\documentclass[11pt,a4paper]{article}

\usepackage[utf8]{inputenc}
\usepackage[T1]{fontenc}
\usepackage{xcolor}
\usepackage{graphicx}
\usepackage{amsmath,amssymb,amsthm,mathtools,geometry,booktabs,
            array,enumitem,tcolorbox}
\tcbuselibrary{skins}

\usepackage{float}

\theoremstyle{definition}

\newtcolorbox{resultbox}{enhanced,sharp corners,
  colback=blue!4!white,colframe=blue!40!black,
  boxrule=0.6pt,left=6pt,right=6pt,top=4pt,bottom=4pt}
\newtcolorbox{warnbox}{enhanced,sharp corners,
  colback=orange!5!white,colframe=orange!60!black,
  boxrule=0.6pt,left=6pt,right=6pt,top=4pt,bottom=4pt}

\newcommand{\SO}{\mathrm{SO}}

\title{\textbf{Towards Wedge Construction of Four-Dimensional Non-Supersymmetric Theories and Torsion Classes}}
\author{
Keshav Dasgupta${}^1$, \, Radu Tatar${}^2$\\[1ex]
\small ${}^1$ Department of Physics, McGill University, Montreal QC Canada\\
\small ${}^2$ Department of Mathematical Sciences, University of Liverpool, Liverpool UK\\
\tt{~~keshav@hep.physics.mcgill.ca; Radu.Tatar@liverpool.ac.uk }
}
\date{\today}

\begin{document}

\maketitle

\begin{abstract}
Motivated by recent proposals relating non-supersymmetric Type 0A theory to
M-theory compactified on a singular wedge geometry, we study an M-theory
compactification on a seven-manifold with \(G_2\) structure, realized as a
deformed K3 fibration over a compact three-manifold. In the Morrison--Vafa
limit, the deformed K3 may be described locally as a non-trivial torus fibration
over a base that is itself a pinched circle fibered over an interval. Once the
doubled-spectrum decomposition and the local pinched structure are specified, we
show that the \(G_2\) torsion classes provide a natural and efficient way to
characterize both the torsion of the seven-manifold and the resulting
supersymmetry breaking in four dimensions. Reducing the system to ten dimensions
in two inequivalent ways leads respectively to Type 0A and Type 0 heterotic
theories compactified on two different non-K\"ahler manifolds, for which the
\(SU(3)\) torsion classes furnish the appropriate mathematical description. In
particular, we argue that the pinching deformation lies in the \(\mathbf{27}\) of
\(G_2\), and that under the two reductions it is distributed differently into the
\(W_2\) and \(W_3\) torsion classes of the corresponding \(SU(3)\) structures. In
the supersymmetric limit, and under suitable assumptions, the two resulting
theories may become U-dual to one another. Away from that limit, however, we
argue that any such duality should be treated with considerable caution.
\end{abstract}

\newpage

\tableofcontents
\bigskip

\section{Introduction and summary}
\label{sec:intro}

A major open problem in string theory is to construct controlled
non-supersymmetric compactifications whose geometry, fluxes, gauge sectors, and
localized degrees of freedom can all be analyzed within a single framework.
Supersymmetric compactifications benefit from the powerful constraints of
special holonomy, calibrated cycles, and Killing-spinor equations \cite{candelas}.  By
contrast, once supersymmetry is broken, the compactification data are much less
rigid: intrinsic torsion, localized sources, non-Kähler geometry, and
branch-dependent sectors can all become important.  The goal of this paper is
to develop a controlled language for studying such a class of
non-supersymmetric backgrounds arising from a wedge compactification of
M-theory.

The starting point is the observation that a seven-dimensional M-theory
geometry may be organized using a $G_2$ structure even when the background is
not supersymmetric \cite{CS, luest}.  The use of a $G_2$ structure should not be confused with
the assumption of $G_2$ holonomy.  A torsion-free $G_2$ structure would imply
strong restrictions and, in the appropriate setting, supersymmetry.  Here we use
a more general torsional $G_2$ structure simply as a geometric bookkeeping
device.  Its intrinsic torsion records the failure of the defining forms to be
closed, and this failure is precisely where the effects of fibration, flux, and
pinching $-$ to be elaborated below $-$ appear.

The geometric input is a K3-fibered M-theory background, viewed locally in an
elliptic presentation.  The ordinary elliptically fibered K3 admits a
Weierstrass description over a $\mathbb P^1$ base, with the elliptic fiber
degenerating over the zeroes of the discriminant \cite{vafaF, MV}.  In the setting of interest
we deform the local fiber geometry so that, instead of an ordinary smooth
circle direction, one encounters a wedge circle
$S^1_+\vee S^1_-$,
motivated by the recent wedge compactification of Baykara--Dudas--Vafa
\cite{BDV} and Altavista et al \cite{Altavista}.
This wedge has two branches meeting at a node.  The two branch radii are
denoted by $R_+$ and $R_-$.  It is then natural to introduce the branch-even
and branch-odd combinations such that 
the scalar $T$ measures the asymmetry of the two branches.  In the effective
Type $0$ description it is the geometric representative of the branch-odd
tachyonic direction, although the statement that this direction is genuinely
tachyonic requires information about the effective potential, not merely the
definition of $T$.

 There is, however, an important difference between the
Baykara--Dudas--Vafa proposal \cite{BDV} and the way we organize the
fields here. In the BDV picture, the resolution property is itself
field-dependent: different fields are assigned to different effective
resolutions of the singular wedge $S^1_+\vee S^1_-$. In particular,
some fields have the disconnected resolution property (DRP), while
others have connected resolution properties (CRP or CRP$'$). In the
notation used below, we reserve the term SSP for the connected
CRP/CRP$'$ sectors, for which the corresponding fields merge into a
single ten-dimensional field. By contrast, DRP fields remain doubled,
because they are effectively supported on the two disconnected
branches. Thus SSP/DRP should not be understood merely as geometric
limits of a single smooth background, but as field-dependent resolution
assignments. The central physical idea of \cite{BDV} is that strong
quantum effects allow different fields to probe different effective
resolutions of the same singular space.

In our approach, which differs from BDV, we take the pinched wedge configuration embedded in the seven-dimensional compactification space as the primary M-theory input. We then organize the degrees of freedom arising from dimensional reduction over the pinched circle together
with possible localized degrees of freedom at the wedge node. In our
description, SSP and DRP are not imposed as fundamental geometric
limits of separate resolved spaces. Rather, SSP-like and DRP-like
behaviour {\it emerges} as a way of organizing the bulk and localized
M-theory degrees of freedom. Fields that are insensitive to the
branch distinction organize into single connected-resolution sectors,
whereas fields sensitive to branch or junction data organize into
doubled DRP-like sectors. In this sense, the field-dependent resolution
properties of the Baykara--Dudas--Vafa construction are translated, in
our framework, into different effective organizations of bulk and
junction-localized M-theory degrees of freedom (DOFs), as made explicit in
section 2.1 and in {\bf Table \ref{bdvvsours}}.

A central theme of the paper is that, once one chooses to describe the wedge
geometry through a branchwise effective dictionary, the $G_2$ and $SU(3)$
torsion classes \cite{CS, luest} become the natural variables for analyzing the resulting
background.  In particular, the branchwise decomposition of the M-theory
four-form flux,
together with the pinched coframe, converts the singular wedge data into
geometric data to which intrinsic torsion is sensitive.  The even and odd
combinations of the branch fluxes then enter naturally into the reduced
$SU(3)$ torsion classes.  In this sense, the torsion classes do not by
themselves derive the doubled spectrum microscopically; rather, once the
doubled branchwise effective description is adopted, they provide a systematic
and powerful way to organize the resulting fibration, flux, and pinch data.

The local seven-dimensional $G_2$ structure also induces a six-dimensional
$SU(3)$ structure after reduction along the circle direction.  The five
$SU(3)$ torsion classes \cite{CS, luest}:
\begin{equation}
W_1,\quad W_2,\quad W_3,\quad W_4,\quad W_5
\end{equation}
then provide the natural language for describing the non-Kähler geometry of
the six-dimensional internal spaces that arise in the two reduction routes.
The fibration contributes to these classes through the non-closure of the
coframe.  The fluxes contribute through their projections onto the appropriate
$SU(3)$ representation channels\footnote{The six-dimensional \(SU(3)\)-structure torsion classes allow a clear tracking of where the non-Kähler geometry comes from. The fibration contributes
because the internal space is twisted rather than a direct product: the fibers
change as one moves over the base. The structure forms \(J\) and \(\Omega\) fail to be closed. The fluxes
contribute after their reduced components are decomposed into \(SU(3)\)
representation pieces, which then enter the corresponding torsion classes
\(W_i\). The wedge pinch supplies the genuinely singular contribution, through
junction-localized terms and smooth branch-odd corrections to the reduced
structure forms.}.  The wedge pinch contributes both through
localized junction-supported terms and through smooth branch-odd backreaction
of the reduced structure forms.

One important refinement is that the localized pinch data should be understood
upstairs as a contribution to the $\mathbf{27}$ component of the $G_2$
intrinsic torsion.  After reduction to an $SU(3)$ structure, this representation
branches into several $SU(3)$ components.  In particular, the component in the
$\mathbf 8$ contributes to the primitive $(1,1)$ torsion class $W_2$, while the
components in the $\mathbf 6\oplus\overline{\mathbf 6}$ contribute to the
primitive $(2,1)+(1,2)$ torsion class $W_3$.  Thus the direct localized
descendants of the parent pinch source can appear in both $W_2$ and $W_3$.
The remaining torsion classes receive pinch-dependent corrections more
indirectly, through the deformation of the reduced $SU(3)$ structure and through
regularized node-resolution data.

The wedge construction leads naturally to two related but distinct
nine-dimensional descriptions, depending on the order of compactification.  In
one route, M-theory is first compactified on the wedge circle, giving a
Type 0A-like branch \cite{type0}, and is then further reduced on the Ho\v{r}ava--Witten
interval \cite{Horava}.  In the other route, M-theory is first compactified on the
Ho\v{r}ava--Witten interval, giving a Type $0{\rm HW}$-like branch, and is then
further compactified on the wedge circle.  Schematically,
\begin{equation}
{\rm M\ theory}
\longrightarrow
0{\rm A}
\longrightarrow
9d,
\qquad
{\rm M\ theory}
\longrightarrow
0{\rm HW}
\longrightarrow
9d.
\end{equation}
These two routes organize the same parent data in different non-supersymmetric local frames.  The
Type 0A route sees the wedge data already in the first step, whereas the
Type $0{\rm HW}$ route sees them only after the second step.  Consequently, the
torsion classes on the two sides should not be compared term by term.  The
correct comparison is instead a representation-theoretic comparison after both
routes have been reduced to the same dimension.

This distinction also affects the gauge sector.  On the Type 0A side, the
gauge data are tied to the orientifold/D-brane system and to the doubled
Type 0A RR gauge fields.  The resulting lower-dimensional gauge algebra is
obtained by imposing bundle holonomy, fibration monodromy, flux-induced
St\"uckelberg, and wedge-deformation conditions.  On the Type $0{\rm HW}$ side,
the natural starting point is the Ho\v{r}ava--Witten wall gauge sector.  After
branch resolution, the effective wall algebra is organized in terms of
branchwise $SO(16)$ factors.  The same kinds of projections appear, but they
act on different parent gauge data.  Thus the gauge-sector comparison has the
same logical form as the torsion-class comparison: the two descriptions have
parallel structures, but the microscopic origins and the detailed
representatives are different.

A related issue is the role of duality.  At the supersymmetric endpoints one
expects a strong--weak relation between the appropriate type-IIA and
Ho\v{r}ava--Witten or heterotic descriptions \cite{Horava, DRS}.  Away from those endpoints,
however, one should be more cautious.  The comparison of spectra, torsion
classes, and gauge sectors is not by itself sufficient to establish a U-duality
between the two non-supersymmetric descriptions.  Indeed, the localized degrees
of freedom at the pinch need not organize themselves in manifestly equivalent
ways in the two frames.  It is possible that a more subtle duality persists
away from the supersymmetric limits, but such a claim would require additional
checks, including the matching of localized junction modes, interaction terms,
anomaly constraints, flux quantization, and the precise map between moduli and
gauge-sector data.

Accordingly, the goal of this paper is to provide a precise effective
description of the two branches and to identify which aspects of their
torsion, flux, spectrum, and gauge data can be compared in a controlled way. We do not claim to
derive a complete microscopic duality between the two non-supersymmetric
branches.  Instead, we construct an effective geometric framework in which the
two routes can be analyzed and compared.  The main output is a systematic
organization of the fibration, flux, pinch, and gauge data in terms of
$G_2$ and $SU(3)$ torsion classes, together with the corresponding commutant
description of the gauge sectors.  This framework makes clear which structures
match between the two descriptions, which are frame-dependent, and which
localized data require further microscopic input. 

At the same time, this effective parallel has a clear boundary.  The torsion
analysis and the commutant analysis identify the representation channels and
projection conditions that can be compared reliably, but they do not determine
the complete microscopic dynamics at the wedge junction.  In particular, the
local degrees of freedom supported at the pinch may include modes whose
normalizations, interactions, and transformation properties are invisible in
the bulk torsion data alone.  Establishing a genuine duality away from the
supersymmetric endpoints would therefore require going beyond the present
effective description and constructing an explicit map of the local junction
theory itself.  Until such a map is available, the safest conclusion is that
the two routes exhibit a controlled structural correspondence, rather than a
demonstrated microscopic equivalence.

\subsection{Organization and brief summaries of the various sections}

One of the main aim of our work here is to motivate the readers towards 
the construction of controlled non-supersymmetric
models using wedge compactifications of M-theory. Our main goal is to explain why
torsional $G_2$ and $SU(3)$ structures are useful even without supersymmetry,
and emphasize that the Type 0A and Type $0{\rm HW}$ descriptions should be
compared as effective nine-dimensional frames rather than as manifestly
identical microscopic duals.  We can also state the main caution of the
paper: spectra, torsion classes, and gauge-sector data provide strong
structural evidence, but do not by themselves prove a full U-duality away from
the supersymmetric endpoints. With this framework in place, we now summarize the organization of the paper. We have also tried to make the presentation reasonably self-contained, so that readers who are less familiar with G-structure techniques can still follow the main ideas and extract the relevant physical conclusions.

Section \ref{sec2} sets up the M-theory geometry as a K3-fibered seven-manifold with
$G_2$ structure.  It explains how the wedge construction modifies the local
fiber geometry and how the resulting fibration, flux, and pinch data are
encoded in intrinsic torsion.  The section introduces the two main branches:
an interval--circle branch and a wedge branch.

Subsection \ref{sec001} explains the proposed dictionary between M-theory on the
singular wedge circle and the effective Type 0A field content.  The main
point is that the NSNS sector is not doubled, while the RR sector is effectively
doubled through branch- or junction-resolved data.  The branch-odd radius
modulus is identified as the geometric representative of the Type $0$ tachyonic
direction, while several branch-odd or mixed components are treated as
auxiliary junction data rather than independent bulk fields. Herein we also propose a way to connect our construction with the one from \cite{BDV}.

Our main journey starts from subsection \ref{sec002} where we introduce the K3-fibered $G_2$-structure background and
explain the use of fiberwise M-theory/heterotic duality over a compact
three-manifold.  The discussion is phrased in an adiabatic regime, where the
K3 fibers vary slowly and the familiar heterotic or type-IIA duality frames can
be used locally.  It also introduces the elliptic and orientifold descriptions
of the K3 fiber that are used later.

In subsection \ref{intervalbranch} we describe the first deformation of the local base geometry,
where the orientifold base is replaced by an interval fibered with an ordinary
circle.  This branch is generally non-K\"ahler, but it need not be
non-supersymmetric.  Supersymmetry must instead be tested using the appropriate
$G_2$ or heterotic $SU(3)$ torsion-class conditions.

The subsection \ref{sec2.2.1} analyzes how the fibration of the K3 hyperk\"ahler structure
over the three-dimensional base generates $G_2$ intrinsic torsion.  The four
$G_2$ torsion classes are interpreted as the singlet, Lee-form, adjoint, and
primitive traceless parts of the non-closure of the defining forms.  The
subsection explains how base twisting, variation of the K3 triplet, and
background fluxes source these torsion classes.

The subsection \ref{wedgebranch} introduces the genuinely non-supersymmetric wedge branch, in
which the ordinary circle is replaced by the singular wedge
$S^1_+\vee S^1_-$.  The two branch radii define a branch-even mode and a
branch-odd imbalance mode.  The wedge node introduces localized data and
motivates the use of torsion classes to organize the resulting
non-supersymmetric geometry.

In subsection \ref{g2pinch1} we explain how the total $G_2$ torsion receives contributions
from three sources: the non-trivial fibration, the internal fluxes, and the
wedge pinch.  The branchwise flux decomposition and the pinched coframe convert
the singular wedge information into geometric data to which intrinsic torsion
is sensitive.  The subsection also clarifies that the torsion formalism is an
effective organizational tool, not a microscopic derivation of the doubled
spectrum.

Subsection \ref{naturalpinch} is crucial. 
Here we argue, hopefully convincingly, that the leading localized pinch deformation is most
naturally placed in the primitive traceless component of the $G_2$ torsion.
The reason is that the pinch is not primarily an overall singlet deformation or
a Lee-form deformation, but a localized shape deformation of the $G_2$
structure.  This observation later explains how the same parent pinch data can
descend into different $SU(3)$ torsion channels.

One of the lingering question in out set-up is what happens when the tachyon itself vanishes. In 
subsection \ref{susyresto} we clarify that setting the branch-odd modulus to zero does not
automatically restore supersymmetry.  Vanishing tachyon removes one source of
branch asymmetry, but supersymmetry also depends on the remaining fibration,
flux, and torsion data.  In particular, a symmetric wedge configuration can
still be torsional or singular.

Section \ref{sec3NK} starts our first descend to ten-dimensions and from there subsequently to four-dimensions.
This section studies the Type 0A reduction obtained by first compactifying
M-theory on the wedge circle.  Since the wedge appears in the first reduction,
the Type 0A torsion classes already contain the branch-even, branch-odd, and
pinch-induced data.  The section develops the $SU(3)$ torsion-class language
for the resulting non-Kähler six-manifold.

Subsection \ref{sec301} discusses the dependence of the $SU(3)$ torsion classes on flux, fibration and pinch.
This subsection decomposes the Type 0A torsion classes into contributions
from fibration, flux, and pinch effects.  The purpose is to show how different
geometric and field-theoretic ingredients populate the five $SU(3)$ torsion
channels.  It also prepares the notation used later for comparing with the
Type $0{\rm HW}$ route.

Subsection \ref{sec311} explains that the fibration contributes to the torsion
classes through the non-closure of the internal coframe.  Since the defining
$SU(3)$ forms are built from this coframe, their exterior derivatives
naturally contain fibration data.  These terms give the purely geometric
contribution to the torsion classes.

The flux contribution to the torsion classes is discussed in subsection \ref{sec312}. Here 
we describe how the branch-even and branch-odd flux data enter
the $SU(3)$ torsion classes through appropriate representation projections.
The even flux sector behaves like the ordinary Type IIA-like contribution,
while the odd sector carries branch-dependent information associated with the
wedge geometry.  The discussion also explains how reduced $G_4$ data can
contribute to the primitive torsion channels.

One of the important question is how does the pinch contributes to the torsion classes. This is studied in subsection \ref{pinchu2} where 
we analyze in detail the torsion induced by the wedge pinch.  The pinch
produces localized junction-supported terms and also smooth branch-odd
corrections.  These effects are encoded through the branch-odd modulus, its
gradients, and possible regularized node-resolution data.

Another pertinent question is as to why the pinch can affect more than just $W_3$? This is answered in subsection \ref{sec314}.
This subsubsection basically explains that although the parent $G_2$ pinch source is
placed in the primitive traceless torsion class, its reduction to an
$SU(3)$ structure can distribute the information into more than one torsion
channel.  Direct localized descendants can enter $W_2$ and $W_3$, while
induced smooth corrections can affect the other classes through the deformation
of $J$ and $\Omega$.

Finally we come to the issue of combining fibration, flux and pinch contributions together. This is achieved in section \ref{sec315}.
The aim of this subsection is to collect the fibration, flux, and pinch terms into a single
Type 0A torsion-class formula.  The key point is that the pinch does not
replace the odd flux sector; it adds additional branch-odd and localized data.
The resulting formula gives a compact summary of the Type 0A 
non-K\"ahler
geometry.

One of the question that we asked in subsection \ref{susyresto} is whether we can restore supersymmetry when the tachyon vanishes. The answer was no because vanishing tachyon is not the supersymmetric IIA point. In subsection \ref{sec32} we compare
symmetric wedge regime, where the tachyon vanishes, with the actual Type IIA endpoints, where the tachyon condenses to non-zero values. In fact,
this subsection distinguishes the branch-symmetric point from the true
Type IIA endpoints.  The symmetric point has equal branch radii, while the
Type IIA endpoints occur when one branch collapses.  This distinction is
important for interpreting the branch-odd scalar and for understanding which
limits restore ordinary Type IIA behavior.

We further analyze the supersymmetry breaking and gauge sector from flux, fibration and pinch in subsection \ref{sosiebrake}.
In here we explain how fibration, flux, and wedge data break
supersymmetry and affect the gauge sector in the Type 0A frame.  The
four-dimensional gauge group is described through commutant conditions
involving D8-brane bundle holonomy, fibration monodromy, flux-induced
St\"uckelberg couplings, and wedge-deformation data.  The result is presented
as the sharpest controlled gauge-sector description without a full microscopic
junction model.

Section \ref{sec40} is our second descend to ten-dimensions and from there subsequently to four-dimensions via
Type $0$ Heterotic theory compactified on a non-K\"ahler 
six-manifold.
This section develops the alternative reduction route, where M-theory is first
compactified on the Ho\v{r}ava--Witten interval and only later on the wedge
circle.  The resulting Type $0{\rm HW}$ description has a different natural
organization of the same parent data.  The section focuses on the induced
$SU(3)$ torsion classes and their comparison with the Type 0A frame.

Subsection \ref{sec41} discusses the reduction to the type $0$ Ho\v{r}ava--Witten branch and torsion classes.
In here we introduce the Type $0{\rm HW}$ internal geometry and explain
how the $G_2$ torsion data descends after reduction along the interval.  Before
the wedge reduction, the torsion classes have a simpler form, built from
fibration and flux contributions.  The wedge-dependent corrections appear only
after the later compactification on the wedge circle.

The subsection \ref{sec411} resolves an apparent mismatch between the Type 0A and
Type $0{\rm HW}$ treatments of the parent $G_2$ pinch data.  The point is that
the two reductions project the same parent torsion information in different
ways.  The Type $0{\rm HW}$ description keeps certain localized primitive
torsion sources more explicit, whereas the Type 0A description distributes
the same information among branch-resolved flux and pinch terms.

We now perform the same set of computatations that we did for the Type 0A case, namely, compute the 
fibration, flux and pinch contributions to $SU(3)$ torsion classes. This is achieved in subsection \ref{sec412}.
This subsection writes the Type $0{\rm HW}$ torsion classes after including
the wedge reduction.  The fibration and flux terms come from the
interval-reduced ten-dimensional frame, while the branch-odd, localized, and
resolved-node terms arise from the subsequent wedge compactification.  This
gives the Type $0{\rm HW}$ analogue of the Type 0A torsion-class
decomposition.

In subsection \ref{sec420} we perform a series of quantitative comparisons of Type $0{\rm HW}$ with the Type $0A$ reduction.
This subsection compares the two reduction routes after both have been brought
to nine dimensions.  It emphasizes that the two frames should not be equated
term by term.  Instead, their torsion classes and spectra should be compared
through common representation channels and effective projection structures.

One of the first comparison is the spectrum for both these theories in nine-dimensions. This is performed in detail in subsection \ref{sec421}.
Herein we compare the field content obtained by the two orders of
compactification.  It explains which bulk zero modes are kept in the common
sector and which branch-odd or mixed components should be treated as localized
or auxiliary junction data. The main goal is to exhibit the precise matching of the common
nine-dimensional massless bosonic sector, while emphasizing that this matching
should not be interpreted as a proof of a full U-duality between the two descriptions away from the supersymmetric end-points. 

Before making comparison of the torsion classes, we first quantify the Type $0{\rm HW}$ torsion classes in ten and nine dimensions in subsection \ref{comparisoon0}.
This subsubsection distinguishes carefully between the ten-dimensional
Type $0{\rm HW}$ torsion classes and the nine-dimensional torsion classes after
the wedge compactification.  The ten-dimensional expressions contain only the
interval-reduced fibration and flux data, while the nine-dimensional
expressions include the wedge-induced localized and branch-odd corrections.

Subsection \ref{comparisoon} discusses the refined comparison of the Type $0A$ and Type $0{\rm HW}$ torsion sectors.
This subsubsection gives the detailed comparison of the five torsion classes in
the two frames.  It explains how the same parent $G_2$ data can appear as
different-looking $SU(3)$ torsion representatives after the two reductions.
The comparison is therefore structural and representation-theoretic, not a
literal equality of coefficients.

In subsection \ref{sec424} we study the Type $0{\rm HW}$ gauge theory sector and compare it to Type $0A$ gauge sector studied in section \ref{sosiebrake}. Here we analyze the branch-resolved Ho\v{r}ava--Witten wall gauge
sector and compare it with the Type 0A gauge-sector construction.  The
Type $0{\rm HW}$ side starts from branchwise $SO(16)$ wall factors, while the
Type 0A side is tied to orientifold/D-brane and doubled RR gauge data.  Both
gauge sectors are organized by holonomy, monodromy, mass-kernel, and wedge
commutant projections, but they act on different parent gauge algebras.

In subsection \ref{sec4343} we study the 
endpoint torsion-frame map which should be understood as a controlled comparison
available only at the supersymmetric Type IIA endpoints, where one of the two
wedge branches has collapsed.  In that regime the Type 0A and
Type $0{\rm HW}$ $SU(3)$-structure data may be related by the usual
strong--weak change of frame, including the appropriate moduli map, pullback of
the internal coframe, Weyl rescaling, and projection onto the relevant torsion
representations.  Away from these endpoints, however, the same formulae should
be regarded only as a formal bookkeeping device for comparing representation
channels.  They {\it do not} by themselves establish a U-duality in the
non-supersymmetric interior of the wedge moduli space, where the localized
junction degrees of freedom and their interactions must be matched separately.

Finally in subsection \ref{fate} we explain how the minimal geometric-pinch contribution is
modified once one allows for genuinely localized degrees of freedom at the
junction.  While the geometric pinch naturally feeds the torsion classes through
the descendants of the parent $G_2$ primitive traceless channel, additional
junction modes may carry other $SU(3)$ representation components.  The
torsion-class framework provides a systematic way to organize these possible
localized contributions and to identify which torsion channels they affect.

In section \ref{sec5} we conclude. We 
emphasize that the Type 0A and Type $0{\rm HW}$ descriptions
are best viewed as two effective organizations of the same parent M-theory
data.  The torsion-class formalism provides a powerful bookkeeping device once
the branchwise flux decomposition and pinched coframe are accepted as part of
the effective description. Our analysis shows that the two routes exhibit a
controlled structural correspondence, while a full microscopic duality would
require further checks involving localized junction modes, interactions,
anomalies, flux quantization, and moduli maps.

\section{M-theory compactified on a $G_2$-structure seven-manifold}
\label{sec2}

In this section we set up the M-theory compactification that underlies the
rest of the paper.  The starting point is a compact seven-manifold with
$G_2$ structure, locally realized as a K3 fibration over a compact
three-manifold \cite{CS}.  We use the familiar seven-dimensional M-theory/heterotic
duality as a local fiberwise guide, in an adiabatic regime where the K3 fiber
varies slowly over the base.  This provides a useful framework for organizing
the geometry, but it should not be interpreted as a proof of a global duality
for arbitrary torsional or singular fibrations.

Our main goal in this section is to explain the geometric proposal, the
assumptions entering the fiberwise description, and the two branches that arise
from deforming the local K3 fiber geometry.  The first branch is the
interval--circle branch, which is generally non-Kähler but may still preserve
supersymmetry if the induced torsion satisfies the appropriate first-order
conditions \cite{luest, strominger, DRS}.  The second branch is the wedge branch, in which an ordinary circle
is replaced by the singular space $S^1_+\vee S^1_-$ \cite{BDV, Altavista}.  This branch introduces
branch-dependent data, a pinch locus, and a natural branch-odd scalar
associated with the Type $0$ direction.

The perspective adopted here is deliberately effective.  The wedge geometry is
treated as the primary M-theory input, while the Type 0A and Type $0{\rm HW}$
descriptions are used as candidate local frames for organizing the resulting
degrees of freedom.  In particular, the branchwise flux decomposition, the
pinched coframe, and the corresponding $G_2$ torsion classes provide a
controlled language for analyzing the local geometry.  The detailed reduction
to Type 0A and Type $0{\rm HW}$ frames, together with the limitations of
their possible duality interpretation, will be discussed in later sections.

\subsection{M-theory on $S^1_+\vee S^1_-$ and the effective type 0A field dictionary \label{sec001}}

We will start by summarizing the M-theory to Type 0A reduction on the singular wedge circle
$S^1_+\vee S^1_-$.
The guiding principle is the following: the NSNS sector is \emph{not} doubled, whereas the RR sector \emph{is} doubled.
More precisely, the perturbative type-$0$A spectrum contains:
\begin{equation}
\big(g_{\mu\nu}, B_{\mu\nu}, \Phi, T\big)_{\rm NS} ~~\oplus ~~\big(C_1^{(\pm)}, C_3^{(\pm)}\big)_{\rm RR}
\end{equation}
in the NSNS sector and the RR sectors respectively. The branch-odd scalar modulus $T$ of the wedge geometry, as we shall describe soon, is identified with the tachyon.
A spectrum counting will tell us that the number of bosonic degrees of freedom in ten-dimensions can be easily computed to be:
\begin{equation}
\mathcal N_{\rm bos}^{({\rm 0A})} = 35+28+1+1+8+8+56+56=193 ,
\label{0A_193_again}
\end{equation}
compared to ${128}$ bosonic bulk degrees of freedom in M-theory in the supergravity limit. Clearly to get an increase in the degrees of freedom, from 128 to 193, the compactification manifold cannot be a smooth one. The claim in \cite{BDV} is that compactification on a pinched geometry $S^1_+ \vee S^1_-$ can reproduce Type 0A from M-theory.

To see how this may appear from eleven-dimensional field decomposition, we will let the eleven-dimensional coordinates be split as
$x^M=(x^\mu,y^+,y^-)$, 
where \(x^\mu\) denote the ten common directions and \(y^\pm\) parametrize the two branches \(S^1_\pm\). The eleven-dimensional fields are
$G_{MN}$ and $C_{MNP}$. Their component decomposition is:
\begin{equation}
C_{MNP}
\;\longrightarrow\;
\Big\{
C_{\mu\nu\rho},
\;
C_{\mu\nu +},
\;
C_{\mu\nu -}, \; C_{\mu + -}
\Big\} \nonumber
\label{C3decomp0A}
\end{equation}
\begin{equation}
G_{MN}
\;\longrightarrow\;
\Big\{
G_{\mu\nu},
\;
G_{\mu +},\,G_{\mu -},
\;
G_{++},\,G_{--},\,G_{+-}
\Big\},
\label{metricdecomp0A}
\end{equation}
from here the ten-dimensional NSNS fields are identified as follows.
The physical ten-dimensional metric is the common branch-even metric, {\it i.e.}
$g_{\mu\nu}\sim G_{\mu\nu}$,
whereas the physical NSNS two-form comes from the branch-even and branch-odd combinations:
\begin{equation}
B^{(e)}_{\mu\nu} = B_{\mu\nu} 
\equiv
\frac12\big(C_{\mu\nu +}+C_{\mu\nu -}\big), \qquad
B^{(o)}_{\mu\nu}
\equiv
\frac12\big(C_{\mu\nu +}-C_{\mu\nu -}\big), 
\label{B2odd0A}
\end{equation}
respectively, where the non-KK nature of the reduction may be justified in the following way. Since the components $C_{\mu\nu\pm}$ are not natural KK reductions, we can  assume that the information about the pinch DOFs are already included in all the field components that have legs along the $\pm$ directions of $S^1_+\vee S^1_-$ in \eqref{metricdecomp0A}. This is also the reason why there is no literal {\it equality} between the left and right hand sides of the decomposition in \eqref{metricdecomp0A}. Now comes our first proposal: the branch-odd combination
is \emph{not} a second physical NSNS two-form of Type 0A. It should instead be viewed as constrained branch/junction data, or as an auxiliary quantity entering effective local parametrizations. We will elaborate more on this below as well as provide a reason soon. The other two 
branch-even and branch-odd scalar combinations:
\begin{equation}
R_B \equiv \frac12\big(\sqrt{G_{++}}+\sqrt{G_{--}}\big),
\qquad
T \equiv \frac12\big(\sqrt{G_{++}}-\sqrt{G_{--}}\big) ,
\label{RBTdef0A}
\end{equation}
where $\sqrt{G_{++}} = R_+$ and $\sqrt{G_{--}} = R_-$ are now identified  
with the branch-even radion-like scalar $\Phi$, while
$T=\frac{R_+-R_-}{2}$ is the branch-odd scalar modulus, identified with the Type $0$ tachyon. (In fact the dilaton \(\Phi\) arises from the above branch-even scalar combination after the usual Weyl rescaling to ten-dimensional string frame.)

The three RR one-forms come directly from the two branchwise metric vectors:
$C^{(+)}_{1,\mu}\sim G_{\mu +}$ and 
$C^{(-)}_{1,\mu}\sim G_{\mu -}$; and from the three-form $C_{\mu+-}$. There is a small subtlety here. At the symmetric point
$R_+=R_-$ with vanishing tachyon at leading order,
the branch-exchange \(\mathbb Z_2\) symmetry is restored, and therefore branch-odd
quantities vanish at leading order. However this does not mean that the wedge
\(S^1_+\vee S^1_-\) has become a smooth two-dimensional internal space with
independent \(+\) and \(-\) tangent directions. It remains a singular one-dimensional
space, namely two circles meeting at a point. Therefore mixed components such as
\(G_{+-}\) and \(C_{\mu+-}\) do not automatically become ordinary propagating bulk
fields even at the symmetric point. At most, they may be allowed as
junction-symmetric or auxiliary data in a resolved or effective description, but they
are not forced to appear in the strict branchwise bulk decomposition lending support to our earlier proposal that these components contain the information of the localized pinch DOFs.
This mean we can keep $G_{\mu +}$ and $G_{\mu -}$ as the dynamical fields and relegate $C_{\mu+-}$ to be an auxiliary field, or equivalently, pass to even/odd combinations: 
\begin{equation}\label{gaugoram}
A_\mu^{(e)}\equiv \frac12\big(G_{\mu +}+G_{\mu -}\big),
\qquad
A_\mu^{(o)}\equiv \frac12\big(G_{\mu +}-G_{\mu -}\big) ,
\end{equation}
implying that there are two independent RR one-form sectors, exactly as required in Type $0$A with the dotted terms coming from the pinch contributions. The NS two-forms are more tricky. Before discussing them, it is useful to decompose the M-theory \(4\)-form flux branchwise. Let
$\eta_\pm$ denote the one-forms along the two branches and
$v_\pm$ their dual vectors. Then the eleven-dimensional \(4\)-form can be written locally as:
\begin{equation}\label{g4c3h3}
\begin{aligned}
G_4 \to &~ G_4^{\rm br}
=
\eta_+\wedge H_3^{(+)}
+
\eta_-\wedge H_3^{(-)}
+
F_4^{(+)}
+
F_4^{(-)}\\
& H_3^{(\pm)}\equiv \iota_{v_\pm}G_4,
\qquad
F_4^{(\pm)}\equiv G_4-\eta_\pm\wedge H_3^{(\pm)} ,
\end{aligned}
\end{equation}
where we notice that we have {\it two} copies for each three-forms $H_3^{(\pm)}$ and four-forms $F_4^{(\pm)}$ from direct decomposition (and the corrections from the pinch DOFs are already assumed to be incorporated in their definitions). We also denote the branch-corrected four-form as $G_4^{\rm br}$.
However, these are only \emph{branchwise local representatives} of the same eleven-dimensional field. Since there is only one M-theory \(C_3\), the branchwise quantities are not independent lower-dimensional fields. Passing to even/odd combinations:
\begin{equation}
H_3^{(e)}\equiv \frac12\big(H_3^{(+)}+H_3^{(-)}\big),
\qquad
H_3^{(o)}\equiv \frac12\big(H_3^{(+)}-H_3^{(-)}\big),
\label{H3evenodd0A}
\end{equation}
one should identify
$H_3^{(e)}$
with the physical NSNS \(3\)-form field strength
$H_3=dB_2$, 
whereas
$H_3^{(o)}$
is not a second physical NSNS field. Rather, it is constrained branch-odd data associated with the wedge geometry and the tachyonic sector.
A useful schematic parametrization is:
\begin{equation}
H_3^{(+)}=H_3^{\rm phys}+\Delta H_3,
\qquad
H_3^{(-)}=H_3^{\rm phys}-\Delta H_3,
\label{H3constraint0A}
\end{equation}
where \(H_3^{\rm phys}\) is the single physical NSNS \(3\)-form, while \(\Delta H_3\) is branch-odd auxiliary data with the assumption that both may contain the pinch contributions. Plugging \eqref{H3constraint0A} in \eqref{H3evenodd0A}, we can easily show that:
\begin{equation}\label{physh3}
H_3^{(e)}=H_3^{\rm phys},
\qquad
H_3^{(o)}=\Delta H_3 ,
\end{equation}
justifying that we have only {\it one} physical two-form field whose field strength is $H_3^{(e)}$, and the branch-odd field strength is an auxiliary data.  Moreover, because the only natural branch-odd \emph{scalar} modulus is \(T\), the odd \(3\)-form should not be counted as a new propagating field. At most it should be treated as a derived branch-odd deformation in the following way:
\begin{equation}
H_3^{(o)}
=
T\,{\cal H}_3^{(1)}
+
dT\wedge {\cal H}_2^{(1)}
+
{\cal O}(T^2),
\label{H3oddderived0A}
\end{equation}
built from the physical fields and the local wedge geometry. In particular,
$T=0$ implies that 
$H_3^{(o)}=0$
at leading order. However once we take the tachyonic fluctuations $\delta T$, we do get non-zero $H_3^{(o)}$. Under exchange $S^1_+\leftrightarrow S^1_-$ one has
$T\mapsto -T$ and $H_3^{(o)}\mapsto -H_3^{(o)}$. Since also
$dT\mapsto -dT$, the parametrization~\eqref{H3oddderived0A} is compatible
with branch exchange provided the coefficient forms are branch-even:
\begin{equation}
\mathcal H_3^{(1)}\big|_{T=0}
=
\text{branch-even},
\qquad
\mathcal H_2^{(1)}\big|_{T=0}
=
\text{branch-even}.
\label{H_brancheven}
\end{equation}
Thus $\mathcal H_3^{(1)}$ and $\mathcal H_2^{(1)}$ are built from the
branch-even background geometry, the physical fields, and the local
junction data. Since $H_3^{(o)}$ is auxiliary branch-odd data rather than an independent
propagating NSNS field strength, its exterior derivative must be determined
by the physical fields and by the wedge geometry. Taking the exterior
derivative of~\eqref{H3oddderived0A} gives, to leading order:
\begin{equation}
dH_3^{(o)}
=
dT\wedge
\left(
\mathcal H_3^{(1)}
-
d\mathcal H_2^{(1)}
\right)
+
T\,d\mathcal H_3^{(1)}
+
O(T^2) ,
\label{dH3o}
\end{equation}
where the coefficient of \(dT\),
namely $\mathcal H^{(1)}_3-d\mathcal H^{(1)}_2$,
is not an independent source term.  It is fixed by the chosen branch-odd
deformation ans\"atze and by the requirement that the odd sector remain auxiliary
rather than becoming a second physical NSNS three-form. At the genuine Type IIA endpoints one of the two wedge branches collapses,
with either \(R_+\to 0\) or \(R_-\to 0\).  In such a limit the two-branch
description is no longer the correct set of variables.  There is only one
surviving circle branch, and hence only one ordinary IIA NSNS three-form
field strength. For example, if the \(+\) branch survives, the physical field strength is
the surviving branch combination
$H^{\rm surv}_3
=
H^{(+)}_3
=
H^{(e)}_3+H^{(o)}_3$.
If the \(-\) branch survives, it is instead
$H^{\rm surv}_3
=
H^{(-)}_3
=
H^{(e)}_3-H^{(o)}_3$.
The orthogonal, collapsed-branch combination is not a second IIA field and
must decouple. Therefore the auxiliary branch-odd deformation \eqref{H3oddderived0A}
cannot leave behind an independent contribution at the endpoint.  Equivalently,
the effective endpoint contribution of the coefficient forms
$\mathcal H^{(1)}_3$ and $\mathcal H^{(1)}_2$, including any switching functions that multiply
them, must vanish or be reabsorbed into the single surviving IIA field strength.
Thus (12) should be regarded as an interior-wedge expansion.  Near
\(R_+=0\) or \(R_-=0\), it must be reorganized in endpoint-adapted variables
rather than interpreted as a second NSNS sector.

At leading order the coefficient forms are determined by the branchwise
decomposition~\eqref{g4c3h3} and \eqref{H3constraint0A}. Thus $\mathcal H_3^{(1)}$ measures the response of the odd flux imbalance
to the branch-odd radius modulus, while $\mathcal H_2^{(1)}$ measures the
response to gradients of that modulus. These coefficient forms are
therefore fixed only after the local wedge geometry and background flux
configuration have been specified.

The subtle part of the reduction is the doubled RR \(3\)-form sector whose field strengths appear as $F_4^{(\pm)}$ in \eqref{g4c3h3}. From the naive component decomposition in \eqref{metricdecomp0A}, one only sees
$C_{\mu\nu\rho}$
which would give a single RR \(3\)-form, as in ordinary type IIA. But perturbative Type $0$A is known to contain two RR \(3\)-forms:
\begin{equation}
C_3^{(+)},\qquad C_3^{(-)}.
\label{RR3double0A}
\end{equation}
Therefore the second RR \(3\)-form cannot be obtained by a naive smooth Kaluza--Klein component count alone. The correct interpretation, and as we have been emphasizing all along, is that the wedge compactification is \emph{not} an ordinary smooth-circle reduction. Rather, the singular branch/junction structure provides two RR sectors associated with the single eleven-dimensional \(C_3\). Thus one should view
$C_{\mu\nu\rho}$
as giving rise to two branch/junction-resolved RR \(3\)-form sectors
$C_3^{(+)}$ and $C_3^{(-)}$,
even though these are not two independent smooth eleven-dimensional \(3\)-form fields\footnote{In other words, if the
wedge reduction gives two two-form slots and two three-form slots, then the
natural branchwise decomposition of the eleven-dimensional three-form is:
\begin{equation}
C_3
\;\leadsto\;
C_3^{\rm br}
=
\eta_+\wedge B_2^{(+)}
+
\eta_-\wedge B_2^{(-)}
+
C_3^{(+)}
+
C_3^{(-)} ,\nonumber
\label{C3_branch_decomposition}
\end{equation}
where $\eta_\pm$ are the one-forms along the two wedge branches that appeared earlier in \eqref{g4c3h3}, $B_2^{(\pm)}$
are the two branchwise two-form fields, and $C_3^{(\pm)}$ are the two
branchwise three-form fields. Taking an exterior derivative reproduces \eqref{g4c3h3}.}. A possible local effective parametrization may be expressed in the following way:
\begin{equation}\label{C3pmansatz0A}
\begin{aligned}
C_3^{(\pm)}
& =
C_3
\pm
\alpha\,B_2^{(o)}\wedge dG_{+-}
\pm \,\,\Xi_3(T, G_{+-}, B_2^{(o)}, ..)\\ 
& = C_3
\pm
\alpha\, \frac12\big(C_{\mu\nu +}-C_{\mu\nu -}\big) \wedge dG_{+-}
\pm\Xi_3(T, G_{+-}, C_{\mu\nu\pm}, ..),
\end{aligned}
\end{equation}
where the dotted terms in $\Xi_3(T, G_{+-}, B_2^{(o)}, ..)$ include all possible combinations of the fields allowed from M-theory, not necessarily the low-energy ones, that include the DOFs from the pinch. However \eqref{C3pmansatz0A} should be regarded only as a speculative effective ans\"atze for branch/junction corrections but it makes sense because $C_3$ doesn't have any legs along the $\pm$ directions and therefore any corrections to it {\it must be additive ones}. Although this appears slightly speculative at this stage, we can make the following interesting observation. In the limit $T = G_{+-} = B_2^{(o)} = 0$, we see that the second term in \eqref{C3pmansatz0A} vanishes, but the third term in general doesn't. On the other hand, the Type IIA limit is when {\it either} $R_+$ or $R_-$ vanishes, and there the two three-forms $C_3^{(\pm)}$ should merge making $C_3^{(+)} = C_3^{(-)} = C_3$. To facilitate this, consider two functions defined as:
\begin{equation}\label{zetaabeta}
\mathcal F_i(\zeta) \equiv \mathcal F_i\left({R_+\over R_-}e^{-{R_+\over R_-}}\right), \qquad  \mathcal G_i(\eta) \equiv \mathcal G_i\left({R_-\over R_+}e^{-{R_-\over R_+}}\right),    
\end{equation}
such that $\mathcal F_i(0) = \mathcal G(0) = 0$ and $i \in \mathbb{Z}$. The exact form of these functions is not needed for the present discussion but the vanishing condition at the origin is necessary\footnote{It is also important that \(\zeta,\eta\to 0\) at the Type IIA
endpoints, and the profile functions vanish there. This guarantees that the
extra branch-resolved correction is switched off when the wedge degenerates
to a single ordinary circle branch. Thus $\mathcal F_i(\zeta)$  and $\mathcal G_i(\eta)$ should be regarded as effective switching functions.
They allow the branch/junction correction to be non-zero in the interior
wedge regime, where both \(R_{+}\) and \(R_{-}\) are finite, while forcing it
to disappear in the ordinary Type IIA limits. Their precise form would depend
on the microscopic dynamics of the localized pinch degrees of freedom, which
is not fixed by the present effective analysis.}. Now assuming $\alpha$ appearing in \eqref{C3pmansatz0A} is itself a function of the form $\alpha = \alpha(\{\mathcal F_i\}, \{\mathcal G_j\})$, we can combine the two terms in \eqref{C3pmansatz0A} into one function $\widehat{\Xi}_3$ and express it as:

{\footnotesize
\begin{equation}
\alpha\,B_2^{(o)}\wedge dG_{+-}
+ \,\,\Xi_3(T, G_{+-}, B_2^{(o)}, ..)  ~\to ~ \widehat{\Xi}_3(T, G_{+-}, B_2^{(o)}, ..)\nonumber
\end{equation}
\begin{equation}
\widehat{\Xi}_3(T, G_{+-}, B_2^{(o)}, ..)~= ~\sum_{i = 1}^\infty \mathcal F_i(\zeta) \,\Xi^{(i)}_3((T, G_{+-}, B_2^{(o)}, ..) + \sum_{j = 1}^\infty\mathcal G_j(\eta) \,\widetilde{\Xi}^{(j)}_3(T, G_{+-}, B_2^{(o)}, ..)\nonumber
\end{equation}
\begin{equation}\label{chotolookai}
C_{\mu\nu\rho}^{(\pm)} \equiv C_{\mu\nu\rho} \pm (\widehat{\Xi}_3)_{\mu\nu\rho} ,
\end{equation}}
then the three-form $C_3^{(\pm)}$ are clearly distinguished by $\pm \widehat{\Xi}_3$. In the IIA limit where $R_+ \to 0$ or $R_- \to 0$, the $\widehat{\Xi}_3$ term cleanly decouples (or becomes a pure gauge) and we get back the standard one-set three-form gauge fields. Note that, as one would have expected, $\widehat{\Xi}_3$ clearly depends on the details of the wedge structure and the tachyon $T$, including other fields, but for it to contribute to the IIA three-form it is required to have the right gauge transformation like $C_3$. (In a similar vein, we could express $\mathcal H_3^{(1)}$ and $\mathcal H_2^{(1)}$ in \eqref{H3oddderived0A}
as a series in $\mathcal F_i(\zeta)$ and $\mathcal G_j(\eta)$ so that the IIA limit will not have any superfluous degrees of freedom.) 
Thus the safest statement is then the following: the doubled RR \(3\)-forms of type-$0$A should be viewed as branch/junction-resolved sectors of the single M-theory $C_3$ by the inclusion of terms like $\pm\widehat{\Xi}_3$ or their generalizations thereof. This way we can account for both the Type 0A and the Type IIA spectra correctly. 

The preceding discussion, and in particular the organization of the degrees of freedom, provides a useful framework for analyzing the spectrum. Nevertheless, despite this success, the manner in which the field doubling has been implemented remains somewhat unsatisfactory.
For example, while the even and odd combinations in \eqref{gaugoram} lead to {\it two} gauge fields, a similar even and odd combinations in \eqref{H3evenodd0A} lead to {\it one} anti-symmetric B-field. Moreover a single three-form is duplicated as in \eqref{chotolookai}, whereas the metric remains as a single copy. This looks highly arbitrary so the question is: how do we justify the spectrum?

This is where the arguments presented in \cite{BDV} provide some hints. As discussed in section \ref{sec:intro}, in
the Baykara--Dudas--Vafa construction, SSP and DRP are field-dependent
resolution properties.  Fields with SSP propagate on the connected resolution
and therefore give a single ten-dimensional field, whereas fields with DRP live
on the two disconnected resolutions and therefore give two ten-dimensional
fields. Our analysis presented here differs slightly in which we have formulated our fields: all field components are doubled, plus there are extra localized fields at the pinch. Some of the doubled fields become physical and propagate in the bulk, while others remain as localized pinch DOFs. However the fact that in \cite{BDV} the fields themselves see different resolutions may be reconciled in the following way. In the DRP scenario, a set of arguments presented in \cite{BDV} justifies that the three-form should be {\it doubled}. In our case, following \eqref{chotolookai}, we argued for the existence of two copies of the three-forms due to the presence of $\pm \widehat{\Xi}_3$ which depend on $T$ and other DOFs at the pinch. 
Since in our case the two circles cannot separate, the DRP like scenario can emerge in the limit where the tachyon and additional {\it massless} DOFs dominate over the pinch DOFs, thus keeping $\widehat{\Xi}_3 \ne 0$  and $C_3^{(+)} \ne C_3^{(-)}$. These two three-forms then exactly match the two three-forms at the DRP point. In a similar vein, when the pinch singularity is blown-up we are at the SSP point of \cite{BDV}. In our case this is achieved when $H_3^{(o)}$ in \eqref{H3oddderived0A} becomes highly subdominant and localized compared to $H_3^{\rm phys}$ in \eqref{physh3}, which in turn is effectively insensitive to the pinch DOFs. This is where $H_3^{\rm phys}$ exactly matches with the expected one copy of the three-form at the SSP point. 
In a similar vein the doubling of the $U(1)$ gauge fields and the presence of a single copy of the graviton can be easily matched with the corresponding results from \cite{BDV}. (See {\bf Table \ref{bdvvsours}} for more details.) Beyond these necessary conditions, the exact form of
$\mathcal{F}_i$ and $\mathcal{G}_j$ is not determined by the present
analysis. Of course a complete microscopic derivation would require a detailed model
of the junction degrees of freedom and their coupling to the bulk fields,
which goes beyond the scope of this work. What the construction does
establish is that functions satisfying the aforementioned conditions are consistent
with the wedge geometry, and that any such choice reproduces the correct IIA
limit and branch-exchange behavior. The arbitrariness in the choice of
$\mathcal{F}_i,\mathcal{G}_j$ within this class reflects the microscopic
ambiguity in how the junction degrees of freedom couple to the bulk
three-form. This then resolves the apparent arbitrariness of procedure adopted in \eqref{gaugoram}, \eqref{H3evenodd0A} and \eqref{chotolookai}.

{At this point we should again compare our analysis with the proposal from \cite{BDV}. The proposal does not derive the resolution property of the M-theory three-form
\(C_3\) from an ordinary KK reduction; rather, it assumes that \(C_3\) has the opposite
resolution property from the metric, namely that \(C_{\pm\mu\nu}\) has SSP while
\(C_{\mu\nu\rho}\) has DRP. This choice is made precisely so that the ten-dimensional spectrum
matches Type \(0\)A: the former gives a single \(B_{\mu\nu}\), while the latter
gives the doubled RR three-forms \(C^{(\pm)}_{\mu\nu\rho}\). The paper further argues
that this assignment is compatible with the parity and circle-exchange symmetries,
and that the SSP of \(C_{\pm\mu\nu}\) implies that an M2-brane wrapped on the
internal direction produces only one $10d$ string rather than two. Thus the
justification is a combination of spectrum matching, symmetry consistency, and the
wrapped-M2-brane interpretation, rather than a first-principles derivation. What we have used here is more specific to the junction data and additional degrees of freedom residing there.}

Therefore to conclude, 
the physical Type $0$A fields descending from the wedge reduction may now be summarized in the following way: 
\begin{equation}
\text{NSNS sector: }
g_{\mu\nu},\quad
B_{\mu\nu}=\frac12(C_{\mu\nu +}+C_{\mu\nu -}),\quad
\Phi,\quad
T=\frac{R_+-R_-}{2} \nonumber
\label{NSNSfinal0A}
\end{equation}
\begin{equation}
\text{RR sector: }
C_1^{(+)}\sim G_{\mu +},\qquad
C_1^{(-)}\sim G_{\mu -},\qquad
C_3^{(+)}\sim C_3 + \hat{\Xi}_3,\qquad
C_3^{(-)}\sim C_3 - \hat{\Xi}_3 ,
\label{RRfinal0A}
\end{equation}
which matches precisely with the fact that the NS sector fields are not doubled but the RR sector fields are in 0A, plus with the presence of a tachyon $T$. By contrast, the following objects should \emph{not} be counted as extra perturbative Type $0$A fields:
\begin{equation}
G_{+-},\qquad C_{\mu + -},
\qquad
B_2^{(o)}=\frac12(C_{\mu\nu +}-C_{\mu\nu -}),
\qquad
H_3^{(o)}=\frac12(H_3^{(+)}-H_3^{(-)}).
\label{nonphys0A}
\end{equation}
They should instead be interpreted as constrained wedge/junction data or as auxiliary branch-odd quantities determined by the local singular geometry and the tachyon background. This is consistent with the present status of the wedge-circle proposal: the doubled RR sector is part of the proposed Type \(0\)A interpretation, while the precise microscopic realization of all doubled fields from the singular M-theory geometry is still not fully understood. It will be interesting to see if with appropriate definition of the function in \eqref{chotolookai} one could justify the presence of the two RR three-forms from M-theory dimensional reduction.

\begin{table}[t]
\centering
\renewcommand{\arraystretch}{1.25}
\newcommand{\thirdcell}[1]{\multicolumn{1}{|c|}{#1}}
\newcommand{\thirdsubcell}[2]{%
\multicolumn{1}{|@{}p{0.38\textwidth}@{}|}{%
\begin{tabular}{@{}p{0.38\textwidth}@{}}
\centering #1 \tabularnewline
\hline
\centering #2 \tabularnewline
\end{tabular}%
}%
}
\begin{tabular}{|c|c|@{}p{0.38\textwidth}@{}|}
\hline
\textbf{Fields} & \textbf{BDV} & \centering\arraybackslash\textbf{Our Approach} \\
\hline
$g_{\mu\nu}$ & SSP & \thirdcell{Bulk DOF} \\
\hline
$H_3^{\rm phys}$ & SSP & \thirdcell{$H_3^{(e)}$} \\
\hline
$H_3^{(o)}$ & $\cdots$ & \thirdcell{$\Delta H_3$} \\
\hline
$A_\mu^{(\pm)}$ & DRP & \thirdsubcell{$A_\mu^{(e)}$}{$A_\mu^{(o)}$} \\
\hline
$C_3^{(\pm)}$ & DRP & \thirdsubcell{$C_3+\widehat{\Xi}_3$}{$C_3-\widehat{\Xi}_3$} \\
\hline
$\varphi$ & SSP &
\thirdcell{$f\!\left({\sqrt{G_{++}}+\sqrt{G_{--}}\over 2}\right)$} \\
\hline
$T$ & SSP &
\thirdcell{${1\over 2}\left(\sqrt{G_{++}}-\sqrt{G_{--}}\right)$} \\
\hline
$\varphi_2$ & $\cdots$ & \thirdcell{$G_{+-}$} \\
\hline
$A_\mu^{(2)}$ & $\cdots$ & \thirdcell{$C_{\mu+-}$} \\
\hline
$\Phi_{\rm massive}$ & $\cdots$ & \thirdcell{Localized pinch UV DOFs} \\
\hline
\end{tabular}
\caption{Comparison between the BDV field assignment and the branchwise effective variables used in our approach. Here $f(x)$ denotes an appropriate function of $x$ such that at the Type IIA end-points $f(x) = {3\over 4}\log(4x^2)$. The highly localized UV DOFs at the pinch are denoted by $\Phi_{\rm massive}$. The $\cdots$ denotes the absence of such fields in the BDV prescription. {Note that away from the common node, each wedge branch is locally
smooth, and the ordinary Wilsonian procedure may be applied, provided that
the retained fields are separated parametrically from the massive modes
that have been integrated out. At the node itself, however, the smooth-bulk
Wilsonian action is not sufficient. The highly localized degrees of freedom $\Phi_{\rm massive}$
may contain both modes that become light near the
degeneration and genuinely short-distance modes associated with the
microscopic completion of the pinch.}}
\label{bdvvsours}
\end{table}

Let us close this section by clarifying the sense in which torsion classes are used below.
Because the compactification involves the singular wedge \(S^1_+\vee S^1_-\), the relevant geometry should not be viewed as a smooth seven-manifold with globally smooth \(G_2\) holonomy.  Instead, it should be treated as a singular, or stratified, space: away from the junction it has smooth branch components carrying ordinary local \(G_2\)-structure data, while the common node contributes additional localized information.

Where do these localized pinch DOFs come from? A useful way to interpret the localized pinch sector is to distinguish between
two possible kinds of junction degrees of freedom.  The first consists of
genuinely heavy short-distance modes associated with the microscopic completion
of the singularity.  The second consists of modes that become light, or even
massless, as the pinch is approached.  Both kinds of modes can contribute to
the local physics at the node, but they enter the low-energy description in
different ways.

In an ordinary smooth compactification, heavy short-distance degrees of
freedom are integrated out in the Wilsonian or exact-renormalization-group
description.  Their effects survive through higher-derivative operators,
threshold corrections, renormalized couplings, and other local terms in the
effective action.  In a singular compactification this logic has to be refined.
Near the node of
$S^1_+\vee S^1_-$,
the local geometry is not smooth, the usual Kaluza--Klein expansion may break
down, and the branch matching conditions can produce localized contributions
that are invisible in a purely smooth-bulk description.  Thus some of the
short-distance information that would normally be encoded in smooth bulk
Wilson coefficients may instead appear as localized defect data at the pinch.

At the same time, the conifold example \cite{conifold} teaches us that not every singular
localized contribution should be interpreted as a UV effect.  In the conifold
case, the apparent singularity of the low-energy theory arises because states
that become massless at the singular point have been integrated out.  Once
these light states are included explicitly, the effective description is
repaired.  A similar possibility should be kept open for the wedge pinch:
some of the degrees of freedom localized near the node may be light states
whose mass scale is controlled by the local degeneration, rather than heavy UV
modes.

Thus the localized junction sector should be interpreted broadly.  It may
contain light states that become important as the pinch is approached, as well
as genuinely short-distance degrees of freedom that remain localized at the
singular node and are not part of the ordinary smooth-bulk Wilsonian spectrum.
We denote this collective junction sector by $\chi_{\rm jct}$ and write the
effective action schematically as:
\begin{equation}
S_{\rm eff}[g, C_3, \chi_{\rm jct}]
=
S_{\rm bulk}^{\rm light}[g,C_3]
+
S_{\rm jct}
\big[
g|_{\Sigma_{\rm jct}},
C_3|_{\Sigma_{\rm jct}},
\chi_{\rm jct}
\big]
+
S_{\rm bulk}^{\rm hd}[g, C_3],
\label{Seff_bulk_jct_general}
\end{equation}
where we have only showed the bosonic DOFs,
$S_{\rm bulk}^{\rm light}$ contains the ordinary low-energy bulk fields
away from the pinch, $S_{\rm jct}$ contains the localized degrees of freedom
supported at the node, and $S_{\rm bulk}^{\rm hd}$ denotes the usual
higher-derivative and threshold corrections in the smooth bulk.
More explicitly, one may separate the junction sector into a part containing
light or near-light localized modes and a part containing intrinsically
short-distance junction degrees of freedom:
\begin{equation}
S_{\rm jct}\big[
g|_{\Sigma_{\rm jct}},
C_3|_{\Sigma_{\rm jct}},
\chi_{\rm jct}
\big]
=
S_{\rm jct}^{\rm light}[\chi_{\rm light}]
+
S_{\rm jct}^{\rm UV}[\chi_{\rm UV}]
+
S_{\rm coupl}
\big[
g|_{\Sigma_{\rm jct}},
C_3|_{\Sigma_{\rm jct}},
\chi_{\rm light},
\chi_{\rm UV}
\big] ,
\label{Sjct_light_UV_split}
\end{equation}
where $\chi_{\rm jct} \equiv (\chi_{\rm light}, \chi_{\rm UV})$.
The fields $\chi_{\rm light}$ denote modes whose mass becomes small near the
pinch and which should therefore be retained explicitly in the low-energy
description of the junction.  The fields $\chi_{\rm UV}$ denote genuinely
short-distance degrees of freedom associated with the microscopic completion
of the singular node.  These are not ordinary propagating bulk fields, but they
need not be integrated out in the same way as heavy modes in the smooth bulk;
the singularity can make them visible as localized defect data.
If one further integrates out some subset of the localized UV sector, then its
effects may be represented by local operators supported at the junction:
\begin{equation}
S_{\rm jct}^{\rm UV,eff}
=
\int_{\Sigma_{\rm jct}}
{\cal L}_{\rm loc}^{\rm UV}
\big(
g|_{\Sigma_{\rm jct}},
C_3|_{\Sigma_{\rm jct}},
G_4|_{\Sigma_{\rm jct}},
\chi_{\rm light},
\ldots
\big).
\label{UV_localized_action}
\end{equation}
However, this is an additional effective step, not the basic assumption.  The
basic point is that the smooth-bulk Wilsonian separation between retained IR
fields and integrated-out UV modes need not apply uniformly at the singular
pinch.  The node may support localized UV-sensitive degrees of freedom, and
these can affect the matching conditions, Bianchi identities, torsion-source
equations, and possibly the branchwise field dictionary, without becoming
ordinary propagating fields throughout the eleven-dimensional bulk\footnote{{The separation between the two descriptions may be stated more precisely
by regulating the node with a small neighborhood
${\cal U}_{\epsilon}(\Sigma_{\rm jct})$. On
${\cal M}_{11}\setminus{\cal U}_{\epsilon}$, the bulk fields satisfy the
ordinary Wilsonian equations of motion. The limit
$\epsilon\rightarrow0$ is well defined only when the canonical fluxes of
the bulk fields obey junction conditions of the schematic form:
\begin{equation}
\lim_{\epsilon\rightarrow0}
\left[
\Pi_{\Phi}^{(+)}
-
\Pi_{\Phi}^{(-)}
\right]_{\partial{\cal U}_{\epsilon}}
=
\frac{\delta S_{\rm jct}}
{\delta\Phi|_{\Sigma_{\rm jct}}} \nonumber
\end{equation}
where $\Pi^{(\pm)}_\Phi$ are the canonical momenta of the massive field $\Phi$ on two sides of the junction.
A complete microscopic solution would require the explicit junction action
and a verification of these conditions. In the present work we instead
identify the representation channels in which these localized terms can
enter. Accordingly, our construction is a controlled effective ansatz on
the smooth strata, supplemented by parametrized defect data at the pinch.}}.

The light junction modes have a different role.  If a mode becomes light at the
pinch, it can mix with the branchwise bulk fields and modify the effective
lower-dimensional field dictionary.  This is the sense in which one may write,
for example:
\begin{equation}
C_3^{(\pm)}
=
C_3
\pm
\widehat{\Xi}_3 .
\label{chotolook}
\end{equation}
which exactly explains the origin of the term in \eqref{chotolookai}.
The correction $\widehat{\Xi}_3$ should then be viewed as a branch- or
junction-sensitive contribution generated by light degrees of freedom at the
pinch, or by modes that become light as the wedge singularity is approached.
Heavy modes may renormalize the coefficients entering $\widehat{\Xi}_3$, or
generate localized operators that source it, but they should not be counted as
independent propagating components of $C_3^{(\pm)}$. In this sense, the propagating doubled RR three-form sector is most naturally associated with the light or near-light branch/junction data, while the genuinely UV sector contributes through localized defect operators and matching conditions.

The heavy localized sector must be confined to the pinch for this interpretation to
be consistent.  If the heavy short-distance degrees of freedom reappeared as
ordinary propagating bulk fields throughout the compactification, the
Wilsonian description would fail globally.  Instead, the more precise
statement is that the singular junction can trap, expose, or require
additional localized data.  Some of this data may consist of heavy UV
remnants encoded in localized operators; some may consist of light states that
must be included explicitly near the pinch. Away from the pinch, the low-energy theory may be described after integrating
out the modes whose masses are nonzero.  In that regime one can write
schematically:
\begin{equation}
S_{\rm eff}^{\rm away}(g, C_3^{\rm br})
=
S_{\rm bulk}^{\rm light}(g, C_3)
+
\Delta S_{\rm loc}^{\rm induced}(g, C_3, \chi_{\rm light}) ,
\label{Seff_away_pinch_rewrite}
\end{equation}
which precisely explains how $C_3 \to C_3^{\rm br}$, or $G_4 \to G_4^{\rm br}$ in \eqref{g4c3h3}, thus giving credence to the doubled spectrum discussed earlier.
Near the pinch, however, one should introduce back the localized heavy modes. 
Then the appropriate near-pinch description is not obtained by keeping these
modes integrated out.  Instead one should use an enlarged local theory:
\begin{equation}
S_{\rm eff}^{\rm near}(g, C_3^{\rm br}, \chi_{\rm UV})
=
S_{\rm bulk}^{\rm light}(g, C_3)
+
\underbrace{S_{\rm jct}^{\rm light}[\chi_{\rm light}]
+
S^{\rm light}_{\rm coupl}[g,C_3,\chi_{\rm light}]}_{\Delta S_{\rm loc}^{\rm induced}(g, C_3, \chi_{\rm light})}
+
S_{\rm jct}^{\rm heavy,eff} ,
\label{Seff_near_pinch_rewrite}
\end{equation}
which combines the wedge analogue of the conifold lesson, {\it i.e.} a singular effective
description signaling that light localized states have been omitted, and a more generic singularity that could host localized heavy UV modes with:
\begin{equation}\label{haramibob}
S_{\rm jct}^{\rm heavy,eff} \equiv S_{\rm jct}^{\rm heavy}(\chi_{\rm heavy}) + S_{\rm coupl}^{\rm heavy}(g, C_3, \chi_{\rm light}, \chi_{\rm heavy}) ,
\end{equation}
where $\chi_{\rm heavy} \subseteq \chi_{\rm UV}$ and equivalently 
$ S_{\rm jct}^{\rm heavy}(\chi_{\rm heavy}) \subseteq S_{\rm jct}^{\rm UV}(\chi_{\rm UV})$ from \eqref{Sjct_light_UV_split}. It should also be clear that the two coupling interactions from \eqref{Seff_near_pinch_rewrite} and \eqref{haramibob} should at least be a subset of the interactions terms in \eqref{Sjct_light_UV_split}. These short distance interactions do not change any of the bulk physics, so for all practical purposes, \eqref{Seff_away_pinch_rewrite} suffices.

The torsion-class framework does not decide which microscopic mechanism is
responsible.  Its role is to organize the possible localized sources into
$G_2$ and, after reduction, $SU(3)$ representation channels.  The detailed
spectrum, masses, and interactions of $\chi_{\rm jct}$ require a microscopic
model of the pinch, or equivalently a localized effective action for the
junction.

With this interpretation there is no contradiction in using \(G_2\) and \(SU(3)\) torsion
classes.  On the smooth locus the standard forms \(\varphi\), \(\psi\), and, after reduction, \(J\), \(\Omega\), are defined branchwise, and their exterior derivatives encode the usual intrinsic torsion sourced by fibration and flux.  The genuinely new feature of the wedge is that these branchwise structures need not extend smoothly across the node.  This failure of smooth extension is precisely what is represented by localized, junction-supported torsion terms.

The torsional analysis should therefore be understood as an effective branchwise and
distributional analysis, not as a claim of smooth \(G_2\) holonomy.  Its purpose is to
organize the smooth fibration and flux contributions together with the singular pinch data
in a common representation-theoretic language.

In this sense the torsion analysis proceeds on
$M_7^{\rm sm}
=
M_7\setminus \Sigma_{\rm jct}$,
where $\Sigma_{\rm jct}$ denotes the locus over which the wedge branches meet.
On $M_7^{\rm sm}$ one may define the usual $G_2$ forms
$\varphi$, and 
$\psi=*_{7}\varphi$, 
and decompose their exterior derivatives into the standard torsion classes, as we shall describe soon.
The singularity appears through the fact that the forms and coframes are only
piecewise smooth.  Their exterior derivatives therefore contain both smooth
branchwise pieces and distributional junction-supported pieces:
\begin{equation}\label{smoothc}
d\varphi
=
(d\varphi)_{\rm sm}
+
(d\varphi)_{\rm jct},
\qquad
d\psi
=
(d\psi)_{\rm sm}
+
(d\psi)_{\rm jct}.
\end{equation}
The smooth terms describe ordinary fibration and flux-induced intrinsic torsion,
while the junction-supported terms encode the pinch contribution.
The junction pieces can receive contributions from both the geometric
non-closure of the pinched coframe and the localized short-distance sector:
\begin{equation}
(d\varphi)_{\rm jct}
=
(d\varphi)_{\rm pinch}
+
{\cal J}_{\varphi}(\chi_{\rm jct})\nonumber
\end{equation}
\begin{equation}\label{hathimerasathi}
(d\psi)_{\rm jct}
=
(d\psi)_{\rm pinch}
+
{\cal J}_{\psi}(\chi_{\rm jct}),
\end{equation}
where ${\cal J}_{\varphi}(\chi_{\rm jct})$ and ${\cal J}_{\psi}(\chi_{\rm jct})$ should in-principle be interpreted broadly as 
containing contributions from light junction fields, from localized
operators induced by heavy modes, or from both. However we could transfer the contributions from the light junction modes to $(d\varphi)_{\rm sm}$ in \eqref{smoothc} so that it could be related to the doubled fields, and leave the junction currents exclusively from the localized heavy UV modes. After projecting into $G_2$ irreducible representations, these terms feed the
localized torsion classes.  Similarly, after reduction to an $SU(3)$ structure,
they can contribute to the localized parts of
$W_1,\ldots,W_5$. For example, if the wedge one-form is written schematically as:
\begin{equation}
e^4_{\rm wedge}
=
R_B(\text{branch-even one-form})
+
T(\text{branch-odd one-form}),
\end{equation}
then its exterior derivative contains ordinary smooth contributions involving
$dR_B$ and $dT$, together with possible distributional contributions supported
at the node.  These are exactly the kinds of terms to which intrinsic torsion
is sensitive.  Therefore the use of torsion classes is not obstructed by the
singularity; rather, the singularity appears as a localized source in the
torsion decomposition. To keep the analysis manageable, we will focus primarily on the contribution of
the geometric pinch terms in \eqref{hathimerasathi} to the torsion classes.  The
additional effects of the localized junction currents
${\cal J}_{\varphi}(\chi_{\rm jct})$ and
${\cal J}_{\psi}(\chi_{\rm jct})$, that contain exclusively the localized UV modes, will be discussed separately in
section~\ref{fate}, except where they are explicitly needed.

The same logic applies after reducing to a six-dimensional $SU(3)$-structure
description.  On each smooth branch, the reduced forms $J$ and $\Omega$ define
the usual $SU(3)$ structure, and their exterior derivatives determine
$W_1,.., W_5$.
At the junction, the reduction produces localized or distributional
contributions to these torsion classes.  Thus the $SU(3)$ torsion classes
should be understood as effective branchwise torsion classes, supplemented by
localized node data. The correct interpretation is therefore, we have 
smooth $G_2/SU(3)$ torsion on each branch, plus
junction-supported torsion at the node.
This is a perfectly meaningful effective framework, provided one does not
pretend that all forms are globally smooth across the wedge point.

The error would only arise if one imposed smooth-manifold identities globally
without accounting for the node.  For example, one should not treat
$S^1_+\vee S^1_-$ as an ordinary smooth circle, nor should one assume that
the branchwise one-forms are globally smooth across the junction.  Instead, one
must allow distributional terms, matching conditions, and possible auxiliary
junction degrees of freedom.  With these included, the torsion classes provide
a consistent way to organize the geometry.

\subsection{Starting point and fiberwise duality frame \label{sec002}}

After this detour, let us now come back to the model that we are interested in. We will start by taking 
$X_7$ to be a compact seven-manifold with $G_2$-structure, equipped
with a fibration by $K3$ surfaces over a compact three-manifold
$\Sigma_3$:
\begin{equation}
K3 \hookrightarrow X_7 \to \Sigma_3 .
\end{equation}
If the $G_2$-structure is torsion-free, compactification of M-theory on
$X_7$ yields a four-dimensional theory with $\mathcal N=1$
supersymmetry \cite{luest}. More generally, if the $G_2$-structure has intrinsic
torsion, one obtains a flux or torsion compactification, and
supersymmetry must be analyzed using the corresponding $G_2$ torsion
classes \cite{CS, luest, strominger, DRS}.

However the motivation for the $G_2$ structure goes beyond supersymmetry. It is a representation-theoretic and geometric framework for organizing the
intrinsic torsion of a seven-dimensional space, irrespective of whether or not the system has any residual supersymmetry or not. As we shall see soon, the non-supersymmetric pinch
case is precisely a case where this torsionful $G_2$ framework is useful. On the other hand, the motivation for considering a $K3$ fibration is the standard
seven-dimensional duality: 
\begin{equation}
\text{M-theory on } K3
\qquad \longleftrightarrow \qquad
\text{heterotic on } T^3 .
\end{equation}
Our proposal is to use this duality locally over $\Sigma_3$. This is the so-called fiber-wise duality conjecture \cite{senduality} that works remarkably well when there is some residual supersymmetry and an underlying adiabaticity.

Throughout this paper we assume an adiabatic regime in which the $K3$
fibration varies slowly over $\Sigma_3$, and the relevant monodromies
act through the part of the $K3$ moduli compatible with a heterotic
$T^3$ description. Thus, and as mentioned above, the duality is employed here $-$ at least in the supersymmetric case $-$ as a
\emph{local fiberwise duality frame}, not as an already-established
global theorem for arbitrary $K3$-fibered $G_2$ backgrounds.
To make contact with the familiar F-theory/heterotic dictionary, we
further assume that the $K3$ fibers admit an elliptic presentation
\begin{equation}
T^2_f \hookrightarrow K3 \to \mathbb{P}^1 ,
\end{equation}
and that we are in a Morrison--Vafa type \cite{MV} regime in which the elliptic
fiber is small compared to the base. In this regime the $K3$ fiber is
well approximated by a thin elliptic surface, and the fiberwise
duality map is most transparent.

We are also interested in the F-theory framework that has some orientifold description. As shown in \cite{sen} this allows a quantitative control of the theory. 
Near the weak-coupling orientifold region of the base, we use the
standard local Sen-type model \cite{sen}:
\begin{equation}\label{oribaba}
\mathbb{P}^1 \;\leadsto\; T^2/\mathbb{Z}_2 ,
\end{equation}
where the double cover is a torus and the four fixed points of the
$\mathbb{Z}_2$ action represent the local orientifold loci. Again, this
should be understood as a \emph{local weak-coupling description} of
the base geometry, not as a literal global equality of complex
structures.

We now consider two deformations of this local base geometry. The
first is designed to leave the Kähler regime while potentially
preserving supersymmetry. The second is the non-supersymmetric wedge
branch.

\subsection{Branch I: the interval--circle branch \label{intervalbranch}}

An elliptically fibered K3 surface may be presented in Weierstrass form as a
family of cubic curves
$y^2=x^3+f(z)x+g(z)$,
parametrized by $z\in\mathbb P^1$, where $f$ is a section of
$\mathcal O(8)$ and $g$ is a section of $\mathcal O(12)$ \cite{vafaF, MV}.  For each value of
$z$ with nonzero discriminant, the equation defines a smooth elliptic curve. For a generic elliptic K3, the elliptic fiber degenerates over the
twenty-four zeroes of $\Delta$ on the base $\mathbb P^1$, counted with
multiplicity. As mentioned earlier, we are in the orientifold limit \eqref{oribaba}, wherein 
the first local replacement is:
\begin{equation}
T^2/\mathbb{Z}_2
\qquad \longrightarrow \qquad
S_a^1/\mathbb{Z}_2 \rtimes S^1_b ,
\end{equation}
where $S^1_b$ denotes the base-side circle inherited from the elliptic
presentation, and $\rtimes$ indicates a nontrivial local fibration or
monodromy of $S^1_b$ over the interval direction. Since:
\begin{equation}
S_a^1/\mathbb{Z}_2 \cong I ,
\end{equation}
the deformed local model is
$I \rtimes S^1_b$.
This is not an ordinary complex-geometric deformation of the base
$\mathbb{P}^1$. Rather, it should be interpreted as a local real or
non-geometric deformation of the fiberwise duality frame. In
particular, the original elliptic-complex description of the $K3$
fiber is no longer preserved in the usual sense, so the resulting
compactification is not naturally described as Kähler in the original
complex-geometric frame.

Nevertheless, supersymmetry need not be lost. On the M-theory side the
relevant condition is the existence of a $G_2$-structure satisfying
the appropriate first-order supersymmetry equations, possibly with
torsion. Thus the correct test is not Kählerity, but compatibility of
the deformed geometry with the $G_2$ Killing spinor equations. There
are many known examples of supersymmetric non-Kähler
$G_2$-structure backgrounds, so this branch is not obviously excluded.

Applying Ho\v rava--Witten reduction along the interval factor
$S^1/\mathbb{Z}_2$ then suggests the local heterotic duality frame:
\begin{equation}
\text{heterotic on } T^2_f \times S^1_b
\quad \text{fibered over } \Sigma_3 .
\end{equation}
Equivalently, the local heterotic three-torus is
$T^3_{\rm het}=T^2_f \times S^1_b$.
This should be understood as a \emph{local bookkeeping identification}
of torus factors in the duality frame, not as a literal
complex-geometric decomposition inside the original $K3$.

On the heterotic side supersymmetry is tested using the
$\mathrm{SU}(3)$-structure torsion classes $(W_1,\dots,W_5)$ of the
internal six-manifold. Writing the fundamental form and holomorphic
three-form as $(J,\Omega)$, the Strominger-type supersymmetry
conditions are \cite{luest, strominger, DRS}:
\begin{equation}
W_1=0,\qquad W_2=0,\qquad 2W_4+W_5=0 .
\end{equation}
Thus Deformation 1 gives a candidate non-Kähler but supersymmetric
branch provided the geometry, fluxes, and gauge bundle satisfy the
heterotic Bianchi identity and the Strominger system. A subsequent U-duality gives Type IIA theory on $T^2_f \rtimes {S_a^1/Z_2} \rtimes \Sigma_3$ at the orientifold points.

\begin{table}[t]
\centering
\renewcommand{\arraystretch}{1.35}
\begin{tabular} {|p{0.14\textwidth}|p{0.38\textwidth}|p{0.38\textwidth}|}
\hline
\textbf{$G_2$ torsion data} & \textbf{Reduced $SU(3)$ torsion data} & \textbf{Interpretation} \\
\hline
$\tau_0$ &
$W_1$ &
Singlet torsion; contributes to the scalar/non-complex part of the
$SU(3)$ structure. \\
\hline
$\tau_1$ &
$W_4,\;W_5$ &
Lee-form-type torsion; controls conformal/non-balanced behavior after
reduction. \\
\hline
$\tau_2$ &
$W_2$ and Lee-form data &
Adjoint-type torsion; contributes to primitive $(1,1)$ obstructions and
can mix with one-form torsion depending on the reduction direction. \\
\hline
$\tau_3$ &
$W_2,\;W_3$ &
Primitive traceless torsion; its $\mathbf 8$ component feeds $W_2$, while
its $\mathbf 6\oplus\overline{\mathbf 6}$ component feeds $W_3$. \\
\hline
\end{tabular}
\caption{Schematic relation between the intrinsic torsion classes of a
seven-dimensional $G_2$ structure and the torsion classes of the induced
six-dimensional $SU(3)$ structure.}
\label{g2su3maap}
\end{table}

\subsubsection{Analysis of the branch using $G_2$ torsion classes \label{sec2.2.1}}

Let $\{e^i\}$, $i=1,2,3$, be an orthonormal coframe on $\Sigma_3$, and let $J^i$, $i=1,2,3$, be the hyperk\"ahler triplet on the $K3$ fiber, normalized by
$J^i\wedge J^j = 2\,\delta^{ij}\,\mathrm{vol}_4$
with 
$*_{4} J^i = J^i$. Then a natural $G_2$ structure on $M_7$ and its dual is given by:

{\footnotesize
\begin{equation}
\varphi
=
 e^{123} + e^1\wedge J^1 + e^2\wedge J^2 + e^3\wedge J^3,~~~~
\psi \equiv *_7\varphi
=
\mathrm{vol}_4 + e^{23}\wedge J^1 + e^{31}\wedge J^2 + e^{12}\wedge J^3 ,
\label{psiG2def}
\end{equation}}
which basically is the standard $SU(2)$-structure-to-$G_2$ construction.
Here $e^{abc} \equiv e^a \wedge e^b \wedge e^c$.
The intrinsic torsion is encoded by four torsion classes:
\begin{equation}\label{taui}
\tau_0\in\Lambda^0,
\qquad
\tau_1\in\Lambda^1_{7},
\qquad
\tau_2\in\Lambda^2_{14},
\qquad
\tau_3\in\Lambda^3_{27},
\end{equation}
where they provide a convenient
way to measure how far the seven-dimensional geometry is from having
torsion-free $G_2$ holonomy.  The scalar class $\tau_0$ captures the nearly
parallel component of the torsion, while the one-form class $\tau_1$ is the
Lee-form-type contribution and measures an overall conformal twisting of the
$G_2$ structure.  The two-form class $\tau_2$ lies in the adjoint
representation of $G_2$ and encodes the part of the torsion associated with
the non-integrability of the coassociative structure.  Finally, the three-form
class $\tau_3$ is the primitive traceless component and is often the most
sensitive to localized geometric deformations, flux backreaction, and pinch
or junction effects (to be discussed soon).  In a torsion-free $G_2$ background all four classes
vanish, whereas in the non-supersymmetric geometries to be considered here these
classes provide a systematic language for organizing fibration, flux, and
localized singular contributions. They may be quantified
through the following form decomposition that are taken from $\varphi$ and $\psi$ described in  \eqref{psiG2def} above:
\begin{equation}
d\varphi
=
\tau_0\,\psi + 3\,\tau_1\wedge\varphi + *_{{}_7}\tau_3,~~~~~
d\psi
=
4\,\tau_1\wedge\psi + \tau_2\wedge\varphi ,
\label{torsiondecomp2}
\end{equation}
where the exterior derivatives of the defining $G_2$ forms decompose into irreducible
$G_2$ representations. This may be explained more quantitatively in the following way. The failure of the three-form $\varphi$ to be closed is
split into three independent pieces: a singlet component proportional to the
dual four-form $\psi$, a Lee-form component obtained by wedging the one-form
torsion class with $\varphi$, and a primitive symmetric-traceless component
encoded by the Hodge dual of $\tau_3$.  Similarly, the failure of the dual
four-form $\psi$ to be closed is governed by two pieces: the same Lee-form
torsion class wedged with $\psi$, and an adjoint-valued two-form torsion class
wedged with $\varphi$.    
To make the torsion generation explicit, we can parametrize the failure of the base and fiber forms to be covariantly constant by:
\begin{equation}
de^i = -\frac12 f^i{}_{jk}\,e^j\wedge e^k,~~~~~~~
dJ^i = M^i{}_{j}\wedge J^j + P^i ,
\label{dJgeneral}
\end{equation}
where
$M^i{}_{j} = M^i{}_{jk}\,e^k$
is an $SO(3)$-valued connection on the bundle of self-dual two-forms over the $K3$ fiber, while:
\begin{equation}
P^i \in \Omega^1(\Sigma_3)\otimes \Lambda^2_{-}(K3)
\end{equation}
contains the components of the variation of the hyperk\"ahler structure that lie outside the span of the $J^i$. In other words, $P^i$ captures the primitive/anti-self-dual part of the twisting. Using \eqref{psiG2def} and \eqref{dJgeneral}, one finds:
\begin{equation}
\begin{aligned}
d\varphi
&=
\sum_{i=1}^3 de^i\wedge J^i - \sum_{i=1}^3 e^i\wedge dJ^i \\
&=
-\frac12 f^i{}_{jk}\,e^{jk}\wedge J^i
-
 e^i\wedge M^i{}_{j}\wedge J^j
-
 e^i\wedge P^i ,
\end{aligned}
\label{dphifibration}
\end{equation}
where note that $d(e^{123}) = 0$, {\it i.e.} vanishes identically using the first equation in \eqref{dJgeneral}. (If $de^i$ also depends on the K3 fiber, then $d(e^{123})$ should be kept\footnote{In other words, if the one-forms $e^i$ also had components along the four-dimensional fiber directions, which could happen due to a non-trivial fibration from the fact that the projection
$\pi: M_7 \to \Sigma_3$
induces 
$e^i=\pi^*(\hat e^i)$
implying the pullback of a base coframe $\hat e^i$ on $\Sigma_3$, then: 
\begin{equation}
de^i
=
-\frac12 f^i{}_{jk}e^{jk}
+
Q^i{}_{Aj}\,E^A\wedge e^j
+
\frac12 R^i{}_{AB}\,E^A\wedge E^B, \nonumber
\end{equation}
with $E^A$ being the K3 fiber one-forms. In that case 
$d(e^{123})$
would no longer vanish in general and should be kept. For our fibration
where 
$M_7 = K3 \rtimes \Sigma_3$,
the cleanest notation is to keep the split explicit with 
$\{E^A\}_{A=1}^4$ for the K3 fiber,
and $\{e^i\}_{i=1}^3$ for the base $\Sigma_3$,
with
$e^i(E_A)=
E^A(\partial_j)=0$, 
then the non-trivial fibration is encoded by the structure equations:
\begin{equation}
dE^A
=
\Theta^A{}_{Bi}\,e^i\wedge E^B
+
\frac12\,\Pi^A{}_{BC}\,E^B\wedge E^C
+
\frac12\,\Xi^A{}_{ij}\,e^i\wedge e^j ,\nonumber
\label{dEAclean}
\end{equation}
and the first equation in \eqref{dJgeneral} if the base itself is anholonomic. Thus in an adapted local frame, the one-forms $e^i$ on 
$\Sigma_3$ are purely horizontal and do not themselves carry K3 fiber components. The non-trivial fibration is instead encoded in the structure equations for the fiber one-forms $E^A$, namely through mixed terms of the form $e^i\wedge E^B$ in $dE^A$. This is the viewpoint we will adopt here to avoid the aforementioned subtle nuances. \label{einuances}}.)
Similarly, from \eqref{psiG2def} one gets schematically:
\begin{equation}
\begin{aligned}
d\psi
&=
d\mathrm{vol}_4
+
d(e^{23})\wedge J^1 + d(e^{31})\wedge J^2 + d(e^{12})\wedge J^3 \\
&\qquad
+
e^{23}\wedge dJ^1 + e^{31}\wedge dJ^2 + e^{12}\wedge dJ^3.
\end{aligned}
\label{dpsifibration}
\end{equation}
Even if the fiber is hyperk\"ahler pointwise, the base dependence of the triplet $J^i$ and the twisting of the base frame are enough to make both $d\varphi$ and $d\psi$ non-zero. The right-hand sides of \eqref{dphifibration} and \eqref{dpsifibration} decompose into $G_2$ irreducible pieces. It is useful to separate the \(\mathfrak{gl}(3)\)-valued coefficient acting on the
hyperk\"ahler triplet into trace, antisymmetric, and symmetric traceless
parts\footnote{The \(SO(3)\) appears because the three self-dual \(2\)-forms
$J^i$ for $i=1,2,3$,
on the \(K3\) fiber form an orthonormal basis of the rank-\(3\) bundle of self-dual two-forms. Since the metric on the fiber preserves the inner product
$\langle J^i,J^j\rangle \propto \delta^{ij}$,
any local change of basis among the \(J^i\) that preserves this normalization must be an orthogonal rotation of the form
$J^i \;\longrightarrow\; O^i{}_{j}\,J^j$
with 
$O\in SO(3)$.
Thus the triplet \(\{J^1,J^2,J^3\}\) is naturally acted on by \(SO(3)\).
Equivalently, on a hyperk\"ahler four-manifold the self-dual \(2\)-forms span a \(3\)-dimensional real vector space. The wedge-product normalization
$J^i\wedge J^j = 2\,\delta^{ij}\,\mathrm{vol}_4$
defines a Euclidean metric on this space, so its structure group is
$SO(3)\simeq SU(2)/\mathbb Z_2$.
Therefore, when the triplet varies over the base \(\Sigma_3\), its variation is described by a matrix-valued one-form
$\mathcal M^i{}_{j}=\mathcal M^i{}_{jk}\,e^k$.
Its antisymmetric part
$A_{ij}\equiv \frac12\big(\mathcal M_{ij}-\mathcal M_{ji}\big)$
is the genuine \(SO(3)\) connection, while the trace and symmetric traceless parts
measure departures from a pure \(SO(3)\) rotation inside the span of the triplet.
More concretely, if the \(K3\) fiber changes as one moves along \(\Sigma_3\), then the self-dual basis \(\{J^i\}\) can rotate into itself. This is exactly what the term
$dJ^i = \mathcal M^i{}_{j}\wedge J^j + P^i$
means: $\mathcal M^i{}_{j}\wedge J^j$
is the part of the variation of \(J^i\) that stays inside the span of the hyperk\"ahler triplet, while:
\begin{equation}
P^i \in \Omega^1(\Sigma_3)\otimes \Lambda^2_{-}(K3)\nonumber
\end{equation}
is the part orthogonal to that span, namely the anti-self-dual or primitive deformation.
If the hyperk\"ahler normalization is preserved exactly along the base, then one must
set:
\begin{equation}
\mathrm{tr}\,\mathcal M=0,
\qquad
S_{ij}=0,
\qquad
\mathcal M_{ij}=A_{ij}\in \mathfrak{so}(3)\nonumber
\end{equation}
and recover the usual \(SO(3)\)-valued connection. So the logic is the following:
self-dual two-forms on $K3
\;\cong\;
\mathbb R^3$ implies 
basis changes preserving $\delta^{ij}$
are $SO(3)$,
and hence the base-dependence of the hyperk\"ahler triplet is governed by an
\(SO(3)\)-valued connection, namely the antisymmetric part of the more general
matrix-valued coefficient \(\mathcal M^i{}_{j}\).}:
\begin{equation}
\mathcal M_{ij} = \frac13 (\mathrm{tr}\,\mathcal M)\,\delta_{ij} + A_{ij} + S_{ij},
\qquad
A_{ij}=-A_{ji},
\qquad
S_{ij}=S_{ji},
\qquad
S_i{}^i=0 ,
\label{Mdecomp}
\end{equation}
from where we see that the projections onto $\mathbf 1\oplus \mathbf 7\oplus \mathbf{27}$ in $d\varphi$ and onto $\mathbf 7\oplus \mathbf{14}$ in $d\psi$ determine the torsion classes. A convenient schematic summary is:
\begin{equation}
\tau_0 \sim \Pi_{\mathbf 1}(d\varphi),
\qquad
\tau_1 \sim \Pi_{\mathbf 7}(d\varphi) \sim \Pi_{\mathbf 7}(d\psi),
\qquad
\tau_2 \sim \Pi_{\mathbf{14}}(d\psi),
\qquad
\tau_3 \sim \Pi_{\mathbf{27}}(d\varphi).
\label{projectorsummary}
\end{equation}
More explicitly the scalar torsion $\tau_0$, or alternatively 
the singlet part, is generated by the part of \eqref{dphifibration} proportional to:
\begin{equation}
\psi = \mathrm{vol}_4 + e^{23}\wedge J^1 + e^{31}\wedge J^2 + e^{12}\wedge J^3.
\end{equation}
Therefore an isotropic, trace-like twisting of the hyperk\"ahler triplet can
contribute to \(\tau_0\).  Likewise, any singlet component of the base twisting
\(f^i{}_{jk}\) contributes to \(\tau_0\).  If this singlet projection vanishes,
then the \(G_2\) structure has no nearly-parallel component, even though the
other torsion classes may still be non-zero.

The one-form torsion vector $\tau_1$ is generated by the $\mathbf 7$ component in both $d\varphi$ and $d\psi$. In the present fibration language it is sourced by the non-uniform warping of the fibration, by the antisymmetric part $A_{ij}$ of $M_{ij}$, and by any one-form part of the flux decomposition below. Thus the failure of the hyperk\"ahler triplet to be parallel transported uniformly over $\Sigma_3$ produces $\tau_1$. On the other hand, the adjoint torsion $\tau_2$ class is special because it appears only in
$d\psi$ defined via the following relation to both $\tau_1$ and $\tau_2$:
\begin{equation}
d\psi = 4\tau_1\wedge\psi + \tau_2\wedge\varphi.
\end{equation}
It is therefore controlled by the $\mathbf{14}$ projection of the terms in \eqref{dpsifibration}. Geometrically, $\tau_2$ is generated by the part of the fibration curvature that preserves the metric but rotates the $G_2$ structure inside the adjoint of $G_2$. In the present setup, this comes from the mismatch between the $d(e^{ij})$ terms and the $dJ^i$ terms after projection onto the $\mathbf{14}$.

Finally for the traceless symmetric torsion $\tau_3$, {\it i.e.}
the $\mathbf{27}$ piece in $d\varphi$, comes from the symmetric traceless part $S_{ij}$ of \eqref{Mdecomp}, from primitive pieces in the base twist, and from the primitive anti-self-dual components $P^i$ in \eqref{dJgeneral}. In particular, the terms:
\begin{equation}
-e^i\wedge P^i
\end{equation}
are automatically orthogonal to the singlet and vector pieces and therefore naturally feed\footnote{Another way to see why the $P^i$ term lands in the ${\bf 27}$: it is anti-self-dual on the K3 fiber, while the singlet and vector pieces are built from the self-dual 
$J_i$ structure. This matches the role of $P^i$ in contribution to $G_2$ torsion.} the $\mathbf{27}$ component, i.e. $\tau_3$.
So the fibration itself already generates torsion. In a product geometry with covariantly constant $J^i$ and untwisted base frame one would have:
\begin{equation}
f^i{}_{jk}=0,
\qquad
M^i{}_j=0,
\qquad
P^i=0,
\end{equation}
and hence all torsion classes would vanish. The non-trivial fibration $K3\rtimes\Sigma_3$ precisely turns on these quantities, thereby generating non-zero torsion classes.

What happens when we add the internal $G_4$ flux? 
Including an internal four-form flux on $M_7$ would imply that 
$G_4 \in \Lambda^4(M_7)$. 
Under $G_2$ we can decompose $\Lambda^4$ as
$\Lambda^4 = \Lambda^4_{\mathbf 1}\oplus \Lambda^4_{\mathbf 7}\oplus \Lambda^4_{\mathbf{27}}$, 
so one may write \cite{CS, behrndt}:
\begin{equation}
G_4 = q\,\psi + v\wedge\varphi + \Xi^{(4)}_{27},
\label{G4decomp}
\end{equation}
where $q\in\Lambda^0,
v\in\Lambda^1_{\bf 7}$ and 
$\Xi^{(4)}_{27}\in\Lambda^4_{\mathbf{27}}$, with $*_7:\Lambda^3_{\bf 27}(M_7)\longrightarrow \Lambda^4_{\bf 27}(M_7)$ and consequently 
$\Xi_{27}^{(3)}
\equiv
*_7^{-1}\Xi_{27}^{(4)} \in\Lambda^3_{27}(M_7)$.
The $G_2$ decomposition itself does not change; what changes is that the supersymmetry equations and the $11d$ equations of motion now relate $d\varphi$ and $d\psi$ to the flux pieces in \eqref{G4decomp}. As a first trial one may schematically write:
\begin{equation}
\tau_0 = \tau_0^{\rm geom} + c_0\,q,
\qquad
\tau_1 = \tau_1^{\rm geom} + c_1\,v,
\qquad
\tau_3 = \tau_3^{\rm geom} + c_3\,*^{-1}\Xi^{(4)}_{27},
\label{taufluxschem}
\end{equation}
with numerical coefficients $c_i$ fixed by the precise BPS/equations-of-motion system. In an actual M-theory compactification these coefficients depend on the precise supersymmetry equations, warp factors, external spacetime ansatz, and possibly source terms. A pure internal $G_4$ contains no $\mathbf{14}$ irrep, so there is no direct, independent $G_4$ source for $\tau_2$; however, once the geometry is itself twisted, the back-reacted $d\psi$ can still have a $\mathbf{14}$ part and hence a non-zero $\tau_2$.
Therefore, in the present fibred background, a better alternative to \eqref{taufluxschem} is:
\begin{equation}\label{g2key}
\tau_i = \tau_i^{\rm geom}(f,M,P) + \tau_i^{\rm flux}(q,v,\Xi^{(4)}_{27}),
\qquad i=0,1,2,3,
\end{equation}
with the important caveat that $\tau_2^{\rm flux}$ is indirect rather than a standalone $G_4$ irrep source. (In other words, the flux contributions are determined by the detailed eleven-dimensional supersymmetry conditions and generally depend on warp factors and the external spacetime ansatz.) The equation \eqref{g2key} is the key equation for us: it explains precisely how the torsion is generated from both the fibration and the background fluxes. In fact for the local limit:
\begin{equation}
K3 \ \leadsto\ T^2_f \rtimes \mathbb P^1,
\qquad
\mathbb P^1 \ \leadsto\ \frac{S^1_a}{\mathbb Z_2} \rtimes S^1_b ,
\end{equation}
which should be viewed as an adiabatic/fiberwise limit rather than an exact global equality, we can use \eqref{g2key} to discuss how the subsequent non-K\"ahlerity arises. 

Reducing along the orbifold circle $S^1_a/\mathbb Z_2$ gives the Ho\v rava--Witten description, hence a heterotic $E_8\times E_8$ theory. Compactifying further on the remaining six-manifold:
\begin{equation}\label{manifold1}
X_6^{(\mathrm{het})} \sim T^2_f \rtimes S^1_b \rtimes \Sigma_3,
\end{equation}
with gauge bundle/background data $V_L\oplus V_R$ on the two walls, the unbroken four-dimensional gauge group is:
\begin{equation}\label{commutant}
\mathcal G^{(4d)}_{\mathrm{het}}
=
\mathrm{Com}_{E_{8,L}}\!\big(\mathrm{Hol}(V_L)\big)
\times
\mathrm{Com}_{E_{8,R}}\!\big(\mathrm{Hol}(V_R)\big),
\end{equation}
where
$\mathrm{Com}_{E_8}(H)$
means the commutant of the subgroup $H\subset E_8$ inside $E_8$.
Physically, this says the following. On the two Ho\v{r}ava--Witten walls one starts with
$E_{8,L}\times E_{8,R}$.
If one turns on gauge bundles $V_L$ and $V_R$, with structure groups:
\begin{equation}
\mathrm{Hol}(V_L)\subset E_{8,L},
\qquad
\mathrm{Hol}(V_R)\subset E_{8,R},
\end{equation}
then the surviving four-dimensional gauge symmetry is the subgroup of $E_{8,L}\times E_{8,R}$ that commutes with those structure groups as expressed in \eqref{commutant}.

Reducing instead along the ordinary circle $S^1_b$ gives type IIA on the interval $S^1_a/\mathbb Z_2$, i.e. the type $I'$ frame. In the perturbative type-$I'$ description, the gauge sector comes from the interval endpoints with O8-planes and D8-branes. For the standard symmetric distribution of eight D8-branes on each endpoint, the perturbative endpoint gauge algebra is:
\begin{equation}
\mathfrak g_{\mathrm{pert}} = \mathfrak{so}(16)_L \oplus \mathfrak{so}(16)_R ,
\label{pertSO16}
\end{equation}
which is not exactly the expected gauge group on the heterotic side. The reason comes from the fact that these two theories are U-dual to each other (as we shall discuss soon). This means 
at strong coupling each endpoint can enhance to $E_8$, and therefore the full non-perturbative identification is:
\begin{equation}
SO(16)_L \times SO(16)_R
\quad \rightsquigarrow \quad
E_{8,L}\times E_{8,R} ,
\label{SO16toE8}
\end{equation}
which is how the gauge group matching could be 
attained\footnote{The above equation is standard for the strong-coupling enhancement of the type-$I'$ endpoints, but in a fluxed compactification with additional fibrations and Wilson lines the unbroken four-dimensional gauge group can be further reduced. In other words, at the level of the local endpoint algebra, $SO(16)$ enhances non-perturbatively to $E_8$,
after which one must still quotient by the effects of fluxes, Wilson lines, and bundle holonomy.}. 
A similar story has recently been developed in sections 4.2 and 4.3 of \cite{transient} (see figures 7 and 8 therein).
Upon further compactification on the six-manifold, locally represented by:
\begin{equation}\label{manifold2}
X_6^{(\mathrm{IIA})} \sim T^2_f \rtimes \frac{S^1_a}{\mathbb Z_2} \rtimes \Sigma_3,
\end{equation}
with fluxes and Wilson lines, the unbroken $4d$ gauge group is the commutant of the brane/bundle data inside \eqref{SO16toE8}. Thus, whenever the heterotic frame is at an unbroken $E_8\times E_8$ point, the type-$I'$ frame reproduces the same algebra only after the standard endpoint enhancement.

However two questions still remain: (1) How is supersymmetry restored on the two non-K\"ahler six manifolds \eqref{manifold1} and \eqref{manifold2}? (2) Are the two four-dimensional theories strong/weak dual? Let us start by answering the issue of supersymmetry. Here the $SU(3)$ torsion classes are best suited, so we will directly use the $G_2 \to SU(3)$ structure decomposition. (See {\bf Table \ref{g2su3maap}}.)

Let the seven-dimensional $G_2$-structure manifold be locally written as a circle or interval fibration over a six-manifold,
$M_7 \longrightarrow X_6$,
with $X_6$ equal to either \eqref{manifold1} or \eqref{manifold2}.
If $v$ denotes the one-form along the M-theory circle/interval direction, then the $G_2$ structure $(\varphi,\psi)$ may be decomposed in terms of an $SU(3)$ structure $(J,\Omega)$ on $X_6$ as:
\begin{equation}
\varphi = J \wedge v + {\rm Re}\,\Omega,
\qquad
\psi = \frac{1}{2}J\wedge J - {\rm Im}\,\Omega \wedge v .
\label{G2SU3decomp}
\end{equation}
The $G_2$ torsion classes are defined by \eqref{torsiondecomp2}.
Upon reduction to six dimensions, these induce the standard $SU(3)$ torsion classes $W_1,\ldots,W_5$ through:
\begin{equation}
dJ
=
-\frac32\,{\rm Im}(W_1 \overline{\Omega})
+W_4\wedge J
+W_3 \nonumber
\label{dJtorsion1}
\end{equation}
\begin{equation}
d\Omega
=
W_1\,J\wedge J
+
W_2\wedge J
+
\overline{W}_5\wedge \Omega .
\label{dJtorsion}
\end{equation}
Thus the six-dimensional torsion is inherited from two sources:
(i) non-trivial fibration data of $T^2_f,\ S^1_b,\ S^1_a/\mathbb Z_2,\ \Sigma_3$,
and (ii) the internal M-theory flux $G_4$,
both of which make the $G_2$ classes $\tau_i$ non-zero and hence induce non-zero $W_i$. Schematically, from \eqref{G2SU3decomp} and \eqref{torsiondecomp2}, one finds:
\begin{equation}
W_i \sim \widetilde{\cal F}_i(\tau_0,\tau_1,\tau_2,\tau_3; dv,\iota_v G_4,\ldots),
\end{equation}
where $(\iota_v G_4)_{PQR} = v^M(G_4)_{MPQR}$; 
so that whenever the circle/interval is non-trivially fibered, $dv\neq 0$, or the flux is non-zero, the six-manifold is generically torsional.
The non-K\"ahlerity is then immediate. A K\"ahler manifold requires
$dJ=0$ which equivalently translates into 
$W_1=W_2 = W_3=W_4=0$.
But here the fibration twists and fluxes generically induce
$W_3 \neq 0,
W_4 \neq 0$
and often also
$W_1 \neq 0$ or $W_2 \neq 0$,
so neither \eqref{manifold1} nor \eqref{manifold2} is K\"ahler in general.

For the heterotic compactification on \eqref{manifold1}
supersymmetry does not require K\"ahlerity. Instead one needs a complex conformally balanced geometry, which in torsion-class language is:
\begin{equation}
W_1=W_2=0,
\qquad
2W_4 + W_5 = 0 .
\label{hetSUSYcond}
\end{equation}
The first condition says that the manifold is complex, while the second is the standard Strominger-system relation. Therefore
heterotic supersymmetry survives only if the torsion induced from the $G_2$ data arranges itself so that \eqref{hetSUSYcond} is satisfied.
In the present backgrounds the fibration and fluxes generically generate
\begin{equation}
W_3\neq 0,
\qquad
W_4\neq 0,
\qquad
W_5\neq 0 ,
\end{equation}
which is compatible with heterotic supersymmetry only when \eqref{hetSUSYcond} holds. This is actually not hard to construct as there have been numerous examples of heterotic compactifications on non-K\"ahler six-manifolds with non-zero supersymmetries \cite{DRS}.

For the type IIA compactification on \eqref{manifold2}
the same inherited torsion data imply that the manifold is again generically non-K\"ahler. In an $SU(3)$-structure language, four-dimensional supersymmetry requires additional algebraic relations among the $W_i$ and the RR/NS fluxes; in particular one again needs a very special alignment of the Lee-form data. Thus, if the reduction of the M-theory $G_2$ classes gives a generic non-zero combination of:
\begin{equation}
W_1,\ W_2,\ W_3,\ W_4,\ W_5,
\end{equation}
but arranged such that \eqref{hetSUSYcond} is being satisfied, then the IIA vacuum is supersymmetric. This is again not too hard to construct as can be inferred from \cite{luest, DRS, DEM}.



Let us now answer the other question, namely: Are the two four-dimensional theories dual in the strong/weak sense?
In the ordinary supersymmetric adiabatic limit, the answer is: \emph{yes, in the standard heterotic/type-$I'$ sense, fiberwise and using the usual T--S--T chain}. The logic is the familiar chain:
\begin{equation}
E_8\times E_8\ \text{heterotic on }S^1
\xrightarrow{T}
SO(32)\ \text{heterotic on }S^1
\xrightarrow{S}
\text{type I on }S^1
\xrightarrow{T}
\text{type }I'.
\label{TSTchain}
\end{equation}
At the level of the $11d$ radii $R_a$ and $R_b$, which correspond to the two cycles $S^1_a$ and $S^1_b$ appearing in \eqref{manifold2} and \eqref{manifold1} respectively, we have: 
\begin{equation}
g_{\mathrm{IIA}} \sim \left(\frac{R_b}{\ell_{11}}\right)^{3/2},
\qquad
g_{\mathrm{HW/het}}^2 \sim \left(\frac{R_a}{\ell_{11}}\right)^3,
\label{couplings11d}
\end{equation}
so the weakly coupled regime of one frame corresponds to the opposite regime of the dual frame after the T--S--T reparametrization. 
Interestingly, even in the presence of generic fluxes and genuinely non-K\"ahler, non-integrable fibrations, the exact four-dimensional duality may continue to hold, provided the background preserves an underlying supersymmetry and the fluxes obey the necessary restrictive consistency conditions \cite{DRS}.
The duality is then automatic. Once supersymmerty is broken the story gets more complicated and the duality 
is no longer automatic. The most conservative statement then is
that the $S_a$ and $S_b$ reductions are expected to remain dual only in a controlled adiabatic/BPS regime.
Outside that regime one must check the full flux quantization, Bianchi identities, tadpoles, and St\"uckelberg masses explicitly. We will come back to this issue soon.




\subsection{Branch II: the wedge branch \label{wedgebranch}}

The second deformation corresponds to the non-supersymmetric wedge branch, which is particularly interesting in view of the discussion above. Locally we replace the base of the K3 fiber to a more singular manifold via the following series of transformations:
\begin{equation}\label{basegeom}
 \mathbb{P}^1 \qquad\longrightarrow\qquad  T^2/\mathbb{Z}_2
  \qquad\longrightarrow\qquad
  S_a^1/\mathbb{Z}_2\rtimes(S^1\vee S^1)_b.
\end{equation}
Here $(S^1\vee S^1)_b$ denotes the wedge of two circles of radii
$R_\pm$ meeting at a junction point $v$, and we will call it a {\it pinch} singularity.  Since
$S_a^1/\mathbb{Z}_2\cong I$, the local model becomes
$I\times(S^1\vee S^1)_b$.
The tachyon modulus and mean radius are\footnote{One needs to be careful here. The quantity $T$ is tachyonic only if there exist an effective potential $V$ such that 
$m^2_T = 
\frac{\partial^2 V}{\partial T^2}
\big|_{T=0}
<0$. In the wedge model, the branch exchange symmetry only suggests that the effective potential near the symmetric point has an even expansion, {\it i.e.} odd powers of $T$ are absent, but the symmetry alone does not fix the sign of $m^2_T$: that sign is dynamical.
However, the usual type-$0$ interpretation supplies the dynamical expectation.  In
ordinary type-$0$ string theory there is a closed-string tachyon in the
NS-NS sector.  In the wedge construction, the branch-odd radius modulus is
identified with the geometric representative of this type-$0$ tachyonic
direction.  Therefore, if the wedge construction is to reproduce the type-$0$
tachyon, the effective potential must have $m^2_T < 0$. This is the only reason why $T$ is regarded as a tachyon here.\label{tachiooon}}:
\begin{equation}\label{eq:TRB}
  T = {R_+-R_-\over 2},
  \qquad
  R_B = {R_++R_-\over 2}.
\end{equation}
This is no longer a smooth complex curve but a stratified real
space with singular locus $I\times\{v\}$.  This branch should be
viewed as an intrinsically non-geometric or quantum deformation of
the fiberwise duality frame.
Motivated by the wedge compactification programme and its
Ho\v rava--Witten/Type-0 interpretation~\cite{BDV,Altavista},
we are led not to an ordinary supersymmetric heterotic
compactification but to a non-supersymmetric \emph{heterotic-like}
background of the form
\begin{equation}\label{hetbg}
  \text{Type 0 Heterotic/HW on }
  T^2_f\times(S^1\vee S^1)_b\quad\text{fibered over }\Sigma_3.
\end{equation}
On the other hand, compactifying along the pinched circle $(S^1\vee S^1)_b$ leads to a non-supersymmetric Type IIA theory on a background of the form
\begin{equation}\label{0abg}
  \text{Type 0A on }
  T^2_f\times S^1/Z_2\quad\text{fibered over }\Sigma_3.
\end{equation}
which is defined at the orientifold points. While these two theories, at the supersymmmetric Type IIA end points, could be locally U-dual to each other under specific conditions by using fiberwise duality arguments, it is not clear at this stage whether such duality arguments could be extended once we are in the symmetric point with non-trivial tachyon fluctuation. We will come back to this discussion soon.

Moreover the wedge branch should \emph{not} be identified with a standard
perturbative $E_8\times E_8$ heterotic compactification.  On the
seven-dimensional side the $\SO(16)^4$ structure is best understood
as a \emph{sectoral} Ho\v rava--Witten/Type-0 structure, not as a
single perturbative heterotic gauge algebra of rank~32.  Similarly in the Type 0A side the local orthogonal gauge symmetry would arise from the orientifold points that are being {\it doubled} by the pinched compactifying circle $(S^1 \vee S^1)_b$. In both cases however the six-dimensional compact manifolds are non-K\"ahler and could even by non-complex.

\subsubsection{Fluxes, fibration and wedge sourcing the $G_2$ torsion classes \label{g2pinch1}}

Consider M-theory on the seven-manifold
$M_7 = K3 \rtimes \Sigma_3$,
with a non-trivial fibration of the $K3$ fiber over the compact three-manifold $\Sigma_3$. We further take a local degeneration limit in 
which:
\begin{equation}\label{degen}
K3 \;\leadsto\; T^2_f \rtimes \frac{S^1_a}{\mathbb Z_2} \rtimes (S^1\vee S^1)_b ,
\end{equation}
where again we can assume that we are at the Morrison-Vafa \cite{MV} limit in which the torus $T^2_f$ is fibered over the base \eqref{basegeom} defined using a pinched circle $(S^1 \vee S^1)_b$ over an interval $I$ defined as $S_a^1/\mathbb{Z}_2\cong I$. We can also 
switch on internal M-theory flux:
\begin{equation}
G_4=dC_3 \in \Lambda^4(M_7).
\end{equation}
Because of the pinching of the $S^1_b$ circle, 
the $G_2$ torsion classes is now generated by {\it three} distinct effects: (i) pinch-induced internal $G_4$-flux,
(ii) non-trivial fibration over $\Sigma_3$, and 
(iii) pinching of $S_b^1 \to S_+^1 \vee S_-^1$, compared to what we had earlier.

The intrinsic torsion of the $G_2$ structure is defined in terms of the associative $3$-form $\varphi$ and coassociative $4$-form
$\psi \equiv *_7 \varphi$
through \eqref{torsiondecomp2} and \eqref{taui}. Let $e^1,e^2$ span the torus fiber $T^2_f$, let $e^3$ denote the local one-form on $S^1_a/\mathbb Z_2$, and let $e^4$ denote the one-form along the circle direction $S_b^1$ before pinching. Similarly, let
$e^5,e^6,e^7$
be a local orthonormal coframe on the base $\Sigma_3$. A convenient local $G_2$ structure is:
\begin{equation}
\varphi
=
e^{127}+e^{347}+e^{567}
+e^{135}-e^{146}-e^{236}-e^{245},
\label{phi_standard}
\end{equation}
with $\psi = *_7\varphi$ and $e^{abc} \equiv e^a \wedge e^b \wedge e^c$. Equivalently, one may view $M_7$ locally as a six-manifold times one direction and write $\varphi$ and $\psi$ in the following way:
\begin{equation}
\varphi = J\wedge e^4 + {\rm Re}\,\Omega,
\qquad
\psi = \frac12 J\wedge J - {\rm Im}\,\Omega\wedge e^4,
\end{equation}
but for present purposes the $G_2$ language in \eqref{phi_standard} is sufficient. Because the fiber is non-trivially fibered over $\Sigma_3$, the one-forms $e^A$ are not closed. Thus for 
$I=1,2,3,4$,
and $m=5,6,7$,
we can express the exterior derivative of $e^I$ as:
\begin{equation}
de^I
=
\Theta^I{}_{Jm}\,e^m\wedge e^J
+
\frac12\,\Xi^I{}_{mn}\,e^m\wedge e^n
+
\frac12\,\Pi^I{}_{JK}\,e^J\wedge e^K ,
\label{deIclean}
\end{equation}
where
$\Theta^I{}_{Jm}$
measures how the fiber directions twist as one moves along the base,
$\Xi^I{}_{mn}$ would describe a purely base-induced non-integrability contribution, and
$\Pi^I{}_{JK}$
would describe intrinsic anholonomy inside the fiber itself\footnote{  The first term is the mixed fibration term and measures how the fibre
directions rotate as one moves along the base. The second term is the
vertical component of the curvature of the horizontal distribution and it
appears when the lift of two base directions fails to close horizontally.
The third term describes twisting internal to the fibre. It is present when
the fibre one-forms \(e^I\) are not closed along the fibre directions
themselves. In the simplified discussion below we isolate only the mixed fibration
effect, namely the variation of the fibre frame along the base. 
and suppress the purely horizontal curvature term  and the
intrinsic fibre anholonomy term. These additional terms can
be restored in a more complete treatment, but they are not needed for the
basic point that a non-trivial fibration already makes the adapted coframe
non-closed and hence sources intrinsic torsion.}. However since we are only looking at the fibration piece and ignoring other nuances, we expect:
\begin{equation}\label{deidei}
de^I \Big|_{\rm fib}
=
\Theta^I{}_{Jm}\,e^m\wedge e^J.
\end{equation}
Therefore, even before introducing fluxes or pinching, one already has
$d\varphi \neq 0$ and 
$d\psi \neq 0$
and hence non-vanishing $G_2$ torsion. To introduce fibration, fluxes and pinching all together, 
it is useful to decompose the torsion classes into three contributions:
\begin{equation}
\tau_i
=
\tau_i^{\rm fib}
+
\tau_i^{\rm flux}
+
\tau_i^{\rm pinch},
\qquad i=0,1,2,3.
\label{tausplit}
\end{equation}
Here
$\tau_i^{\rm fib}$ comes from the non-trivial fibration data \eqref{deIclean},
$\tau_i^{\rm flux}$ comes from the internal $G_4$ flux, and
$\tau_i^{\rm pinch}$ comes from the singular degeneration
$S_b^1 \to S_+^1 \vee S_-^1$.

To study the flux contributions we can decompose the internal four-form fluxes under $G_2$ as
$\Lambda^4
\cong
\mathbf{1}\oplus\mathbf{7}\oplus\mathbf{27}$. However the localized DOFs at the pinch should also influence the form of the fluxes.
Hence one may write \cite{CS, behrndt}:
\begin{equation}
G^{\rm br}_4
=
f\,\psi
+
X\wedge \varphi
+
G_4^{(27)},
\label{G4decomp_full}
\end{equation}
where
$f\in \Lambda^0$, 
$X\in \Lambda^1_7$, and 
$G_4^{(27)}\in \Lambda^4_{27}$. (Note that this is the same $G_4^{\rm br}$ from \eqref{g4c3h3} as we are counting the {\it total} four-form fluxes in the background.)
Comparing the torsional equations and the $G_2$ decomposition \eqref{torsiondecomp2} and \eqref{taui}, one obtains schematically:
\begin{equation}
\tau_0^{\rm flux} \propto f,
\qquad
\tau_1^{\rm flux} \propto X,
\qquad
\tau_3^{\rm flux} \propto *_7 G_4^{(27)},
\label{flux_to_tau}
\end{equation}
while
$\tau_2^{\rm flux}$
is not directly sourced by a generic internal $G^{\rm br}_4$ irrep in the same way. Thus the generic effect of $G_4$ is simply one-on-one mapping of the form:
\begin{equation}
G_4^{(\mathbf 1)} \longleftrightarrow \tau_0,
\qquad
G_4^{(\mathbf 7)} \longleftrightarrow \tau_1,
\qquad
G_4^{(\mathbf{27})} \longleftrightarrow \tau_3.
\end{equation}
Now consider the effect of the non-trivial fibration
$K3 \rtimes \Sigma_3$.
Because the $K3$ fiber varies over $\Sigma_3$, its metric moduli, hyperk\"ahler two-forms, and local coframe all acquire dependence on the base coordinates. Inserting \eqref{deidei} into $d\varphi$ and $d\psi$ generates the fibration-induced torsion classes. At the representation-theoretic level one finds:

\begin{equation}
\text{trace part of the twisting}
\;\longrightarrow\;
\tau_1^{\rm fib},\nonumber
\end{equation}
\begin{equation}
\text{singlet scalar contraction of the twisting}
\;\longrightarrow\;
\tau_0^{\rm fib},\nonumber
\end{equation}
\begin{equation}
\text{traceless symmetric / primitive part of the twisting}
\;\longrightarrow\;
\tau_3^{\rm fib},\nonumber
\end{equation}
or alternatively, the singlet projection of the fibration contributes to $\tau_0$, the vector/Lee-form projection contributes to $\tau_1$, and the primitive traceless projection contributes to $\tau_3$.
If the induced $G_2$ connection fails to lie in $G_2$ even after removing the $\mathbf{1}$, $\mathbf{7}$ and $\mathbf{27}$ pieces, one may also generate
$\tau_2^{\rm fib}\neq 0$.
Thus the non-trivial fibration can in principle source all four torsion classes:
\begin{equation}
\tau_i^{\rm fib}\neq 0,
\qquad i=0,1,2,3 ,
\end{equation}
which we can alternatively represent as the following: The scalar or isotropic part of the fibration twisting contributes to the scalar torsion class $\tau_0$.
The vector-like part of the twisting contributes to $\tau_1$.
The adjoint ${\bf 14}$-dimensional part that appears in $d\psi$ contributes to $\tau_2$.
The primitive or symmetric-traceless part of the twisting contributes to $\tau_3$. In particular, even when the flux vanishes, the twisting of the $K3$ fiber over $\Sigma_3$ already makes the $G_2$ structure generically non-torsion-free.

What about the pinch contribution that converts $S^1_b$ to a pinched circle? This is much more non-trivial so we will spend some time clarifying the extra effect of pinching the circle $S_b^1$. Before pinching, let
$e^4 = R_B\, d\theta_b$
be the one-form along the smooth circle. After pinching,
$S_b^1 \;\longrightarrow\; S_+^1 \vee S_-^1$,
the one-form becomes piecewise defined\footnote{A clean way to justify the ansatz \eqref{e4wedge} is to first note that on an ordinary smooth circle $S_b^1$ with angular
coordinate $\theta_b$, the natural one-form along the circle is simply
$e^4 = R_B\,d\theta_b$,
with 
$\theta_b\sim \theta_b+2\pi$,
because locally every one-dimensional Riemannian manifold is described by a line
element of the form
$ds_b^2 = (e^4)^2 = R_B^2\, d\theta_b^2$.
So $e^4$ is just the orthonormal coframe for the one-dimensional metric on the
circle. After pinching, the topology changes from a single smooth circle to a wedge of two
circles,
$S_b^1 \;\longrightarrow\; S_+^1 \vee S_-^1$,
which means that away from the common junction the geometry is still locally just a
disjoint union of two one-dimensional smooth branches. Hence on each branch
separately there is still a natural local angular coordinate and a natural local
orthonormal one-form that may be succinctly expressed as:
\begin{equation}
e^4_{(+)} = R_+\,d\theta_+,
\qquad
e^4_{(-)} = R_-\,d\theta_-.\nonumber
\label{branchwise_e4}
\end{equation}
Since the singularity is concentrated only at the junction, there is no reason for
the one-form to become more complicated \emph{away from} that point. Therefore the
most economical global way to write the branchwise form is precisely \eqref{e4wedge}, namely as the piecewise union of the two smooth branchwise coframes. This justifies our choice of \eqref{e4wedge}. Note however that \eqref{e4wedge} is not meant to be the most general
one-form compatible with the singular wedge geometry. Rather, it is the
\emph{minimal local coframe ansatz} that captures the branchwise radius data of
the pinched circle while keeping the singularity localized at the junction. In
particular, it is the direct branchwise continuation of the smooth-circle form
$e^4 = R_B\,d\theta_b$
to the case where the single circle degenerates into two branches.
The pinch does not force $e^4$ itself to acquire a more complicated bulk form,
because the nontriviality of the pinch is encoded not in the local expression on
each branch, but in the \emph{failure of smooth gluing at the junction}. In other
words, the complication is distributional and localized, rather than spread over
each branch. This is exactly why one expects
$de^4_{\rm wedge}$
to contain singular support at the pinching point, even though $e^4_{\rm wedge}$
looks simple on each branch separately. Indeed, differentiating \eqref{e4wedge} schematically gives:
\begin{equation}
de^4_{\rm wedge}
=
dR_+\wedge d\theta_+\,\mathbf{1}_{S_+^1}
+
dR_-\wedge d\theta_-\,\mathbf{1}_{S_-^1}
+
R_+\,d\theta_+\wedge d\mathbf{1}_{S_+^1}
+
R_-\,d\theta_-\wedge d\mathbf{1}_{S_-^1} ,\nonumber
\label{de4_distributional}
\end{equation}
where the first two terms are ordinary smooth branchwise contributions; and the last two
terms are distributional and are supported at the junction because
$d\mathbf{1}_{S_\pm^1}$
is concentrated where the branch starts or ends. Thus the pinch data are already
encoded in \eqref{e4wedge}: they appear through the singular derivatives of
the characteristic functions, not through a more elaborate smooth expression for
$e^4$ on each branch. Equivalently, one may say that \eqref{e4wedge} is the correct leading
ansatz under the following assumptions:
(1) each branch away from the junction is locally an ordinary smooth circle; (2) the only new data introduced by the pinch are the two branch radii
$R_\pm$ and the singular gluing at the common point; (3)
all additional complications of the pinched geometry are localized at the
junction and therefore enter through distributional terms in derivatives, torsion,
or curvature, rather than through a more complicated branchwise coframe itself.
This is also why it is natural to define $R_B$ and $T$
so that the wedge one-form may be rewritten as \eqref{e4def}
showing explicitly that the smooth-circle mode $R_B$ is supplemented by the
branch-odd pinching modulus $T$. The pinch therefore does make the geometry more
complicated, but it does so by splitting the single circle data into branch-even
and branch-odd pieces and by introducing singular gluing at the junction, rather
than by forcing a more complicated local one-form on each branch.}:
\begin{equation}
e^4_{\rm wedge}
=
R_+\,d\theta_+\,\mathbf{1}_{S_+^1}
+
R_-\,d\theta_-\,\mathbf{1}_{S_-^1},
\label{e4wedge}
\end{equation}
where the characteristic functions $\mathbf{1}_{S_\pm^1}$ localize the two branches. This is easy to argue from looking at the local topology of the pinched circle. We can now introduce:
\begin{equation}
R_B=\frac{R_++R_-}{2},
\qquad
T=\frac{R_+-R_-}{2} ,
\label{RBandTagain}
\end{equation}
where $R_B$ measures the symmetric breathing mode of the two circles, whereas $T$ measures their antisymmetric imbalance. The wedge-circle proposal \cite{BDV} identifies precisely this antisymmetric deformation with the closed-string tachyon modulus of the type-$0$ theory. This can be stated more physically as follows. If
$R_+=R_-$,
then
$T=0$,
so the two branches are symmetric. Turning on
$T\neq 0$
moves the system away from that symmetric point by making one branch larger than the other. In the proposed dictionary \cite{BDV}, this is the geometric avatar of exciting the type-$0$ tachyon. Moreover, tachyon condensation is argued to correspond to shrinking one of the two circles, thereby flowing to the supersymmetric type-IIA limit. With this in mind, we can rewrite \eqref{e4wedge} as:
\begin{equation}\label{e4def}
e^4_{\rm wedge}
=
R_B\Big(d\theta_+\,\mathbf{1}_{S_+^1}+d\theta_-\,\mathbf{1}_{S_-^1}\Big)
+
T\Big(d\theta_+\,\mathbf{1}_{S_+^1}-d\theta_-\,\mathbf{1}_{S_-^1}\Big).
\end{equation}
The second term is odd under exchange of the two branches and is controlled by the tachyon modulus $T$, thus perfectly consistent with the fact that in the wedge-circle/type-$0$ dictionary \cite{BDV}, this antisymmetric mode is identified with the effective tachyon modulus. Now:
\begin{equation}\label{bndsign}
d\mathbf{1}_{S_+^1}=+\delta_v,
\qquad
d\mathbf{1}_{S_-^1}=-\delta_v,
\end{equation}
where $\delta_v$ is the current supported at the junction $v$. The minus sign comes from the \emph{orientation convention} for the two branches at the junction. The point is that $\mathbf{1}_{S_+^1}$ and $\mathbf{1}_{S_-^1}$ are not ordinary smooth functions, but characteristic functions of the two branches of the pinched circle
$S^1_+ \vee S^1_- $.
Their exterior derivatives are distributions supported at the common junction point $v$. The sign is fixed by how the oriented boundary of each branch is defined\footnote{A simple way to see this is to think of each branch locally as an interval meeting at $v$. Let $x$ be a local coordinate transverse to the junction, with
$x>0$ on $S^1_+$,
and $x<0$ on $S^1_-$. 
Let
$\mathbf{1}_{S_+^1} = \mathbb{H}(x)$,
and 
$\mathbf{1}_{S_-^1} = \mathbb{H}(-x)$,
where $\mathbb{H}$ is the Heaviside step function. Then:
\begin{equation}
d\mathbf{1}_{S_+^1}
=
dH(x)
=
\delta(x)\,dx
\equiv
+\delta_v ,~~~~~~~~
d\mathbf{1}_{S_-^1}
=
dH(-x) =H'(-x)\,d(-x)=
-\delta(x)\,dx
\equiv
-\delta_v. \nonumber
\end{equation}
Equivalently, in terms of oriented boundaries,
$\partial S^1_+ = +\,v$
and 
$\partial S^1_- = -\,v$,
so the distributional derivative of the characteristic function of an oriented branch is the delta-current Poincar\'e dual to its boundary, {\it i.e.}
$d\mathbf{1}_{S^1_\pm} = \delta_{\partial S^1_\pm}$.
This justifies \eqref{bndsign}.}. Taking the exterior derivative of \eqref{e4def}, and plugging \eqref{bndsign} in \eqref{e4def}, one finds:
\begin{equation}
de^4_{\rm wedge}
=
T\,\delta_v\wedge \eta_1
+
(\text{regular terms}),
\label{de4pinch}
\end{equation}
where \(\eta_1\) denotes the limiting one-form tangent to the wedge-circle
direction at the junction.  Since the wedge is singular, \(\eta_1\) should not
be interpreted as a globally smooth one-form on \(S^1_+\vee S^1_-\).  Rather,
it is a distributional shorthand for the common branch-direction obtained from
the limits of \(d\theta_+\) and \(d\theta_-\) at the node.  Thus
\(\delta_v\wedge\eta_1\) is a junction-supported two-current measuring the
failure of the branchwise one-form \(e^4_{\rm wedge}\) to extend smoothly
through the node. We have also absorbed any $dT$ dependent factors in the definition of the regular terms. Thus the pinching contributes a distributional piece localized at the junction.
Substituting \eqref{de4pinch} into \eqref{phi_standard}, every term in $\varphi$ containing $e^4$ produces a localized contribution to $d\varphi$. Therefore:
\begin{equation}
d\varphi\big|_{\rm pinch}
=
T\,\delta_v\wedge \rho_3^{\rm loc} ,
\label{dphi_pinch}
\end{equation}
for some smooth local $3$-form $\rho_3^{\rm loc}$ built out of the remaining local coframe. Equivalently, on the full seven-manifold this current is supported on the six-dimensional locus:
\begin{equation}
\Sigma_6^{\rm jct}
=
T^2_f \rtimes \frac{S_a^1}{\mathbb Z_2} \rtimes \Sigma_3 ,
\end{equation}
at the node $v$ precisely because of the fact that the $G_2$ structure manifold is at least locally of the form $\Sigma^{\rm jct}_6  \rtimes (S^1\vee S^1)_b$ away from the degeneration points. Hence:
\begin{equation}\label{pinchN}
d\varphi\big|_{\rm pinch}
=
T\,\delta_{\Sigma_6^{\rm jct}}\,n\wedge \rho_3 ,
\end{equation}
where 
$n \propto \eta_1$ denotes the local unit normal one-form to the six-dimensional junction locus \(\Sigma^{\rm jct}_6\subset M_7\), so that
$\delta_{\Sigma^{\rm jct}_6}\,n$
is the Poincar\'e-dual current localizing the pinch contribution on the junction. Here \(\rho_3\) is the smooth three-form tangent to \(\Sigma_6^{\rm jct}\) obtained by projecting the local singular form \(\rho_3^{\rm loc}\) to the tangent bundle of the junction locus\footnote{There is a simple way to see how $\rho_s^{\rm loc}$ converts to $\rho_3$. The key point is that \(e^4\) is a one-form on the full seven-manifold
$M_7 \simeq \Sigma_6 \rtimes (S^1_+\vee S^1_-)$,
not merely on the singular one-dimensional wedge fiber by itself. Therefore its exterior derivative
$de^4$
must be understood as the exterior derivative on the full seven-dimensional space. In particular, \(de^4\) is a two-form on \(M_7\), and hence its two legs need not both lie in the fiber sector.
So the right statement is not that \(\mathbf 1_{S^1_\pm}\) themselves are one-forms on the base, but rather that their variation must be understood as taking place over the full seven-dimensional geometry. This means that the characteristic functions \(\mathbf 1_{S^1_\pm}\) cannot be treated as depending only on an isolated one-dimensional fiber coordinate. Instead, they must be regarded as distributions on the total space whose support is localized at the degeneration locus and whose exterior derivatives naturally produce forms with one singular leg in the wedge/fiber sector and one leg that is meaningful relative to the embedding into the full seven-manifold.
A clean way to say this is the following.
Starting from \eqref{e4def}
it is easy to argue that
$de^4_{\rm wedge}\in \Lambda^2(T^*M_7)$.
Now \(\Lambda^2(T^*M_7)\) decomposes schematically as:
\begin{equation}
\Lambda^2(T^*M_7)
=
\Lambda^2(T^*\Sigma_6)
\oplus
\Big(T^*\Sigma_6\wedge T^*(S^1\vee S^1)\Big)
\oplus
\Lambda^2\big(T^*(S^1\vee S^1)\big) ,\nonumber
\label{lambda2split}
\end{equation}
where the last term is too small in degree or singular to carry the whole pinch information by itself, because the fiber is only one-dimensional away from the node and singular at the node. Therefore the singular contribution to \(de^4\) is naturally expected to lie in the mixed sector
$T^*\Sigma_6\wedge T^*(S^1\vee S^1)$.
So the pinch contribution should indeed have one leg along the wedge/fiber direction and one leg along the six-dimensional side.
The characteristic functions \(\mathbf 1_{S^1_\pm}\) distinguish the two branches of the degenerate fiber. But in the full geometry, the location of the pinch is not just an abstract point of an isolated fiber: it sits over the six-dimensional junction locus
$\Sigma_6^{\rm jct}\subset M_7$.
Hence the distributions \(\mathbf 1_{S^1_\pm}\) should really be regarded as distributions on the total space, adapted to the fibration structure, rather than as functions only of a single fiber coordinate.
Their exterior derivatives therefore produce currents supported at the degeneration locus in the total space, {\it i.e.}
$d\mathbf 1_{S^1_\pm}
\sim
\delta_{\Sigma_6^{\rm jct}}\; n_\pm$,
where \(n_\pm\) denotes the appropriate singular one-form direction associated with the pinched branch. (Previously we took the simpler case where $n_+ \sim n_- = n$.) In this form, the support of the singularity is already spread over the six-dimensional junction locus, rather than being a point in an isolated one-dimensional fiber.
Differentiating \eqref{e4def} gives schematically \eqref{deeff}, from where it is easy to argue that 
the singular pinch terms in \(de^4\) are naturally of the form:
\begin{equation}
de^4_{\rm wedge}\big|_{\rm pinch}
\in
T^*\Sigma_6\wedge T^*(S^1\vee S^1) \sim
T\,\delta_{\Sigma_6^{\rm jct}}\, \hat{n}\wedge \eta_1,\nonumber
\label{de4_mixed_statement}
\end{equation}
thus justifying what we said above. (Plus it identifies $n$ in say \eqref{pinchN} to be proportional to $\eta_1$.) Therefore the singular term should be interpreted schematically not as
$\delta_v\wedge \eta_1$
with both factors living purely in the one-dimensional fiber, but rather as an element of the mixed sector with 
$\hat{n}\in T^*\Sigma_6$ and 
$\eta_1\in T^*(S^1\vee S^1)$
schematically denote the base-side and fiber-side legs of the singular two-form. More invariantly, one should simply view the singular term as a mixed two-form on the total space, supported at the degeneration locus. From here we see that in \eqref{dphi_pinch} we simply swap the forms to convert $\rho_3^{\rm loc}$ to $\rho_3$ in \eqref{pinchN}.}. Now comes the crucial point: 
The full pinch contribution to $d\varphi$
contains two structurally distinct terms:
\begin{equation}
d\varphi\big|_{\rm pinch}
=
\underbrace{T\,\delta_v\wedge \rho_3^{\rm loc}}_{\text{singular, localized}}
+
\underbrace{dT\wedge \widetilde\rho_3^{\rm loc}}_{\text{smooth, branch-odd}}
+
(\text{regular terms}),
\label{dphi_twoterms}
\end{equation}
where
$\rho_3^{\rm loc}
=
\eta_1\wedge\Sigma_2$ and 
$\widetilde\rho_3^{\rm loc}
=
\alpha_1\wedge\Sigma_2$.
These two terms have different $G_2$ representation content and must be
treated separately. The singular localized term is the one whose representation content is
controlled by the primitivity argument. Lifting the node $v$ to the
six-dimensional junction locus $\Sigma_6^{\rm jct}\subset M_7$, we 
write:
\begin{equation}
K_{\rm sing}
\equiv
d\varphi\big|_{\rm pinch,sing}
=
T\,\delta_{\Sigma_6^{\rm jct}}\,n\wedge\rho_3 ,
\end{equation}
where $n$ is the local normal one-form to the junction locus and $\rho_3$ is
the tangential three-form induced from $\rho_3^{\rm loc}$. 
Because this is a localized primitive deformation of the $G_2$ structure rather than a scalar or Lee-form type deformation, its natural home is the $\mathbf{27}$ piece. Looking at the definition of $d\varphi$ from \eqref{torsiondecomp2} we see that it is now proportional to $\ast_{{}_7}\tau_3$. Therefore the pinch primarily contributes to:
\begin{equation}
\tau_3^{\rm pinch}\neq 0,
\qquad
\tau_3^{\rm pinch}\propto T\,\delta_{\Sigma_6^{\rm jct}}\,\Xi^{(3)}_{27},
\label{tau3pinch}
\end{equation}
with $\Xi^{(3)}_{27} \in \Lambda^3_{27}$, 
while at leading order
$\tau_0^{\rm pinch}=
\tau_1^{\rm pinch}=
\tau_2^{\rm pinch}=0$
unless additional warping or singular backreaction converts the junction source into $\mathbf{1}$, $\mathbf{7}$, or $\mathbf{14}$ components as well. An alternative way to justify this is to take \eqref{dphi_pinch}, and show that $\pi_1(d\varphi|_{\rm pinch}) = \pi_7(d\varphi|_{\rm pinch}) = 0$ and therefore the pinch does not source $\tau_0$ and $\tau_7$ from \eqref{taui} to this order. This gives:
\begin{equation}\label{pinchu}
d\varphi|_{\rm pinch}
= *_{{}_7}\tau_3 = 
T\,\delta_{\Sigma_6^{\rm jct}}\, n\wedge \rho_3 ~~ \implies ~~
\tau_3^{\rm pinch}
=
T\,\delta_{\Sigma^{\rm jct}_6}\,*_7(n\wedge \rho_3)
=
T\,\delta_{\Sigma^{\rm jct}_6}\,*_6\rho_3 ,
\end{equation}
again modulo an overall sign depending on conventions; 
and we see that taking the Hodge dual immediately gives us \eqref{tau3pinch} because $\Xi^{(3)}_{27} \equiv *_6\rho_3 \in \Lambda^3_{27}$.
Thus the direct leading effect of the wedge asymmetry is that the asymmetric mode of the two circles $S^1_+$ and $S^1_-$, which is in fact related to the tachyon $T$, is directly given by the $\tau_3$ torsion class of the $G_2$ structure manifold:
\begin{equation}\label{takibeta}
T \longleftrightarrow \tau_3^{\rm pinch} ,
\end{equation}
implying a deeper connection to geometry.
Thus the singular localized pinch source contributes directly to  $\tau_3$ and not to $\tau_0$ or $\tau_1$ at leading localized order. It also gives no direct contribution to $\tau_2$, since $\tau_2$ appears in $d\psi$, not in $d\varphi$. On the other hand, the smooth branch-odd term:
\begin{equation}
K_{\rm smooth}
\equiv
dT\wedge\widetilde\rho_3^{\rm loc}
=
dT\wedge\alpha_1\wedge\Sigma_2
\end{equation}
is different. It is not localized at the junction, and therefore the
distributional $\mathbf{27}$ argument above does not apply to it. In the
wedge-adapted $G_2\to SU(3)$ splitting, the normal direction is the pinched
circle direction $\tilde{n}$ (similar to $n$ before), and from
$\varphi\big|_{e^4}
=
e^4\wedge\Sigma_2$
one locally identifies
$J=\Sigma_2$.
If $\alpha_1$ is identified with the normal one-form $\tilde{n}$ in this split, then
the smooth term takes the schematic form:
\begin{equation}
K_{\rm smooth}
\sim
dT\wedge \tilde{n}\wedge J .
\label{smoothpinch_form}
\end{equation}
This term has vanishing singlet projection, which may be easily verified by computing $\pi_{\bf 1}$, but it is not generically a pure
$\mathbf{7}$ form. Rather, after projecting into $G_2$ irreducible
representations, one has:
\begin{equation}
K_{\rm smooth}
=
\pi_{\mathbf 7}(K_{\rm smooth})
+
\pi_{\mathbf{27}}(K_{\rm smooth}),
\qquad
\pi_{\mathbf 1}(K_{\rm smooth})=0 .
\label{smoothpinch_reps}
\end{equation}
Therefore the smooth branch-odd pinch term can source both a Lee-form-type
$\tau_1$ correction and a smooth $\tau_3$ correction typically of the form
$\pi_{\mathbf 7}(K_{\rm smooth})
=
3\,\delta_{\rm smooth}\tau_1\wedge\varphi$, and 
$\pi_{\mathbf{27}}(K_{\rm smooth})
=
*_7\delta_{\rm smooth}\tau_3$.
There is no direct $\tau_2$ contribution from this term, because $\tau_2$ is
read from $d\psi$. Thus the precise statement is:
\begin{equation} 
\underbrace{T\,\delta_v\wedge\rho_3^{\rm loc}}_{\text{singular}}
\longrightarrow
\mathbf{27}
\longrightarrow
\tau_3^{\rm pinch,sing}\nonumber
\end{equation}
\begin{equation} \label{pinch_rep_summary}
\underbrace{dT\wedge\alpha_1\wedge\Sigma_2}_{\text{smooth}}
\longrightarrow
\mathbf{7}\oplus\mathbf{27}
\longrightarrow
\delta_{\rm smooth}\tau_1+\delta_{\rm smooth}\tau_3 .
\end{equation}
However if one wants the smooth term to source only $\tau_1$, this must be imposed as
an extra truncation condition, {\it i.e.}
$\pi_{\mathbf{27}}(K_{\rm smooth})=0$,
rather than inferred automatically from the geometry.
Thus the direct localized effect of the wedge asymmetry is that the
branch-odd mode of the two circles $S^1_+$ and $S^1_-$, which is related to
the tachyon $T$, appears as the coefficient of a localized $\mathbf{27}$
torsion source as in \eqref{takibeta}.
And on the other hand, the smooth $dT$-dependent part of the same branch-odd deformation is instead
a non-localized torsion correction with $\mathbf{7}\oplus\mathbf{27}$
content.

Two important remarks follow immediately. First, if
$R_+=R_-$, then $T=0$, and the antisymmetric localized source vanishes.
In that symmetric wedge limit the singular pinch-induced contribution to
the torsion disappears at leading order. Second, the pinching does not
replace the flux- or fibration-induced torsion; rather, it adds both a
singular localized piece and smooth branch-odd corrections. Thus the total
$\tau_3$ is more accurately written as
\begin{equation}
\tau_3
=
\tau_3^{\rm fib}
+
\tau_3^{\rm flux}
+
\tau_3^{\rm pinch,sing}
+
\delta_{\rm smooth}\tau_3,
~~~~ {\rm with}~~~~
\tau_3^{\rm pinch,sing}
\propto
T\,\delta_{\Sigma_6^{\rm jct}} .
\label{tauteen}
\end{equation}
Collecting the various contributions, the $G_2$ torsion classes on
$M_7=K3\rtimes\Sigma_3$, with $K3$ locally given by the degeneration limit
\eqref{degen}, take the form:
\begin{equation}\label{torwedge}
\begin{aligned}
\tau_0
&=
\tau_0^{\rm fib}
+
c_0\,f,
\qquad
\tau_1
=
\tau_1^{\rm fib}
+
c_1\,X +
\delta_{\rm smooth}\tau_1,
\qquad
\tau_2
=
\tau_2^{\rm fib} + \tau_2^{\rm sing} \\
\tau_3
& =
\tau_3^{\rm fib}
+
c_3\,*_7 G_4^{(27)}
+
T\,\delta_{\Sigma_6^{\rm jct}}\,\Xi^{(3)}_{27}  +
\delta_{\rm smooth}\tau_3  + \mathcal O(T^2), \qquad
G_4 = f\,\psi + X\wedge \varphi + G_4^{(27)} ,
\end{aligned}
\end{equation}
where the constants $c_0, c_1, c_3$ depend on the normalization conventions, and $\mathcal O(T^2)$ denotes beyond-the-leading-order corrections.
The singular localized pinch term
$T\,\delta_{\Sigma_6^{\rm jct}}\,\Xi^{(3)}_{27}$ lies in the $\mathbf{27}$ and
therefore contributes directly to $\tau_3$. The smooth branch-odd term
$dT\wedge\alpha_1\wedge\Sigma_2$ has a $\mathbf{7}$ projection and, in
general, a $\mathbf{27}$ projection; these are denoted by
$\delta_{\rm smooth}\tau_1$ and $\delta_{\rm smooth}\tau_3$ respectively.
The pinch does not directly source $\tau_0$ at this order because the singlet
projection of both the singular and smooth pinch terms vanishes. It also does
not directly source $\tau_2$, since $\tau_2$ is determined by the $\mathbf{14}$
component in $d\psi$. Indirect back-reaction may still generate a non-zero
$\tau_2$,which we denote here\footnote{$\tau_2^{\rm sing}$ may not necessarily be singular because the pinch directly does not produces a singular $\tau_2$. Nevertheless it is included here for completeness and may be regarded as an indirect $\tau_2$ correction or backreaction-induced $\tau_2$ correction.} 
by $\tau_2^{\rm sing}$, but that requires solving the coupled torsion equations that we do not perform here. In conclusion, we see that the tachyon 
modulus:
\begin{equation}
T=\frac{R_+-R_-}{2}
\end{equation}
has a precise geometric meaning on the M-theory side: it measures the asymmetric pinching of the wedge circle and appears directly as the coefficient of the localized $\tau_3$-torsion source at the junction. In this sense the total $G_2$ torsion on the singular branch is the sum of three effects: (1) bulk flux $G_4$, (2) non-trivial fibration over $\Sigma_3$, and 
(3) localized wedge pinching controlled by $T$.

Finally, one should emphasize that torsion classes and $G_4$ flux listed in \eqref{torwedge} are the most precise universal statements that can be made without specifying an explicit metric, connection, and singular local model for the pinched geometry. With a fully explicit local ansatz, the tensors $\tau_i^{\rm fib}$ and the localized form $\Xi_{27}$ can be computed component by component.

\subsection{Why is the pinch naturally in the ${\bf 27}$ of $G_2$ at leading order in $T$? \label{naturalpinch}}

We would like to revisit the question as to why the pinch lies in the ${\bf 27}$ of $G_2$ using an alternative proof, compared to what we had in the previous section {\it i.e.} section \ref{g2pinch1}, via $G_2 \to SU(3)$ structure decomposition.

Recall that, before pinching, the circle one-form is
$e^4 = R_B\,d\theta_b$.
After the degeneration
$S_b^1 \;\longrightarrow\; S_+^1 \vee S_-^1$,
the one-form becomes piecewise defined as given in \eqref{e4wedge}, which when expressed in terms of $R_B$ and $T$, takes the form \eqref{e4def}. From here one can easily show that the 
exterior derivative of the wedge one-form becomes:
\begin{equation}\label{deeff}
\begin{aligned}
de^4_{\rm wedge}
 & =
dR_B\wedge \sigma_1
+
R_B\,d\sigma_1
+
dT\wedge \alpha_1
+
T\,d\alpha_1\\
& =
dT\wedge \alpha_1
+
T\,\delta_v\wedge \eta_1
+
(\text{regular terms}) ,
\end{aligned}
\end{equation}
where 
$\sigma_1
\equiv
d\theta_+\,\mathbf{1}_{S^1_+}
+
d\theta_-\,\mathbf{1}_{S^1_-}$ and 
$\alpha_1
\equiv
d\theta_+\,\mathbf{1}_{S^1_+}
-
d\theta_-\,\mathbf{1}_{S^1_-}$. The relation \eqref{deeff} shows that the pinch contains
a smooth branch-odd term $dT\wedge \alpha_1$
and a singular localized term $T\,\delta_v\wedge \eta_1$. Now 
taking the standard local associative form for $\varphi$ in \eqref{phi_standard}, the terms containing $e^4$ are of the form
$\varphi\big|_{e^4}
=
e^4\wedge \Sigma_2$,
where 
$\Sigma_2 \equiv -e^{37}+e^{16}+e^{25}$ denotes a two-form on the six-dimensional base.
Therefore the pinch-sensitive part of \(d\varphi\) comes from:
\begin{equation}
d\big(e^4\wedge \Sigma_2\big)
=
de^4\wedge \Sigma_2
-
e^4\wedge d\Sigma_2.
\end{equation}
The second term contains no \(de^4\), so it contributes only to regular background terms. Hence the pinch contribution is entirely from
$de^4_{\rm wedge}\wedge \Sigma_2$, such that:
\begin{equation}
d\varphi\big|_{\rm pinch}
=
dT\wedge \widetilde\rho_3^{\rm loc}
+
T\,\delta_v\wedge \rho_3^{\rm loc}
+
(\text{regular terms}),
\label{dphi_pinch_master}
\end{equation}
where
$\widetilde\rho_3^{\rm loc} \equiv \alpha_1\wedge \Sigma_2$
and 
$\rho_3^{\rm loc} \equiv \eta_1\wedge \Sigma_2$. The term
$dT\wedge \widetilde\rho_3^{\rm loc}$
is smooth. It is not supported on the node. After reduction to the six-dimensional \(SU(3)\)-structure side, it becomes the source of the \(dT\)-dependent corrections in \(W_2,W_3,W_4,W_5\) as we saw earlier. On the other hand, 
the term
$T\,\delta_v\wedge \rho_3^{\rm loc}$
is singular and localized at the node. This is the term that should be identified with the singular \(G_2\)-torsion source.
Lifting the node \(v\) to the six-dimensional junction locus \(\Sigma_6^{\rm jct}\subset M_7\), one writes:
\begin{equation}
d\varphi\big|_{\rm pinch,\,sing}
=
T\,\delta_{\Sigma_6^{\rm jct}}\,n\wedge \rho_3,
\label{dphi_sing_master}
\end{equation}
where \(n\) is the local normal one-form to \(\Sigma_6^{\rm jct}\). Thus \(n\) is a one-form in the fiber \(S^1_+ \vee S^1_-\), while \(\rho_3\) is a three-form tangent to the six-dimensional junction locus, obtained by projecting the local form \(\rho_3^{\rm loc}\) to \(T^*\Sigma_6^{\rm jct}\). We can also quantify this in the following way. Locally writing
$M_7 \simeq \mathbb R_n \times \Sigma_6^{\rm jct}$, we see that 
$n \in \Lambda^1(N^*\Sigma_6^{\rm jct})$
is the one-form along the direction transverse to the junction hypersurface, while
$\rho_3 \in \Lambda^3(T^*\Sigma_6^{\rm jct})$
is a \(3\)-form tangent to the six-dimensional junction locus. In fact this is the base where the seven-dimensional $G_2$ structure induces a $SU(3)$ structure. Using local splitting:
\begin{equation}
\varphi = J\wedge n + {\rm Re}\,\Omega,
\qquad
\psi = \frac12 J\wedge J - {\rm Im}\,\Omega\wedge n.
\label{SU3split_master}
\end{equation}
In such a space a real three-form decomposes as $(3, 0) + (0, 3) + (2, 1) + (1, 2)$. We can represent this using the $SU(3)$ structure data $(J, \Omega)$ using the following decompositions:
\begin{equation}\label{robeta}
\rho_3  \in \Lambda^3
=
\langle {\rm Re}\,\Omega\rangle
\oplus
\langle {\rm Im}\,\Omega\rangle
\oplus
\Lambda^3_{\rm prim,(2,1)+(1,2)} ~~\implies ~~
\rho_3 = a\,{\rm Re}\,\Omega
+
b\,{\rm Im}\,\Omega
+
J\wedge \lambda_1
+
\rho_3^{\rm prim},
\end{equation}
where $a,b\in \Lambda^0,
\lambda_1\in \Lambda^1(T^*\Sigma_6^{\rm jct})$
and 
$\rho_3^{\rm prim}\in \Lambda^3_{\rm prim,(2,1)+(1,2)}$. It is now immediately obvious that imposing primitivity of $\rho_3$  implies 
$J\wedge \rho_3 = 0$, {\it i.e.} $\lambda_1 = 0$,
and the absence of \((3,0)+(0,3)\) piece implies
$\rho_3\wedge \Omega = 
\rho_3\wedge \overline\Omega = 0$, {\it i.e.} $a = b = 0$. Therefore demanding a primitive $(2, 1) + (1,2)$ form would mean $a = b = \lambda_1 = 0$. 

The aforementioned analysis however {\it does not} explain why are we demanding primitivity of the three-form $\rho_3$ in the first place. Note that primitivity is neither implied topologically nor geometrically, so the reason should lie elsewhere. In fact there are three main reasons for demanding primitivity here.

\vskip.1in

\noindent (1) The pinch is supposed to be a shape deformation, not a volume deformation. The \({\rm Re}\,\Omega\) and \({\rm Im}\,\Omega\) pieces correspond to singlet deformations of the \(SU(3)\) structure. In the \(G_2\) language, these feed the scalar torsion \(\tau_0\), i.e. the nearly-parallel/singlet component. But the wedge asymmetry
$T=\frac12(R_+-R_-)$
is naturally interpreted as a \emph{traceless branch-odd distortion} of the pinched circle geometry, not as an overall scalar rescaling of the local \(SU(3)\) structure. So one expects it to avoid the singlet sector. Quantitatively, this means setting $a = b = 0$ in \eqref{robeta}.

\vskip.1in

\noindent (2) The pinch is localized and anisotropic, so it should be traceless rather than Lee-form type.
The \(J\wedge \lambda_1\) piece is the non-primitive component of a \(3\)-form and corresponds to Lee-form-type data. This is associated with the \(\mathbf 7\) piece in the \(G_2\) decomposition, hence \(\tau_1\). But the local pinch source is built from the singularity of one specific fiber direction \(e^4\), through \eqref{deeff}
and then inserted into
$\varphi\big|_{e^4}=e^4\wedge\Sigma_2$.
This produces a localized anisotropic defect with one transverse leg and three tangential legs. Such a source is naturally interpreted as a traceless distortion of the \(G_2\) structure, not as a Lee-form deformation. Quantitatively, that suggests 
$\lambda_1=0$ in \eqref{robeta}.

\vskip.1in

\noindent (3) The pinch should have some consistency with the intended physical interpretation. If one wants the pinch to be the geometric origin of the tachyonic branch-odd mode, then one usually wants the \emph{leading localized source} to sit in one distinguished torsion class, namely \(\tau_3\), rather than simultaneously exciting \(\tau_0\), \(\tau_1\), and possibly \(\tau_2\). This is a modeling choice, but a very natural one. It is precisely enforced by taking
$\rho_3=\rho_3^{\rm prim}$ in \eqref{robeta}.

\vskip.1in

\noindent The above three reasons provide reasonable arguments to choose a primitive $\rho_3$. However there does exist a slightly stronger reason for this choice.  Note that, 
from
$\rho_3^{\rm loc}
=
\eta_1\wedge\Sigma_2$ with $\Sigma_2$ being a two-form defined earlier, 
one can at least say that the induced tangential form \(\rho_3\) is built from the same local frame that defines the \(G_2\) structure. So its representation content is highly constrained by the local geometry. If, in addition, the local frame is chosen so that \(\Sigma_2\) is primitive as a two-form with respect to the induced \(SU(3)\) structure, namely
$J\wedge \Sigma_2 = 0$
and the tangential projection of \(\eta_1\) is orthogonal to the distinguished complex directions singled out by \(\Omega\), then
\(\rho_3\) indeed lies in the primitive \((2,1)+(1,2)\) sector. This means a
stronger version of the assumption is that we can always 
choose the local \(SU(3)\) structure so that the induced pinch form is primitive.

Thus primitivity of the three-form is more natural in our setting because of the aforementioned reasoning, although we will speculate later what happens when we go away from the primitivity requirement. With this in mind, we can ask as to 
why the singular piece has no \(\mathbf{1}\) part. To see this, let us 
define:
\begin{equation}
K_{\rm sing}
\equiv
d\varphi\big|_{\rm pinch,\,sing}
=
T\,\delta_{\Sigma_6^{\rm jct}}\,n\wedge \rho_3 ,
\end{equation}
where the subscript sing means that we are ignoring all the smooth and regular terms from \eqref{dphi_pinch_master}.
The \(\mathbf 1\)-piece of a four-form is its projection onto the line spanned by \(\psi\), namely:
\begin{equation}
\pi_{\mathbf 1}(K_{\rm sing})
=
\frac{\langle K_{\rm sing},\psi\rangle}{\langle \psi,\psi\rangle}\,\psi ,
\label{pi1_master}
\end{equation}
where note that a typical four-form generically may be decomposed as \eqref{G4decomp_full} under $G_2$ structure. We have also defined the contraction as
$\langle \alpha_p,\beta_p\rangle
=
\frac{1}{p!}\,\alpha_{m_1\cdots m_p}\beta^{m_1\cdots m_p}$
giving the pointwise form inner product. Taking the split form of $\psi$
from \eqref{SU3split_master} we can easily infer that $-$
since \(K_{\rm sing}\) contains one factor of \(n\) $-$ it is automatically orthogonal to the purely tangential term \(\frac12J\wedge J\). Therefore:
\begin{equation}\label{ksing}
\langle K_{\rm sing},\psi\rangle
=
-\,T\,\delta_{\Sigma_6^{\rm jct}}\,
\langle n\wedge \rho_3,\ {\rm Im}\,\Omega\wedge n\rangle 
=
-\,T\,\delta_{\Sigma_6^{\rm jct}}\,
\langle \rho_3,\ {\rm Im}\,\Omega\rangle ,
\end{equation}
where we used the identity
$\langle n\wedge \alpha_p,\ n\wedge \beta_p\rangle
=
\langle \alpha_p,\beta_p\rangle$, and we are typically ignoring the $\langle \psi, \psi\rangle$ piece as it forms an inconsequential multiplicative factor.
But if \(\rho_3\) has no \((3,0)+(0,3)\) piece, then it is orthogonal to \({\rm Im}\,\Omega\). This means:
\begin{equation}\label{pinchoo1}
\langle \rho_3,{\rm Im}\,\Omega\rangle=0 ~~~\implies ~~~~
\pi_{\mathbf 1}(K_{\rm sing})=0 ,
\end{equation}
justifying the fact that $K_{\rm sing}$ cannot have a ${\bf 1}$ 
part\footnote{A more accurate statement would be that 
the singlet part is removed by requiring
$\Pi_{(3,0)+(0,3)}(\rho_3)=0$,
not by primitivity alone.}. On the other hand, if $\rho_3$ is not primitive then it takes the form 
given in \eqref{robeta} with $(a, b) \ne 0$, with the bracket condition \eqref{ksing} still remaining the same. Using the inner-product argument already discussed, one now finds:
\begin{equation}
\pi_{\mathbf 1}(K_{\rm sing})
\propto
\langle \rho_3,{\rm Im}\,\Omega\rangle\,\psi
=
b\,\|{\rm Im}\,\Omega\|^2\,\psi ,
\label{pi1_from_b}
\end{equation}
implying that the coefficient \(b\) sources the singlet part, thus making \(\tau_0^{\rm pinch}\) non-zero. However from the three reasoning given earlier, $b \ne 0$ would in turn imply that the pinching deforms the entire volume of the six-dimensional base, thus unfortunately going against the very idea of the local pinching that converts $S^1_b \to (S^1_+ \vee S^1_-)_b$.

We can also ask why the singular piece has no \(\mathbf{7}\) part. To see this note that for a \(G_2\) structure seven-manifold, one has
$\Lambda^4
=
\Lambda^4_1\oplus \Lambda^4_7\oplus \Lambda^4_{27}$,
with
$\Lambda^4_7
=
\{\, X\wedge \varphi \;:\; X\in \Lambda^1 \,\}$.
So the \(\mathbf 7\) projection of a \(4\)-form \(K\) is the unique form of the type
$\pi_{\mathbf 7}(K)=X_K\wedge \varphi$
for some \(1\)-form \(X_K\). Equivalently, after Hodge duality,
$*_7:\Lambda^4_7 \longrightarrow \Lambda^3_7$,
and
$\Lambda^3_7
=
\{\, Y\lrcorner \psi \;:\; Y\in \Lambda^1 \,\}$.
So a cleaner way to test whether a \(4\)-form has a \(\mathbf 7\) piece is to Hodge dualize it and ask whether the resulting \(3\)-form has a component in \(\Lambda^3_7\). Therefore taking the Hodge dual of $K_{\rm sing}$ we get:
\begin{equation}
\chi_{\rm sing}
\equiv *_7 K_{\rm sing}
=
T\,\delta_{\Sigma_6^{\rm jct}}\,*_7(n\wedge \rho_3)
=
T\,\delta_{\Sigma_6^{\rm jct}}\,*_6\rho_3 ,
\label{chi_master}
\end{equation}
where, for the 
local metric split as
$ds_7^2 = n^2 + ds_6^2$ and
with orientation chosen so that
${\rm vol}_7 = n\wedge {\rm vol}_6$,
any \(p\)-form \(\alpha_p\) tangent to the six-dimensional subspace satisfy 
$*_7(n\wedge \alpha_p)=*_6\alpha_p$
and
$*_7\alpha_p = (-1)^p\,n\wedge *_6\alpha_p$. (The sign in the second relation depends on conventions, but the first equation is the convention used in the discussion above.) Therefore if \(\rho_3\) is primitive \((2,1)+(1,2)\), then \(*_6\rho_3\) is also primitive \((2,1)+(1,2)\). But the \(\mathbf 7\) inside \(\Lambda^3\) corresponds to the non-primitive sector. Therefore \(*_6\rho_3\) is orthogonal to the \(\mathbf 7\)-piece, and so:
\begin{equation}\label{pinchoo2}
\pi_{\mathbf 7}(K_{\rm sing})=0 ,
\end{equation}
implying no ${\bf 7}$ piece either. However when $\rho_3$ is not primitive and therefore satisfy \eqref{robeta} then the non-primitive piece
$J\wedge \lambda_1$
is precisely the type of \(SU(3)\) component that feeds the \(\mathbf 7\) inside \(\Lambda^3\), and therefore contributes to the \(\mathbf 7\) piece of 
$K_{\rm sing}=n\wedge \rho_3$.
So \(\lambda_1\neq 0\) generically gives
non-zero $\pi_{\mathbf 7}(K_{\rm sing})$
hence a nonzero \(\tau_1^{\rm pinch}\). Unfortunately such a choice clashes with the second reason that we gave earlier, and so allowing 
nonzero \(\tau_1^{\rm pinch}\) becomes unnatural in the pinch-deformed scenario. Therefore, following \eqref{pinchoo1} and \eqref{pinchoo2}, and the fact that $d\varphi$ for the $G_2$ manifold decomposes as in \eqref{torsiondecomp2}, it follows that:
\begin{equation}
K_{\rm sing}=*_7\tau_3^{\rm pinch} ~~~ \implies ~~~
\tau_3^{\rm pinch}
=
T\,\delta_{\Sigma_6^{\rm jct}}\,*_7(n\wedge \rho_3)
=
T\,\delta_{\Sigma_6^{\rm jct}}\,*_6\rho_3 = T\,\delta_{\Sigma_6^{\rm jct}}\,\Xi^{(3)}_{27} ,
\label{tau3_final_master}
\end{equation}
where 
$\Xi^{(3)}_{27}\equiv *_6\rho_3 \in \Lambda^3_{27}$. Note that, 
from the singular part of \(d\varphi\) and imposing some reasonable arguments, one proves directly that 
$\tau_3^{\rm pinch}\neq 0$
and $\tau_0^{\rm pinch}=\tau_1^{\rm pinch}=0$.
To also conclude
$\tau_2^{\rm pinch}=0$, 
one must separately assume that no singular \(\mathbf{14}\)-valued term is generated in \(d\psi\) at this order. Such an assumption is in-built in our set-up both in our choice of G-fluxes and in our choice of the $G_2$ structure. Therefore 
$T$ controls the singular localized \(\mathbf{27}\)-valued pinch torsion source, while
$dT$ controls the smooth branch-odd torsion deformations that descend after reduction.

\subsection{Does supersymmetry get restored when the tachyon vanishes?\label{susyresto}}

There is however one crucial question regarding supersymmetry when the tachyon modulus $T = 0$. Is the supersymmetry restored? The answer in general is {\it no}, but the story has some subtleties that we want to elucidate in the following.  
We start by collecting the discussion of the branch-asymmetry modulus:
\begin{equation}
R_B=\frac{R_++R_-}{2},
\qquad
T=\frac{R_+-R_-}{2},
\label{RTdef_summary}
\end{equation}
and explain carefully what is meant by the symmetric point
$T=0$, 
how it differs from the type-IIA endpoint, and under what conditions supersymmetry can or cannot be restored.

The first important point is that one must distinguish the \emph{background value} of the branch-odd modulus from its \emph{fluctuations}. Writing the tachyon field as:
\begin{equation}
T(x)=T_0+\delta T(x),
\label{Texpand_summary}
\end{equation}
the condition
$T_0=0$
implies 
$R_+=R_-$,
means only that the \emph{classical background} is at the symmetric wedge point. It does \emph{not} imply that the fluctuation
$\delta T(x)$
vanishes. Thus the correct interpretation of
$T_0=0$
is that the expectation value of the branch-odd deformation vanishes, not that the corresponding mode has disappeared from the spectrum.
Equivalently, one should distinguish
$\langle T\rangle =0$
from
$\delta T =0$. Only the first is implied by the condition $R_+ = R_-$.

One may also see it from an effective potential and the tachyonic mode.
(See also footnote \ref{tachiooon}.) Suppose the branch-odd mode has an effective potential of the form:
\begin{equation}
V(T)=V_0+\frac12\,m_T^2\,T^2+\lambda T^4+\cdots.
\label{VTeff_summary}
\end{equation}
Then
$\left.\frac{\partial V}{\partial T}\right|_{T=0}=0$,
so \(T=0\) is a stationary point. However, whether this point is stable depends on the sign of the mass term $m_T$ appearing above. The mass is defined as:
\begin{equation}
m_T^2=\left.\frac{\partial^2 V}{\partial T^2}\right|_{T=0} ,
\end{equation}
so if
$m_T^2<0$,
then \(\delta T\) is tachyonic, and the point \(T=0\) is unstable. In that case the statement
$T_0=0$
does \emph{not} mean that supersymmetry has been restored; it only means that one is sitting at the symmetric point of the wedge geometry.

We can also ask how the pinch contribution disappears at {$T_0=0$} and how it affects the torsion classes. From the earlier analysis of the $G_2$ torsion classes, the pinch contribution enters through:
\begin{equation}
\tau_3^{\rm pinch}
=
T\,\delta_{\Sigma_6^{\rm jct}}\,\Xi^{(3)}_{27}
+
{\cal O}(T^2),
\qquad
\Xi^{(3)}_{27}\in \Lambda^3_{27}.
\label{tau3pinch_summary}
\end{equation}
Therefore, for the \emph{background} geometry,
$T_0=0$ we expect 
$\tau_3^{\rm pinch}|_{\rm background}=0$.
So at the symmetric point, the branch-odd \emph{background} contribution to the $G_2$ torsion disappears at leading order. However, fluctuations still induce:
\begin{equation}
\delta \tau_3^{\rm pinch}
\sim
\delta T\,\delta_{\Sigma_6^{\rm jct}}\,\Xi^{(3)}_{27},
\label{deltatau3_summary}
\end{equation}
so if \(\delta T\) remains as a tachyonic mode, then the supersymmetry-breaking channel associated with the wedge branch remains open at the level of fluctuations.

The story remains similar from the {$SU(3)$}-structure side too. After reducing to the six-dimensional space with \(SU(3)\) structure, the torsion classes take the schematic form:
\begin{equation}
W_i
=
W_i^{\rm fib}
+
W_i^{\rm flux}
+
\delta_T W_i,
\qquad i=1,\dots,5,
\label{Wi_master_summary}
\end{equation}
where \(\delta_T W_i\) contains the branch-odd pinch contribution. At the background level,
$T_0=0$ and therefore we expect
$\delta_T W_i|_{\rm background}=0$
at leading order in the pinch sector. But the fluctuations still generate:
\begin{equation}
\delta W_i \neq 0
\end{equation}
whenever \(\delta T\neq 0\). Thus the condition \(T_0=0\) removes only the \emph{background} pinch-induced torsion; it does not guarantee that the full fluctuation spectrum is supersymmetric.

With the above discussion, we can now quantify the reason as to why {$T_0=0$} does not automatically restore supersymmetry. The point is that, 
it is crucial to distinguish the pinch contribution from the flux- and fibration-induced contributions. On the $G_2$ side, we found the torsion classes as in \eqref{torwedge}.
Hence
$T_0=0$
implies only:
\begin{equation}
\tau_3^{\rm pinch}=0,
\end{equation}
but in general
$\tau_0,\ \tau_1,\ \tau_2,\ \tau_3$ and the corresponding $\delta\tau_i$
need not vanish because the flux and fibration pieces remain.
Likewise, on the reduced \(SU(3)\)-structure side, one may still have nonzero:
\begin{equation}
W_i^{\rm fib},
\qquad
W_i^{\rm flux},
\end{equation}
as well as $\delta W_i$ even when
$T_0=0$.
Therefore supersymmetry is \emph{not} automatically restored at the symmetric wedge point.

So when can supersymmetry be restored?
Supersymmetry can be restored only if \emph{all} remaining supersymmetry-breaking sources are also absent or satisfy the appropriate constraints. On the {$G_2$} side a
sufficient condition for strict $G_2$ holonomy is:
\begin{equation}
\tau_0=\tau_1=\tau_2=\tau_3=0.
\label{G2holonomy_summary}
\end{equation}
More generally, in flux backgrounds one may allow nonzero torsion provided the Killing-spinor equations are satisfied. Thus \(T_0=0\) can at most remove the pinch contribution; the remaining pieces must still satisfy the supersymmetry constraints. On the {$SU(3)$} side and  
in a heterotic/Strominger-type language, supersymmetry requires:
\begin{equation}
W_1=W_2=0,
\qquad
2W_4+W_5=0,
\label{strom_summary}
\end{equation}
together with the appropriate flux and Bianchi identities. Hence, even when \(T_0=0\), supersymmetry is restored only if
$W_i^{\rm fib}+W_i^{\rm flux}$ and the corresponding $\delta W_i$
satisfy \eqref{strom_summary}. If these conditions fail, the background remains non-supersymmetric even though the branch-odd background deformation has been switched off.

However a more pertinent question is: when can supersymmetry \emph{not} be restored? There are two distinct obstructions. First is the obvious one coming from the flux/fibration obstruction. The point is that, even if
$T_0=0$,
supersymmetry is not restored if the remaining torsion classes fail to satisfy the required $G_2$ or \(SU(3)\) constraints. Concretely,
$T_0=0$
but $\tau_i^{\rm fib}\neq 0$
or 
$\tau_i^{\rm flux}\neq 0$
may still lead to a non-supersymmetric background, notwithstanding the fact that $\delta \tau_i$ still remains a potential source of susy breaking.

The second one is slightly more tricky and comes from the tachyonic fluctuation. Even if the \emph{background} torsion constraints are satisfied at \(T_0=0\), supersymmetry is not restored if the fluctuation \(\delta T\) remains genuinely tachyonic:
\begin{equation}
m_{\delta T}^2<0.
\label{mneg_summary}
\end{equation}
In that case the point \(T_0=0\) is unstable and cannot be a supersymmetric vacuum. Thus a true supersymmetric restoration at \(T_0=0\) requires both:
(1) the remaining flux/fibration torsion satisfies the SUSY constraints,
and
(2) the branch-odd fluctuation $\delta T$ is absent, stabilized, or no longer tachyonic.

The above discussion also answers the question as to why the type-IIA limit is different from {$T_0=0$}. The type-IIA limit is not characterized by
$T_0=0$.
Rather, it is characterized by the collapse of one branch of the wedge:
$R_+=0$ or 
$R_-=0$.
In that case the wedge
$S^1_+\vee S^1_-$
reduces to a \emph{single} circle, and one recovers the ordinary type-IIA reduction. Notice that at this endpoint
$T=\pm R_B$,
so in general one might think that 
$T\neq 0$. This is not exactly the right interpretation. The tachyon $T$ is only identified by the asymmetry between $R_+$ and $R_-$. So when $R_+ = R_-$, that tachyon vanishes but the fluctuation survive. When $R_+ \ne R_-$, both the tachyon and its fluctuation survive. However when $R_- = 0$, we are at the point in the moduli space where we have one circle of radius $R_+$ where $T = R_B = {R_+\over 2}$ is now identified with $R_B$, {\it i.e.} with the IIA dilaton. 
Therefore the IIA limit is \emph{not} the symmetric point \(T_0=0\); it is instead a boundary of the wedge moduli space where one branch disappears. This resolves the apparent paradox:
$T_0=0$
means symmetric wedge,
whereas
$R_\pm=0$
means single-circle IIA endpoint.
They are different points in moduli space. Thus we see that 
$T_0=0$ is at most a necessary condition for removing the background pinch deformation,
but it is not sufficient for supersymmetry restoration unless both the remaining torsion constraints and their fluctuations are satisfied and $\delta T$ is no longer 
tachyonic\footnote{{It is useful to separate the recovery of the Type IIA spectrum from the
restoration of supersymmetry. When one branch collapses, the doubled RR
fields reduce to their single surviving combinations, the independent
branch-odd data disappear, and the wedge becomes an ordinary circle. This
establishes the kinematical recovery of the Type IIA field content.
Supersymmetry is restored only if the surviving metric, flux, and bundle
data also solve the Type IIA Killing-spinor equations and Bianchi
identities. Thus tachyon condensation removes the Type $0$ instability and
takes the system to a single-circle endpoint, but the endpoint is
supersymmetric only when
$\delta\psi_M=0$ and 
$\delta\lambda=0$,
or, equivalently, when the corresponding endpoint torsion and flux
conditions are satisfied. This distinction will be assumed throughout the
subsequent discussion.}}.

Therefore, there is no contradiction: the presence of a tachyonic mode in the original type-\(0\) regime is exactly what allows the system to flow to a supersymmetric endpoint where that instability is no longer present. The choice $R_+=0$ with $R_->0$ or 
$R_-=0$ with $R_+>0$ is a different vacuum where the old unstable fluctuation has driven the system to this new background. In this background the original scalar field $T$ now gets identified with the IIA dilaton, the doubled RR fields get merged into one another and we recover the Type IIA spectrum.

{Before ending this section let us make one more observation. The single-circle Type IIA endpoints discussed above should be
distinguished from the perturbative orbifold construction at the symmetric
wedge point. At
$R_{+}=R_{-}$, we have vanishing tachyon, {\it i.e.}
$T=0$,
and the branch-exchange operation acts as a $\mathbb Z_2$ symmetry. The
branch-even and branch-odd combinations transform as
\begin{equation}
\Phi^{(e)}
\longmapsto
+\Phi^{(e)},
\qquad
\Phi^{(o)}
\longmapsto
-\Phi^{(o)}.
\end{equation}
Identifying this operation with $(-1)^{G_L}$ in the Type $0$A description,
the orbifold projection removes the branch-odd sectors and retains the
diagonal RR combinations:
\begin{equation}
C_1^{\rm IIA}
=
\frac{C_1^{(+)}+C_1^{(-)}}{\sqrt{2}},
\qquad
C_3^{\rm IIA}
=
\frac{C_3^{(+)}+C_3^{(-)}}{\sqrt{2}} ,
\end{equation}
where $C_1^{(\pm)}$ and $C_3^{(\pm)}$ are defined in \eqref{RRfinal0A}. These 
are precisely the one- and the three-forms in the RR sector. 
Together with the invariant NSNS fields, the surviving bosonic spectrum is:
\begin{equation}
\left\{
g_{\mu\nu},
B_2,
\Phi;
C_1^{\rm IIA},
C_3^{\rm IIA}
\right\},
\end{equation}
giving rise to the ordinary Type IIA bosonic spectrum. Note that, in the parametrization
$C_3^{(\pm)}=C_3\pm\widehat\Xi_3$, the projection simply sets
$\widehat\Xi_3=0$. Similarly, the branch-odd torsion and flux pieces are
removed. The resulting background is supersymmetric only if the remaining
branch-even data satisfy the Type IIA Killing-spinor equations.}

{This perturbative projection is conceptually different from the
tachyon-condensation endpoint. In the former, both branches remain at equal
radius and one projects onto the invariant Hilbert space. In the latter,
one branch disappears dynamically and the wedge becomes a single circle.}

\section{Type 0A theory compactified on a non-K\"ahler six-manifold \label{sec3NK}}

Our analysis of the second kind of deformation in section \ref{wedgebranch} produced two theories in four-dimensions: a Type 0 Heterotic/HW compactification from \eqref{hetbg} and a Type 0A compactification from \eqref{0abg}. Both are compactified on six-dimensional non-K\"ahler (or torsional) manifolds with zero supersymmetries in four-dimensions. The pertinent question is: Are these two theories U-dual to each other because their supersymmetric cousins $-$ studied in section \ref{intervalbranch} $-$ are U-dual? To answer this question we will have to study these two compactifications in details and infer the consequences from there. In this section we will study the Type 0A theory, and in the next section we will address the Type 0 Heterotic theory.

The Type 0A theory comes from reduction on the pinched circle \cite{BDV} and the resulting four-dimensional theory is induced from the $SU(3)$-structure on the six-manifold \eqref{0abg}. Recall that 
we consider M-theory on
$M_7 = K3 \rtimes \Sigma_3$,
where locally:
\begin{equation}\label{locK3}
K3 \;\leadsto\; T^2_f \rtimes \frac{S_a^1}{\mathbb Z_2} \rtimes (S^1_+\vee S^1_-)_b ,
\end{equation}
and where internal M-theory flux
$G_4=dC_3$ is switched on. As discussed above, the seven-dimensional $G_2$-structure torsion classes receive three distinct 
contributions from the flux components, the fibrations and the pinching
as shown in \eqref{torwedge}, with possible $\mathcal O(T^2)$ corrections beyond the leading order. Reducing along the pinched circle direction leads to a six-dimensional non-K\"ahler (or torsional) manifold whose local geometry takes the form:
\begin{equation}\label{m6def}
M_6 \;\sim\; T^2_f \rtimes \frac{S_a^1}{\mathbb Z_2} \rtimes \Sigma_3 ,
\end{equation}
which is an orientifold because of the $\mathbb Z_2$ action on $S_a^1$.
Even for this local geometry we can ask how the 
$SU(3)$ structure is induced from the $G_2$ structure studied earlier. 
This is easily seen from \eqref{G2SU3decomp}, and for the present convenience we can replace $\eta$ therein by 
$\eta_b$ to denote the one-form along the pinched circle direction prior to reduction, with $v_b$ representing the dual vector. Taking exterior derivatives gives the exact identities:
\begin{equation}
d\varphi
=
(dJ)\wedge \eta_b
+
J\wedge d\eta_b
+
d\,{\rm Re}\,\Omega \nonumber
\label{dphi_exact_0A}
\end{equation}
\begin{equation}
d\psi
=
J\wedge dJ
-
d\,{\rm Im}\,\Omega\wedge \eta_b
+
{\rm Im}\,\Omega\wedge d\eta_b ,
\label{dpsi_exact_0A}
\end{equation}
where $(J,\Omega)$ defines an $SU(3)$ structure on $M_6$. The $SU(3)$ torsion classes are defined by \eqref{dJtorsion}, such that 
the six-dimensional torsion classes are determined by the three basic inputs appearing in \eqref{dpsi_exact_0A}, namely:
\begin{equation}
dJ,
\qquad
d\Omega,
\qquad
d\eta_b.
\end{equation}
These three basic ingredients will shape our present discussion of the non-K\"ahler six-manifold $M_6$ from \eqref{m6def} in the Type 0A side. For example, we can start by asking how the $SU(3)$ torsion classes depend on flux, fibration and pinch.

\subsection{Dependence of the $SU(3)$ torsion classes on flux, fibration and pinch \label{sec301}}

The most efficient way to organize the answer is to expand the reduced $SU(3)$-structure data in powers of the branch-odd parameter, {\it i.e.} the tachyon, $T$ in the following way:
\begin{equation}
J
=
J^{(0)}
+
T\,J^{(1)}
+
O(T^2),
\qquad
\Omega
=
\Omega^{(0)}
+
T\,\Omega^{(1)}
+
O(T^2) ,
\label{JOmega_expand_T}
\end{equation}
where to zeroth order, and by definition, we expect the flux and the fibration contributions to $J$ and $\Omega$. Following our earlier strategy, we will analyze only to the first order in $T$ and keep the $\mathcal O(T^2)$ for future works. Accordingly:
\begin{equation}\label{torson}
W_i
=
W_i^{\rm fib}
+
W_i^{\rm flux}
+
\delta_T W_i
+
O(T^2),
\end{equation}
where
$\delta_T W_i$
denotes the correction induced by the branch-asymmetry modulus $T$ that we shall quantify in the following. The equation \eqref{torson} is a useful book-keeping device to track the torsion classes, but their explicit form depend on $dJ$ and $d\Omega$ from \eqref{JOmega_expand_T}. This can be written as:
\begin{equation}
dJ
=
dJ^{(0)}
+
T\,dJ^{(1)}
+
dT\wedge J^{(1)}
+
O(T^2) \nonumber
\label{dJexpandT}
\end{equation}
\begin{equation}
d\Omega
=
d\Omega^{(0)}
+
T\,d\Omega^{(1)}
+
dT\wedge \Omega^{(1)}
+
O(T^2) ,
\label{dOmegaexpandT}
\end{equation}
from where we see that the terms
$dT\wedge J^{(1)}$ and 
$dT\wedge \Omega^{(1)}$
are precisely the origin of the derivative corrections to the torsion classes. We can also see this class by class. For \(W_1\), which is a scalar, the correction can only be of the form:
\begin{equation}
W_1
=
W_1^{\rm fib}
+
W_1^{\rm flux}
+
T\,W_1^{(T)}
+
\mathcal O(T)
+
\mathcal O(T^2),
\end{equation}
where 
$W_1^{(T)}=\left.\frac{\partial W_1}{\partial T}\right|_{T=0}$
and the $\mathcal O(T)$ is 
a schematic way to indicate a possible \emph{derivative correction} to the scalar \(W_1\) built from \(dT\). Since \(W_1\) is a scalar, one cannot write a term like \(dT\wedge(\cdots)\). Instead one would need to contract \(dT\) with some one-form built using $J, \Omega$, flux and the fibration tensors. Details are however not important so we can put all the $T$ dependent terms in $\delta_T W_1$ to specify direct dependence on $T$ and $dT$. Similarly for 
\(W_2\in \Lambda^2\) and $W_3 \in \Lambda^3$, one may have:
\begin{equation}\label{w2andw3}
\begin{aligned}
& W_2
=
W_2^{\rm fib}
+
W_2^{\rm flux}
+
T\,W_2^{(T)}
+
dT\wedge \widetilde W_2^{(T)}
+
O(T^2) \nonumber\\
& W_3
=
W_3^{\rm fib}
+
W_3^{\rm flux}
+
T\,W_3^{(T)}
+
dT\wedge \widetilde W_3^{(T)}
+
O(T^2) ,
\end{aligned}
\end{equation}
with 
$\widetilde W_2^{(T)}\in \Lambda^1$ and 
$\widetilde W_3^{(T)}\in \Lambda^2$.
For the Lee-form classes \(W_4,W_5\in \Lambda^1\), the cleanest possibility is actually:
\begin{equation}
W_4
=
W_4^{\rm fib}
+
W_4^{\rm flux}
+
T\,W_4^{(T)}
+
c_4\,dT
+
O(T^2)\nonumber
\label{W4dT}
\end{equation}
\begin{equation}
W_5
=
W_5^{\rm fib}
+
W_5^{\rm flux}
+
T\,W_5^{(T)}
+
c_5\,dT
+
O(T^2),
\label{W5dT}
\end{equation}
because these are already one-forms. Note that writing
$dT\wedge \widetilde W_{4, 5}^{(T)}$
would force \(\widetilde W_{4, 5}^{(T)}\) to be a scalar, in which case it is simpler just to write a coefficient times \(dT\).
Thus the notation
$\widetilde W_i^{(T)}$
should be understood as the lower-degree form whose wedge product with \(dT\) gives the derivative contribution to the torsion class \(W_i\), when such a contribution is allowed by degree.

\subsubsection{Fibration contribution to the torsion classes \label{sec311}}

With the above formulation of the torsion classes, we can now quantify the piecewise contributions from the fibration structure. 
Let $M_6$ from \eqref{m6def}
be the six-manifold obtained after reduction along the pinched circle direction. Choosing an adapted local coframe
$\{E^A\}_{A=1,2,3}$
on $T^2_f \rtimes \frac{S_a^1}{\mathbb Z_2}$
and
$\{e^m\}_{m=4,5,6}$ on
$\Sigma_3$, the base one-forms $e^m$ are purely horizontal, and the non-trivial fibration is encoded in the non-closure of the fiber one-forms:
\begin{equation}
dE^A\Big|_{\rm fib}
=
\Theta^A{}_{Bm}\,e^m\wedge E^B
+
\frac12\,\Pi^A{}_{BC}\,E^B\wedge E^C
+
\frac12\,\Xi^A{}_{mn}\,e^m\wedge e^n ,
\label{dEAfib_correct}
\end{equation}
which also appeared in \eqref{deIclean} and the parameters are defined therein.
The $SU(3)$ structure on $M_6$ is specified by $(J,\Omega)$, and the torsion classes are defined in \eqref{dJtorsion}.
The geometric twisting \eqref{dEAfib_correct} therefore induces
$dJ^{\rm fib}\neq 0$ and 
$d\Omega^{\rm fib}\neq 0$, where:
\begin{equation}\label{djneo}
\begin{aligned}
& dJ^{\rm fib}
=
-\frac{3}{2}\,{\rm Im}\!\big(W_1^{\rm fib}\,\overline\Omega\big)
+
W_3^{\rm fib}
+
W_4^{\rm fib}\wedge J\\
& d\Omega^{\rm fib}
=
W_1^{\rm fib}\,J\wedge J
+
J\wedge W_2^{\rm fib}
+
\Omega\wedge W_5^{\rm fib} ,
\end{aligned}
\end{equation}
which is easy to see from the coframe choice $\mathcal E = (E^A, e^m)$  and the definition $\mathcal E^2 \equiv \mathcal E \wedge \mathcal E$ that $J \in \mathcal E^2$ and $\Omega \in (\mathcal E + i\mathcal E)^3$ induces $dJ^{\rm fib}\sim (dE^A)\wedge(\cdots)$
and $d\Omega^{\rm fib}\sim (dE^A)\wedge(\cdots)$. Since $dE^A \ne 0$ from \eqref{dEAfib_correct}, we get \eqref{djneo}. Putting everything together, we therefore expect them to contribute to the torsion classes. More precisely\footnote{Equation (2.7) of Cardoso et al.\ (see the first reference in \cite{luest}) and the projection formulae used in \eqref{g2torsionformula} are the
same $SU(3)$-structure torsion-class decomposition.  Their equation gives the
representation-theoretic identification of the five classes inside $dJ$ and
$d\Psi$, while our formulae give explicit projection operators that extract the
corresponding components. More quantitatively, equation (2.7) therein
writes the torsion class decomposition with the subscript $0$ denoting the primitive part.  This is exactly the
representation-theoretic content used in our extraction formulae: $W_1$ is the
singlet part, $W_2$ is the primitive $(1,1)$ class read from the primitive
$(2,2)$ part of $d\Omega$, $W_3$ is the primitive $(2,1)+(1,2)$ part of $dJ$,
and $W_4,W_5$ are the one-form Lee-type classes. 
The only caveats are convention-dependent numerical
normalizations, the choice of $\Psi$ versus $\Omega$, and the precise definition
of the contraction operator. These can be easily adjusted.}:
\begin{equation}\label{g2torsionformula}
W_1^{\rm fib}
=
-\frac{i}{12}\,\overline\Omega\lrcorner\, dJ^{\rm fib},
\qquad
J\wedge W_2^{\rm fib}
=
\Pi_{\rm prim}^{(2,2)}\!\big(d\Omega^{\rm fib}\big)
\nonumber
\end{equation}
\begin{equation}
W_3^{\rm fib}
=
\Pi_{\rm prim}^{(2,1)+(1,2)}\!\big(dJ^{\rm fib}\big)
\nonumber
\end{equation}
\begin{equation}\label{toruequ}
W_4^{\rm fib}
=
\frac12\,J\lrcorner\, dJ^{\rm fib},
\qquad
W_5^{\rm fib}
=
\frac12\,{\rm Re}\!\left(\overline\Omega^{-1}\!\lrcorner\, d\Omega^{\rm fib}\right) ,
\end{equation}
where \(\Pi_{\rm prim}\) denotes projection onto the primitive part with respect to the fundamental form \(J\), while \(\lrcorner\) denotes interior contraction of forms. For example,
$(J\lrcorner \gamma)_m=\frac12 J^{pq}\gamma_{pqm}$,
so \(J\lrcorner dJ\) is a one-form, whereas
$\overline\Omega\lrcorner dJ$
is obtained by contracting all three indices of \(\overline\Omega\) into the three-form \(dJ\) and is therefore a scalar\footnote{A short introduction to {\it interior contraction} and {\it primitive forms} is as follows. The interior contraction between two forms $\alpha$ and $\beta$, denoted by the symbol
$\alpha \lrcorner \beta$,
means contraction of the indices of \(\alpha\) into those of \(\beta\). For example, if \(J\) is a \(2\)-form and \(\gamma_3\) is a \(3\)-form, then
$J\lrcorner \gamma_3$
is a \(1\)-form obtained by contracting the two indices of \(J\) into two of the indices of \(\gamma_3\). In components,
$(J\lrcorner \gamma)_m
=
\frac12\,J^{pq}\,\gamma_{pqm}$.
This is why
$W_4^{\rm fib}
=
\frac12\,J\lrcorner dJ^{\rm fib}$
is a \(1\)-form. Likewise,
$\overline\Omega\lrcorner\, dJ^{\rm fib}$
means contract the three indices of \(\overline\Omega\) into the three indices of the \(3\)-form \(dJ^{\rm fib}\), producing a scalar. That is why
$W_1^{\rm fib}
=
-\frac{i}{12}\,\overline\Omega\lrcorner\, dJ^{\rm fib}$
is a scalar. In a similar vein,
$\overline\Omega^{-1}\!\lrcorner\, d\Omega^{\rm fib}$
means contract the \((0,3)\) indices of \(\overline\Omega\) against the \((3,1)\)-type object inside \(d\Omega^{\rm fib}\), giving a \(1\)-form. The notation \(\overline\Omega^{-1}\) is shorthand for the inverse tensor obtained by raising indices with the metric. In a similar vein, on an \(SU(3)\)-structure manifold with fundamental form \(J\), a \(k\)-form is called \emph{primitive} if its contraction with \(J\) vanishes in the appropriate sense. For example
$\alpha_2$ primitive implies
$J\lrcorner \alpha_2 = 0$, 
and for a \(3\)-form \(\beta_3\), primitiveness means that the part proportional to \(J\wedge(\cdots)\) has been removed.
Thus
$\Pi_{\rm prim}^{(1,1)}(d\Omega^{\rm fib})$
means: take the \((1,1)\) part of \(d\Omega^{\rm fib}\), and then project out the non-primitive piece, keeping only the primitive \((1,1)\) part. Similarly,
$\Pi_{\rm prim}^{(2,1)+(1,2)}(dJ^{\rm fib})$
means: take the \((2,1)+(1,2)\) part of \(dJ^{\rm fib}\), and remove any component proportional to \(J\wedge(\text{one-form})\), leaving only the primitive part.  This hopefully explains all the equations appearing in \eqref{toruequ}. \label{primzak}}.
Thus the non-trivial fibration by itself can in principle generate all five $SU(3)$ torsion classes\footnote{There is a small notational shortcut in the formula for $W_1$ quoted in \eqref{g2torsionformula}.
Strictly speaking, the scalar torsion class $W_1$ appears in both
$dJ$ and $d\Omega$.  Indeed, with the convention \eqref{djneo}
the same scalar $W_1$ is visible in two different projections:
$W_1$ is the singlet component of $dJ$,
and also
$W_1$ is the $J\wedge J$ component of $d\Omega$.
Therefore one can extract $W_1$ either from $dJ$ or from $d\Omega$, provided
the normalization conventions for $J$ and $\Omega$ are fixed consistently. More explicitly, from \eqref{djneo}, the singlet part of $dJ$ is
$(dJ)_{\bf 1}
=
-\frac{3}{2}\,
{\rm Im}\!\left(W_1\,\overline\Omega\right)$.
Contracting this with $\overline\Omega$ isolates the scalar torsion class produces $W_1$ in \eqref{g2torsionformula}
up to the chosen normalization of $\Omega$.  The other pieces in $dJ$ do not
contribute to this contraction because
$\overline\Omega\lrcorner W_3=0$,
and 
$\overline\Omega\lrcorner (W_4\wedge J)=0$.
Thus the contraction with $\overline\Omega$ picks out only the singlet
component of $dJ$. On the other hand, from \eqref{djneo}, the $J\wedge J$ component of
$d\Omega$ is
$(d\Omega)_{\bf 1}
=
W_1\,J\wedge J$. Since $W_2$ is primitive,
$J\lrcorner W_2=0$,
and since the $\Omega\wedge W_5$ term lies in a different $SU(3)$
representation, the double contraction with $J$ isolates the scalar
coefficient.  Schematically,
$W_1
=
\frac{1}{\mathcal N_J}\,
(J\wedge J)\lrcorner d\Omega$,
where $\mathcal N_J$ is the normalization factor defined by
$(J\wedge J)\lrcorner (J\wedge J)=\mathcal N_J$.
With the common convention used in many $SU(3)$-structure decompositions, this
can be written as
$W_1
=
\frac{1}{12}\,
J\lrcorner\big(J\lrcorner d\Omega\big)$,
again up to normalization conventions.
Thus the two extraction formulae are equivalent:
\begin{equation}
-\frac{i}{12}\,\overline\Omega\lrcorner dJ
=
\frac{1}{12}\,
J\lrcorner\big(J\lrcorner d\Omega\big)
=
W_1 ,\nonumber
\label{W1_two_extractions}
\end{equation}
where the equality above should be understood as a
consistency condition on the $SU(3)$-structure decomposition.  In other words,
$dJ$ and $d\Omega$ are not independent arbitrary forms; they are both built
from the same intrinsic torsion.  Therefore the $W_1$ extracted from $dJ$ must
agree with the $W_1$ extracted from $d\Omega$.}.

\subsubsection{Flux contribution to the torsion classes \label{sec312}}

For the flux contribution, there is a subtlety that we need to consider. For example, the discussion written purely in terms of
$G_4=\eta_b\wedge H_3+F_4^{\rm int}$
is the \emph{single-circle} decomposition appropriate to an ordinary IIA reduction. For the wedge reduction
$(S^1_+\vee S^1_-)_b$
one should instead keep the two branches separate. This is precisely how the type $0$ doubling is reflected.
A convenient local description is to introduce branchwise one-forms
$\eta_\pm = R_\pm\,d\theta_\pm$
and 
$R_\pm = R_B \pm T$, so that the M-theory $4$-form is decomposed as:
\begin{equation}\label{GHFetc}
\begin{aligned}
 G_4 \to & ~G_4^{\rm br}
=
\eta_+\wedge H_3^{(+)}
+
\eta_-\wedge H_3^{(-)}
+
F_4^{(+)}
+
F_4^{(-)} \\
&H_3^{(\pm)} \equiv \iota_{v_\pm} G_4,
\qquad
F_4^{(\pm)}
\equiv
G_4-\eta_\pm\wedge H_3^{(\pm)} ,
\end{aligned}
\end{equation}
where in the second line we have denoted the branchwise six-dimensional fluxes. The point is that, in the wedge-circle proposal \cite{BDV}, M-theory on
$S^1\vee S^1$
is interpreted as type $0$A, and the two branches encode the doubled weak-coupling data rather than a single IIA sector. In particular, the recent proposal identifies the wedge reduction with type $0$A and interprets the two-circle asymmetry mode as the tachyon modulus, while type $0$ strings have a doubled RR sector compared to type II. In general therefore, we can define:
\begin{equation}
 F_4^{(\pm)} = a^{(\pm)}F_4 \pm \widetilde{\Xi}_4(T, ...) + \cdots  \nonumber 
\end{equation}
\begin{equation}
H_3^{(\pm)} = a_1^{(\pm)} H_3 + a_2^{(\pm)}\ast_6 H_3 \pm \widetilde{\Xi}_3(T, ...) + \cdots ,
\end{equation}
where the dotted terms are additional DOFs at the pinch, $a_i^{(\pm)}, a^{(\pm)}$ are model dependent coefficients; and $\widetilde{\Xi}_j(T, ...)$ for $j = 3, 4$ are functions similar to 
$\widehat{\Xi}_3(T, ...)$ that we encountered in \eqref{chotolookai} and depend on the tachyon $T$ and other DOFs at the pinch. Note that our definition of $H_3^{(\pm)}$ is slightly more general than \eqref{H3constraint0A} as it puts both $H_s^{\rm phys}$ and $\Delta H_3$ on a firmer footing. However as in our analysis in section \ref{sec001}, 
it is natural to pass from the branch basis to even/odd combinations:
\begin{equation}\label{HFfatu}
\begin{aligned}
& H_3^{(e)} \equiv \frac12\Big(H_3^{(+)}+H_3^{(-)}\Big),
\qquad
H_3^{(o)} \equiv \frac12\Big(H_3^{(+)}-H_3^{(-)}\Big)\\
& F_4^{(e)} \equiv \frac12\Big(F_4^{(+)}+F_4^{(-)}\Big),
\qquad
F_4^{(o)} \equiv \frac12\Big(F_4^{(+)}-F_4^{(-)}\Big).
\end{aligned}
\end{equation}
where note that we have taken the {\it full} doubled field structure from \eqref{H3evenodd0A}, \eqref{H3constraint0A} and 
\eqref{C3pmansatz0A}, including the non-physical 0A fields from \eqref{nonphys0A}. This is more useful then taking on the restrictive 0A fields from \eqref{RRfinal0A}, as will become clear soon. 
Similarly, defining 
$\eta_e \equiv \frac12(\eta_+ + \eta_-)$ and 
$\eta_o \equiv \frac12(\eta_+ - \eta_-)$, 
the M-theory four-form $G_4$ from \eqref{GHFetc} becomes:
\begin{equation}
G^{\rm br}_4
=
\underbrace{\eta_e\wedge H_3^{(e)}
+
\eta_o\wedge H_3^{(o)}
+
F_4^{(e)}}_{G_4^{(e)}}
+
\underbrace{\eta_e\wedge H_3^{(o)}
+
\eta_o\wedge H_3^{(e)}
+
F_4^{(o)}}_{G_4^{(o)}} ,
\label{G4_even_odd_full}
\end{equation}
where using
$\eta_\pm=(R_B\pm T)\,d\theta_\pm$,
one sees that the branch-odd part is controlled by the same parameter
$T=\frac{R_+-R_-}{2}$
that plays the role of the type-$0$ tachyon modulus in the wedge-circle proposal.

However, the \emph{intrinsic} $SU(3)$ torsion classes
$W_1,\dots,W_5$ are defined from
$dJ$ and 
$d\Omega$, 
so only those combinations of fluxes that backreact on the reduced metric, $B$-field, and the $SU(3)$-structure forms enter them directly.
A way to facilitate this is to define the following map:
\begin{equation}
{\cal Q}_j^{H^{(s)}}:\ H^{(s)}_3\longrightarrow \Lambda^j,
\qquad
{\cal Q}_j^{G^{(s)}}:\ G^{(s)}_4\longrightarrow \Lambda^j,
\qquad
{\cal Q}_j^{F^{(s)}}:\ F^{(s)}_2\longrightarrow \Lambda^j,
\end{equation}
where $s = (e, o)$. Note that we have included the abelian gauge fields strengths coming from the doubled fields \eqref{gaugoram}, but since they are two-forms their contributions to $dJ$ and $d\Omega$ would be more non-trivial. In the following we will briefly mention them but come back to a more detailed study later. The flux-induced $dJ$ and $d\Omega$ can be written as:
\begin{equation}
(dJ)_{{\rm flux},s}
=
\alpha_{H_s}{\cal Q}_3^{H^{(s)}}(H^{(s)}_3)
+
\alpha_{F_s}{\cal Q}_3^{F^{(s)}}(F^{(s)}_2)
+
\alpha_{G_s}{\cal Q}_3^{G^{(s)}}(G^{(s)}_4) \nonumber
\end{equation}
\begin{equation}\label{mapwa}
(d\Omega)^{\rm tot}_{{\rm flux}, s}
=
\underbrace{\beta_{H_s}{\cal Q}_4^{H^{(s)}}(H^{(s)}_3)
+
\beta_{F_s}{\cal Q}_4^{F^{(s)}}(F^{(s)}_2)}_{(d\Omega)_{{\rm flux}, s}}
+
\beta_{G_s}{\cal Q}_4^{G^{(s)}}(G^{(s)}_4) ,
\end{equation}
where note that only after these maps are specified
does it make sense to project the flux-induced pieces onto the torsion classes; and $(\alpha_{\zeta_s}, \beta_{\zeta_s})$ are model-dependent coefficients. The branch-odd sector is more constrained than not arbitrary, but it is also not fully
determined without an explicit local junction model. For example, under branch exchange
\(S^1_+\leftrightarrow S^1_-\), the odd fluxes change sign, so the odd
coefficients must be organized as odd functions of the branch-asymmetry
modulus:
\begin{equation}
\alpha_{\zeta_o}
=
T\,\widehat\alpha_{\zeta_o}
+
O(T^3),
\qquad
\beta_{\zeta_o}
=
T\,\widehat\beta_{\zeta_o}
+
O(T^3),
\label{odd_coeff_parity_short}
\end{equation}
with branch-even hatted coefficients. Whereas in the true IIA endpoint limit, the even and odd branch contributions
must recombine into the single surviving IIA flux sector as
$\alpha_{\zeta_e}+\alpha_{\zeta_o}
\longrightarrow
\alpha_{\rm IIA}$, and
$\beta_{\zeta_e}+\beta_{\zeta_o}
\longrightarrow
\beta_{\rm IIA}$, thus giving us the requisite torsion classes for the supersymmetric case from the maps \eqref{mapwa}. Note that these maps could be motivated from EOMs, Bianchi-identities or consistency conditions, but \eqref{mapwa} contains the full information of the connection between the geometric quantities and the DOFs of Type 0A theory. 
Thus the doubled branchwise reduction that we studied in section \ref{sec001} and also above, should be incorporated in the torsion classes in the following way:
\begin{equation}
W_i^{\rm flux}
=
W_i^{\rm flux,(e)}
+
W_i^{\rm flux,(o)},
\qquad i=1,\dots,5,
\label{Wi_flux_even_odd_split}
\end{equation}
in \eqref{torson},
where the branch-even part is the ordinary IIA-like contribution, and the branch-odd part is the genuinely type-$0$ branch-odd contribution tied to the wedge asymmetry. To see how this works with the map \eqref{mapwa}, we will try the following strategy.
Taking the even combination, the corresponding flux contribution to $W_1$ may be quantified in the following way:
\begin{equation}
\begin{aligned}
W_1^{\rm flux,e}
 & =
-\frac{i}{12}\,
\overline{\Omega}\lrcorner
(dJ)_{\rm flux,e}\\
& =
-\frac{i}{12}\,
\overline{\Omega}\lrcorner
\left[
\alpha_{H_e}\,\mathcal Q_3(H_3^{(e)})
+
\alpha_{F_e}\,{\cal Q}_3(F_2^{(e)})
+
\alpha_{G_e}\,{\cal Q}_3(G_4^{(e)})
\right]
\sim
\Pi_{(3,0)+(0,3)}
\left(\mathcal H_3^{(e)}\right)
+\cdots ,
\end{aligned}
\end{equation}
where $\mathcal H_j^{(s)} \equiv (\delta_{j3}\alpha_{H_s} + \delta_{j4}\beta_{H_s})\mathcal Q_j(H_3^{(s)})$, and similarly 
$\mathcal F_j^{(s)} \equiv (\delta_{j3}\alpha_{F_s} + \delta_{j4}\beta_{F_s})\mathcal Q_j(F_2^{(s)})$ and 
$\mathcal G_j^{(s)} \equiv (\delta_{j3}\alpha_{G_s} + \delta_{j4}\beta_{G_s})\mathcal Q_j(G_4^{(s)})$. To lowest order we expect $\mathcal H^{(s)}_3 \sim H_3^{(s)}$ and $\mathcal G_3^{(s)} \sim 0$; and $\mathcal H_4^{(s)} \sim 0$ but $\mathcal G_4^{(s)}$ is slightly involved (see \eqref{Q4G_definition}). (For gauge fields this is more non-trivial so we will discuss this later).
It is therefore no surprise that 
the branch-even combinations behave like the ordinary IIA reduction and contribute as:
\begin{equation}
W_1^{\rm flux,(e)}
\sim
\Pi_{(3,0)+(0,3)}\!\big(\mathcal H_3^{(e)}\big)\nonumber
\end{equation}
\begin{equation}
W_2^{\rm flux,(e)}
\sim
{\rm L}_J^{-1}\left[\Pi_{\rm prim}^{(2,2)}\!\big(d\Omega\big)_{\rm flux,e}\right]
+
\widetilde a_2\,
\Pi_{\rm prim}^{(1,1)}\!\big(\iota_{v_e} *_7 G_4^{(27,e)}\big)\nonumber
\end{equation}
\begin{equation}
W_3^{\rm flux,(e)}
=
\Pi_{\rm prim}^{(2,1)+(1,2)}\!\big(\mathcal H_3^{(e)}\big)\nonumber
\end{equation}
\begin{equation}
W_4^{\rm flux,(e)} = a_4\,X^{(e)}_\perp,
\qquad
W_5^{\rm flux,(e)} = a_5\,X^{(e)}_\perp ,
\label{W45flux_even}
\end{equation}
where the meaning of all the symbols used here is defined in footnote \ref{primzak}, and ${\rm L}_J$ is the map  $L_J:\Lambda_{\rm prim}^{1,1}\longrightarrow \Lambda_{\rm prim}^{2,2}$ such that
$L_J(\alpha)=J\wedge \alpha$. A more detailed explanation of the various terms appearing in \eqref{W45flux_even} is as follows. Recall that on the \(G_2\) side we decomposed the internal M-theory flux as \eqref{G4decomp_full}
where
$f\in \Lambda^0,
X\in \Lambda^1_{7}$ and 
$G_4^{(27)}\in \Lambda^4_{27}$.
When we pass to the wedge-circle reduction, every branch-dependent quantity can be split into even and odd parts under
$+\leftrightarrow -$. This means the \emph{branch-even part} of the \(\mathbf{27}\)-component of the M-theory \(4\)-form flux becomes:
\begin{equation}\label{g4oodd}
G_4^{(27,e)}
\equiv
\frac12\Big(G_{4,+}^{(27)}+G_{4,-}^{(27)}\Big) ,
\end{equation}
and consequently use this as an input in the construction of $\mathcal G_4^{(27, e)}$. 
Equivalently, it is the part of the primitive/traceless \(\mathbf{27}\) flux that survives in the symmetric combination of the two branches.
However in the following definition:
\begin{equation}\label{MBFlina}
\Pi_{\rm prim}^{(1,1)}\!\big(\iota_{v_e} *_7  G_4^{(27,e)}\big) = 
\Pi_{\rm prim}^{(1,1)}\!\left[{v_e}^M \big(*_7  G_4^{(27,e)}\big)_{MNP}\right],
\end{equation}
one first takes the branch-even \(\mathbf{27}\) flux, then Hodge dualizes it on the \(G_2\) manifold, contracts along the even circle direction \(v_e\), and finally projects onto the primitive \((1,1)\) part on the reduced \(SU(3)\)-structure manifold. This works because 
the two-form can then be decomposed into $SU(3)$ representations\footnote{To see why this is the case, let us elaborate the story a little bit. Let $Y_7$ be the $G_2$-structure manifold and let $v_s$ be the vector associated
with the branch-$s$ reduction direction.  The contraction
$\iota_{v_s}*_7G_4^{(27,s)}$
is a two-form on the six-dimensional space $X_6$ transverse to $v_s$ and therefore belongs to 
$\Lambda^2(X_6)$.
Once the structure group on $X_6$ is reduced to $SU(3)$, every two-form
decomposes into irreducible $SU(3)$ representations.
Using the almost-complex structure, a general two-form $\alpha\in\Lambda^2(X_6)$
decomposes by complex type as
$\alpha
=
\alpha^{(1,1)}
+
\alpha^{(2,0)}
+
\alpha^{(0,2)}$. The $(1,1)$ part further decomposes into a trace part proportional to $J$ and a
primitive part, {\it i.e.}
$\alpha^{(1,1)}
=
\alpha_{\bf 1}
+
\alpha_{\bf 8}$,
where
$\alpha_{\bf 1}
=
\frac{1}{3}(J\lrcorner\alpha)\,J$,
and
$\alpha_{\bf 8}
=
\alpha^{(1,1)}
-
\frac{1}{3}(J\lrcorner\alpha)\,J $.
The primitive condition is
$J\lrcorner\alpha_{\bf 8}=0$.
Thus
$\Lambda^{1,1}
=
\Lambda^{1,1}_{\bf 1}
\oplus
\Lambda_{\rm prim}^{1,1}$.
Under $SU(3)$,
$\Lambda^{1,1}_{\bf 1}\cong {\bf 1}$, and 
$\Lambda_{\rm prim}^{1,1}\cong {\bf 8}$.
The remaining pieces are the $(2,0)$ and $(0,2)$ components.  In complex
dimension three,
$\Lambda^{2,0}\cong \overline{\bf 3}$,
and $\Lambda^{0,2}\cong {\bf 3}$.
Equivalently, the real part
$\Lambda^{2,0}\oplus \Lambda^{0,2}$
is a six-real-dimensional representation, often denoted
by 
${\bf 3}\oplus \overline{\bf 3}$.
Therefore a general two-form decomposes as
$\Lambda^2(X_6)
=
\Lambda^2_{\bf 1}
\oplus
\Lambda^2_{\bf 8}
\oplus
\Lambda^2_{{\bf 3}\oplus\overline{\bf 3}}$.
Applying this to
$\alpha
=
\iota_{v_s}*_7G_4^{(27,s)}$
gives \eqref{su3decompg4}.
Note that 
this is not a decomposition of the original seven-dimensional
$G_2$ representation by itself.  It is the decomposition of the
{six-dimensional two-form}
after reducing from the $G_2$ structure on $Y_7$ to the $SU(3)$ structure on
$X_6$. The dimensions also match: for example
$\dim_{\mathbb R}\Lambda^2(X_6)=15$,
while
$\dim_{\mathbb R}{\bf 1}
+
\dim_{\mathbb R}{\bf 8}
+
\dim_{\mathbb R}({\bf 3}\oplus\overline{\bf 3})
=
1+8+6
=
15$. Thus there are no missing components.}:
\begin{equation}\label{su3decompg4}
\iota_{v_s}*_7G_4^{(27,s)}
=
\left(\iota_{v_s}*_7G_4^{(27,s)}\right)_{\bf 1}
+
\left(\iota_{v_s}*_7G_4^{(27,s)}\right)_{\bf 8}
+
\left(\iota_{v_s}*_7G_4^{(27,s)}\right)_{{\bf 3}\oplus \overline{\bf 3}} ,
\end{equation}
for $s = (e, o)$, where the primitive $(1,1)$ part is exactly the ${\bf 8}$ component, giving credence to \eqref{MBFlina}.
Notice however that we are using $G_4^{(27, e)}$ and {\it not} $\mathcal G_4^{(27, e)}$. The reason is because the map ${\cal Q}_4^{G^{(s)}}:
G_4^{(s)}
\longrightarrow
\Lambda^4(X_6)$  in \eqref{mapwa} is a degree-matching map.  Its purpose is to convert the seven-dimensional
$G_4$-flux data into a four-form on the reduced six-dimensional
$SU(3)$-structure space $X_6$, because $d\Omega$ is a four-form on $X_6$. For the primitive/traceless $G_2$ component
$G_4^{(27,s)}\in \Lambda^4_{27}(Y_7)$,
a natural leading-order contribution to this map is\footnote{First note that $ {\cal Q}_4^{G^{(s)}}\left(G_4^{(27,s)}\right)\equiv \mathcal G_4^{(27, s)}\ne G_4^{(27, s)}$. This is simply because   
$\Omega\in \Lambda^{3,0}(X_6)$,
and therefore
$d\Omega\in \Lambda^4(X_6)$.
By contrast, before reduction the M-theory flux is a four-form on the
seven-dimensional $G_2$-structure manifold $Y_7$, such that 
$G_4\in \Lambda^4(Y_7)$.
Thus $d\Omega$ and $G_4$ are not the same kind of object.  One is the exterior
derivative of the $SU(3)$-structure form on $X_6$; the other is a physical
four-form flux on $Y_7$. Only the pieces of $G_4$ that land in the correct $SU(3)$ representation
can contribute to a given torsion class.  Since $W_2$ is a primitive $(1,1)$
form,
$W_2\in \Lambda_{\rm prim}^{1,1}(X_6)$,
the corresponding piece of $d\Omega$ must be of the form
$J\wedge W_2
\in
\Lambda_{\rm prim}^{2,2}(X_6)$. We should also take into consideration that in the non-supersymmetric type 0A case, the relation
$(d\Omega)_{G_4,s}
=
{\cal Q}_4^{G^{(s)}}\!\left(G_4^{(s)}\right)$
should be presented as a \emph{flux-induced torsion ansatz} or as the
leading-order result of solving the backreacted equations of motion.  It is not
a universal identity.  More explicitly, one should write
$(d\Omega)_{{\rm flux},s}$ as in \eqref{mapwa}
where the maps ${\cal Q}_4$ are fixed either by an explicit dimensional
reduction of the equations of motion, or by a controlled leading-order
backreaction ansatz. For the primitive $G_2$ component $G_4^{(27,s)}$, a natural leading
representation-compatible choice is \eqref{Q4G_definition}.
This formula is justified at the level of representation theory and degree
matching.  It says that the $G_4^{(27,s)}$ flux component contributes to the
primitive $(2,2)$ part of $d\Omega$, and therefore to $W_2$.  But the numerical
coefficient $\gamma_G^{(s)}$, and even whether this contribution is present in
a particular compactification, must be determined from the explicit reduction or
from the backreacted equations of motion. All in all it means that the symbol ${\cal Q}_4^{G^{(s)}}$ is not trivial. It encodes the sequence of operations required to map the seven-dimensional
$G_4$ flux to a four-form on $X_6$ in the appropriate $SU(3)$ representation
channel contributing to $d\Omega$, with the understanding that such a map is
not merely kinematical but must be supported by the underlying equations of
motion or by a controlled reduction ansatz.}:
\begin{equation}
{\cal Q}_4^{G^{(s)}}\left(G_4^{(27,s)}\right)
=
\gamma_G^{(s)}\,
J\wedge
\Pi_{\rm prim}^{(1,1)}
\left(
\iota_{v_s}*_7G_4^{(27,s)}
\right),
\label{Q4G_definition}
\end{equation}
where $\gamma_G^{(s)}$ is a normalization coefficient that may be absorbed into
$\beta_{G_s}$ or into $\widetilde a_2^{(s)}$. In the language of primitive $(1, 1)$ form, this at least provides a reason for the choice of $G_4^{(27, e)}$ in the $W_2$ torsion class \eqref{W45flux_even}. We can make this more precise in the following way.
In the chain of maps represented via:
\begin{equation}
G_4^{(27,s)}
\in \Lambda^4_{27}(Y_7)
\xrightarrow{\ *_7\ }
*_7G_4^{(27,s)}
\in \Lambda^3_{27}(Y_7)
\xrightarrow{\ \iota_{v_s}\ }
\iota_{v_s}*_7G_4^{(27,s)}
\in \Lambda^2(X_6) ,
\end{equation}
where the last bit comes from the contraction operation $\iota_{v_s}$ for any vector $v_s^M$, we can see easily that we can 
projects this two-form onto the primitive $(1,1)$ component in the following way:
\begin{equation}
\Pi_{\rm prim}^{(1,1)}
\left(
\iota_{v_s}*_7G_4^{(27,s)}
\right)
\in
\Lambda_{\rm prim}^{1,1}(X_6).
\end{equation}
Since $W_2$ is a primitive $(1,1)$ form, this is precisely the correct
representation channel for $W_2$.  However, $d\Omega$ is a four-form.  Therefore
the corresponding contribution to $d\Omega$ is obtained by wedging with $J$:
\begin{equation}\label{maychokh}
J\wedge
\Pi_{\rm prim}^{(1,1)}
\left(
\iota_{v_s}*_7G_4^{(27,s)}
\right)
\in
\Lambda_{\rm prim}^{2,2}(X_6) ,
\end{equation}
which explains the appearance of the factor $J\wedge$ in
\eqref{Q4G_definition}. Thus the $G_4^{(27,s)}$ contribution to $d\Omega$ is not independent of the
map ${\cal Q}_4^{G^{(s)}}$.  Rather, it is the explicit leading-order form of
that map
$(d\Omega)_{G_4,s}
=
\beta_{G_s}\,
{\cal Q}_4^{G^{(s)}}\left(G_4^{(27,s)}\right)$ in \eqref{mapwa}
with
${\cal Q}_4^{G^{(s)}}\left(G_4^{(27,s)}\right)$
exactly given by \eqref{maychokh} with a proportionality constant 
$\gamma_G^{(s)}$.
Equivalently, defining
$\widetilde a_2^{(s)}
\equiv
\beta_{G_s}\gamma_G^{(s)}$, 
one now obtains
$(d\Omega)_{G_4,s}$ to be proportional to \eqref{maychokh} with the proportionality constant
$\widetilde a_2^{(s)}$.
Now recall the $W_2$ extraction formula.  Since $d\Omega$ takes the form \eqref{dJtorsion}, 
the primitive $(2,2)$ part of $d\Omega$ is $J \wedge W_2$. This immediately gives us:
\begin{equation}\label{capquebetag}
W_2^{G_4,s}
=
{\rm L}_J^{-1}
\left[
\Pi_{\rm prim}^{(2,2)}
\left(
\widetilde a_2^{(s)}
J\wedge
\Pi_{\rm prim}^{(1,1)}
\left(
\iota_{v_s}*_7G_4^{(27,s)}
\right)
\right)
\right] ,
\end{equation}
where the ${\rm L}_J$ map has been defined earlier in \eqref{W45flux_even}. Looking at the nested form structure in \eqref{capquebetag}, some simplifications easily follow. The primitive (1, 1) form inside the first bracket belongs to 
$\Lambda_{\rm prim}^{1,1}(X_6)$,
so wedging it with $J$ already gives a primitive $(2,2)$ form belonging to 
$\Lambda_{\rm prim}^{2,2}(X_6)$.
Hence the primitive projector acts trivially on this term:
\begin{equation}\label{primitmagic}
\Pi_{\rm prim}^{(2,2)}
\left[
J\wedge
\Pi_{\rm prim}^{(1,1)}
\left(
\iota_{v_s}*_7G_4^{(27,s)}
\right)
\right]
=
J\wedge
\Pi_{\rm prim}^{(1,1)}
\left(
\iota_{v_s}*_7G_4^{(27,s)}
\right).
\end{equation}
Plugging \eqref{primitmagic} in \eqref{capquebetag}, and using the fact that the ${\rm L}^{-1}_j$ map naturally satisfy 
${\rm L}_J^{-1}(J\wedge \alpha)=\alpha$
for 
$\alpha\in\Lambda_{\rm prim}^{1,1}(X_6)$, 
we immediately get the second contribution to $W_2$ in \eqref{W45flux_even} expressed in terms of $G_4^{(27, e)}$.

For \(X_\perp^{(e)}\) we are taking the \emph{horizontal branch-even part} of the \(\mathbf{7}\)-component \(X\). More explicitly, \(X\) is a \(1\)-form in the \(\mathbf 7\) of \(G_2\). Relative to the reduction direction, it can be decomposed into a piece along the circle and a piece tangent to the six-dimensional space, {\it i.e.}
$X = X_{\parallel}\,\eta + X_\perp$,
where
$X_\perp$ is the component orthogonal to the reduction one-form \(\eta\), i.e. the part tangent to the six-dimensional internal manifold.
Then:
\begin{equation}\label{xee}
X_\perp^{(e)}
\equiv
\frac{(X_\perp)_+ + (X_\perp)_-}{2}
\end{equation}
is its branch-even part.
So \(X_\perp^{(e)}\) is simply the symmetric horizontal \(1\)-form descending from the \(\mathbf 7\) flux sector, and it contributes to the Lee-form torsion classes:
\begin{equation}
W_4^{\rm flux,(e)} = a_4\,X_\perp^{(e)},
\qquad
W_5^{\rm flux,(e)} = a_5\,X_\perp^{(e)}.
\end{equation}
On the other hand, the branch-odd combinations are absent in an ordinary smooth-circle reduction and are the new ingredient of the wedge compactification. Since they are odd under
$+\leftrightarrow -$
they must vanish at the symmetric point
$R_+=R_-$ where $T=0$.
Therefore their contribution necessarily starts at linear order in the branch-odd parameter \(T\) or its derivatives:
\begin{equation}
W_i^{\rm flux,(o)}
=
T\,{\cal W}_i^{(o)}
+
(\partial T)\cdot {\cal Y}_i^{(o)}
+
O(T^2) ,
\label{Wi_flux_odd_general}
\end{equation}
which is a generic ans\"atze so we have carefully implement the second term $\mathcal Y_i^{(o)}$ for $i = 1, 2, .., 5$. 
More explicitly, the degree-compatible form is:
\begin{equation}
\begin{aligned}
& W_1^{\rm flux,(o)}
=
T\,{\cal W}_1^{(o)}
+
\iota_{V_T}{\cal Y}_1^{(o)}
+
O(T^2)\\
& W_2^{\rm flux,(o)}
=
T\,{\cal W}_2^{(o)}
+
dT\wedge {\cal Y}_2^{(o)}
+
O(T^2)\\
& W_3^{\rm flux,(o)}
=
T\,{\cal W}_3^{(o)}
+
dT\wedge {\cal Y}_3^{(o)}
+
O(T^2),
\end{aligned}
\end{equation}
\begin{equation}
W_4^{\rm flux,(o)}
=
T\,{\cal W}_4^{(o)}
+
c_4^{(o)}\,dT
+
O(T^2),
\qquad
W_5^{\rm flux,(o)}
=
T\,{\cal W}_5^{(o)}
+
c_5^{(o)}\,dT
+
O(T^2) , \nonumber
\label{W45flux_odd_general}
\end{equation}
which keeps track of the fact that $W_i \in \Lambda^i$ for $i = 1, .., 3$ and $W_4, W_5 \in \Lambda^1$ (or alternatively \(W_1\) is a scalar, \(W_2\) is a primitive \((1,1)\)-form,
\(W_3\) is a primitive \((2,1)+(1,2)\)-form, and \(W_4,W_5\) are
one-forms). The symbol
$\iota_{V_T}$
denotes the \emph{interior product} (or contraction) with the vector field \(V_T\), which is the vector dual to the one-form \(dT\), namely
$(V_T)^\mu \equiv g^{\mu\nu}\,\partial_\nu T$.
So if
${\cal Y}_1^{(o)}$
is a one-form, then
$\iota_{V_T}{\cal Y}_1^{(o)}$
means contract the vector \(V_T\) into that one-form, producing a scalar: 
$\iota_{V_T}{\cal Y}_1^{(o)}
=
(V_T)^\mu\,({\cal Y}_1^{(o)})_\mu$.
This is why it can appear in \(W_1\), since
$W_1$
is a scalar torsion class and cannot contain a wedge product with \(dT\). Note also that 
the coefficient forms \({\cal W}_i^{(o)}\) and \({\cal Y}_i^{(o)}\) are built from the branch-odd flux combinations:
\begin{equation}
H_3^{(o)},\qquad F_4^{(o)},\qquad G_4^{(27,o)},\qquad X_\perp^{(o)} ,
\end{equation}
defined, for example, in \eqref{g4oodd} and \eqref{xee} by taking the minus combinations. In particular, the primitive branch-odd pieces provide additional contributions to the $W_2$ and $W_3$ torsion as:
\begin{equation}
W_{\rm 2, add}^{\rm flux, (o)} = \widetilde a_2^{(o)}\,\Pi_{\rm prim}^{(1,1)}\!\big(\iota_{v_o} *_7 G_4^{(27,o)}\big), ~~~W_{\rm 3, add} ^{\rm flux,(o)}
=
\Pi_{\rm prim}^{(2,1)+(1,2)}\!\big(\mathcal H_3^{(o)}\big),
\label{W3flux_odd_precise}
\end{equation}
where $\iota_{v_o} *_7 G_4^{(27,o)} =v_o^M\big(*_7 G_4^{(27,o)}\big)_{MNP}$ is defined in somewhat similar way as before, namely  one first takes the branch-odd \(\mathbf{27}\) flux, then Hodge dualizes it on the \(G_2\) manifold, contracts along the odd circle direction \(v_0\), and finally projects onto the primitive \((1,1)\) part on the reduced \(SU(3)\)-structure manifold. Finally, the odd \(\mathbf 7\) component contributes to the Lee-form classes:
\begin{equation}
W_4^{\rm flux,(o)}\sim a_4^{(o)} X_\perp^{(o)},
\qquad
W_5^{\rm flux,(o)}\sim a_5^{(o)} X_\perp^{(o)}.
\label{W45flux_odd_precise}
\end{equation}
Therefore the correct type-$0$ version of the earlier IIA-like discussion is not
$G_4=\eta_b\wedge H_3+F_4^{\rm int}$
with a \emph{single} six-dimensional flux sector as in IIA, but rather the branchwise decomposition \eqref{GHFetc} together with the even/odd recombination \eqref{HFfatu}. The flux contribution to the $SU(3)$ torsion should then be written as \eqref{Wi_flux_even_odd_split}
with:
\begin{equation}\label{toruequ2}
\begin{aligned}
 W_4^{\rm flux}
&=
a_4\,X_\perp^{(e)}
+
a_4^{(o)}\,X_\perp^{(o)}
+
c_4^{(o)}\,dT
+
O(T^2)\\[4pt]
 W_5^{\rm flux}
&=
a_5\,X_\perp^{(e)}
+
a_5^{(o)}\,X_\perp^{(o)}
+
c_5^{(o)}\,dT
+
O(T^2)\\[4pt]
 W_1^{\rm flux}
&=
\Pi_{(3,0)+(0,3)}\!\big(\mathcal H_3^{(e)}\big)
+
T\,{\cal W}_1^{(o)}
+
\iota_{V_T}{\cal Y}_1^{(o)}
+
O(T^2)\\[4pt]
 W_3^{\rm flux}
&=
\Pi_{\rm prim}^{(2,1)+(1,2)}\!\big(\mathcal H_3^{(e)}\big)
+
\Pi_{\rm prim}^{(2,1)+(1,2)}\!\big(\mathcal H_3^{(o)}\big)
+
O(T^2)\\[4pt]
& + 
T\,{\cal W}_3^{(o)}
+
dT\wedge {\cal Y}_3^{(o)}\\[4pt]
 W_2^{\rm flux}
&=
{\rm L}_J^{-1}\left[\Pi_{\rm prim}^{(2,2)}\!\big(d\Omega\big)_{\rm flux,e}\right]
+
\widetilde a_2\,\Pi_{\rm prim}^{(1,1)}\!\big(\iota_{v_e} *_7 G_4^{(27,e)}\big) +
dT\wedge {\cal Y}_2^{(o)}\\
&+
T\,{\cal W}_2^{(o)} +
\widetilde a_2^{(o)}\,\Pi_{\rm prim}^{(1,1)}\!\big(\iota_{v_o} *_7 G_4^{(27,o)}\big)
+
O(T^2),
\end{aligned}
\end{equation}
by combining the even and odd cases from \eqref{W45flux_even}, \eqref{W45flux_odd_general} and \eqref{W3flux_odd_precise} respectively.
The even combinations reproduce the ordinary IIA-like torsion contributions, while the odd combinations are the genuinely type-$0$ data tied to the branch asymmetry and the tachyon modulus \(T\). This is the closed-string flux reflection of the type-$0$ doubling. 

A final caveat is important. The wedge-circle/type-$0$A proposal is recent, and the follow-up orientifold analysis explicitly notes that some type-$0$A open-string sectors do not yet have a complete microscopic M-theory derivation. So the branchwise doubled closed-string flux analysis above is the natural quantitative implementation of the proposal, but it should still be read as the appropriate conjectural refinement of the ordinary IIA formulas, not as a fully settled microscopic derivation.

\subsubsection{Pinch contribution to the torsion classes \label{pinchu2}}

Looking at the torsion classes from the fibration and the flux contributions in \eqref{toruequ} and \eqref{toruequ2}, we notice that once the flux sector is treated correctly on the type-\(0\)A side, the pinch contribution should also be written in a slightly more careful way. The main conceptual point is the following. The pinch is still controlled by the branch-odd scalar:
\begin{equation}
T=\frac{R_+-R_-}{2},
\end{equation}
so its \emph{direct} effect on the reduced \(SU(3)\)-structure torsion classes still comes through \(T\) and \(dT\), exactly as before. However, after taking into account the Type \(0\)A interpretation of the wedge reduction, the pinch contribution should \emph{not} be viewed as replacing the branch-odd flux sector. Rather, it provides an additional branch-odd source with the same degree structure. Therefore the clean decomposition is no longer \eqref{torson} with \(W_i^{\rm flux}\) understood as a single ordinary IIA-like flux sector, but instead of the following two equivalent ways:
\begin{equation}
W_i
=
W_i^{\rm fib}
+
W_i^{\rm flux,(e)}
+
W_i^{\rm flux,(o)}
+
W_i^{\rm pinch},
\qquad i=1,\dots,5 \nonumber
\label{Wi_master_corrected_min}
\end{equation}
\begin{equation}
W_i
=
W_i^{\rm fib}
+
W_i^{\rm flux+pinch},
\qquad
W_i^{\rm flux+pinch}
\equiv
W_i^{\rm flux,(e)}+W_i^{\rm flux,(o)}+W_i^{\rm pinch}.
\label{Wi_master_corrected_min2}
\end{equation}
In other words, after the Type \(0\)A reinterpretation, the odd-flux sector should \emph{still} be retained, while the pinch sector adds a new contribution on top of it. This point will be important for us.

To proceed, first let us recall the M-theory origin of the pinch term. This is elaborated in section \ref{sec001}, where the pinch term appears from the odd combination of \(R_+\) and \(R_-\). On the seven-dimensional \(G_2\)-structure side, the direct effect of the wedge asymmetry is more precisely captured by:
\begin{equation}
\tau_3^{\rm pinch}
=
T\,\delta_{\Sigma_6^{\rm jct}}\,\Xi^{(3)}_{27}
+
O(T^2),
\qquad
\Xi^{(3)}_{27}\in \Lambda^3_{27},
\label{tau3pinch_minimal}
\end{equation}
whose explicit form appears in \eqref{torwedge}; see also section \ref{naturalpinch}. Thus the pinch contribution is distributional on the M-theory side, supported on the junction locus \(\Sigma_6^{\rm jct}\subset M_7\). This part does \emph{not} change: the branch-odd scalar \(T\) is still the unique pinch modulus, and its direct descendant still feeds the reduced torsion through the primitive \((2,1)+(1,2)\) sector. Only after reducing along the pinched direction is the localized current integrated out into an effective \(T\)-dependent deformation of the \(SU(3)\) structure.

After reduction along the pinched direction, the localized current on the wedge fiber is integrated out, so what survives on \(M_6\) is a deformation of the reduced \(SU(3)\) structure given by \eqref{JOmega_expand_T}, such that \(dJ\) and \(d\Omega\) satisfy  \eqref{dOmegaexpandT}. The \(dT\)-dependent terms in these expressions are simply the chain-rule contributions coming from the \(T\)-dependence of the reduced structure.

The next question is how the pinch term relates to the odd-flux effects. The correct statement is now slightly different from what was analyzed earlier from M-theory. In the wedge/Type \(0\)A interpretation, the branchwise split of the M-theory \(G^{\rm br}_4\) flux, given by \eqref{GHFetc}, provides only one physical NSNS flux \(H_3^{(e)}\), while the branch-odd combination \(H_3^{(o)}\) is not an independent physical NSNS field. Rather, it is constrained by the wedge geometry and the tachyon background. Thus one may still regard it schematically as in \eqref{H3oddderived0A}, and similarly for the branch-odd pieces of the \(\mathbf 7\) and \(\mathbf{27}\) flux sectors:
\begin{equation}
X_\perp^{(o)}
=
T\,{\cal X}_1^{(1)}
+
c_X\,dT
+
{\cal O}(T^2),
\qquad
G_4^{(27,o)} 
=
T\,{\cal G}_{4,27}^{(1)}
+
dT\wedge {\cal G}_{3,27}^{(1)}
+
{\cal O}(T^2),
\label{G27odd_as_Tdata_min}
\end{equation}
where \(X_\perp^{(o)}\) can be defined from \eqref{xee} by taking the odd combination of \((X_\perp)_\pm\); and $\mathcal G_{j,27}^{(1)}$ appear from non-trivial mappings. However, this does \emph{not} imply that the pinch term absorbs and replaces the odd-flux sector. What it implies is that both sectors are controlled by the same branch-odd data \((T,dT)\), so that the pinch contribution adds to the odd-flux contribution in a compatible way.

With this understood, the corrected pinch/\(T\)-dependent pieces may be collected as follows. We start with the scalar class \(W_1\). Since \(W_1\in \Lambda^0\), one cannot write a universal wedge product with \(dT\). The pinch contribution therefore takes the form
\begin{equation}
W_1^{\rm pinch}
=
T\,W_1^{(T,{\rm pinch})}
+
\iota_{V_T}Y_1^{({\rm pinch})}
+
{\cal O}(T^2),
\label{deltaW1_corrected_min}
\end{equation}
where
$(V_T)^m \equiv g^{mn}\partial_n T$
is the vector dual to \(dT\), \(\iota_{V_T}Y_1^{({\rm pinch})}=(V_T)^\mu (Y_1^{({\rm pinch})})_\mu\), and \(Y_1^{({\rm pinch})}\in \Lambda^1(M_6)\) is a one-form built from the reduced geometry and the pinch-induced branch-odd data. This should be combined additively with the odd-flux contribution from \eqref{toruequ2}
so that the total odd contribution to \(W_1\) may be expressed succinctly as:
\begin{equation}
W_1^{\rm odd,total}
=
T\,{\bf W}_1^{(o)}
+
\iota_{V_T}{\bf Y}_1^{(o)}
+
O(T^2) \nonumber
\label{W1oddtotal_min}
\end{equation}
\begin{equation}
{\bf W}_1^{(o)}
=
{\cal W}_1^{(o)}+W_1^{(T,{\rm pinch})},
\qquad
{\bf Y}_1^{(o)}
=
{\cal Y}_1^{(o)}+Y_1^{({\rm pinch})} ,
\label{W1coeffmap_min}
\end{equation}
where the superscript odd,total simply stands for the combined odd and pinch terms from \eqref{Wi_master_corrected_min2}. (We will follow this nomenclature throughout but the meaning should be clear from the master equation \eqref{Wi_master_corrected_min2}.)
In a similar vein, the pinch contribution to the two-form class \(W_2\) may be expressed in the following way:
\begin{equation}
W_2^{\rm pinch}
=
T\,W_2^{(T,{\rm pinch})}
+
dT\wedge \widehat W_2^{({\rm pinch})}
+
{\cal O}(T^2),
\qquad
\widehat W_2^{({\rm pinch})}\in \Lambda^1(M_6).
\label{deltaW2_corrected_min}
\end{equation}
Here \(\widehat W_2^{({\rm pinch})}\) is not a new independent torsion class; it is the one-form coefficient multiplying \(dT\) in the derivative correction to \(W_2\). More concretely, since \(W_2\in\Lambda^2(M_6)\), any term linear in \(dT\) must be of the form \(dT\wedge(\text{one-form})\). Starting from the \(T\)-dependent \(SU(3)\)-structure forms \(J(T),\Omega(T)\) from \eqref{JOmega_expand_T}, one compares
$d\Omega
=
d\Omega^{(0)}
+
T\,d\Omega^{(1)}
+
dT\wedge \Omega^{(1)}
+
{\cal O}(T^2)$
with the standard decomposition
$d\Omega
=
W_1\,J\wedge J
+
W_2\wedge J
+
\overline W_5\wedge \Omega$.
Since \(W_2\) is extracted from the primitive \((1,1)\) part of \(d\Omega\), the \(dT\)-dependent contribution to \(W_2\) comes from the primitive \((1,1)\) projection of \(dT\wedge \Omega^{(1)}\). By definition:
\begin{equation}
dT\wedge \widehat W_2^{({\rm pinch})}
\end{equation}
is the part of this expression that contributes to \(W_2\wedge J\). Equivalently, \(\widehat W_2^{({\rm pinch})}\) may be viewed as the one-form extracted from the pinch-induced derivative correction. This should now be added to the odd-flux part of $W_2$ from \eqref{toruequ2}
so that:
\begin{equation}
W_2^{\rm odd,total}
=
T\,{\bf W}_2^{(o)}
+
dT\wedge {\bf Y}_2^{(o)}
+
\widetilde a_2^{(o)}\,
\Pi_{\rm prim}^{(1,1)}\!\big(\iota_{v_o} *_7  G_4^{(27,o)}\big)
+
O(T^2) \nonumber
\label{W2oddtotal_min}
\end{equation}
\begin{equation}
{\bf W}_2^{(o)}
=
{\cal W}_2^{(o)}+W_2^{(T,{\rm pinch})},
\qquad
{\bf Y}_2^{(o)}
=
{\cal Y}_2^{(o)}+\widehat W_2^{({\rm pinch})}.
\label{W2coeffmap_min}
\end{equation}
Notice that the reduced odd \(\mathbf{27}\)-flux term remains as part of the flux sector; it has no direct pinch analogue at leading order. There is a simple reason behind it: The odd \(\mathbf{27}\)-flux term is a genuine \emph{flux} contribution coming from the branch-odd part of the M-theory \(4\)-form
$G_4^{(27,o)}$. After Hodge duality and contraction, it contributes to \(W_2\) through the primitive \((1,1)\) projection as shown above.
So this is a smooth bulk flux effect in the odd \(\mathbf{27}\)-sector.
By contrast, the pinch source comes from the localized torsion current
\eqref{tau3pinch_minimal}
whose natural descendant after reduction lies in the primitive \((2,1)+(1,2)\) channel, namely the \(W_3\) sector that we discuss in the following.

We now come to the three-form class \(W_3\). This is the most important one, because it contains the direct descendant of \(\tau_3^{\rm pinch}\) that we studied in \eqref{tauteen} and \eqref{torwedge} for M-theory on the \(G_2\)-structure manifold; see also section \ref{naturalpinch}. For the present case, the pinch contribution is:
\begin{equation}
W_3^{\rm pinch}
=
b_3\,T\,
\Pi_{\rm prim}^{(2,1)+(1,2)}
\!\big(\iota_{v_b}\Xi^{(4)}_{27}\big)
+
dT\wedge \widehat W_3^{({\rm pinch})}
+
{\cal O}(T^2),
\label{deltaW3_corrected_min}
\end{equation}
where $\Xi_{27}^{(4)}\in \Lambda^4_{27}(M_7),~~
\Xi_{27}^{(3)}=*_{7}\Xi_{27}^{(4)}$; and 
$\widehat W_3^{({\rm pinch})}\in \Lambda^2(M_6)$
contains the chain-rule contribution from \(J(T)\) together with the pinch-induced derivative correction\footnote{As a reminder to the readers, the projection \(\Pi_{\rm prim}^{(2,1)+(1,2)}\) means the following: first take the \((2,1)+(1,2)\) part with respect to the reduced \(SU(3)\) structure, and then project to the primitive piece. For a \(3\)-form \(\beta_3\) on an \(SU(3)\)-structure manifold, the decomposition is schematically:
\begin{equation}
\beta_3
=
\beta_3^{(3,0)+(0,3)}
+
\beta^{(2,1)+(1,2)}_{\rm prim}
+
J\wedge \lambda_1, \nonumber
\end{equation}
where \(\lambda_1\) is a one-form. Then
$\Pi_{\rm prim}^{(2,1)+(1,2)}(\beta_3)
=
\beta^{(2,1)+(1,2)}_{\rm prim}$.
Equivalently, it is the part of \(\beta_3\) satisfying
$J\wedge \Pi_{\rm prim}^{(2,1)+(1,2)}(\beta_3)=0$,
with no \((3,0)\) or \((0,3)\) component. Thus
$\Pi_{\rm prim}^{(2,1)+(1,2)}\!\big(\iota_{v_b}\Xi_{27}\big)$
means: take the \(3\)-form \(\iota_{v_b}\Xi_{27}\), decompose it under \(SU(3)\), and keep only its primitive \((2,1)+(1,2)\) part.
The contraction \(\iota_{v_b}\) denotes interior contraction with the vector field \(v_b\), which is dual to the one-form \(\eta_b\) along the pinched-circle reduction direction. Concretely, if \(\Xi^{(4)}_{27}\) is written as a \(4\)-form on the \(G_2\) side, then
$(\iota_{v_b}\Xi^{(4)}_{27})_{mnp}
=
(v_b)^q\,(\Xi^{(4)}_{27})_{qmnp}$.
If instead \(\Xi^{(3)}_{27}\) is already understood as a reduced \(3\)-form descendant, then \(\iota_{v_b}\Xi^{(3)}_{27}\) is simply a schematic notation for extracting the component of the \(G_2\) \(\mathbf{27}\)-source relevant to the six-dimensional \(W_3\) torsion.
Earlier, one also writes the singular part of \(d\varphi\) as
$d\varphi\big|_{\rm pinch,sing}
=
T\,\delta_{\Sigma_6^{\rm jct}}\,n\wedge \rho_3$,
which is a \(4\)-form. Since
$d\varphi\big|_{\rm pinch,sing}=*_7\tau_3^{\rm pinch}$,
we get
$\tau_3^{\rm pinch}
=
T\,\delta_{\Sigma_6^{\rm jct}}\,*_7(n\wedge \rho_3)$.
Defining
$\Xi_{27}^{(3)}
\equiv *_7(n\wedge \rho_3)=*_6\rho_3$,
one obtains
$\tau_3^{\rm pinch}
=
T\,\delta_{\Sigma_6^{\rm jct}}\,\Xi_{27}^{(3)}$ with 
$\Xi_{27}^{(3)}\in \Lambda^3_{27}$.
This is the convention used in \eqref{tau3pinch_minimal}.
But one could instead keep the \(4\)-form
$\Xi_{27}^{(4)}
\equiv n\wedge \rho_3
\in \Lambda^4_{27}(M_7)$,
so that
$d\varphi\big|_{\rm pinch,sing}
=
T\,\delta_{\Sigma_6^{\rm jct}}\,\Xi_{27}^{(4)}$.
Then
$\tau_3^{\rm pinch}
=
T\,\delta_{\Sigma_6^{\rm jct}}\,*_7\Xi_{27}^{(4)}$.
In this convention, if one reduces along the \(b\)-direction, one may write expressions involving
$\iota_{v_b}\Xi_{27}^{(4)}$,
which is indeed a \(3\)-form. \label{3or4}}. 
Here the explicit \(\delta_{\Sigma_6^{\rm jct}}\) has disappeared because the reduction has already integrated out the localized current on the pinched direction; what survives is its effective \(T\)-dependent descendant on \(M_6\).

Finally, \(\widehat W_3^{({\rm pinch})}\) is the coefficient of the derivative-of-\(T\) correction to \(W_3\). Since \(W_3\in\Lambda^3(M_6)\) and \(dT\in\Lambda^1(M_6)\), the object multiplying \(dT\) must be a \(2\)-form, {\it i.e.}
$\widehat W_3^{({\rm pinch})}\in \Lambda^2(M_6)$.
Quantitatively, one may define it by:
\begin{equation}
dT\wedge \widehat W_3^{({\rm pinch})}
=
\Pi_{\rm prim}^{(2,1)+(1,2)}\!\big(dT\wedge J^{(1)}\big)
+
\cdots,
\label{W3hatdef_min}
\end{equation}
where the dotted terms denote further derivative corrections.
This pinch piece must then be added to the odd-flux contribution from \eqref{toruequ2} so that the total odd contribution becomes:
\begin{equation}
W_3^{\rm odd,total}
=
\Pi_{\rm prim}^{(2,1)+(1,2)}\!\big(H_3^{(o)}\big)
+
T\,{\bf W}_3^{(o)}
+
dT\wedge {\bf Y}_3^{(o)}
+
O(T^2) \nonumber
\label{W3oddtotal_min}
\end{equation}
\begin{equation}
{\bf W}_3^{(o)}
=
{\cal W}_3^{(o)}
+
b_3\,\Pi_{\rm prim}^{(2,1)+(1,2)}\!\big(\iota_{v_b}\Xi^{(4)}_{27}\big),
\qquad
{\bf Y}_3^{(o)}
=
{\cal Y}_3^{(o)}
+
\widehat W_3^{({\rm pinch})}.
\label{W3coeffmap_min}
\end{equation}
For the Lee-form classes \(W_4,W_5\), we can use the parametrization that we advocated in \eqref{zetaabeta}, namely $\zeta =\frac{R_+}{R_-}e^{-R_+/R_-}$ and $\eta = \frac{R_-}{R_+}e^{-R_-/R_+}$ to construct a scalar and from there define a one-form. We will also need a branch-odd parameter. A possible candidate is the following\footnote{ Before using $\varepsilon$ and $T$ simultaneously in the torsion formulas,
we should clarify their relation. Recall that
$T=\frac{R_+-R_-}{2}$,
and $R_B=\frac{R_++R_-}{2}$,
so that near the symmetric point one may write
$R_+=R_B+T,
R_-=R_B-T$, and 
$\delta\equiv \frac{T}{R_B}$.
The regular branch-odd parameter in~\eqref{epsregdef_min} is odd under
$R_+\leftrightarrow R_-$, or equivalently under $T\mapsto -T$. Therefore
its Taylor expansion around $T=0$ at fixed $R_B$ contains only odd powers
of $T$. Let:
\begin{equation}
f(x)=x e^{-x},
\qquad
x=\frac{R_+}{R_-}=\frac{1+\delta}{1-\delta} ,\nonumber
\end{equation}
which vanishes for both $x \to 0$ and $x \to \infty$.
Then
$\varepsilon
=
f(x)-f(x^{-1})$.
Since $f'(1)=e^{-1}(1-1)=0$, the linear term in the expansion around
$x=1$ vanishes. Expanding to third order gives:
\begin{equation}
\varepsilon
=-{4\over e}\left({T\over R_B}\right)
-\frac{8}{3e}
\left(\frac{T}{R_B}\right)^3
+
O\!\left(\frac{T^5}{R_B^5}\right). \nonumber
\label{eps_vs_T}
\end{equation}
Thus $\varepsilon$ is linearly related to $T$ near the symmetric wedge
point. To next order, it is suppressed by two extra powers of $T/R_B$ relative to $T$:
$\varepsilon = O\!\left(\left(\frac{T}{R_B}\right)^3\right)$.
Consequently, the simultaneous appearance of $T$ and $\varepsilon$ in the
torsion formulas is not redundant. The parameter $T$ organizes the leading
branch-odd expansion around the symmetric wedge point, while $\varepsilon$
captures a regular branch-odd combination that is especially useful when
discussing the approach to the Type IIA endpoints.}:
\begin{equation}
\varepsilon
\equiv \zeta - \eta = 
\frac{R_+}{R_-}e^{-R_+/R_-}
-
\frac{R_-}{R_+}e^{-R_-/R_+},
\label{epsregdef_min}
\end{equation}
which is odd under
$R_+\leftrightarrow R_-$ because 
$\varepsilon\mapsto -\,\varepsilon$,
and vanishes at the symmetic point
$R_+=R_-$
as well as remaining finite at the Type IIA endpoints. Using this, we can construct a one-form $d\epsilon$, so that we can parametrize the pinch contributions to the Lee-form classes as:
\begin{equation}
W_4^{\rm pinch}
=
b_4\,d\varepsilon
+
\varepsilon\,W_4^{(\varepsilon,{\rm pinch})}
+
O(\varepsilon^2),
\label{W4pinch_min}
\end{equation}
\begin{equation}
W_5^{\rm pinch}
=
b_5\,d\varepsilon
+
\varepsilon\,W_5^{(\varepsilon,{\rm pinch})}
+
O(\varepsilon^2).
\label{W5pinch_min}
\end{equation}
These should again be \emph{added} to the odd-flux pieces rather than used to replace them. Thus the total odd Lee-form contributions become
\begin{equation}
W_4^{\rm odd,total}
=
a_4^{(o)}X_\perp^{(o)}
+
c_4^{(o)}\,dT
+
T\,{\cal W}_4^{(o)}
+
b_4\,d\varepsilon
+
\varepsilon\,W_4^{(\varepsilon,{\rm pinch})}
+
O(T^2,\varepsilon^2),
\label{W4oddtotal_min}
\end{equation}
\begin{equation}
W_5^{\rm odd,total}
=
a_5^{(o)}X_\perp^{(o)}
+
c_5^{(o)}\,dT
+
T\,{\cal W}_5^{(o)}
+
b_5\,d\varepsilon
+
\varepsilon\,W_5^{(\varepsilon,{\rm pinch})}
+
O(T^2,\varepsilon^2).
\label{W5oddtotal_min}
\end{equation}
Therefore to conclude, 
the pinch contribution does not absorb and replace the odd-flux contribution;
instead,
it adds a new branch-odd source with the same degree structure.
Equivalently, the total odd sector is:
\begin{equation}
W_i^{\rm odd,total}
=
W_i^{\rm flux,(o)}
+
W_i^{\rm pinch},
\qquad i=1,\dots,5.
\end{equation}
This is the form that should be combined smoothly with the even sector contributions and with the fibration contributions. In the following 
subsection we will elaborate on the  combined fibration, flux and pinch contributions to the \(SU(3)\)-structure torsion classes, but before turning to that, let us first address an important question: why can the pinch affect more than just $W_3$?

\subsubsection{Why the pinch can affect more than just \(W_3\)?\label{sec314}}

The short point is that $\tau_3^{\rm pinch} \ne 0$ 
as a statement on the parent $G_2$
-manifold does not imply that, after reduction to six dimensions, 
only $W_3$ can feel the pinch. What it implies is only that the direct localized singular source sits in the ${\bf 27}$ of $G_2$, hence in 
the $\tau_3$ 
 channel. Once one reduces to the $SU(3)$
structure manifold, the pinch can affect the reduced structure forms 
$J \equiv J(T)$ and $\Omega \equiv \Omega(T)$, and through their 
$T$ dependence it can induce smooth corrections to several 
$W_i$.
The direct singular descendant is special to $W_3$, but the induced smooth backreaction need not be. In fact there are really two different statements, and they should not be conflated. 

The first is the direct statement on the parent \(G_2\)-structure manifold. On the seven-dimensional side, the localized pinch source appears in the torsion decomposition as $d\varphi$ as given in \eqref{torsiondecomp2}.
The singular pinch source to the lowest order in $T$ is:
\begin{equation}
\tau_3^{\rm pinch}
=
T\,\delta_{\Sigma_6^{\rm jct}}\,\Xi_{27}
+
O(T^2),
\label{tau3pinch_parent_expl}
\end{equation}
with no corresponding singular \(\tau_0^{\rm pinch}\), \(\tau_1^{\rm pinch}\), or \(\tau_2^{\rm pinch}\) at leading order. Therefore the \emph{direct localized pinch source} lies entirely in the \(\tau_3\) channel. This is a statement about the irreducible \(G_2\)-representation content of the singular piece on the parent seven-manifold.

The second statement is the reduced statement on the \(SU(3)\)-structure manifold. After reducing along the pinched circle direction, one no longer works directly with
\(\tau_0,\tau_1,\tau_2,\tau_3\),
but with the six-dimensional torsion classes \(W_1,\dots,W_5\), defined through $dJ$ and $d\Omega$ as in \eqref{dJtorsion}.
Now the crucial point is that the pinch does not only descend as the explicit localized source \eqref{tau3pinch_parent_expl}; it also deforms the reduced \(SU(3)\)-structure itself:
\begin{equation}
J(T)=J^{(0)}+T\,J^{(1)}+O(T^2),
\qquad
\Omega(T)=\Omega^{(0)}+T\,\Omega^{(1)}+O(T^2) ,
\label{JOmega_T_dependence_again}
\end{equation}
such that $dJ$ and $d\Omega$ now takes the form \eqref{dOmegaexpandT}.
Therefore, even if the \emph{direct singular source} came only through \(\tau_3^{\rm pinch}\), the \emph{induced reduced deformation} can contribute to all \(SU(3)\) torsion classes that are extracted from \(dJ\) and \(d\Omega\).

Therefore the correct distinction is the following: 
Direct singular descendant of the pinch implies 
$W_3$ only.
And therefore the direct localized source \(\tau_3^{\rm pinch}\) is a primitive \((2,1)+(1,2)\)-type object after reduction, so it feeds the \(W_3\) channel:
\begin{equation}
W_3^{\rm pinch,\,direct}
=
b_3\,T\,
\Pi_{\rm prim}^{(2,1)+(1,2)}\!\big(\iota_{v_b}\Xi_{27}\big).
\label{W3_direct_only}
\end{equation}
By contrast, indirect smooth backreaction of the pinch on $J(T)$ and $\Omega(T)$ implies that 
$W_i$ with $i = 1,..., 5$ may all receive corrections.
These are not new singular descendants of \(\tau_3\). Rather, they are the ordinary \(SU(3)\)-torsion components extracted from the \(T\)-dependent deformations \eqref{dOmegaexpandT}. This means that the 
schematic formulas that we quoted earlier should be interpreted carefully. In other words, one should really write:
\begin{equation}
W_i^{\rm pinch}
=
W_i^{\rm pinch,\,direct}
+
W_i^{\rm pinch,\,induced}\nonumber
\end{equation}
\begin{equation}
W_i^{\rm pinch,\,direct}
=
b_i\,T\,
\Pi_{\rm prim}^{(2,1)+(1,2)}\!\big(\iota_{v_b}\Xi^{(4)}_{27}\big)\, \delta_{i3} ,
\end{equation}
with $i = 1, ..., 5$, implying that it is only $W_3^{\rm pinch,\,direct}$ that
is nonzero at leading direct order. Here $\Xi_{27}^{(4)}\in \Lambda^4_{27}(M_7)$ such that $(\iota_{v_b}\Xi^{(4)}_{27})_{NPQ} = v_b^M (\Xi_{27}^{(4)})_{MNPQ}$ and the operation $\Pi_{\rm prim}^{(2, 1) + (1, 2)}$ projects into the requisite primitive form. (See also footnote \ref{3or4}.)
The remaining terms should then be understood as induced smooth corrections:
\begin{equation}
W_1^{\rm pinch,\,induced}
=
T\,W_1^{(T,{\rm pinch})}
+
\iota_{V_T}Y_1^{({\rm pinch})}
+
O(T^2)\nonumber
\end{equation}
\begin{equation}
W_2^{\rm pinch,\,induced}
=
T\,W_2^{(T,{\rm pinch})}
+
dT\wedge \widehat W_2^{({\rm pinch})}
+
O(T^2) \nonumber
\end{equation}
\begin{equation}
W_3^{\rm pinch,\,induced}
=
dT\wedge \widehat W_3^{({\rm pinch})}
+
O(T^2)\nonumber
\end{equation}
\begin{equation}
W_4^{\rm pinch,\,induced}
=
b_4\,d\varepsilon
+
\varepsilon\,W_4^{(\varepsilon,{\rm pinch})}
+
O(\varepsilon^2)\nonumber
\end{equation}
\begin{equation}
W_5^{\rm pinch,\,induced}
=
b_5\,d\varepsilon
+
\varepsilon\,W_5^{(\varepsilon,{\rm pinch})}
+
O(\varepsilon^2).
\end{equation}
Thus previously, our interpretation that 
the pinch contributes to $W_1,W_2,W_3,\dots$
should be understood in the following refined sense:
only $W_3$ receives the \emph{direct} descendant of the localized $\tau_3^{\rm pinch}$ source,
whereas
$W_1,\ W_2,\ W_4$ and  $W_5$
receive only \emph{induced smooth} corrections because the pinch deforms $J(T)$ and $\Omega(T)$.

\subsubsection{Combining fibration, flux and pinch contributions\label{sec315}}

We now combine the fiber, flux and the pinch contributions into a single formulation. As we saw in section \ref{pinchu2} the main point is that the pinch contribution does \emph{not} replace the odd part of the flux contribution. Rather, it adds a new branch-odd source with the same degree structure. Thus the total torsion classes may be organized as:
\begin{equation}
W_i
=
W_i^{\rm fib}
+
W_i^{\rm flux}
+
W_i^{\rm pinch} ~~~~ {\rm with} ~~~~
W_i^{\rm flux}
=
W_i^{\rm flux,(e)}
+
W_i^{\rm flux,(o)} ,
\label{Wi_flux_even_odd_decomp_combined}
\end{equation}
where $i=1,\dots,5$.
The fibration piece \(W_i^{\rm fib}\) is unchanged by the present discussion, so our focus will be on how the odd flux sector and the pinch sector combine. In fact, the branch-even part behaves in the same way as the ordinary IIA-like reduction and contributes as:
\begin{equation}
W_1^{\rm flux,(e)}
\sim
\Pi_{(3,0)+(0,3)}\!\big(\mathcal H_3^{(e)}\big)\nonumber
\label{W1_flux_even_combined}
\end{equation}
\begin{equation}
W_2^{\rm flux,(e)}
\sim
{\rm L}_J^{-1}\left[\Pi_{\rm prim}^{(2,2)}\!\big(d\Omega\big)_{\rm flux,e}\right]
+
\widetilde a_2\,\Pi_{\rm prim}^{(1,1)}\!\big(\iota_{v_e} *_7 G_4^{(27,e)}\big)\nonumber
\label{W2_flux_even_combined}
\end{equation}
\begin{equation}
W_3^{\rm flux,(e)}
=
\Pi_{\rm prim}^{(2,1)+(1,2)}\!\big(\mathcal H_3^{(e)}\big)\nonumber
\label{W3_flux_even_combined}
\end{equation}
\begin{equation}
W_4^{\rm flux,(e)}
=
a_4\,X_\perp^{(e)},
\qquad
W_5^{\rm flux,(e)}
=
a_5\,X_\perp^{(e)} ,
\label{W45_flux_even_combined}
\end{equation}
where all the parameters and notations appearing above have been defined earlier. The main new contributions start from the branch-odd flux sector which we express around the symmetric $T = 0$ point. They are then expressed as:
\begin{equation}
W_1^{\rm flux,(o)}
=
T\,{\cal W}_1^{(o)}
+
\iota_{V_T}{\cal Y}_1^{(o)}
+
O(T^2)\nonumber
\label{W1_flux_odd_combined}
\end{equation}
\begin{equation}
W_2^{\rm flux,(o)}
=
T\,{\cal W}_2^{(o)}
+
dT\wedge {\cal Y}_2^{(o)}
+
\widetilde a_2^{(o)}\,
\Pi_{\rm prim}^{(1,1)}\!\big(\iota_{v_o} *_7 G_4^{(27,o)}\big)
+
O(T^2)\nonumber
\label{W2_flux_odd_combined}
\end{equation}
\begin{equation}
W_3^{\rm flux,(o)}
=
\Pi_{\rm prim}^{(2,1)+(1,2)}\!\big(\mathcal H_3^{(o)}\big)
+
T\,{\cal W}_3^{(o)}
+
dT\wedge {\cal Y}_3^{(o)}
+
O(T^2)\nonumber
\label{W3_flux_odd_combined}
\end{equation}
\begin{equation}
W_4^{\rm flux,(o)}
=
a_4^{(o)}X_\perp^{(o)}
+
c_4^{(o)}\,dT
+
T\,{\cal W}_4^{(o)}
+
O(T^2)\nonumber
\label{W4_flux_odd_combined}
\end{equation}
\begin{equation}
W_5^{\rm flux,(o)}
=
a_5^{(o)}X_\perp^{(o)}
+
c_5^{(o)}\,dT
+
T\,{\cal W}_5^{(o)}
+
O(T^2).
\label{W5_flux_odd_combined}
\end{equation}
These terms are built from the branch-odd flux data
$H_3^{(o)},
X_\perp^{(o)}$ and 
$G_4^{(27,o)}$,
together with their induced \(T\)- and \(dT\)-dependent corrections so that they would vanish exactly at the symmetric point. Recall that the symmetric point is not the supersymmetric IIA point, but we will come back to the IIA point soon.

The pinch contributions can also be quantified directly from the parent M-theory seven-manifold. Due to various reason given earlier, the pinch sector is constructed using a primitive three-form on the six-dimensional base. Putting these facts together, we can show that
the pinch sector originates from the localized \(G_2\)-torsion source:
\begin{equation}
\tau_3^{\rm pinch}
=
T\,\delta_{\Sigma_6^{\rm jct}}\,\Xi_{27}
+
O(T^2),
\label{tau3_pinch_combined}
\end{equation}
where the details of the various terms are provided earlier. The reduction of \eqref{tau3_pinch_combined}
gives new odd contributions to the \(SU(3)\)-structure torsion classes. These take the typical schematic form:
\begin{equation}
W_1^{\rm pinch}
=
T\,W_1^{(T,{\rm pinch})}
+
\iota_{V_T}Y_1^{({\rm pinch})}
+
O(T^2)\nonumber
\label{W1_pinch_combined}
\end{equation}
\begin{equation}
W_2^{\rm pinch}
=
T\,W_2^{(T,{\rm pinch})}
+
dT\wedge \widehat W_2^{({\rm pinch})}
+
O(T^2)\nonumber
\label{W2_pinch_combined}
\end{equation}
\begin{equation}
W_3^{\rm pinch}
=
b_3\,T\,
\Pi_{\rm prim}^{(2,1)+(1,2)}\!\big(\iota_{v_b}\Xi_{27}\big)
+
dT\wedge \widehat W_3^{({\rm pinch})}
+
O(T^2) ,
\label{W3_pinch_combined}
\end{equation}
where the appearance of the primitive contribution to $W_3$ is directly the consequence of \eqref{tau3_pinch_combined}.
For the Lee-form classes it is more convenient to use a regular branch-odd parameter instead of \(d\log(R_+/R_-)\), because the latter diverges at the actual Type IIA endpoints. A convenient regular choice is the one motivated from $\zeta$ and $\eta$ in \eqref{zetaabeta}, namely:
\begin{equation}\label{epsregdef_combined}
\varepsilon
\equiv
\frac{R_+}{R_-}\,e^{-R_+/R_-}
-
\frac{R_-}{R_+}\,e^{-R_-/R_+}.
\end{equation}
This satisfies the two requisite conditions: 
(1) $R_+\leftrightarrow R_-~\Longrightarrow~
\varepsilon\mapsto -\,\varepsilon$
and
(2) $R_+=R_-
~\Longrightarrow~
\varepsilon=0$, 
while remaining finite at the Type IIA endpoints. The Lee-form pinch contributions may therefore be written as:
\begin{equation}
W_4^{\rm pinch}
=
b_4\,d\varepsilon
+
\varepsilon\,W_4^{(\varepsilon,{\rm pinch})}
+
O(\varepsilon^2)\nonumber
\label{W4_pinch_combined}
\end{equation}
\begin{equation}
W_5^{\rm pinch}
=
b_5\,d\varepsilon
+
\varepsilon\,W_5^{(\varepsilon,{\rm pinch})}
+
O(\varepsilon^2) ,
\label{W5_pinch_combined}
\end{equation}
which interestingly vanishes exactly at the symmetric and is regular at the IIA points. This is because, near the two Type IIA end-points, we have  
$ d\varepsilon\big|_{R_\pm=0}
\sim \pm
\frac{dR_\pm}{R_\mp}$. They vanish only if one restricts to the endpoint loci themselves, where the
collapsed branches are held fixed, {\it i.e.} when $dR_\pm = 0$ for $R_\pm = 0$ and $R_\mp > 0$. Otherwise they remain regular functions. (The standard logarithmic choice, like \(d\log(R_+/R_-)\), would work well for the symmetric point but would be a bad choice for the IIA point.) However we must distinguish two different scenarios here.
First, if one restricts exactly to the type-IIA endpoint, then the collapsed
branch is removed from the physical configuration.  For example, at
$R_+=0$
one should impose
$dR_+=0$
for variations tangent to the endpoint theory.  Then
$d\varepsilon\big|_{\rm IIA}=0$,
and hence:
\begin{equation}
\left.
W_4^{\rm pinch}
\right|_{\rm IIA}
=
0,
\qquad
\left.
W_5^{\rm pinch}
\right|_{\rm IIA}
=
0.
\label{W45_pinch_vanish_endpoint}
\end{equation}
This is the physically expected result: at the type-IIA endpoint there is only
one surviving ordinary circle branch, so the wedge-node pinch has disappeared
from the bulk type-IIA geometry. Second, if one studies fluctuations away from the endpoint, then
$dR_+\neq 0$
or
$dR_-\neq 0$
is allowed.  In that case $d\varepsilon$ need not vanish.  The term
$b_{4,5}\,d\varepsilon$
then describes the leading response of the Lee-form torsion to moving away
from the type-IIA endpoint back into the wedge interior.  It is not a residual
pinch effect inside the endpoint theory; it is the first normal variation away
from the endpoint. Equivalently, one may write the endpoint behavior as
$W_{4,5}^{\rm pinch}
=
O(\varepsilon,d\varepsilon)$,
with the endpoint boundary condition:
\begin{equation}
\varepsilon\big|_{\rm IIA}=0,
\qquad
d\varepsilon\big|_{T_{\rm IIA}\mathcal M}=0 ,
\end{equation}
where $T_{\rm IIA}\mathcal M$ denotes the tangent space to the type-IIA endpoint
subspace of the full wedge moduli space.  In this restricted sense, the pinch
contribution vanishes exactly at the IIA endpoints.
Therefore the formulae \eqref{W5_pinch_combined} 
should be interpreted as describing the neighborhood of the endpoint in the
wedge moduli space.  On the endpoint itself, where the collapsed branch is
removed and the pinch no longer exists, these terms must be set to zero.  Their
nonzero value only measures an infinitesimal departure from the endpoint back
toward the non-supersymmetric wedge geometry.

The pinch contribution is therefore an \emph{additive refinement} of the odd flux contribution and not just a redefinition of them. For the scalar class \(W_1\), one combines the coefficients as:
\begin{equation}
T\,{\cal W}_1^{(o)}
\;\longrightarrow\;
T\,{\bf W}_1^{(o)}
\equiv
T\Big({\cal W}_1^{(o)}+W_1^{(T,{\rm pinch})}\Big)\nonumber
\label{mapW1a_combined}
\end{equation}
\begin{equation}
\iota_{V_T}{\cal Y}_1^{(o)}
\;\longrightarrow\;
\iota_{V_T}{\bf Y}_1^{(o)}
\equiv
\iota_{V_T}\Big({\cal Y}_1^{(o)}+Y_1^{({\rm pinch})}\Big) ,
\label{mapW1b_combined}
\end{equation}
where we see how the pinch part {\it adds} to the existing odd contribution from the flux sector. This is the key point. Similarly,
for the two-form class \(W_2\), one has:
\begin{equation}
T\,{\cal W}_2^{(o)}
\;\longrightarrow\;
T\,{\bf W}_2^{(o)}
\equiv
T\Big({\cal W}_2^{(o)}+W_2^{(T,{\rm pinch})}\Big)\nonumber
\label{mapW2a_combined}
\end{equation}
\begin{equation}
dT\wedge {\cal Y}_2^{(o)}
\;\longrightarrow\;
dT\wedge {\bf Y}_2^{(o)}
\equiv
dT\wedge\Big({\cal Y}_2^{(o)}+\widehat W_2^{({\rm pinch})}\Big) ,
\label{mapW2b_combined}
\end{equation}
where we again see similar behavior. The hatted variables are the direct descendant from \eqref{W3_pinch_combined} and they have been defined earlier. Interestingly, one crucial observation here is that 
the odd reduced \(\mathbf{27}\)-flux term:
\begin{equation}
\widetilde a_2^{(o)}\,
\Pi_{\rm prim}^{(1,1)}\!\big(\iota_{v_o} *_7 G_4^{(27,o)}\big)
\end{equation}
has no direct pinch analogue at leading order, because the pinch \(\mathbf{27}\)-source contributes naturally to \(W_3\), not to \(W_2\).
On the other hand, if the M-theory $G_2$ structure pinch description involved a {\it non-primitive} three-form, then the story would be different. However as explained earlier, the non-primitive choice is likely to clash with the very construction of the pinch itself, so we will not worry about it here. For the three-form class \(W_3\), the map is:
\begin{equation}
T\,{\cal W}_3^{(o)}
\;\longrightarrow\;
T\,{\bf W}_3^{(o)}
\equiv
T\Big({\cal W}_3^{(o)}+b_3\,\Pi_{\rm prim}^{(2,1)+(1,2)}(\iota_{v_b}\Xi_{27})\Big)\nonumber
\label{mapW3a_combined}
\end{equation}
\begin{equation}
dT\wedge {\cal Y}_3^{(o)}
\;\longrightarrow\;
dT\wedge {\bf Y}_3^{(o)}
\equiv
dT\wedge\Big({\cal Y}_3^{(o)}+\widehat W_3^{({\rm pinch})}\Big).
\label{mapW3b_combined}
\end{equation}
Thus the direct descendant of the localized pinch source shifts the coefficient of the \(T\)-linear odd piece in \(W_3\). Finally, for the Lee-form classes \(W_4,W_5\), it is better to keep the odd flux and pinch pieces separate, because they involve different regular odd one-forms. One therefore writes
\begin{equation}
W_4^{\rm odd,total}
=
a_4^{(o)}X_\perp^{(o)}
+
c_4^{(o)}\,dT
+
T\,{\cal W}_4^{(o)}
+
b_4\,d\varepsilon
+
\varepsilon\,W_4^{(\varepsilon,{\rm pinch})}
+
O(T^2,\varepsilon^2)\nonumber
\label{mapW4_combined}
\end{equation}
\begin{equation}
W_5^{\rm odd,total}
=
a_5^{(o)}X_\perp^{(o)}
+
c_5^{(o)}\,dT
+
T\,{\cal W}_5^{(o)}
+
b_5\,d\varepsilon
+
\varepsilon\,W_5^{(\varepsilon,{\rm pinch})}
+
O(T^2,\varepsilon^2).
\label{mapW5_combined}
\end{equation}
Thus the pinch contribution adds a new regular odd one-form \(d\varepsilon\) and a new algebraic odd piece proportional to \(\varepsilon\); it does not replace the odd \(\mathbf 7\)-flux contribution. Notice that for the Lee-form classes, the odd flux parameters
$a_{4,5}^{(o)},
c_{4,5}^{(o)}$ and 
${\cal W}_{4,5}^{(o)}$
are supplemented by the new pinch coefficients
$b_{4,5}$
and 
$W_{4,5}^{(\varepsilon,{\rm pinch})}$,
with the branch-odd one-form now naturally encoded by \(d\varepsilon\).
The full torsion classes then take the form:
\begin{equation}
W_i
=
W_i^{\rm fib}
+
W_i^{\rm flux+pinch},
\label{Wi_total_final_combined}
\end{equation}
where $i=1,\dots,5,$ and the fibration contributions may be read up from \eqref{toruequ}. They are not affected by either the flux or the pinch effects and only depend on the fibration. Also since we are not performing any T-dualities, but only dimensional reductions, there is no mixing between the flux and the fibration. 
Collecting everything, the total flux-plus-pinch contribution may be written as:
\begin{equation}
\begin{aligned}
W_1^{\rm flux+pinch}
&=
\Pi_{(3,0)+(0,3)}\!\big(\mathcal H_3^{(e)}\big)
+
T\,{\bf W}_1^{(o)}
+
\iota_{V_T}{\bf Y}_1^{(o)}
+
O(T^2),
\\[4pt]
W_2^{\rm flux+pinch}
&=
{\rm L}_J^{-1}\left[\Pi_{\rm prim}^{(2,2)}\!\big(d\Omega\big)_{\rm flux,e}\right]
+
\widetilde a_2\,\Pi_{\rm prim}^{(1,1)}\!\big(\iota_{v_e} *_7 G_4^{(27,e)}\big)
\\
&\hspace{1.9cm}
+
\widetilde a_2^{(o)}\,\Pi_{\rm prim}^{(1,1)}\!\big(\iota_{v_o} *_7 G_4^{(27,o)}\big)
+
T\,{\bf W}_2^{(o)}
+
dT\wedge {\bf Y}_2^{(o)}
+
O(T^2),
\\[4pt]
W_3^{\rm flux+pinch}
&=
\Pi_{\rm prim}^{(2,1)+(1,2)}\!\big(\mathcal H_3^{(e)}\big)
+
\Pi_{\rm prim}^{(2,1)+(1,2)}\!\big(\mathcal H_3^{(o)}\big)
+
T\,{\bf W}_3^{(o)}
+
dT\wedge {\bf Y}_3^{(o)}
+
O(T^2),
\\[4pt]
W_4^{\rm flux+pinch}
&=
a_4\,X_\perp^{(e)}
+
a_4^{(o)}X_\perp^{(o)}
+
c_4^{(o)}\,dT
+
T\,{\cal W}_4^{(o)}
+
b_4\,d\varepsilon
+
\varepsilon\,W_4^{(\varepsilon,{\rm pinch})}
+
O(T^2,\varepsilon^2),
\\[4pt]
W_5^{\rm flux+pinch}
&=
a_5\,X_\perp^{(e)}
+
a_5^{(o)}X_\perp^{(o)}
+
c_5^{(o)}\,dT
+
T\,{\cal W}_5^{(o)}
+
b_5\,d\varepsilon
+
\varepsilon\,W_5^{(\varepsilon,{\rm pinch})}
+
O(T^2,\varepsilon^2) ,
\end{aligned}
\label{W_flux_pinch_combined_final}
\end{equation}
where we can summarize this in the following way:
the pinch contribution does not replace the odd flux contribution;
it adds new terms with the same degree structure,
and more precisely,
in the \(W_1,W_2,W_3\) channels it shifts the odd \(T\)- and \(dT\)-dependent coefficients,
while
in the Lee-form channels \(W_4,W_5\) it adds a new regular branch-odd one-form $d\varepsilon$
and a new algebraic odd piece proportional to $\varepsilon$. The fibration contribution then simply adds to the flux-plus-pinch contribution.
This gives the desired single formulation of the fibration, flux and pinch contributions to the \(SU(3)\)-structure torsion classes.

Before ending this section let us make a few observations. So far we did not take the doubled $U(1)$ gauge fields from \eqref{gaugoram} and \eqref{RRfinal0A}. To make our construction complete, it will be useful to also take them into account. In type 0A, the doubled bulk $U(1)$ gauge fields are the doubled RR one-form
potentials
$C_1^{(+)}$, and 
$C_1^{(-)}$,
with gauge-invariant RR two-form field strengths
$F_2^{(+)}
=
dC_1^{(+)}$,
and $F_2^{(-)}
=
dC_1^{(-)}$.
Equivalently, we define the branch-even and branch-odd combinations:
\begin{equation}
F_2^{(e)}
=
\frac{1}{2}
\left(
F_2^{(+)}+F_2^{(-)}
\right),
\qquad
F_2^{(o)}
=
\frac{1}{2}
\left(
F_2^{(+)}-F_2^{(-)}
\right) ,
\end{equation}
which, under exchange of the two type 0A branches,
$(+)\longleftrightarrow (-)$,
transform as
$F_2^{(e)}
\longmapsto
F_2^{(e)}$, and 
$F_2^{(o)}
\longmapsto
-
F_2^{(o)}$. Thus $F_2^{(e)}$ belongs to the branch-even bulk flux sector, while
$F_2^{(o)}$ belongs to the branch-odd bulk flux sector. The important point is that the intrinsic torsion classes do not depend
directly on the gauge potentials $C_1^{(\pm)}$.  They can only depend on the
gauge-invariant RR field strengths $F_2^{(\pm)}$.  Therefore the correct
schematic decomposition of the torsion classes is not just \eqref{Wi_total_final_combined} but as additive piece $W_i^{F_2}$.
Equivalently, one may keep the same notation $W_i^{\rm flux+pinch}$, provided
one understands that the flux sector now includes:
\begin{equation}
{\rm Flux}_{0A}
=
\left\{
H_3^{(e)},H_3^{(o)},
F_2^{(e)},F_2^{(o)},
G_4^{(27,e)},G_4^{(27,o)}
\right\} ,
\end{equation}
which includes {\it all} the doubled field at the pinch that we derived in section \ref{sec001}. Thus the generic strategy that all fields, excluding the metric, are literally doubled at the pinch provides a more powerful way to analyze the system. The torsion classes then become:
\begin{equation}
W_i^{\rm flux+pinch}
=
W_i^{H_3}
+
W_i^{F_2}
+
W_i^{G_4}
+
W_i^{\rm pinch} ,
\end{equation}
where both even (e) and odd (o) combinations are taken in the same footing. Away from the pinch, in the DRP and the SSP regime \cite{BDV}, certain combinations become dynamical and others auxiliary, but these details are irrelevant so long as the two $S^1$ in $S^1\vee S^1$ are in contact with each other. Since we are always in this regime, we will assume that the 
two-form flux $F_2$ decomposes under the $SU(3)$ structure as:
\begin{equation}
F_2^{(s)}
=
f^{(s)}J
+
F_{2,{\rm prim}}^{(1,1;s)}
+
F_2^{(2,0;s)}
+
F_2^{(0,2;s)},
\qquad
s=e,o ,
\end{equation}
where
$f^{(s)}
=
\frac{1}{3}
J\lrcorner F_2^{(s)}
=
\frac{1}{6}
J^{mn}F_{mn}^{(s)}$. 
The primitive $(1,1)$ part satisfies
$J\lrcorner F_{2,{\rm prim}}^{(1,1;s)}=0$. Naively therefore it would appear that
the singlet piece $f^{(s)}J$ and the primitive $(1,1)$ piece
$F_{2,{\rm prim}}^{(1,1;s)}$ are the parts that naturally feed the same
representation channels as $W_1$ and $W_2$.  The non-$(1,1)$ pieces
$F_2^{(2,0;s)}+F_2^{(0,2;s)}$ measure the failure of $F_2$ to be compatible
with the almost-complex structure and can be included as an additional
non-integrability source.

However using the decomposition of $dJ$ and $d\Omega$ in \eqref{mapwa}, and the representation of the gauge field as $\mathcal F_j^{(s)} \equiv (\delta_{j3}\alpha_{F_s} + \delta_{j4}\beta_{F_s})\mathcal Q_j(F_2^{(s)})$, we note that to lowest order $\mathcal Q_j(F_2^{(s)}) \sim 0$ simply because $F_2^{(s)}$ is a two-form whereas $dJ$ and $d\Omega$ are three-form and four-form respectively. Thus unless $F_s^{(s)}$ is wedged with another natural form on $X_6$ or $X_7$, the lowest order contributions to the torsion classes typically would vanish. Therefore the best way to introduce the effects from the tw-form fluxes is to rewrite \eqref{W_flux_pinch_combined_final} in the following suggestive way:
\begin{equation}\label{W_flux_pinch_corrected}
\begin{aligned}
W_1^{\rm flux+pinch}
&=
-\frac{i}{12}\,
\overline{\Omega}\lrcorner
\left[
{\cal Q}_3^{H,e}(H_3^{(e)})
+
{\cal Q}_3^{F,e}(F_2^{(e)})
+
{\cal Q}_3^{F,o}(F_2^{(o)})
\right]\\
&\hspace{1.2cm}
+
{T}\,{\bf W}_1^{(o)}
+
\iota_{V_{{T}}}{\bf Y}_1^{(o)}
+
O({T}^2),
\\[6pt]
W_2^{\rm flux+pinch}
&=
{\rm L}_J^{-1}\Big\{\Pi_{\rm prim}^{(2,2)}
\left[
(d\Omega)_{\rm flux,e}
+
{\cal Q}_4^{F,e}(F_2^{(e)})
+
{\cal Q}_4^{F,o}(F_2^{(o)})
\right]\Big\}\\
&\hspace{1.2cm}
+
\widetilde a_2\,
\Pi_{\rm prim}^{(1,1)}
\left(
\iota_{v_e} *_7G_4^{(27,e)}
\right)
+
\widetilde a_2^{(o)}\,
\Pi_{\rm prim}^{(1,1)}
\left(
\iota_{v_o} *_7G_4^{(27,o)}
\right)\\
&\hspace{1.2cm}
+
J\wedge
{T}\,{\bf W}_2^{(o)}
+
J\wedge
d{T}\wedge{\bf Y}_2^{(o)}
+
O({T}^2) \\
W_3^{\rm flux+pinch}
&=
\Pi_{\rm prim}^{(2,1)+(1,2)}
\left[
{\cal Q}_3^{H,e}(H_3^{(e)})
+
{\cal Q}_3^{H,o}(H_3^{(o)})
+
{\cal Q}_3^{F,e}(F_2^{(e)})
+
{\cal Q}_3^{F,o}(F_2^{(o)})
\right]\\
&\hspace{1.2cm}
+
{T}\,{\bf W}_3^{(o)}
+
d{T}\wedge{\bf Y}_3^{(o)}
+
O({T}^2)
\\
W_4^{\rm flux+pinch}
&=
\frac{1}{2}
J\lrcorner
(dJ)_{\rm flux}
+
c_4^{(o)}\,d{T}
+
{T}\,{\cal W}_4^{(o)}
+
b_4\,d\varepsilon
+
\varepsilon\,W_4^{(\varepsilon,{\rm pinch})}
+
O({T}^2,\varepsilon^2)\\
W_5^{\rm flux+pinch}
&=
\frac{1}{2}
{\rm Re}
\left[
\overline{\Omega}^{-1}
\lrcorner
(d\Omega)_{\rm flux}
\right]
+
c_5^{(o)}\,d{T}
+
{T}\,{\cal W}_5^{(o)}
+
b_5\,d\varepsilon
+
\varepsilon\,W_5^{(\varepsilon,{\rm pinch})}
+
O(T^2,\varepsilon^2) ,
\end{aligned}
\end{equation}
which is more precise.  It avoids saying that a two-form flux $F_2$ directly
contributes to a scalar torsion class or to a three-form torsion class.  Instead,
it says that $F_2$ contributes only through the appropriate flux-to-geometry
maps:
\begin{equation}
F_2
\longmapsto
{\cal Q}_3^{F}(F_2)
\in \Lambda^3,
\qquad
F_2
\longmapsto
{\cal Q}_4^{F}(F_2)
\in \Lambda^4.
\end{equation}
Similarly, $G_4$ contributes only after dimensional reduction and projection to
the appropriate six-dimensional form degree. The fibration contribution is still independent of the flux and pinch sectors
because no T-duality is being performed.  The pinch contribution still does not
replace the odd flux contribution.  It adds additional branch-odd $T$,
$dT$, $d\varepsilon$, and $\varepsilon$ dependent terms.  The new point is
that the type 0A bulk $U(1)$ gauge fields add a further RR two-form flux sector,
namely
$F_2^{(s)}$
whose singlet, primitive $(1,1)$, and non-$(1,1)$ components feed the
appropriate $SU(3)$ representation channels of the torsion classes in a way shown in \eqref{W_flux_pinch_corrected}.

Our second observation concerns a subtle but important point that setting $T=0$ in
\eqref{W_flux_pinch_combined_final} does \emph{not} reproduce the supersymmetric
type IIA limit. The reason is twofold. First, the supersymmetric IIA endpoint is
reached at
$T=\pm R_B$, with $T$ and $R_B$ defined as in say \eqref{RBTdef0A},
rather than at $T=0$. Second, even if one chooses a symmetric background value
$T_0=0$,
the fluctuations
$T=T_0+\delta T=\delta T$
still contribute at first order through terms such as:
\begin{equation}
\delta T\,{\bf W}_i^{(o)},
\qquad
d(\delta T)\wedge {\bf Y}_i^{(o)},
\qquad
c_i^{(o)}\,d(\delta T) ,
\end{equation}
where we are not required to keep $\delta T$ very small. This way even when we are infinitesimally away from the symmetric point where the tachyon becomes $T = 0 + \epsilon_T + \delta T$ with $\epsilon_T$ quantifying the distance away from the symmetric point, \eqref{W_flux_pinch_combined_final} continues to provide the torsion classes for this case.
Thus $T_0=0$ describes only the branch-symmetric wedge background, not the
supersymmetric endpoint, and the torsion classes remain sensitive to the odd
sector through the fluctuation $\delta T$ already at linear order. To recover
the genuine type IIA limit, one must instead expand around
$T=\pm R_B$
and require simultaneously that the independent branch-odd wedge data disappear
in that degeneration limit.

There are several practical advantages to expanding about the symmetric point
$T=0$
where 
$R_+=R_-$,
rather than about one of the supersymmetric type IIA endpoints
$T=\pm R_B$
where 
$R_\mp=0$.
The basic reason is that the point $T=0$ is the most natural point for organizing
the full two-branch wedge geometry itself, whereas the type IIA endpoints are
degeneration limits in which one branch disappears.
First, the symmetric point treats the two branches democratically. Since the wedge
geometry is built from the pair $(S^1_+,S^1_-)$, the point
$R_+=R_-$
is the unique point at which the branch exchange symmetry is manifest. Expanding
around this point makes the decomposition into branch-even and branch-odd sectors
completely transparent:
\begin{equation}
(\text{even})\sim \frac12\big(X_+ + X_-\big),
\qquad
(\text{odd})\sim \frac12\big(X_+ - X_-\big).
\end{equation}
In particular, the tachyonic modulus
$T_0=\frac{R_+-R_-}{2}$
is then naturally small, so an expansion in $T$ is precisely an expansion in the
departure from branch symmetry. Second, the symmetric point is the natural place to separate physical bulk data
from auxiliary branch-odd wedge data. Many of the wedge-specific structures,
such as branch-odd corrections to fluxes, torsion classes, or local junction data,
are most cleanly organized as perturbations around the symmetric configuration.
In this language, one can write systematically:
\begin{equation}
W_i = W_i^{(e)} + T\,W_i^{(o)} + dT\wedge Y_i^{(o)}+\cdots,
\end{equation}
and immediately see which pieces are even and which are odd. This bookkeeping is
less transparent near an endpoint where one branch has already collapsed. Third, the expansion around $T=0$ is better suited for describing the genuinely
non-supersymmetric wedge regime. If the goal is to understand the full branch-resolved
geometry, the doubled RR sectors, or the emergence of the tachyon as a branch-odd
modulus, then the symmetric point is the natural reference background. By contrast,
an expansion around $T=\pm R_B$ is tailored to the neighborhood of a supersymmetric
endpoint and therefore tends to hide the intrinsically two-branch nature of the wedge. Fourth, the symmetric point allows a single local expansion that covers both branches
simultaneously. Near the supersymmetric endpoints, one generally needs two different
local expansions,
\begin{equation}
\delta_+ = R_B-T = R_-,
\qquad
\delta_- = R_B+T = R_+,
\end{equation}
depending on which branch is collapsing. Thus the endpoint description is naturally
split into two charts, one near $R_-=0$ and one near $R_+=0$. By contrast, the
expansion around $T=0$ provides one unified expansion valid in the interior of the
two-branch moduli space.
Fifth, from the point of view of effective field theory, the symmetric point is
often technically simpler because the expansion parameter is just the odd modulus
itself, {\it i.e.}
$T_0 \ll R_B$.
This gives a clean perturbative hierarchy between the branch-even background and
the branch-odd deformation. Near the IIA endpoint, the natural small parameter is
instead one of the vanishing radii, {\it i.e.}
$\delta_\pm = R_B\mp T$,
and the expansion must be reorganized in a way that mixes the background value of
$T$ with the radion $R_B$. That is better for endpoint physics, but less natural
for the interior wedge regime.
Therefore the two expansions serve different purposes:
expansion around $T=0$ (or more appropriately $T_0 = 0$) is best for describing the full symmetric two-branch wedge geometry and its odd/even decomposition,
whereas expansion around $T_0=\pm R_B$ is best for describing the approach to the supersymmetric type IIA endpoints.
So the advantage of expanding about the symmetric point is not that it captures the
supersymmetric limit more faithfully; rather, it is that it gives the most natural
and economical local description of the \emph{non-supersymmetric wedge geometry
itself}.

\subsection{Symmetric wedge regime versus the actual Type IIA endpoints \label{sec32}}

In the previous discussion the reduced \(SU(3)\)-structure data were organized as an expansion in the branch-odd parameter 
$T=\frac12(R_+-R_-)$
and 
$R_B=\frac12(R_++R_-)$,
with $J$ and $\Omega$ defined as in \eqref{JOmega_expand_T}.
This is the natural parametrization near the \emph{symmetric wedge point} defined as:
\begin{equation}
R_+\approx R_-,
\qquad
|T|\ll R_B,
\label{symm_wedge_regime_endpointsec}
\end{equation}
where both branches are present and the theory is still in the type-\(0\) regime. However, the actual supersymmetric Type IIA endpoint of the wedge construction is not the symmetric point \(T=0\). Instead, it is obtained when \emph{one} branch shrinks and the other survives as the ordinary M-theory circle. Thus the two genuine Type IIA endpoints are:
\begin{equation}\label{IIAlimit}
\begin{aligned}
& ~~~\left(R_+\to 0,~~ R_->0\right) ~~ \implies ~~ R_+=0 \iff T=-R_B \\
{\rm or}& ~~~ \left(R_-\to 0,~~ R_+>0\right) ~~ \implies ~~ R_-=0 \iff T=+R_B, 
\end{aligned}
\end{equation}
implying that the earlier \(T\)-expansion should be interpreted as the \emph{symmetric wedge/type-\(0\) expansion}, and not as an expansion around the actual Type IIA endpoint. Our aim here is to rewrite the torsion classes in a form adapted to the true Type IIA endpoints. The main conclusions are:

\begin{enumerate}
\item the fibration contribution is unaffected by going to the endpoint;
\item the \emph{direct localized} pinch contribution disappears at the actual Type IIA endpoint, because once one branch has collapsed there is no longer a wedge singularity;
\item the odd-branch corrections, including the \emph{smooth descendants} of the pinch sector, must be reorganized in terms of endpoint variables rather than the symmetric variable \(T\).
\end{enumerate}

\noindent To see how the fibration contribution remains unchanged at the Type IIA endpoint, one has to recall that the fibration contribution is purely geometric and comes from the non-closure of the fiber co-frame \eqref{dEAfib_correct} which only depends only on the twisting of
$T_f^2\rtimes \frac{S_a^1}{\mathbb Z_2}$
over \(\Sigma_3\), and not on whether the wedge has two branches or has already degenerated to a single surviving branch. Therefore the fibration-induced torsion classes remain exactly the same as before, namely the ones from \eqref{toruequ}, implying that the fibration contribution is unaffected when one moves from the symmetric wedge regime to the actual Type IIA endpoint. In other words:
\begin{equation}
W_i^{\rm fib}\Big|_{\rm endpoint}=W_i^{\rm fib}\Big|_{\rm symmetric},
\qquad i=1,\dots,5 ,
\label{fib_unaffected_endpoint}
\end{equation}
where note that if one includes metric backreaction from the collapsing branch, the same
fibration data may be represented in a different normalized coframe, and
the numerical components of $W_i^{\rm fib}$ can receive higher-order
endpoint corrections.
A related question is as to why the direct pinch contribution disappears at the actual Type IIA endpoint. To see this recall that, 
on the M-theory side, the singular pinch source is supported on the junction locus and takes the form:
\begin{equation}
\tau_3^{\rm pinch}
=
T\,\delta_{\Sigma_6^{\rm jct}}\,\Xi^{(3)}_{27}
+
{\cal O}(T^3),
\label{tau3pinch_Mside_endpointsec}
\end{equation}
with \(\Xi^{(3)}_{27}\in \Lambda^3_{27}\). (For detail see section \ref{naturalpinch}.) This expression is meaningful only while the geometry still contains the
two-branch wedge \(S^1_+\vee S^1_-\).  At a genuine Type IIA endpoint one
branch has collapsed completely, so the two-branch wedge description is no
longer the correct set of variables.  There is then only a single surviving
circle, and hence no independent junction singularity.  Consequently the
explicit distributional source
$\delta_{\Sigma_6^{\rm jct}}$
does not survive as an independent term at the exact Type IIA endpoint.
In particular, the direct localized pinch contribution to \(W_3\), for example
the term:
\begin{equation}
b_3 T\,
\Pi_{\rm prim}^{(2,1)+(1,2)}
\left(
\iota_{v_b}\Xi_{27}^{(4)}
\right),
\end{equation}
is a term in the symmetric two-branch wedge expansion.  It should not be
carried unchanged to the exact Type IIA endpoint, because at that endpoint
there is no longer a wedge singularity from which such a localized source
could descend. Near the endpoint, however, the collapsing branch can still leave smooth
non-distributional remnants in the surviving geometry.  We call these
``smooth descendants'' of the pinch sector.  They are not independent
delta-supported sources; rather, they are regular endpoint-adapted corrections
to the \(SU(3)\)-structure forms, flux coefficients, and torsion classes,
controlled by small collapse parameters such as \(\varepsilon_\pm\) and
\(d\varepsilon_\pm\).  Schematically, the endpoint expansion contains terms of
the form:
\begin{equation}
d\varepsilon_\pm\,\omega_i^{(1)}
+
\varepsilon_\pm\,\omega_i^{(0)}
+\cdots,
\end{equation}
rather than the wedge-interior term
$T\,\delta_{\Sigma_6^{\rm jct}}\Xi_{27}$.
Thus the localized pinch source is present only in the genuine two-branch
wedge regime.  Near the Type IIA endpoint it must be reorganized into smooth
endpoint variables, and at the exact endpoint it disappears or is absorbed
into the ordinary single-branch IIA torsion and flux data.

\begin{figure}[htbp]
\centering
\includegraphics[width=0.8\textwidth]{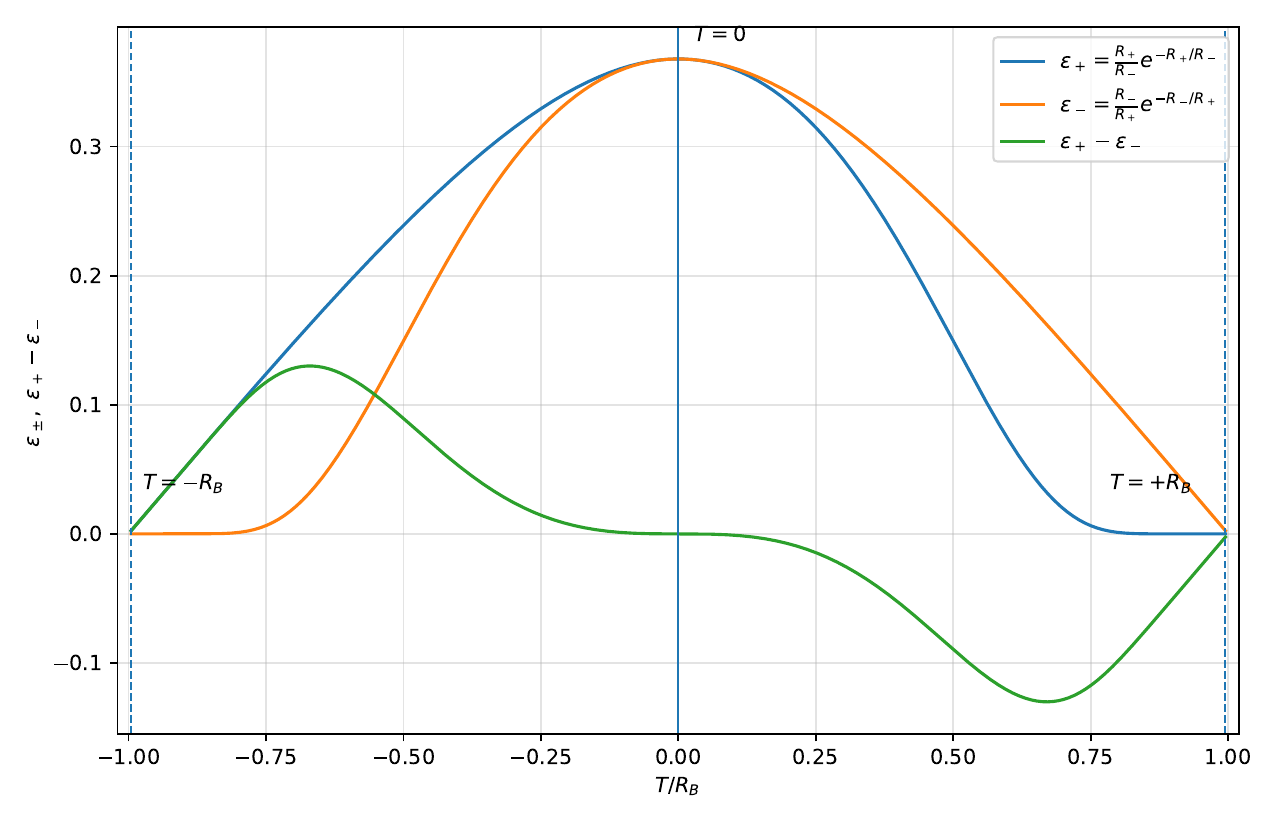}
\caption{The behavior of $\epsilon_+$ and $\epsilon_-$ from \eqref{epsplus_endpointsec} as well as the difference $\epsilon_+ - \epsilon_-$ from \eqref{global_eps_endpoint} at the symmetric point $T = 0$ as well as the two supersymmetric end points $T = \pm R_B = \pm 1$ where Type IIA is realized. Note that all the parameters remain well-defined at the symmetric point as well as the two end points.}
\label{fig:epsilon_curves}
\end{figure}

\begin{figure}[htbp]
\centering
\includegraphics[width=0.8\textwidth]{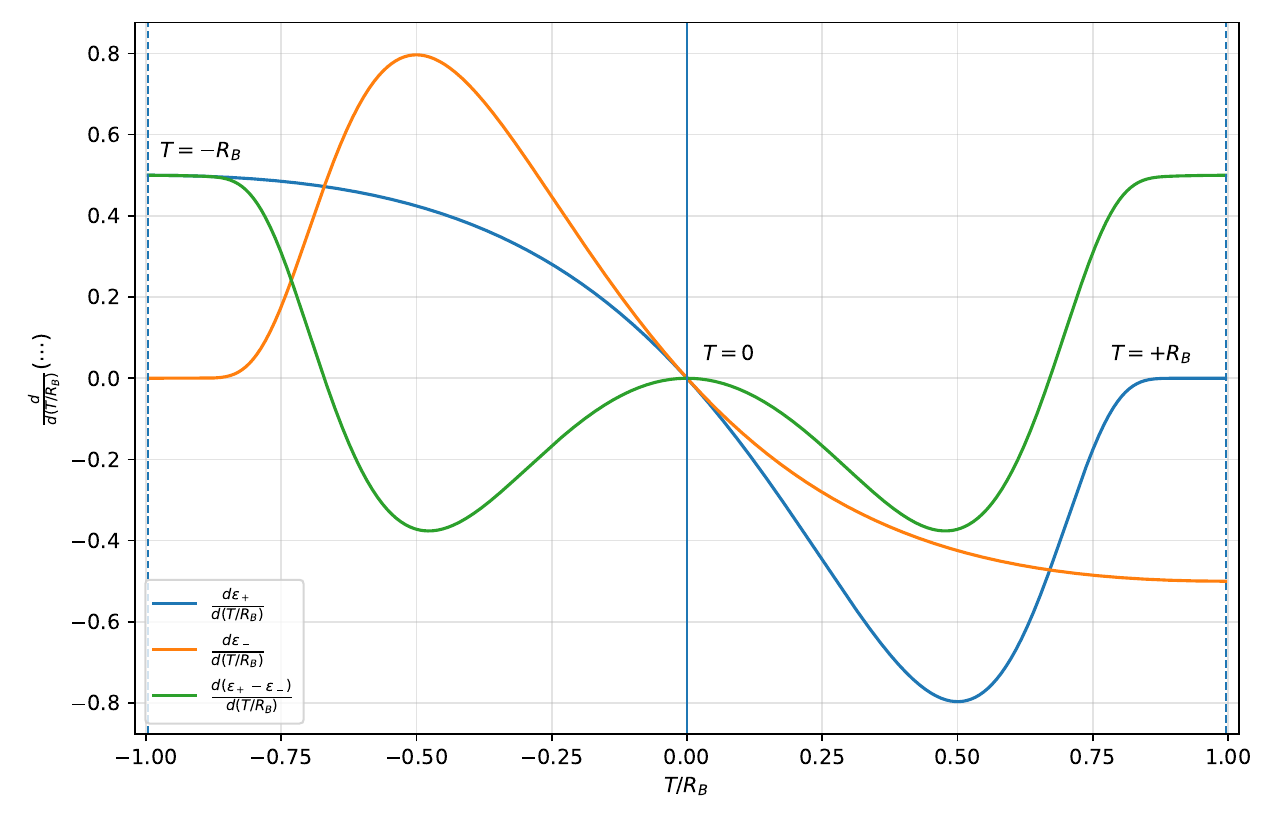}
\caption{Similar plots as the previous ones but now for the derivatives as defined in \eqref{drivvv}. Note again that the functions remain well defined both at the symmetric and at the IIA end-points.}
\label{fig:epsilon_derivatives}
\end{figure}

Our strategy now would be the analyze the system in the following way. We will choose the collapsing branch, say $R_+ \to 0$ or $R_- \to 0$, and arrange the first corrections to the torsion results induced by the corresponding collapsing branch. To make this more efficient it will be useful to choose some endpoint-adapted variables. For example, 
near the endpoints
$R_+\to 0$ with $R_->0$ or $R_-\to 0$ with $R_+>0$, 
the natural small parameters are respectively:
\begin{equation}
\varepsilon_+
\equiv
\frac{R_+}{R_-}\,e^{-R_+/R_-},~~~~~
\varepsilon_+\to 0, ~~~~ {\rm or}~~~~  \varepsilon_-
\equiv
\frac{R_-}{R_+}\,e^{-R_-/R_+},
~~~~~
\varepsilon_-\to 0 \label{epsplus_endpointsec}
\end{equation}
These parameters vanish precisely at the actual Type IIA endpoints. See {\bf figure \ref{fig:epsilon_curves}}.
They are the endpoint-adapted analogues of the regularized odd variable used earlier in the Lee-form sector. Their differentials are also regular:
\begin{equation}\label{drivvv}
\begin{aligned}
& d\varepsilon_+
=
e^{-R_+/R_-}\left(1-\frac{R_+}{R_-}\right)
d\!\left(\frac{R_+}{R_-}\right)\\
& d\varepsilon_-
=
e^{-R_-/R_+}\left(1-\frac{R_-}{R_+}\right)
d\!\left(\frac{R_-}{R_+}\right) ,
\end{aligned}
\end{equation}
which would tell us why logarithmic choices like $\epsilon_\pm = \log\left({R_\pm\over R_\mp}\right)$ are not good end-point variables despite being well-defined at the symmetric point. See {\bf figure \ref{fig:epsilon_derivatives}}.
It is also useful to distinguish these one-sided endpoint variables from the globally odd regularized combination:
\begin{equation}
\varepsilon
\equiv
\frac{R_+}{R_-}e^{-R_+/R_-}
-
\frac{R_-}{R_+}e^{-R_-/R_+},
\label{global_eps_endpoint}
\end{equation}
which is convenient near the symmetric wedge regime. Near a given endpoint, however, the one-sided variables \(\varepsilon_\pm\) are the cleaner expansion parameters. The symmetric wedge expansion and the endpoint expansion should therefore
be understood as two complementary asymptotic regimes. The symmetric
expansion is valid near
$|T|\ll R_B$,
whereas the endpoint expansion is valid near
$\varepsilon_\pm\ll 1$,
that is, when one of the two branches collapses. Consequently, the true
Type-IIA endpoint behavior should not be obtained by directly substituting
\(T=\pm R_B\) into the small-\(T\) formulas. Instead, the endpoint torsion
classes must be written separately in the regular variables
\(\varepsilon_\pm\). The endpoint correction coefficients introduced below
encode the first regular corrections in these endpoint variables and are
the endpoint limits of the same underlying geometric and flux data.

Thus they replace the symmetric-wedge expansion variable \(T\) when one wants the true endpoint behavior. The analysis of the endpoint behavior is however slightly more involved than the symmetric point behavior, so to simplify our analysis we can start by explaining the 
endpoint correction symbols that appear in the endpoint-adapted formulas:
\begin{equation}
\mathfrak W_i^{(+)},\ \mathfrak Y_i^{(+)},\ \mathfrak X_{4,5}^{(+)},
\qquad
\widetilde{\mathfrak W}_i^{(-)},\ \widetilde{\mathfrak Y}_i^{(-)},\ \widetilde{\mathfrak X}_{4,5}^{(-)} ,
\label{fraktur_symbols_list}
\end{equation}
where $i = 1, 2, 3$.
The superscript $\pm$ denotes the endpoint \(R_\pm\to 0\), with the \(\mp\) branch surviving. In other words, near
$R_\pm\to 0$ with  $R_\mp>0$,
the branch \(\mp\) is the surviving IIA-like branch, while the branch \(\pm\) is the collapsing branch. Therefore the fraktur symbols with superscript \((\pm)\) denote the \emph{first smooth corrections induced by the collapsing \(\pm\) branches}, including both the odd-flux descendants and the smooth descendants of the pinch sector. More precisely:
\begin{equation}
\mathfrak W_1^{(+)}\in \Lambda^0(M_6),
\qquad
\mathfrak W_2^{(+)}\in \Lambda^2(M_6),
\qquad
\mathfrak W_3^{(+)}\in \Lambda^3(M_6)\nonumber
\label{frakW_degrees}
\end{equation}
\begin{equation}\label{frak1}
\widetilde{\mathfrak W}_1^{(-)}\in \Lambda^0(M_6),
\qquad
\widetilde{\mathfrak W}_2^{(-)}\in \Lambda^2(M_6),
\qquad
\widetilde{\mathfrak W}_3^{(-)}\in \Lambda^3(M_6),
\end{equation}
are the leading \emph{algebraic} endpoint corrections to \(W_1,W_2,W_3\) at the two ends shown in say {\bf figure \ref{fig:epsilon_curves}}. They are multiplied by one power of the small parameters \(\varepsilon_\pm\), and therefore encode the first smooth deformation of the torsion classes caused by the shrinking \(\pm\) branch. Similarly: 
\begin{equation}
\mathfrak Y_1^{(+)}\in \Lambda^1(M_6),
\qquad
\mathfrak Y_2^{(+)}\in \Lambda^1(M_6),
\qquad
\mathfrak Y_3^{(+)}\in \Lambda^2(M_6)\nonumber
\label{frakY_degrees}
\end{equation}
\begin{equation}\label{frak2}
\widetilde{\mathfrak Y}_1^{(-)}\in \Lambda^1(M_6),
\qquad
\widetilde{\mathfrak Y}_2^{(-)}\in \Lambda^1(M_6),
\qquad
\widetilde{\mathfrak Y}_3^{(-)}\in \Lambda^2(M_6),
\end{equation}
are the coefficient forms of the \emph{derivative} corrections. Since \(d\varepsilon_\pm\in \Lambda^1(M_6)\), they appear in the degree-compatible combinations:
\begin{equation}
\iota_{V_{\varepsilon_+}}\mathfrak Y_1^{(+)},
\qquad
d\varepsilon_+\wedge \mathfrak Y_2^{(+)},
\qquad
d\varepsilon_+\wedge \mathfrak Y_3^{(+)}\nonumber
\end{equation}
\begin{equation}\label{frak3}
\iota_{V_{\varepsilon_-}}\widetilde{\mathfrak Y}_1^{(-)},
\qquad
d\varepsilon_-\wedge \widetilde{\mathfrak Y}_2^{(-)},
\qquad
d\varepsilon_-\wedge \widetilde{\mathfrak Y}_3^{(-)}.
\end{equation}
Thus \(\mathfrak Y_i^{(+)}\) and $\widetilde{\mathfrak Y}_i^{(-)}$ are the two endpoints' analogues of the earlier derivative coefficients \({\cal Y}_i^{(o)}\), \(\widehat W_i^{(T)}\), and the corresponding smooth pinch descendants in the symmetric \(T\)-expansion. In a similar vein:
\begin{equation}\label{frak4}
\left(\mathfrak X_4^{(+)},\ \mathfrak X_5^{(+)}\right)\in \Lambda^1(M_6); ~~~~~~
\left(\widetilde{\mathfrak X}_4^{(-)},\ \widetilde{\mathfrak X}_5^{(-)}\right) \in \Lambda^1(M_6) ,
\end{equation}
are the leading one-form corrections to the Lee-form torsion classes \(W_4\) and \(W_5\) induced by the collapsing \(\pm\) branches respectively. The above set of parameters suggest that we can adapt our analysis to one of the end-point since the behavior at the other end-point will be very similar (given by the tilde parameters). With this in mind, 
we now rewrite the torsion classes near the genuine Type IIA endpoint
$R_+\to 0$ with 
$R_->0$.
Since the direct wedge source is absent there, the full torsion classes take the form:
\begin{equation}
W_i
=
W_i^{\rm fib}
+
W_i^{\rm flux,\,IIA(-)}
+
W_i^{\rm pinch,\,smooth(-)}
+
\delta_{\varepsilon_+}W_i,
\qquad
\varepsilon_+=\frac{R_+}{R_-}e^{-R_+/R_-}\to 0 ,
\label{Wi_endpoint_minus_rewritten_new}
\end{equation}
where as before, the fibration contributions remain unchanged and may be easily extracted from \eqref{toruequ}.
The flux part \(W_i^{\rm flux,\,IIA(-)}\) is just the ordinary IIA-like contribution from the surviving \((-)\) branch, which one may collect from \eqref{W45flux_even}:
\begin{equation}
W_1^{\rm flux,\,IIA(-)}
=
\Pi_{(3,0)+(0,3)}\!\big(H_3^{(-)}\big)\nonumber
\end{equation}
\begin{equation}
W_2^{\rm flux,\,IIA(-)}
=
\Pi_{\rm prim}^{(1,1)}\!\big(d\Omega\big)_{\rm flux,-}
+
\widetilde a_2\,\Pi_{\rm prim}^{(1,1)}\!\big(\iota_{v_-} *_7 G_4^{(27,-)}\big) \nonumber
\end{equation}
\begin{equation}
W_3^{\rm flux,\,IIA(-)}
=
\Pi_{\rm prim}^{(2,1)+(1,2)}\!\big(H_3^{(-)}\big)\nonumber
\end{equation}
\begin{equation}
W_4^{\rm flux,\,IIA(-)}
=
a_4\,X_\perp^{(-)},
\qquad
W_5^{\rm flux,\,IIA(-)}
=
a_5\,X_\perp^{(-)} ,
\label{fluxIIAminus_rewritten_new}
\end{equation}
which {\it does not} clash with supersymmetry as long as the supersymmetry constraints in the presence of the torsion classes are satisfied. We will have more to say about this in the next section. The key point now is the following. The collapsing \(+\) branch contributes only smooth endpoint corrections of the following form:
\begin{equation}
\delta_{\varepsilon_+}W_1
=
\varepsilon_+\,\mathfrak W_1^{(+)}
+
\iota_{V_{\varepsilon_+}}\mathfrak Y_1^{(+)}
+
{\cal O}(\varepsilon_+^2)\nonumber
\end{equation}
\begin{equation}
\delta_{\varepsilon_+}W_2
=
\varepsilon_+\,\mathfrak W_2^{(+)}
+
d\varepsilon_+\wedge \mathfrak Y_2^{(+)}
+
{\cal O}(\varepsilon_+^2)\nonumber
\end{equation}
\begin{equation}
\delta_{\varepsilon_+}W_3
=
\varepsilon_+\,\mathfrak W_3^{(+)}
+
d\varepsilon_+\wedge \mathfrak Y_3^{(+)}
+
{\cal O}(\varepsilon_+^2)\nonumber
\end{equation}
\begin{equation}
\delta_{\varepsilon_+}W_4
=
\varepsilon_+\,\mathfrak X_4^{(+)}
+
\mathfrak c_4\,d\varepsilon_+
+
{\cal O}(\varepsilon_+^2)\nonumber
\end{equation}
\begin{equation}
\delta_{\varepsilon_+}W_5
=
\varepsilon_+\,\mathfrak X_5^{(+)}
+
\mathfrak c_5\,d\varepsilon_+
+
{\cal O}(\varepsilon_+^2) ,
\end{equation}
which are naturally expressed using $\epsilon_+$ and $d\epsilon_+$; and the parameters defined in \eqref{frak1}, \eqref{frak2}, \eqref{frak3} and \eqref{frak4}. The exact form of these parameters are non-trivial to work out because they involve dynamical motion of the pinched circle whose analysis is beyond the scope of this work. 

There is however couple more subtlety. The first one is the following: since in the symmetric regime the pinch contribution did not only produce the direct localized source in \(W_3\), but also smooth descendants that shifted the \(T\)- and \(dT\)-dependent coefficients in the \(W_1\), \(W_2\), and \(W_3\) channels, and added new regular odd one-forms in the \(W_4\) and \(W_5\) channels, one should allow for their endpoint descendants as well. A minimal way to incorporate them is to write:
\begin{equation}
W_i^{\rm pinch,\,smooth(-)}
=
\varepsilon_+\,\mathfrak P_i^{(+)}
+
d\varepsilon_+\wedge \mathfrak Q_i^{(+)} + O(\epsilon_+^2) 
\label{pinch_smooth_endpoint_23}
\end{equation}
\begin{equation}
W_1^{\rm pinch,\,smooth(-)}
=
\varepsilon_+\,\mathfrak P_1^{(+)}
+
\iota_{V_{\varepsilon_+}}\mathfrak Q_1^{(+)}+ O(\epsilon_+^2)\nonumber
\label{pinch_smooth_endpoint_1}
\end{equation}
\begin{equation}
W_4^{\rm pinch,\,smooth(-)}
=
\varepsilon_+\,\mathfrak P_4^{(+)}
+
\mathfrak d_4\,d\varepsilon_++ O(\epsilon_+^2),
~~~~
W_5^{\rm pinch,\,smooth(-)}
=
\varepsilon_+\,\mathfrak P_5^{(+)}
+
\mathfrak d_5\,d\varepsilon_++ O(\epsilon_+^2), \nonumber
\label{pinch_smooth_endpoint_45}
\end{equation}
for $i = 2, 3$.
Here the symbols \(\mathfrak P_i^{(+)}\) and \(\mathfrak Q_i^{(+)}\) denote the smooth endpoint descendants of the pinch sector. They have the same degree structure as the fraktur symbols above and, if desired, may be absorbed into a redefinition of the latter. The second subtlety is the following: near the symmetric wedge regime, the odd pieces were written explicitly in terms of the branch-odd variable \(T\). In particular, one had the odd primitive \((1,1)\) flux contribution to \(W_2\), the odd \(\mathbf 7\)-sector contributions to the Lee-form classes, and, in the pinch sector, the direct primitive descendant in the \(W_3\) channel, expressed respectively in the following way:
\begin{equation}
W_2^{\rm flux,(o)}
\supset
\widetilde a_2^{(o)}\,
\Pi_{\rm prim}^{(1,1)}\!\big(\iota_{v_o} *_7 G_4^{(27,o)}\big)\nonumber
\label{W2odd_symm_short}
\end{equation}
\begin{equation}
W_4^{\rm flux,(o)}
\supset
a_4^{(o)}X_\perp^{(o)},
\qquad
W_5^{\rm flux,(o)}
\supset
a_5^{(o)}X_\perp^{(o)}\nonumber
\label{W45oddX_symm_short}
\end{equation}
\begin{equation}
W_3^{\rm pinch,\,direct}
=
b_3\,T\,
\Pi_{\rm prim}^{(2,1)+(1,2)}\!\big(\iota_{v_b}\Xi_{27}\big).
\label{W3pinch_direct_symm_short}
\end{equation}
Near the true endpoint, however, the localized wedge source is no longer kept as an explicit singular object. What survives are only its \emph{smooth descendants}, and all such \(O(\varepsilon_+)\) and \(O(d\varepsilon_+)\) corrections are packaged into endpoint coefficients. Thus one writes schematically:
\begin{equation}
\delta_{\varepsilon_+}W_2
=
\varepsilon_+\,\mathfrak W_2^{(+)}
+
d\varepsilon_+\wedge \mathfrak Y_2^{(+)}
+
O(\varepsilon_+^2)\nonumber
\end{equation}
\begin{equation}
\delta_{\varepsilon_+}W_4
=
\varepsilon_+\,\mathfrak X_4^{(+)}
+
\mathfrak c_4\,d\varepsilon_+
+
O(\varepsilon_+^2),
\qquad
\delta_{\varepsilon_+}W_5
=
\varepsilon_+\,\mathfrak X_5^{(+)}
+
\mathfrak c_5\,d\varepsilon_+
+
O(\varepsilon_+^2)\nonumber
\end{equation}
\begin{equation}
W_1^{\rm pinch,\,smooth(-)} =
\varepsilon_+\,\mathfrak P_3^{(+)}
+
d\varepsilon_+\wedge \mathfrak Q_3^{(+)}
+
O(\varepsilon_+^2).
\end{equation}
The important point is that the previously explicit terms have \emph{not} disappeared. They are simply absorbed into the coefficients appearing above. In fact what we wanted to show is that, near the Type IIA endpoint, the odd primitive flux term in $W_2$,
the odd $X_\perp$ terms in the Lee forms $W_4$ and $W_5$,
and the primitive pinch descendant in $W_3$
are all absorbed into the endpoint coefficients.
More precisely, this amounts to the following identifications:
\begin{equation}
\mathfrak W_2^{(+)}
=
\widehat{\mathfrak W}_2^{(+)}
+
\widetilde a_2^{(+)}\,
\Pi_{\rm prim}^{(1,1)}\!\big(\iota_{v_+} *_7 G_4^{(27,+)}\big)\nonumber
\label{frakW2_decomp_short}
\end{equation}
\begin{equation}
\mathfrak X_4^{(+)}
=
\widehat{\mathfrak X}_4^{(+)}
+
\alpha_4^{(+)}X_\perp^{(+)},
\qquad
\mathfrak X_5^{(+)}
=
\widehat{\mathfrak X}_5^{(+)}
+
\alpha_5^{(+)}X_\perp^{(+)}\nonumber
\label{frakX45_decomp_short}
\end{equation}
\begin{equation}
\mathfrak P_3^{(+)}
=
\beta_3^{(+)}
+
\widetilde b_3\,
\Pi_{\rm prim}^{(2,1)+(1,2)}\!\big(\iota_{v_b}\Xi_{27}\big).
\label{frakP3_decomp_short}
\end{equation}
Thus the primitive pinch descendant, including other primitive and lee form terms, are still present near the endpoint, but they are no longer written as a separate term; they are hidden inside $\mathfrak W_2^{(+)}$, $\mathfrak X_{4, 5}^{(+)}$ and  \(\mathfrak P_3^{(+)}\).
With this understood, the endpoint-adapted torsion classes are:
\begin{equation}
\begin{aligned}
W_1
&=
W_1^{\rm fib}
+
\Pi_{(3,0)+(0,3)}\!\big(H_3^{(-)}\big)
+
\varepsilon_+\,\mathfrak W_1^{(+)}
+
\iota_{V_{\varepsilon_+}}\mathfrak Y_1^{(+)}
+
\varepsilon_+\,\mathfrak P_1^{(+)}
+
\iota_{V_{\varepsilon_+}}\mathfrak Q_1^{(+)}
+
{\cal O}(\varepsilon_+^2),
\\[4pt]
W_2
&=
W_2^{\rm fib}
+
\Pi_{\rm prim}^{(1,1)}\!\big(d\Omega\big)_{\rm flux,-}
+
\widetilde a_2\,\Pi_{\rm prim}^{(1,1)}\!\big(\iota_{v_-} *_7 G_4^{(27,-)}\big)
\\
&\hspace{2.6cm}
+
\varepsilon_+\,\mathfrak W_2^{(+)}
+
d\varepsilon_+\wedge \mathfrak Y_2^{(+)}
+
\varepsilon_+\,\mathfrak P_2^{(+)}
+
d\varepsilon_+\wedge \mathfrak Q_2^{(+)}
+
{\cal O}(\varepsilon_+^2),
\\[4pt]
W_3
&=
W_3^{\rm fib}
+
\Pi_{\rm prim}^{(2,1)+(1,2)}\!\big(H_3^{(-)}\big)
+
\varepsilon_+\,\mathfrak W_3^{(+)}
+
d\varepsilon_+\wedge \mathfrak Y_3^{(+)}
+
\varepsilon_+\,\mathfrak P_3^{(+)}
+
d\varepsilon_+\wedge \mathfrak Q_3^{(+)}
+
{\cal O}(\varepsilon_+^2),
\\[4pt]
W_4
&=
W_4^{\rm fib}
+
a_4\,X_\perp^{(-)}
+
\varepsilon_+\,\mathfrak X_4^{(+)}
+
\mathfrak c_4\,d\varepsilon_+
+
\varepsilon_+\,\mathfrak P_4^{(+)}
+
\mathfrak d_4\,d\varepsilon_+
+
{\cal O}(\varepsilon_+^2),
\\[4pt]
W_5
&=
W_5^{\rm fib}
+
a_5\,X_\perp^{(-)}
+
\varepsilon_+\,\mathfrak X_5^{(+)}
+
\mathfrak c_5\,d\varepsilon_+
+
\varepsilon_+\,\mathfrak P_5^{(+)}
+
\mathfrak d_5\,d\varepsilon_+
+
{\cal O}(\varepsilon_+^2) ,
\end{aligned}
\label{Wi_endpoint_minus_complete_new}
\end{equation}
which basically collects everything together. A similar set of result with tilde parameters can be attained at the other end-point which we won't show here.

However looking carefully at the set of torsion classes leads to the following puzzle.
Suppose near the endpoint \(R_+\to 0\) with \(R_->0\) we define $\epsilon_+$ as in \eqref{epsplus_endpointsec}.
Then \(\varepsilon_+\to 0\) at the endpoint, but in general
$d\varepsilon_+
\neq 0$
unless the collapse is also stationary. Since the torsion classes \(W_i\) may contain terms such as:
\begin{equation}
d\varepsilon_+\wedge \left(\mathfrak Y_i^{(+)} +\mathfrak Q_i^{(+)}\right),
\qquad
\iota_{V_{\varepsilon_+}}\left(\mathfrak Y_1^{(+)} +\mathfrak Q_1^{(+)}\right),
\qquad
\left(\mathfrak c_{4,5} +\mathfrak d_{4,5}\right)\,d\varepsilon_+,
\label{deps_corrections_problem}
\end{equation}
where we explicitly show their pinch-descendant counterparts,
it seems that even at \(\varepsilon_+=0\) one could still have extra contributions beyond the ordinary IIA torsion classes. That would indeed be unacceptable if one were claiming to be {exactly at the Type IIA endpoint}. The resolution is that the formulas containing \(d\varepsilon_+\) are {endpoint-adapted asymptotic expansions in a neighborhood of the endpoint}, not formulas evaluated on the exact endpoint background itself. Let us clarify this in the following.

Let the family of wedge backgrounds be parameterized by the collapsing branch modulus \(R_+(y)\), where \(y\) denotes coordinates on the internal six-manifold. Then near the endpoint one may write the torsion classes in the following way:
\begin{equation}
W_i(y;\varepsilon_+)
=
W_i^{\rm IIA(-)}(y)
+
\delta_{\varepsilon_+}W_i(y),~~~~~ \delta_{\varepsilon_+}W_i
=
O(\varepsilon_+)
+
O(d\varepsilon_+) ,
\label{Wi_family_expand}
\end{equation}
This is the correct formula for a \emph{family of nearby wedge geometries}. Here \(W_i^{\rm IIA(-)}\) denotes the genuine single-circle Type IIA torsion class, while \(\delta_{\varepsilon_+}W_i\) contains both the smooth odd-flux descendants and the smooth descendants of the pinch sector. The exact Type IIA endpoint is obtained by imposing \(R_+(y)\equiv 0\) as an
identity on the internal space, not merely by setting \(R_+\) to zero at a
single point.  Since \(R_+\) is then the constant zero function, its exterior
derivative vanishes, {\it i.e.}
$R_+(y)\equiv0 ~\Rightarrow ~ dR_+(y)=0$.
If \(dR_+\neq0\), then one is not on the exact endpoint but on a nearby
configuration in which the collapsing branch still varies over the internal
space. Using \(\varepsilon_+\), which is a smooth function of \(R_+/R_-\), this gives:
\begin{equation}
\varepsilon_+(y)\equiv 0
\qquad\Longrightarrow\qquad
d\varepsilon_+(y)=0.
\label{eps_and_deps_zero}
\end{equation}
One may also do a quick quantitative check. From
$\varepsilon_+
=
x\,e^{-x}$,
with 
$x\equiv \frac{R_+}{R_-}$,
we get
$d\varepsilon_+
=
e^{-x}(1-x)\,dx$ 
and 
$dx
=
\frac{1}{R_-^2}(R_-\,dR_+ - R_+\,dR_-)$.
At the endpoint \(R_+=0\), this reduces to
$d\varepsilon_+\big|_{R_+=0}
=
\frac{dR_+}{R_-}$. Therefore 
If
$R_+(y)\equiv 0$,
then automatically
$dR_+=0$,
and hence
$d\varepsilon_+=0$. So no extra torsion survives. However if 
$dR_+\neq 0$,
then
$d\varepsilon_+\to \frac{dR_+}{R_-}\neq 0$.
This simply means one is not yet on the exact Type IIA background, but on a nearby non-supersymmetric deformation of it. In that case extra torsion corrections are perfectly allowed. In particular, they may include smooth descendants of the pinch sector even though the direct localized wedge source has already been removed from the endpoint description.
Therefore on the \emph{exact endpoint background}, we claim that the following condition must hold:
\begin{equation}
\delta_{\varepsilon_+}W_i\Big|_{\rm exact\ endpoint}=0,
\label{deltaWi_zero_endpoint}
\end{equation}
otherwise we will have contradiction with the known Type IIA torsion classes. Imposing \eqref{deltaWi_zero_endpoint} for all paths leading to the either of the two end-points
one recovers precisely:
\begin{equation}
W_i\Big|_{\rm exact\ endpoint}
=
W_i^{\rm fib}
+
W_i^{\rm flux,\,IIA(-)}.
\label{Wi_exact_endpoint}
\end{equation}
So there is no contradiction: the \(d\varepsilon_+\) terms are nonzero only when one is still moving inside the wedge family, not when one has already landed on the exact IIA vacuum.

The ordinary IIA torsion classes describe the intrinsic \(SU(3)\)-structure of the \emph{single-circle} compactification obtained after one branch has fully disappeared. In contrast, the \(\varepsilon_\pm\)-dependent corrections describe how this IIA geometry is deformed when one still remembers the existence of a small collapsing second branch.
Thus the derivative terms \eqref{deps_corrections_problem} are not corrections to the \emph{exact} IIA endpoint. They are corrections to a {near-endpoint wedge configuration}. The same statement applies to the smooth descendants of the pinch sector: they need not vanish in a neighborhood of the endpoint, but they do vanish on the exact endpoint background because there
\begin{equation}
\varepsilon_\pm=d\varepsilon_\pm=0.
\end{equation}
The relation to the symmetric two-branch formulas is now transparent. In the symmetric wedge regime one writes the torsion classes $W_i$ as 
\eqref{Wi_total_final_combined}, 
with 
$W_i^{\rm flux+pinch}$
given by \eqref{W_flux_pinch_combined_final},
so that both the branch-even flux sector and the explicit pinch descendants are present. At the actual Type IIA endpoints one instead writes the torsion classes as \eqref{Wi_endpoint_minus_rewritten_new}
(say in the $-$ branch), and one gets \eqref{Wi_exact_endpoint}
at the IIA point. Equivalently, the symmetric \(T\)-expansion describes the two-branch type-\(0\) regime, whereas the \(\varepsilon_\pm\)-expansion describes a neighborhood of the true Type IIA endpoints in which the direct localized pinch source is absent but its smooth descendants may still be present until one lands exactly on the endpoint.

\subsection{Supersymmetry breaking and gauge sector from flux, fibration and pinch \label{sosiebrake}}

 Although the reduced theory is type \(0\)A and therefore has no unbroken target-space supersymmetry, the intrinsic torsion of the \(SU(3)\)-structure still provides a precise measure of the failure of the \emph{would-be} supersymmetry conditions. A convenient set of diagnostics is:
\begin{equation}\label{diagnostics}
\Delta_{\rm int}
\equiv
(W_1,W_2),
\qquad
\Delta_{\rm Lee}
\equiv
2W_4+W_5 .
\end{equation}
For a Strominger-type supersymmetric \(SU(3)\)-structure one would require
$W_1=W_2=0$
and 
$2W_4+W_5=0$.
In the present background these conditions fail because the torsion receives independent contributions from the fibration, the flux sector, and the pinch sector:
\begin{equation}
W_i
=
W_i^{\rm fib}
+
W_i^{\rm flux+pinch},
\qquad
i=1,\dots,5 
\label{Wi_master_short}
\end{equation}
which we managed to quantify in \eqref{Wi_total_final_combined}
and \eqref{W_flux_pinch_combined_final}. Our aim here is to explicitly work out the diagnostic factors from \eqref{Wi_master_short} using the values of the torsion classes. We can even represent \eqref{Wi_master_short} more finely as: 
\begin{equation}
\Delta_{\rm int}
=
\Delta_{\rm int}^{\rm fib}
+
\Delta_{\rm int}^{\rm flux+pinch},
\qquad
\Delta_{\rm Lee}
=
\Delta_{\rm Lee}^{\rm fib}
+
\Delta_{\rm Lee}^{\rm flux+pinch} ,
\end{equation}
so that there is a neat separation between the fibration and the flux-plus-pinch contributions. This is useful because, 
using the combined flux-plus-pinch structure developed earlier in \eqref{Wi_total_final_combined}
and \eqref{W_flux_pinch_combined_final}, the obstruction data may be written schematically as:

{\footnotesize
\begin{equation}\label{fluxpluspinch}
\begin{aligned}
\Delta_{\rm Lee}^{\rm flux+pinch}
 = &
(2a_4+a_5)\,X_\perp^{(e)}
+
(2a_4^{(o)}+a_5^{(o)})\,X_\perp^{(o)}
+
(2c_4^{(o)}+c_5^{(o)})\,dT\\
 + &
~T\,(2{\cal W}_4^{(o)}+{\cal W}_5^{(o)})
+
(2b_4+b_5)\,d\varepsilon
+
\varepsilon\,(2W_4^{(\varepsilon,{\rm pinch})}+W_5^{(\varepsilon,{\rm pinch})})
+
O(T^2,\varepsilon^2)\\
\Delta_{\rm int}^{\rm flux+pinch}
= &
\Big(
\Pi_{(3,0)+(0,3)}(H_3^{(e)})
+
T\,{\bf W}_1^{(o)}
+
\iota_{V_T}{\bf Y}_1^{(o)},\;\;
{\rm L}_J^{-1}\big[\Pi_{\rm prim}^{(1,1)}(d\Omega)_{\rm flux,e}\big]\\
+ &
~ \widetilde a_2\,\Pi_{\rm prim}^{(1,1)}(\iota_{v_e} *_7 G_4^{(27,e)})
+
\widetilde a_2^{(o)}\,\Pi_{\rm prim}^{(1,1)}(\iota_{v_o} *_7 G_4^{(27,o)})
+
T\,{\bf W}_2^{(o)}
+
dT\wedge {\bf Y}_2^{(o)}
\Big)
+
O(T^2) ,
\end{aligned}
\end{equation}}
where $\epsilon$ is the regular branch-odd parameter introduced earlier in \eqref{global_eps_endpoint}. Thus the obstruction to the would-be supersymmetry conditions is controlled quantitatively by three ingredients: (i) the non-integrable fibration data,
(ii) the \(\mathbf 7\)- and \(\mathbf{27}\)-components of the original \(G_4\)-flux, and (iii) the branch-odd wedge deformation, encoded by \(T\), \(dT\), \(\varepsilon\), and \(d\varepsilon\).
In particular, the pinch is not a spectator effect: it contributes directly to the intrinsic torsion through the odd coefficients \({\bf W}_i^{(o)}\), \({\bf Y}_i^{(o)}\) and through the regular Lee-form terms proportional to \(d\varepsilon\) and \(\varepsilon\). Therefore when there is no pinch, the induced wedge obstruction disappears. These are the exact Type IIA endpoints where one of the two branches collapses completely. Then the diagnostic is captured by:
\begin{equation}
\varepsilon_\pm=0,
\qquad
d\varepsilon_\pm=0,
\end{equation}
where $\epsilon_\pm$ are defined in \eqref{epsplus_endpointsec}; so the smooth endpoint descendants of the pinch sector decouple, and one recovers the ordinary IIA-like torsion built from the surviving branch. Away from the exact endpoints, however, the collapsing branch leaves residual corrections, and these precisely measure the departure from the supersymmetric IIA limit. Therefore, the reduction of M-theory on
$M_7=K3\rtimes \Sigma_3$, with a deformed K3 defined as in \eqref{locK3},
along the pinched circle produces a type-\(0\)A compactification on
a six-dimensional non-K\"ahler manifold that is locally represented by \eqref{m6def}
whose \(SU(3)\)-structure torsion is controlled by the three ingredients from \eqref{Wi_total_final_combined}.
Near the symmetric wedge regime, the combined flux-plus-pinch sector is organized by the branch-even and branch-odd data as in
\eqref{W_flux_pinch_combined_final}; near the genuine Type IIA endpoints, the same data are reorganized in terms of the endpoint variables \(\varepsilon_\pm\), and the smooth descendants of the collapsing branch decouple only on the exact endpoint background.
The failure of the would-be supersymmetry conditions is measured by
\eqref{diagnostics}
and is sourced simultaneously by
the non-integrable fibration,
the \(\mathbf 7\)- and \(\mathbf{27}\)-components of \(G_4\),
and the branch-odd wedge deformation.

We can also talk about the orientifold interval and gauge-sector structure. Because the internal space contains the orientifold interval
$\frac{S_a^1}{\mathbb Z_2}$,
there are two fixed loci at
$x_a=0$, 
and 
$x_a=\pi R_a$,
which in the conservative type-\(I'\)-like description are O8-planes. Spacetime-filling D8-branes may sit on the fixed loci or in the interval. If \(N_L\) and \(N_R\) denote the D8-brane multiplicities at the two ends, then for the ordinary O8\({}^-\) projection the local eight-dimensional gauge algebra is:
\begin{equation}
\mathfrak g_{8d}
=
\mathfrak{so}(2N_L)\oplus \mathfrak{so}(2N_R),
\label{so2N_8d_rewritten}
\end{equation}
while an O8\({}^+\) projection would replace the orthogonal factors by symplectic ones. In the symmetric locally tadpole-free configuration
$N_L=N_R=8$,
one obtains the oft-quoted gauge group of 
$G_{8d}^{\rm sym}
=
SO(16)_L\times SO(16)_R$.
If one keeps the type-\(0\) doubling of open-string sectors, the natural extension is:
\begin{equation}
\mathfrak g_{8d}^{(0A)}
=
\mathfrak g_L^{(+)}
\oplus
\mathfrak g_L^{(-)}
\oplus
\mathfrak g_R^{(+)}
\oplus
\mathfrak g_R^{(-)},
\label{gauge_0A_doubled_rewritten}
\end{equation}
with each factor orthogonal or symplectic depending on the orientifold choice. As emphasized earlier, this doubled open-string structure should be viewed as the natural type-\(0\)A expectation, not as a fully established microscopic M-theory derivation in every sector.
The D8/O8 system is governed by the Bianchi identity for the Romans mass:
\begin{equation}
dF_0
=
\sum_a \left[N_a\,\delta(x_a-x_a^{(a)})
-
8\,\delta(x_a)
-
8\,\delta(x_a-\pi R_a)\right] + \cdots ,
\label{Bianchi_F0_O8D8_rewritten}
\end{equation}
where $\delta(x_a-x_a^{(a)})$ denotes a delta-function localized at the
position $x_a^{(a)}$ of a D8-brane stack, while $\delta(x_a)$ and
$\delta(x_a-\pi R_a)$ denote delta-functions localized on the two O8-planes. The dotted terms are the induced effects from the localized DOFs at the pinch as well as the tachyon $T$. (For the present discussion, we will ignore these effects.) The integer $N_a$ denotes the number of D8-branes in the $a$-th stack.  In the
normalization used in \eqref{Bianchi_F0_O8D8_rewritten}, a single D8-brane has
D8-charge $+1$, while each O8-plane has D8-charge $-8$.  Therefore the two
orientifold planes together carry total D8-charge
$Q_{\rm O8/O8}
=
-8-8
=
-16$.
The total localized D8-charge on the interval is therefore
$Q_{\rm loc}
=
\sum\limits_a N_a -16$.
On a compact interval, the integrated Bianchi identity gives the tadpole
condition
$\int_{S^1_a/\mathbb Z_2} dF_0
=
0$ which implies that
$\sum\limits_a N_a =16$.
Thus, in the standard compact D8/O8 background, the total positive D8-brane
charge cancels the total negative orientifold-plane charge. The Bianchi identity \eqref{Bianchi_F0_O8D8_rewritten} also suggests that 
\(F_0\) is piecewise constant and jumps across D8-brane stacks\footnote{The statement that $F_0$ is piecewise constant follows directly from the
Bianchi identity.  Away from localized D8/O8 sources, and ignoring the induced effects from the pinch, one has
$dF_0=0$,
and hence
$F_0=\text{constant}$
on each open interval between adjacent localized sources.  Across a source,
$F_0$ jumps by the charge carried by that source.  More explicitly, integrating
the Bianchi identity across small intervals surrounding a D8-brane stack at
$x_a=x_a^{(b)}$, O8-planes at $x_a = 0$, and another set of O8-planes at $x_a = \pi R_a$, give:
\begin{equation}
F_0\big(x_a^{(b)}+\epsilon\big)
-
F_0\big(x_a^{(b)}-\epsilon\big)
=
N_b, \;\;
F_0(0^+)-F_0(0^-)
=
-8, \;\;
F_0((\pi R_a)^+)-F_0((\pi R_a)^-)
=
-8 , \nonumber
\end{equation}
respectively.
In the interval description, one often keeps only the physical region
$0\leq x_a\leq \pi R_a$, in which case the orientifold planes appear as boundary
sources.  The same charge information is then encoded in boundary jump
conditions for $F_0$.
If the D8-brane stacks are ordered along the interval as
$0<x_a^{(1)}<x_a^{(2)}<\cdots <x_a^{(k)}<\pi R_a$,
then $F_0$ takes a constant value on each segment.  For example, if the value
just to the right of the first O8-plane is denoted by $m_0$, then
$F_0(x_a)
=
m_0$ for
$0<x_a<x_a^{(1)}$,
while after crossing the first D8 stack one has
$F_0(x_a)
=
m_0+N_1$
for 
$x_a^{(1)}<x_a<x_a^{(2)}$.
More generally, on the interval
$x_a^{(j)}<x_a<x_a^{(j+1)}$,
one has:
\begin{equation}
F_0(x_a)
=
m_0+\sum_{b=1}^{j}N_b .\nonumber
\label{piecewise_F0}
\end{equation}
Thus the Romans mass is quantized and changes only by integer amounts when one
crosses a localized D8-brane stack.  This is the precise meaning of the
statement that $F_0$ is piecewise constant and jumps across D8-brane stacks. Note however that the above analysis ignores the effects from the pinch.}. After further compactification on
$T_f^2\rtimes \Sigma_3$,
the four-dimensional gauge group is the commutant of all geometric and flux data:
\begin{equation}
G_{4d}
=
{\rm Com}_{G_{8d}}
\Big(
{\rm Hol}(V_{\rm D8})
\cup
{\rm Mon}_{\rm fib}
\cup
{\rm Flux}_{\rm stuck}
\cup
{\rm Def}_{\rm wedge}
\Big),
\label{G4d_commutant_0A_rewritten}
\end{equation}
where the various term appearing above are explained in the following way. The notation
${\rm Com}_{G_{8d}}({\cal S})$
means the commutant, or centralizer, of the set ${\cal S}$ inside $G_{8d}$. At the level of the Lie algebras $\mathfrak g_{4d}$ this should be the vanishing of a commutator. 
Explicitly, this means:
\begin{equation}
{\rm Com}_{G_{8d}}({\cal S})
=
\left\{
g\in G_{8d}
\; \big| \;
gs=sg
\quad
\forall
\quad
s\in {\cal S}
\right\} \nonumber
\end{equation}
\begin{equation}\label{commdeff}
\mathfrak g_{4d}
=
\left\{
X_A\in \mathfrak g_{8d}
\; \big| \;
[X,Y]=0
\quad
\forall
\quad
Y\in \mathfrak s
\right\},
\end{equation}
where $\mathfrak s$ denotes the algebraic data associated with the holonomies,
monodromies, flux couplings, and branch-odd deformations. The holonomy
${\rm Hol}(V_{\rm D8})$,
is the holonomy of the gauge bundle living on the D8-branes.  If
$V_{\rm D8}$ is a gauge bundle with connection $A$, then the holonomy around a
closed one-cycle $\gamma$ in the compactification space is:
\begin{equation}
U_\gamma
=
{\cal P}\exp
\left(
i\oint_\gamma A
\right),
\end{equation}
where ${\cal P}$ denotes path ordering.  A four-dimensional gauge boson survives
only if its generator commutes with all such holonomies, {\it i.e.}
for $X_A\in\mathfrak g_{8d}$,
$U_\gamma X_A U_\gamma^{-1}=X_A$, or equivalently
$[U_\gamma,X_A]=0$. 
Thus the bundle holonomy breaks $G_{8d}$ to the subgroup that commutes with the
chosen Wilson lines or gauge bundle background. 

The monodromy
${\rm Mon}_{\rm fib}$,
denotes the monodromy induced by the fibration
$T_f^2\rtimes \Sigma_3$.
If $\gamma$ is a one-cycle in $\Sigma_3$, then transporting the torus fiber
around $\gamma$ may act on the torus one-cycles as:
\begin{equation}
\begin{pmatrix}
A\\
B
\end{pmatrix}
\longmapsto
M_\gamma
\begin{pmatrix}
A\\
B
\end{pmatrix},
\qquad
M_\gamma\in SL(2,\mathbb Z).
\end{equation}
Here $A$ and $B$ denote a basis of one-cycles of the torus fiber $T_f^2$, while
$M_\gamma$ is the geometric monodromy matrix associated with the loop $\gamma$
in the base $\Sigma_3$. More generally, this geometric monodromy may induce an action on the full
charge lattice of the compactification.  This charge lattice can include KK
momenta, winding charges, wrapped-brane charges, and gauge charges.  We denote
the induced action by
$\rho(M_\gamma)$.
Thus $\rho(M_\gamma)$ is the representation of the geometric monodromy on the
space in which the relevant gauge or charge data are defined.
A four-dimensional gauge boson, associated with a generator
$X_A\in \mathfrak g_{8d}$, survives the compactification only if it is globally well-defined
over the fibration.  Equivalently, it must be invariant under the monodromy
action induced by transporting the fiber around any closed loop $\gamma$ in
$\Sigma_3$, {\it i.e.},

{\footnotesize
\begin{equation}
\rho(M_\gamma)\,X_A\,\rho(M_\gamma)^{-1}
=
X_A
~~~
\forall
~~~
\gamma\in H_1(\Sigma_3,\mathbb Z) ~~\implies ~~
\left[\rho(M_\gamma),X_A\right]=0
~~ \forall ~~
\gamma\in H_1(\Sigma_3,\mathbb Z)\nonumber
\end{equation}
\begin{equation}
\mathfrak g_{4d}
\subset
\left\{
X_A\in \mathfrak g_{8d}
\;\big|\;
\left[\rho(M_\gamma),X_A\right]=0
\quad
\text{for all}
\quad
\gamma\in H_1(\Sigma_3,\mathbb Z)
\right\}.
\end{equation}}
Therefore ${\rm Mon}_{\rm fib}$ removes from the four-dimensional gauge group
all generators that are not invariant under the induced monodromy action, with the understanding that ${\rm Mon}_{\rm fib}$ denotes not merely the
geometric matrices $M_\gamma\in SL(2,\mathbb Z)$ acting on the torus cycles, but
their induced action $\rho(M_\gamma)$ on the gauge and charge data.

The third entry,
${\rm Flux}_{\rm stuck}$, in \eqref{G4d_commutant_0A_rewritten} is interesting.
It denotes flux-induced St\"uckelberg couplings.  These couplings can make some
four-dimensional gauge fields massive even if the corresponding symmetry is
present at the level of the higher-dimensional gauge algebra.  A standard
four-dimensional St\"uckelberg coupling has the form:
\begin{equation}
{\cal L}_{\rm stuck.}
=
-\frac{1}{2}
G_{IJ}
\left(
d a^I + k^I_A A^A
\right)
\wedge
*
\left(
d a^J + k^J_B A^B
\right),
\end{equation}
where $a^I$ are axions, $A^A$ are four-dimensional gauge fields, $G_{IJ}$ is the
axion metric, and $k^I_A$ are integer charges determined by the background
fluxes.  The corresponding gauge transformations are standard:
$A^A\longmapsto A^A+d\lambda^A$, and 
$a^I\longmapsto a^I-k^I_A\lambda^A$.
The gauge boson mass matrix is then schematically
$(M^2)_{AB}
=
G_{IJ}k^I_A k^J_B$.
Therefore a four-dimensional gauge field remains massless only if its generator
lies in the kernel of the St\"uckelberg charge matrix:
\begin{equation}
k^I_A v^A=0
~~~
\forall
~~~
I .
\end{equation}
Thus ${\rm Flux}_{\rm stuck}$ should be understood quantitatively as the set of
flux-induced charges $k^I_A$ that determine which gauge bosons acquire
St\"uckelberg masses. 

This brings us to the last entry in \eqref{G4d_commutant_0A_rewritten}, namely
${\rm Def}_{\rm wedge}$.
This term collects the branch-odd deformation data associated with the wedge
asymmetry and with its endpoint descendants.  In the wedge construction, one
has two local branches, which may be denoted schematically by
$S^1_+ \vee S^1_-$.
A symmetric configuration is invariant under exchange of the two branches,
$S^1_+ \longleftrightarrow S^1_-$.
A wedge asymmetry is a deformation that is odd under this exchange.
To avoid confusion with a tachyon field, let us denote the deformation
parameters, or effective branch tensions, by
${\cal T}_+$,
and 
${\cal T}_-$.
These quantities are not tachyon fields.  They are parameters measuring the
relative deformation strength, or effective tension, associated with the two
branches of the wedge.  A convenient branch-odd and branch-even combinations are:
\begin{equation}
{\cal T}_{\rm odd}
=
\frac{1}{2}
\left(
{\cal T}_+ - {\cal T}_-
\right), ~~~~~~~
{\cal T}_{\rm even}
=
\frac{1}{2}
\left(
{\cal T}_+ + {\cal T}_-
\right).
\end{equation}
Under exchange of the two branches,
one has
${\cal T}_+
\longleftrightarrow
{\cal T}_-$.
Therefore
${\cal T}_{\rm even}
\longmapsto
{\cal T}_{\rm even}$, and 
${\cal T}_{\rm odd}
\longmapsto
-
{\cal T}_{\rm odd}$.
Thus ${\cal T}_{\rm odd}$ is the branch-odd deformation parameter, while
${\cal T}_{\rm even}$ is branch-even. In the lower-dimensional theory, such branch-odd data can descend to localized
torsion, localized flux, or localized gauge-bundle deformations at the endpoints
of the interval.  Schematically, one may write the endpoint descendants as a
collection of background deformation data
${\cal W}_r,
\qquad
r=1,\ldots,n_{\rm def}$,
where the label $r$ runs over the different localized descendants of the wedge
asymmetry.  These descendants may include localized torsion classes, localized
flux components, or localized gauge-bundle data.  They act as additional
background structures and can therefore break the gauge group further.
If a wedge descendant is represented by an adjoint-valued element
${\cal W}_r\in \mathfrak g_{8d}$,
then a four-dimensional gauge generator
$X_A\in \mathfrak g_{8d}$
survives only if it commutes with every such descendant:

{\footnotesize
\begin{equation}
[X_A,{\cal W}_r]=0
~~~
\forall
~~~
r ~~ \implies ~~X_A
\in
\ker\left({\rm ad}_{{\rm Def}_{\rm wedge}}\right) ~~ \implies ~~
\left\{
X_A\in\mathfrak g_{8d}
\;\big|\;
[{\cal W}_r,X_A]=0
~~~ \forall
~~~
r
\right\} ,
\end{equation}}
implying that the surviving generators lie in the kernel of the adjoint action
of the wedge-deformation data, and 
${\rm ad}_{{\cal W}_r}(X_A)
\equiv
[{\cal W}_r,X_A]$.

Putting all ingredients together, the four-dimensional gauge algebra is the
subalgebra of $\mathfrak g_{8d}$ that survives the D8-brane bundle holonomies,
the fibration monodromies, the flux-induced St\"uckelberg masses, and the
branch-odd wedge descendants.  A precise way to write this is:
\begin{equation}
\mathfrak g_{4d}
=
\left\{
X_A\in\mathfrak g_{8d}
\;\bigg|\;
[U_\gamma,X_A]=0,\quad
[\rho(M_\eta),X_A]=0,\quad
k^I_A=0,\quad
[{\cal W}_r,X_A]=0
\right\} ,
\end{equation}
where $U_\gamma$ denotes a D8-brane bundle holonomy, $\rho(M_\eta)$ denotes the
induced action of a fibration monodromy on the gauge or charge data, $k^I_A$
are the flux-induced Stückelberg charges, and ${\cal W}_r$ are the localized
wedge descendants.  More explicitly, the four conditions are: (i)
$[U_\gamma,X_A]=0$
for all D8-brane bundle holonomies
$U_\gamma$, (ii)
$[\rho(M_\eta),X_A]=0$
for all induced fibration monodromies
$\rho(M_\eta)$,
(iii) $k^I_A=0$ for all axions
$a^I$,
and (iv)
$[{\cal W}_r,X_A]=0$
for all wedge descendants
${\cal W}_r$.
Equivalently, one may write the surviving algebra as the simultaneous
intersection
\begin{equation}\label{frakg8d}
\mathfrak g_{4d}
=
\mathfrak g_{8d}
\cap
\ker\left({\rm ad}_{{\rm Hol}(V_{\rm D8})}\right)
\cap
\ker\left({\rm ad}_{{\rm Mon}_{\rm fib}}\right)
\cap
\ker\left(M^2_{\rm Stuck}\right)
\cap
\ker\left({\rm ad}_{{\rm Def}_{\rm wedge}}\right) ,
\end{equation}
where the expression should be understood as follows.  The first two kernels impose
ordinary commutant conditions: the gauge generator must commute with the
D8-brane bundle holonomies and with the induced monodromy action of the
fibration.  The third kernel imposes the condition that the gauge boson remains
massless under flux-induced St\"uckelberg couplings.  The fourth kernel imposes
compatibility with the branch-odd wedge-deformation data and with the localized
endpoint descendants of that data.

Therefore to conclude, the local non-abelian gauge sector is that of an O8/D8 system on the orientifold interval, while the four-dimensional gauge group is obtained by taking the commutant with respect to D8-bundle holonomy, geometric monodromy, flux-induced St\"uckelberg couplings, and the wedge-deformation data. Without specifying an explicit metric, a complete flux ansatz, and a fully microscopic orientifold realization, this is the sharpest controlled description of the Type 0A compactification on a non-K\"ahler six-manifold.

\section{Type 0 Heterotic theory compactified on a non-K\"ahler six-manifold \label{sec40}}

Having studied the Type 0A compactified on a non-K\"ahler six-manifold in details,  
we now elaborate on a formal Ho\v rava--Witten-type reduction of M-theory with $G_4$ flux on a seven-manifold of the schematic form:
\begin{equation}\label{77mani}
X_7
=
T^2_f \rtimes S^1_a/\mathbb{Z}_2 \rtimes (S^1\vee S^1)_b \rtimes \Sigma_3 ,
\end{equation}
equipped with a $G_2$ structure and a localized pinch source associated with the wedge factor $(S^1\vee S^1)_b$. Compactifying along $S^1_a/\mathbb{Z}_2$, we obtain a six-dimensional internal space:
\begin{equation}\label{76hw}
\mathbb{X}_6 \equiv X_6^{(0{\rm HW})}
=
T^2_f \rtimes (S^1\vee S^1)_b \rtimes \Sigma_3 ,
\end{equation}
which we interpret as the internal geometry of a type $0$ heterotic Ho\v rava--Witten-like theory. Our aim here would be to derive the induced $SU(3)$-structure torsion classes coming from fibration, flux, and pinch sectors, and compare them quantitatively with the torsion classes obtained earlier from the type $0$A reduction along $(S^1\vee S^1)_b$. We will see that the pinch contribution survives explicitly as a localized primitive torsion source in the $0$HW reduction, whereas in the $0$A reduction it was repackaged into odd/endpoint coefficient deformations in the reduced torsion classes. In the heterotic side, we will provide some explanation as to why the resulting $SU(3)$-structure torsion classes look so different, whether there is a one-to-one relation between them, and why the pinch source appears to affect $W_2$ and $W_3$ directly in the $0$HW description whereas in the $0$A analysis it seemed to contribute directly only to $W_3$. The resolution appears to be that both reductions are different projections of the same parent $G_2$ torsion data, but they keep different structures explicit. The type $0$HW presentation is simpler because it is a compressed leading-order, localized-source description, whereas the type $0$A presentation is an unpacked branch-resolved description in which the same information is distributed among even/odd sectors, endpoint corrections, and flux-plus-pinch combinations. In the following let us elaborate the aforementioned points in details.

\subsection{Reduction to the type $0$ Ho\v rava--Witten  branch and torsion classes \label{sec41}}

We begin with M-theory on a seven-manifold \eqref{77mani}
with background $G_4$ flux and a $G_2$ structure specified by a positive three-form $\Phi$ and its Hodge dual $*_7\Phi$. We assume that the intrinsic torsion of the $G_2$ structure admits the standard decomposition \eqref{torsiondecomp2}
with $\tau_i$ given by \eqref{taui}.
We further assume, in direct analogy with the earlier analysis, that the $G_2$ torsion classes admit three contributions:
\begin{equation}
\tau_i
=
\tau_i^{\rm fib}
+
\tau_i^{\rm flux}
+
\tau_i^{\rm pinch},
\qquad
i=0,1,2,3 .
\label{taudecomp1}
\end{equation}
The first two pieces arise from the non-trivial fibration structure and the background $G_4$ flux, while the pinch piece is localized at the junction locus of the wedge:
\begin{equation}
(S^1\vee S^1)_b
=
S^1_{b,+}\cup S^1_{b,-},
\end{equation}
where the two branches meet at one point.
The crucial assumption inherited from the previous discussion is that, at leading order in the pinch parameter $T$, the pinch source enters only through the primitive $\mathbf{27}$ sector:
\begin{equation}
\tau_0^{\rm pinch} = 0,
\qquad
\tau_1^{\rm pinch} = 0,
\qquad
\tau_2^{\rm pinch} = 0,
\qquad
\tau_3^{\rm pinch}
=
T\, \delta_{\Sigma^{\rm jct}_6}\, \Xi^{(3)}_{27},
\label{pinch27assumption}
\end{equation}
where $\delta_{\Sigma^{\rm jct}_6}$ denotes a localized delta-current supported on the six-dimensional junction locus in the seven-manifold, and $\Xi^{(3)}_{27}$ is a primitive $G_2$ three-form in the $\mathbf{27}$ representation. This is the M-theory input that we now reduce along $S^1_a/\mathbb{Z}_2$.
Formally, this gives a Ho\v rava--Witten-type reduction to a heterotic-like theory with six-dimensional internal space ${X}^{\rm 0HW}_6$ given by \eqref{76hw}.
We assume that $X_6^{(0{\rm HW})}$ admits an $SU(3)$ structure specified by a real two-form $J$ and a complex decomposable three-form $\Omega$ satisfying the usual compatibility conditions:
\begin{equation}
J\wedge \Omega = 0,
\qquad
\frac{i}{8}\,\Omega\wedge \overline{\Omega}
=
\frac{1}{3!}\,J^3 .
\end{equation}
The intrinsic torsion is then encoded in the standard $SU(3)$-structure decomposition \eqref{dJtorsion}
where
$W_1 \in \Lambda^0 \otimes \mathbb{C},
W_2 \in \Lambda^{(1,1)}_{0} \otimes \mathbb{C},
W_3 \in \Lambda^{(2,1)+(1,2)}_{0},
W_4 \in \Lambda^1$ and 
$W_5 \in \Lambda^{(1,0)}$.
As before, we decompose the torsion classes into fibration, flux, and pinch pieces:
\begin{equation}
W_i
=
W_i^{\rm fib}
+
W_i^{\rm flux}
+
W_i^{\rm pinch},
\qquad
i=1,\dots,5 ,
\label{Wdecomp}
\end{equation}
which is generically achieved by projecting the $G_2$ structure to an {$SU(3)$} structure. This is achieved as follows.
To reduce from a $G_2$ structure on $X_7$ to an $SU(3)$ structure on $X_6^{(0{\rm HW})}$, we locally split the seven-dimensional geometry as:
\begin{equation}
X_7
\sim
X_6^{(0{\rm HW})} \rtimes S^1_a/\mathbb{Z}_2 ,
\end{equation}
and write the $G_2$ structure parameters $(\varphi, \psi)$ in terms of the $SU(3)$ structure parameters $(J, \Omega)$ as \eqref{G2SU3decomp}, 
where $v = e^a$
and $e^a$ is the one-form along the orbifold interval $S^1_a/\mathbb{Z}_2$. Note that, when reducing \eqref{torsiondecomp2}, the $G_2$ torsion classes project into the $SU(3)$ torsion classes by separating the pieces tangent and orthogonal to $e^a$. Schematically one obtains:
\begin{equation}
W_i
\sim
{\cal P}_i(\tau_0,\tau_1,\tau_2,\tau_3),
\label{Wtauprojection}
\end{equation}
where ${\cal P}_i$ are projection maps determined by the decomposition of $G_2$ representations under
$G_2 \supset SU(3)$.
In particular, the $\mathbf{27}$ of $G_2$ decomposes as:
\begin{equation}
\mathbf{27}
\to
\mathbf{1}
\oplus
\mathbf{8}
\oplus
\mathbf{6}
\oplus
\overline{\mathbf{6}}
\oplus
\mathbf{3}
\oplus
\overline{\mathbf{3}},
\label{27decomp}
\end{equation}
so a primitive $\tau_3$ source can feed into the primitive $(1,1)$ sector $W_2$ and the primitive $(2,1)+(1,2)$ sector $W_3$, while potentially also inducing trace/vector pieces if the source is not purely primitive relative to the chosen $SU(3)$ structure.
By construction, however, the pinch source \eqref{pinch27assumption} was chosen to be primitive at leading order. Therefore its leading projection obeys $W_i = 0$ for $i = 1, 4, 5$
while $W_2$ and $W_3$ torsion classes take the following form:
\begin{equation}
W_2^{\rm pinch}
=
\kappa_2\, T\, \delta_{\Sigma^{\rm jct}_5}\, \xi_2,
\qquad
W_3^{\rm pinch}
=
\kappa_3\, T\, \delta_{\Sigma^{\rm jct}_5}\, \xi_3 .
\label{W23pinch}
\end{equation}
Here $\delta_{\Sigma^{\rm jct}_5}$ is the delta-current obtained by integrating $\delta_{\Sigma^{\rm jct}_6}$ over the interval direction, $\xi_2$ is a primitive $(1,1)$-form on $X_6^{(0{\rm HW})}$, $\xi_3$ is a primitive real $(2,1)+(1,2)$ form, and $\kappa_2,\kappa_3$ are numerical projection coefficients that depend on the detailed local embedding of the pinch inside the $SU(3)$ structure.

\subsubsection{How is eq. \eqref{W23pinch} compatible with the Type 0A reduction of $\tau_3$? \label{sec411}}

The equation \eqref{W23pinch} immediately raises the following question:
Why did it seem in the type {$0$}A case that {$\tau_3$} contributed directly only to {$W_3$}? In fact this is an important conceptual point.
At the parent $G_2$ level, the primitive $\mathbf{27}$ source
$\tau_3^{\rm pinch}
=
T\,\delta_{\Sigma^{\rm jct}_6}\,\Xi^{(3)}_{27}$
does not project only to one $SU(3)$ torsion class in general. Under
$G_2 \supset SU(3)$,
the $\mathbf{27}$ decomposition \eqref{27decomp} would immediately suggest that 
the primitive $(1,1)$ piece belongs to the $\mathbf{8}$ and contributes to $W_2$, while the primitive $(2,1)+(1,2)$ pieces belong to the $\mathbf{6}\oplus\overline{\mathbf{6}}$ and contribute to $W_3$.
Therefore, in a generic reduction,
$\tau_3$ can contribute to both 
$W_2$
and $W_3$.
So why did the type $0$A discussion make it look as though the direct contribution of $\tau_3$ was primarily to $W_3$?

The question can be made more precise by making the following set of observations. The first one concerns the following. Under \(SU(3)\), the standard torsion classes transform as:
\begin{equation}
W_1 \in \mathbf{1},
\qquad
W_2 \in \mathbf{8},
\qquad
W_3 \in \mathbf{6}\oplus \overline{\mathbf{6}},
\qquad
W_4 \in \mathbf{3}\oplus \overline{\mathbf{3}},
\qquad
W_5 \in \mathbf{3}\oplus \overline{\mathbf{3}}.
\label{SU3torsionreps}
\end{equation}
So once one decomposes ${\bf 27}$ as in \eqref{27decomp}
one immediately sees that (i) the \(\mathbf{8}\) can feed \(W_2\),
(ii) the \(\mathbf{6}\oplus\overline{\mathbf{6}}\) can feed \(W_3\),
(iii) the \(\mathbf{3}\oplus\overline{\mathbf{3}}\) can feed \(W_4\) and/or \(W_5\), and 
(iv) the \(\mathbf{1}\) can feed the singlet torsion \(W_1\).
Therefore, at the purely group-theoretic level, it is incorrect to claim that the
$\mathbf{27}$ gives only \(W_2\) and \(W_3\). So why are we claiming that the $G_2$ torsion only descends to $W_2$ and $W_3$? This is precisely the question that we aim to answer here.

One immediate conclusion, motivated from our earlier arguments, can be the following. 
If the pinch deformation is taken to be 
primitive, traceless, orthogonal to the singlet,
and with no Lee-form-type \(\mathbf{3}\oplus\overline{\mathbf{3}}\) component
at the order under consideration,
then the only nonvanishing \(SU(3)\)-components of the \(\mathbf{27}\) are the
$\mathbf{8}$
and 
$\mathbf{6}\oplus\overline{\mathbf{6}}$.
In that case the pinch contributes only to
$W_2$
and 
$W_3$. So the correct logic is that the representation theory allows  $\mathbf{1},\mathbf{8},\mathbf{6}\oplus\overline{\mathbf{6}},\mathbf{3}\oplus\overline{\mathbf{3}}$, but
the specific pinch ansatz may project out the $\mathbf{1}$ and 
$\mathbf{3}\oplus\overline{\mathbf{3}}$ pieces.

So what happens to the \(\mathbf{1}\) and \(\mathbf{3}\oplus\overline{\mathbf{3}}\)?
The singlet \(\mathbf{1}\) would contribute to \(W_1\), {\it i.e.} to the nearly-K\"ahler-type
scalar torsion. So if one says there is no pinch contribution to \(W_1\), one is
implicitly assuming that the singlet component of the \(\mathbf{27}\) is absent.  Similarly, the \(\mathbf{3}\oplus\overline{\mathbf{3}}\) would contribute to the
Lee-form-type classes \(W_4\) and \(W_5\). So if one says the pinch is entirely in
\(W_2\) and \(W_3\), one is also implicitly assuming the absence of ${\bf 3} \oplus \overline{\bf 3}$. Quantitatively, this means:
\begin{equation}
\Pi_{\mathbf{1}}(\text{pinch})=0,~~~~~~~ \Pi_{\mathbf{3}\oplus\overline{\mathbf{3}}}(\text{pinch})=0,
\label{singlet_zero}
\end{equation}
at the order being studied. This may happen, for example, if the pinch is taken to be primitive and not to
induce an overall conformal rescaling or Lee-form deformation. But this is an
\emph{extra assumption}, not a consequence of the decomposition
\eqref{27decomp} itself. For example, on a six-manifold with \(SU(3)\) structure, any real three-form \(\rho_3\) can be
decomposed uniquely as:
\begin{equation}
\rho_3
=
a\,\Omega
+
\overline{a}\,\overline{\Omega}
+
J\wedge \alpha
+
\rho_3^{\rm prim},~~~~~ a\in \mathbb C,
~~~
\alpha\in \Lambda^1,
~~~
\rho_3^{\rm prim}\in \Lambda_{\rm prim}^{(2,1)+(1,2)} \nonumber
\end{equation}
\begin{equation}
\rho_3 \in \Lambda^3_{\mathbb R}
=
\underbrace{\langle {\rm Re}\,\Omega,\ {\rm Im}\,\Omega\rangle}_{\mathbf{1}\oplus \mathbf{1}}
\;\oplus\;
\underbrace{J\wedge \Lambda^1}_{\mathbf{3}\oplus \overline{\mathbf{3}}}
\;\oplus\;
\underbrace{\Lambda_{\rm prim}^{(2,1)+(1,2)}}_{\mathbf{6}\oplus \overline{\mathbf{6}}} ,
\label{rho3primate}
\end{equation}
where primitivity of the real three-form $\rho_3$ means that 
$J\lrcorner\,\rho_3=0,
\Omega\lrcorner\,\rho_3=0$, and 
$\overline{\Omega}\lrcorner\,\rho_3=0$. Plugging this in \eqref{rho3primate} and using the standard identity $J\lrcorner (J\wedge \alpha) = 2\,\alpha$, immediately gives us $\alpha = 0$. This implies $-$ since $\alpha$ is precisely the coefficient of the
\(\mathbf{3}\oplus\overline{\mathbf{3}}\) piece \(J\wedge \alpha\) $-$ the
vanishing of the \(\mathbf{3}\oplus\overline{\mathbf{3}}\) component.
Similar, using the fact that $ \overline{\Omega}\lrcorner\,\rho_3
=
a\,(\overline{\Omega}\lrcorner\,\Omega)$ and ${\Omega}\lrcorner\,\rho_3
=
b\,({\Omega}\lrcorner\,\overline{\Omega})$, their vanishing would imply $a = b = 0$ in \eqref{rho3primate}. Together they imply \eqref{singlet_zero}. Notice, however, that the above proof applies to the three-form \(\rho_3\) itself.
It shows that \(\rho_3\) lies purely in
$\mathbf{6}\oplus\overline{\mathbf{6}}$,
not that the full pinch deformation lies only there. The \(\mathbf{8}\) relevant for
\(W_2\) can still arise after reducing the full \(G_2\)-structure deformation or after
projecting other pinch-induced data into the primitive \((1,1)\) sector of the
\(SU(3)\) structure. Thus
primitive $\rho_3$ kills the $\mathbf{1}$ and 
$\mathbf{3}\oplus\overline{\mathbf{3}}$ pieces of that three-form, but it does not by itself
eliminate the possibility of an $\mathbf{8}$ contribution elsewhere in the pinch data. Thus the full analysis will involve going beyond the simple group theory analysis.
In the following, we will provide one reason each in Type 0A and Type 0HW theories to justify why ${\bf 27}$ descends to $W_2$ and $W_3$.

The reason in the Type 0A side is that in the $0$A formulation, the $W_2$ projection of the $\mathbf{27}$ source was partly hidden inside the flux-sector notation. 
Consider the torsion classes from \eqref{W_flux_pinch_combined_final}.
Indeed, the term:
\begin{equation}
\widetilde a_2\,\Pi_{\rm prim}^{(1,1)}\!\big(\iota_{v_e} *_7 G_4^{(27,e)}\big)
+
\widetilde a_2^{(o)}\,\Pi_{\rm prim}^{(1,1)}\!\big(\iota_{v_o} *_7 G_4^{(27,o)}\big)
\label{W2flux27}
\end{equation}
coming from the $W_2$ piece in \eqref{W_flux_pinch_combined_final}
already contains the primitive $(1,1)$ projection of the $\mathbf{27}$ sector, but it was being interpreted as part of the reduced flux data. In other words, the $W_2$ contribution from the primitive parent source was not absent; it was simply packaged differently. By contrast, in the $W_3$ sector from \eqref{W_flux_pinch_combined_final} we wrote explicitly:
\begin{equation}
T\,{\bf W}_3^{(o)}
+
dT\wedge {\bf Y}_3^{(o)},
\label{W3oddT}
\end{equation}
and so, comparing with \eqref{mapW3b_combined}, the pinch dependence looked more direct there. Therefore the correct statement here is that 
even in the Type 0A reduction, the parent $\tau_3$ can feed both $W_2$ and $W_3$.
It only looked asymmetric because the $W_2$ piece was absorbed into the projected flux notation, while the $W_3$ piece was displayed more explicitly in the odd/pinch sector. We can also make the following 
quantitative comparison of the two descriptions.
In Type {$0$}HW, 
at leading order in $T$, the explicit pinch corrections are:
\begin{equation}
\Delta W_1^{(0HW)} = 0,
\qquad
\Delta W_4^{(0HW)} = 0,
\qquad
\Delta W_5^{(0HW)} = 0 \nonumber
\end{equation}
\begin{equation}
\Delta W_2^{(0HW)}
=
\kappa_2\,T\,\delta_{\Sigma^{\rm jct}_5}\,\xi_2,
\qquad
\Delta W_3^{(0HW)}
=
\kappa_3\,T\,\delta_{\Sigma^{\rm jct}_5}\,\xi_3.
\label{DeltaHW}
\end{equation}
Thus the pinch appears as a localized primitive source with support on the reduced junction locus. Whereas in the Type {$0$}A
branch, the corresponding leading terms are branch-integrated or branch-resolved. Schematically, from \eqref{W_flux_pinch_combined_final}, we can express this in the following way:
\begin{equation}
\Delta W_3^{(0A)}
=
\Pi_{\rm prim}^{(2,1)+(1,2)}\!\big(H_3^{(o)}\big)
+
T\,{\bf W}_3^{(o)}
+
dT\wedge {\bf Y}_3^{(o)}
+\cdots 
\label{DeltaW30A}
\end{equation}
\begin{equation}
\Delta W_2^{(0A)}
=
\widetilde a_2\,\Pi_{\rm prim}^{(1,1)}\!\big(\iota_{v_e} *_7 G_4^{(27,e)}\big)
+
\widetilde a_2^{(o)}\,\Pi_{\rm prim}^{(1,1)}\!\big(\iota_{v_o} *_7 G_4^{(27,o)}\big)
+
T\,{\bf W}_2^{(o)}
+
dT\wedge {\bf Y}_2^{(o)}
+\cdots , \nonumber
\label{DeltaW20A}
\end{equation}
implying that the same primitive parent source appears through branch-even and branch-odd projected fluxes, plus explicit odd/pinch pieces.
One may also summarize the two branches as follows: in Type 0HW the local source is kept explicit, whereas in Type 0A
local source is integrated and redistributed into projected flux/odd data. In fact
if one integrates the localized $0$HW pinch correction over a tubular neighborhood ${\cal U}_{\rm jct}$ of the junction, one obtains:
\begin{equation}
\int_{\mathcal N(\Gamma_2)} \Delta W_2^{(0HW)}
=
\kappa_2\,T
\int_{\mathcal N(\Gamma_2)}
\delta_{\Sigma^{\rm jct}_5}\,\xi_2 = \kappa_2\,T
\int_{\Gamma_2}\xi_2
\nonumber
\end{equation}
\begin{equation}\label{gam23}
\int_{\mathcal N(\Gamma_3)} \Delta W_3^{(0HW)}
=
\kappa_3\,T
\int_{\mathcal N(\Gamma_3)}
\delta_{\Sigma^{\rm jct}_5}\,\xi_3 = \kappa_3\,T
\int_{\Gamma_3}\xi_3 ,
\end{equation}
where  \(\Gamma_p\subset \Sigma^{\rm jct}_5\) is a \(p\)-cycle and
\(\mathcal N(\Gamma_p)\) is a small tubular neighborhood of 
\(\Gamma_p\) in
the reduced space.
The conditions in \eqref{gam23} should be compared not to a single term in the $0$A formulas, but to the full branch-projected combinations:
\begin{equation}
\int_{{\cal U}_{\rm red}}
\Delta W_2^{(0A)},
\qquad
\int_{{\cal U}'_{\rm red}}
\Delta W_3^{(0A)},
\end{equation}
where ${\cal U}_{\rm red}$ and $\mathcal U'_{\rm red}$ are respectively the appropriate reduced two- and three-cycle neighborhood after compactifying along $(S^1\vee S^1)_b$.
Thus, the matching is not term-by-term, but only after summing the projected pieces in the $0$A description.

The reason in the Type 0HW side as to how the parent $G_2$ $\mathbf{27}$ source reduces to the
localized $W_2$ and $W_3$ sources is the following. The
seven-dimensional pinch source from \eqref{tau3pinch} is
$\tau_3^{\rm pinch}
=
T\,\delta_{\Sigma_6^{\rm jct}}\,\Xi_{27}^{(3)}$,
where $\Sigma_6^{\rm jct} = T^2_f\rtimes\frac{S^1_a}{\mathbb{Z}_2}\rtimes\Sigma_3$
contains the HW interval direction $S^1_a/\mathbb{Z}_2$ as a factor. Since
the support of $\delta_{\Sigma_6^{\rm jct}}$ contains the reduction
direction, the Poincar\'e dual current is pulled back from the reduced
six-manifold rather than acquiring an additional $e^a$ factor. The
push-forward under fiber integration is therefore:
\begin{equation}
\int_{S^1_a/\mathbb{Z}_2}\delta_{\Sigma_6^{\rm jct}}
=
\ell_a\,\delta_{\Sigma_5^{\rm jct}},
\label{delta_pushforward}
\end{equation}
where $\ell_a = \pi R_a$ is the length of the HW interval and
$\Sigma_5^{\rm jct} = T^2_f\rtimes\Sigma_3$ is the reduced junction locus
in $X_6^{(0{\rm HW})}$. We can now decompose the $G_2$ $\mathbf{27}$ three-form with respect to the HW
reduction direction $e^a$ in the following standard way:
\begin{equation}
\Xi_{27}^{(3)}
=
e^a\wedge\xi_2 + \xi_3, \qquad \xi_2\in\Lambda_{\rm prim}^{(1,1)}(X_6^{(0{\rm HW})}),
\qquad
\xi_3\in\Lambda_{\rm prim}^{(2,1)+(1,2)}(X_6^{(0{\rm HW})}),
\label{Xi27_split}
\end{equation}
where $\xi_2$ and $\xi_3$ are the components of $\Xi_{27}^{(3)}$ with and without a leg along $e^a$
respectively, both primitive relative to the induced $SU(3)$ structure on
$X_6^{(0{\rm HW})}$. Primitivity of $\xi_2$ and $\xi_3$ follows from the
primitivity of $\Xi_{27}^{(3)}$ as a $G_2$ $\mathbf{27}$ source together
with the absence of singlet and vector projections assumed
in~\eqref{pinch27assumption}. Applying the Hodge star identities for the product metric split
$ds_7^2 = (e^a)^2 + ds_6^2$ gives us
${*}_7(e^a\wedge\xi_2) = {*}_6\xi_2$, and 
${*}_7\xi_3 = e^a\wedge{*}_6\xi_3$,
up to orientation signs. Therefore the source term
$d\varphi|_{\rm pinch} = {*}_7\tau_3^{\rm pinch}$ becomes:
\begin{equation}
d\varphi\big|_{\rm pinch}
=
T\,\delta_{\Sigma_6^{\rm jct}}
\Big(
{*}_6\xi_2
+
e^a\wedge{*}_6\xi_3
\Big).
\label{dphi_pinch_HW}
\end{equation}
Comparing with the $G_2\to SU(3)$ split along $e^a$, {\it i.e.}
$\varphi = J\wedge e^a + {\rm Re}\,\Omega$, an exterior derivative gives us
$d\varphi = dJ\wedge e^a + d\,{\rm Re}\,\Omega + \cdots$,
from where we can read off the contributions to each torsion channel in the following way.

The tangential four-form $T\,\delta_{\Sigma_6^{\rm jct}}\,{*}_6\xi_2$ has
no $e^a$ leg and therefore contributes to $d\,{\rm Re}\,\Omega$. Since
$W_2$ is extracted from the primitive $(1,1)$ component of $d\Omega$ via
$W_2\wedge J\subset d\Omega$, and $\xi_2$ is primitive $(1,1)$, this gives:
\begin{equation}
W_2^{\rm pinch}
=
\kappa_2\,T\,\delta_{\Sigma_5^{\rm jct}}\,\xi_2.
\label{W2pinch_derived}
\end{equation}
The term $T\,\delta_{\Sigma_6^{\rm jct}}\,e^a\wedge{*}_6\xi_3$ has one
$e^a$ leg and therefore contributes to $dJ$. Since $W_3$ is the primitive
$(2,1)+(1,2)$ component of $dJ$, and since ${*}_6\xi_3$ is again a primitive
real $(2,1)+(1,2)$ form up to the usual $SU(3)$-structure convention-dependent
signs, this gives:
\begin{equation}
W_3^{\rm pinch}
=
\kappa_3\,T\,\delta_{\Sigma_5^{\rm jct}}\,\xi_3.
\label{W3pinch_derived}
\end{equation}
The $\mathbf{1}$, $\mathbf{3}\oplus\overline{\mathbf{3}}$ components of
the $\mathbf{27}$ branching~\eqref{27decomp} would contribute to $W_1$
and to trace pieces respectively, but these are absent at leading order
by the primitivity assumption on $\Xi_{27}^{(3)}$. The contributions to
the Lee-form classes $W_4$ and $W_5$ are also absent at this order because
the localized source~\eqref{tau3pinch} has no $\mathbf{7}$ component.
Therefore
\begin{equation}
W_1^{\rm pinch} = W_4^{\rm pinch} = W_5^{\rm pinch} = 0
\label{W145zero}
\end{equation}
at leading order. The constants $\kappa_2$ and $\kappa_3$ absorb the
interval length $\ell_a$, orientation signs from the Hodge-star identities above,
and the normalization of the primitive projection maps.

This completes the derivation of~\eqref{W23pinch} from the parent $G_2$
torsion data. The channel assignment is summarized as follows: the component
$e^a\wedge\xi_2$ of $\Xi_{27}^{(3)}$ carries the $W_2$ source via
$d\,{\rm Re}\,\Omega$, while the component $\xi_3$ carries the $W_3$
source via $dJ$.

\subsubsection{Fibration, flux and pinch contributions to $SU(3)$ torsion classes \label{sec412}}

We now write the full torsion classes in the reduced theory. The fibration sector is generated by the non-trivial twisting of the internal manifold $X_6^{\rm 0HW}$ in \eqref{76hw}
and is already present even in the absence of background $G_4$ flux in M-theory. (Recall that the twistings are not related to fluxes as no T-dualities have been performed.) The corresponding torsion classes may be parameterized in the following suggestive way:
\begin{equation}
W_1^{\rm fib} = w_1^{\rm fib},
\qquad
W_2^{\rm fib} = w_2^{\rm fib},
\qquad
W_3^{\rm fib} = w_3^{\rm fib},
\qquad
W_4^{\rm fib} = w_4^{\rm fib},
\qquad
W_5^{\rm fib} = w_5^{\rm fib},
\label{Wfibparam}
\end{equation}
where the $w_i^{\rm fib}$ are determined by the connection data of the successive fibrations.

The flux sector arises from the reduction of the M-theory four-form flux $G_4$. Under compactification along $S^1_a/\mathbb{Z}_2$, $G_4$ splits into pieces tangent to $X_6^{(0{\rm HW})}$ and pieces with one leg along $e^a$. Denoting the corresponding reduced fluxes schematically by:
\begin{equation}
G_4
\to
H_3 \wedge e^a,
\end{equation}
where note the absence of the M-theory four-form. This is already understood in the Ho\v rava-Witten  set-up \cite{Horava, Altavista} from the fact that the three-form $C_3$ is odd under the $\mathbb{Z}_2$ action, {\it i.e.} $C_3 \to -C_3$, which projects out those components of $C_3$ whose legs are away from the HW circle. 
The flux-induced torsion classes may be written as:
\begin{equation}
W_1^{\rm flux} = w_1^{\rm flux},
\qquad
W_2^{\rm flux} = w_2^{\rm flux},
\qquad
W_3^{\rm flux} = w_3^{\rm flux},
\qquad
W_4^{\rm flux} = w_4^{\rm flux},
\qquad
W_5^{\rm flux} = w_5^{\rm flux}.
\label{Wfluxparam}
\end{equation}
The precise formulas depend on the detailed flux ansatz, but for the purposes of comparison we only need the fact that these are smooth bulk contributions, unlike the localized pinch piece \eqref{W23pinch}.
Combining \eqref{Wdecomp}, \eqref{Wfibparam}, and \eqref{W23pinch}, the full torsion classes in the type $0$HW branch are therefore:
\begin{equation}
W_1
=
w_1^{\rm fib}
+
w_1^{\rm flux} + .. \nonumber
\label{W1full}
\end{equation}
\begin{equation}
W_2
=
w_2^{\rm fib}
+
w_2^{\rm flux}
+
\kappa_2\, T\, \delta_{\Sigma^{\rm jct}_5}\, \xi_2 + .. \nonumber
\label{W2full}
\end{equation}
\begin{equation}
W_3
=
w_3^{\rm fib}
+
w_3^{\rm flux}
+
\kappa_3\, T\, \delta_{\Sigma^{\rm jct}_5}\, \xi_3 + ..\nonumber
\label{W3full}
\end{equation}
\begin{equation}
W_4
=
w_4^{\rm fib}
+
w_4^{\rm flux} + .., \qquad
W_5
=
w_5^{\rm fib}
+
w_5^{\rm flux} + .. ,
\label{W5full}
\end{equation}
where, to leading order in $T$, the pinch affects only $W_2$ and $W_3$ and the dotted terms will be discussed soon. There is however something a little unusal about the terms appearing in the torsion classes $W_2$ and $W_3$: there is a tachyon $T$ in both of them. This is a bit surprising because in Type 0HW, the tachyon $T$ only kicks in {\it after} we further compactify on $S^1_+ \vee S^1_-$. While this is consistent from the $G_2$ ${\bf 27}$ reduction to ${\bf 8}$ and ${\bf 6} \oplus \overline{\bf 6}$ of $SU(3)$, the result in \eqref{W5full} suggests that it cannot be a ten-dimensional set of torsion classes. It is best understood as a nine-dimensional result. In section \ref{comparisoon}, we will show that this is indeed the case. Nevertheless, and up to the dotted terms, 
the result in \eqref{W5full} looks remarkably simple compared to what we had for the Type 0A case in \eqref{Wi_total_final_combined} and \eqref{W_flux_pinch_combined_final}. To see whether this is really the case, we need to provide quantitative comparison with the Type 0A reduction. In the following we will discuss this, starting with the comparison of the spectra in the two cases. Before that however let us briefly discuss the Bianchi identity in the presence of the wedge.

The torsion-class analysis gives necessary local geometric conditions, but it
does not by itself guarantee a globally consistent heterotic/HW background.
One must also impose the heterotic Bianchi identity:
\begin{equation}
dH_3
=
\frac{\alpha'}{4}
\left(
\mathrm{tr}\,R^+\wedge R^+
-
\mathrm{tr}\,F\wedge F
\right)
+
\mathcal J_4^{\rm loc},
\label{hetBianchi_0HW_short}
\end{equation}
where \(R^+\) is the curvature of the torsionful Hull connection and
\(\mathcal J_4^{\rm loc}\) denotes possible localized source terms at the wedge junction. Away
from the wedge junction, this is the standard Strominger--Hull Bianchi
condition and imposes the usual cohomological constraint in
\(H^4(X_6^{(0{\rm HW})})\). \cite{strominger, DRS}

The wedge branch introduces an additional localized issue. Since $W_2$ and $W_3$ are given by \eqref{W23pinch},
the torsionful curvature \(R^+\) generally acquires a junction-supported
contribution. Consistency therefore requires the localized part of
\(\mathrm{tr}\,R^+\wedge R^+\) to be canceled by a corresponding localized
gauge-bundle contribution or by an allowed junction source:
\begin{equation}
\left[
\left(\mathrm{tr}\,R^+\wedge R^+\right)_{\rm jct}
-
\left(\mathrm{tr}\,F\wedge F\right)_{\rm jct}
+
\mathcal J_4^{\rm jct}
\right]
=0
\qquad
\text{near } \Sigma_5^{\rm jct}.
\label{localized_Bianchi_short}
\end{equation}
A complete verification of this condition requires an explicit local model of
the junction geometry and its gauge sector. Thus the torsion-class analysis in
this section should be read as a necessary local analysis, not as a proof of a
fully global heterotic solution.

\subsection{Quantitative comparison of Type 0HW with the Type {$0$}A reduction \label{sec420}}

The apparent simplicity of the torsion classes from \eqref{W5full} for the Type 0HW case begs the question of the comparison with the Type 0 story. We will then start by first analyzing the spectra on both sides and then compare the torsion classes.

\subsubsection{Comparison of the two reduction orders and the study of the spectra \label{sec421}}

It is important to phrase the comparison in a way that keeps track of the
\emph{branchwise} data of the wedge. If one only keeps the doubly even bulk
components from the start, then the branch-odd scalar modulus is hidden and the
tachyon seems to disappear. This is not correct. In both reduction orders, the
tachyon is the same branch-odd wedge modulus:
\begin{equation}
T \equiv \frac{1}{2}\left(R_+ - R_-\right),
\qquad
R_\pm \equiv \sqrt{G_{\pm\pm}},
\label{tachyon_def_compare}
\end{equation}
or equivalently the odd combination of the two branchwise radii. Thus the
comparison should be organized branchwise first, and only afterwards separated
into even and odd combinations. In what follows we suppress possible boundary gauge sectors and concentrate only
on the universal bosonic fields descending from the eleven-dimensional metric
$G_{MN}$ and three-form $C_{MNP}$. We will also use the notations from section \ref{sec001} so that direct comparison can be made.

Let the eleven-dimensional coordinates be split as
$x^M = (x^\mu, y^a, y^+, y^-)$,
where $x^\mu$ denote the common nine-dimensional directions, $y^a$ is the
Ho\v{r}ava--Witten interval coordinate, and $y^\pm$ parametrize the two branches
of the wedge circle $S^1_+\vee S^1_-$. The relevant eleven-dimensional
components are:
\begin{equation}
C_{MNP}
\;\longrightarrow\;
\Big\{
C_{\mu\nu\rho},
\;
C_{\mu\nu a},
\;
C_{\mu\nu +},\,C_{\mu\nu -},
\;
C_{\mu a +},\,C_{\mu a -},
\;
C_{\mu + -},
\;
C_{a + -}
\Big\} \nonumber
\label{C3decomp_compare0}
\end{equation}
\begin{equation}
G_{MN}
\;\longrightarrow\;
\Big\{
G_{\mu\nu},
\;
G_{\mu a},
\;
G_{\mu +},\,G_{\mu -},\,G_{a+}, \,G_{a-},
\,
G_{aa},
\;
G_{++},\,G_{--},
\;
G_{+-}
\Big\}.
\label{C3decomp_compare}
\end{equation}
For the 
Ho\v{r}ava--Witten $\mathbb Z_2$ action on $y^a$ we will follow the standard prescription, namely $y^a \to -y^a$ and $C_3 \to - C_3$ \cite{Horava}. This means all the metric components that are odd under $\mathbb{Z}_2$ will be projected out whereas all the three-form field contents that are odd under $\mathbb{Z}_2$ will be kept in. Alternatively, the fields are represented as either odd or even: 
\begin{equation}
G_{mn}\ ({\rm even}),
\qquad
G_{m a}\ ({\rm odd}),
\qquad
C_{mnp}\ ({\rm odd}),
\qquad
C_{mn a}\ ({\rm even}),
\label{HW_parity_compare2}
\end{equation}
where $m, n, p$ run over the ten-dimensional directions tangent to
the Ho\v{r}ava--Witten boundary. In the following we will then follow two routes: in one, we first compactify on $S^1_+\vee S^1_-$, then on $S^1_a/\mathbb Z_2$; and in the other, we first compactify on $S^1_a/\mathbb Z_2$, then on $S^1_+\vee S^1_-$.

\begin{table}[t]
\centering
\renewcommand{\arraystretch}{1.3}
\begin{tabular}{|c|c|c|c|}
\hline
\textbf{10d field} & \textbf{9d decomposition} & \textbf{$\mathbb Z_2$ parity} & \textbf{Kept?} \\
\hline
$g_{MN}$ 
& $g_{\mu\nu},\ g_{\mu a},\ g_{aa}$
& $+,\ -,\ +$
& $g_{\mu\nu},\ g_{aa}$
\\
\hline
$B_{MN}$
& $B_{\mu\nu},\ B_{\mu a}$
& $-,\ +$
& $B_{\mu a}$
\\
\hline
$\Phi$
& $\Phi$
& $+$
& $\Phi$
\\
\hline
$T$
& $T$
& $+$
& $T$
\\
\hline
$C_1^{(+)}$
& $C_\mu^{(+)},\ C_a^{(+)}$
& $+,\ -$
& $C_\mu^{(+)}$
\\
\hline
$C_1^{(-)}$
& $C_\mu^{(-)},\ C_a^{(-)}$
& $+,\ -$
& $C_\mu^{(-)}$
\\
\hline
$C_3^{(+)}$
& $C_{\mu\nu\rho}^{(+)},\ C_{\mu\nu a}^{(+)}$
& $-,\ +$
& $C_{\mu\nu a}^{(+)}$
\\
\hline
$C_3^{(-)}$
& $C_{\mu\nu\rho}^{(-)},\ C_{\mu\nu a}^{(-)}$
& $-,\ +$
& $C_{\mu\nu a}^{(-)}$
\\
\hline
\end{tabular}
\caption{Reduction of the ten-dimensional type-$0$A fields on $S^1_a/\mathbb Z_2$ to nine dimensions, showing the decomposition of each field, its inherited $\mathbb Z_2$ parity, and the surviving components.}
\label{0Anine}
\end{table}

\subsubsection*{%
\fbox{Route I: first on $S^1_+\vee S^1_-$, then on $S^1_a/\mathbb Z_2$}
}

If one first reduces on the wedge circle, one obtains the candidate type-$0$A
local frame discussed earlier in section \ref{sec001}. The guiding principle is
that the NSNS sector not doubled, whereas the RR sector doubled.
The branchwise decomposition is:
\begin{equation}
C_{MNP}
\;\longrightarrow\;
\Big\{
C_{\mu\nu\rho},\;C_{\mu\nu a}
\;
C_{\mu\nu +},
\;
C_{\mu\nu -},
\;
C_{\mu a +},
\;
C_{\mu a -},\;C_{\mu + -}, \; C_{a+-}
\Big\}\nonumber
\label{routeI_C3_branchwise}
\end{equation}
\begin{equation}
G_{MN}
\;\longrightarrow\;
\Big\{
G_{\mu\nu},\; G_{\mu a},
\;
G_{\mu +},\;G_{\mu -},\; G_{a+},\;G_{a-},
\;
G_{++},\,G_{--},
\;
G_{+-},
\;
G_{aa}
\Big\} ,
\label{routeI_metric_branchwise}
\end{equation}
which produces the spectrum in ten-dimensions where $(\mu, \nu)$ span nine dimensions. As discussed in section \ref{sec001} this is not exactly the expected spectrum of Type 0A, but can be made to coincide with the 0A spectrum by performing certain redefinitions. We will not repeat those details here, but conclude by saying that the physical type-$0$A fields are then identified as:
\begin{equation}
\big(g_{mn}, B_{mn}, \Phi, T\big)_{\rm NS}
~~\oplus~~
\big(C_1^{(\pm)}, C_3^{(\pm)}\big)_{\rm RR}\nonumber
\label{routeI_0A_spectrum}
\end{equation}
\begin{equation}
B_{mn}
\equiv
\frac{1}{2}\Big(C_{mn +}+C_{mn -}\Big),
\qquad
B^{(o)}_{mn}
\equiv
\frac{1}{2}\Big(C_{mn +}-C_{mn -}\Big)\nonumber
\label{routeI_B_evenodd}
\end{equation}
\begin{equation}
R_B \equiv \frac{1}{2}(R_+ + R_-),
\qquad
T \equiv \frac{1}{2}(R_+ - R_-),
\qquad
R_\pm \equiv \sqrt{G_{\pm\pm}}\nonumber
\label{routeI_RB_T}
\end{equation}
\begin{equation}
C_{1,m}^{(+)} \sim G_{m +},
\qquad
C_{1,m}^{(-)} \sim G_{m -} ,
\label{routeI_RR1}
\end{equation}
where $x^M = (x^m, y^+, y^-)$ now; and
the odd two-form $B^{(o)}_{mn}$ is not counted as a second physical NSNS
two-form; rather it is auxiliary branch-odd data. Likewise, the doubled RR
three-form sector is encoded in the two branch/junction-resolved sectors
$C_3^{(\pm)}$. Note that we have removed other auxiliary fields from \eqref{routeI_metric_branchwise} in writing \eqref{routeI_RR1}.

Now compactifying this type-$0$A frame on $S^1_a/\mathbb Z_2$ removes the fields
with odd Ho\v{r}ava--Witten parity. But the orientifold action also kicks in. This action can be easily inferred from the supersymmetric IIA sector. It tells us metric, dilaton and the gauge fields are even under the O-action whereas the NS B-field and the three-form fields are odd under the O-action. The decomposition of the spectrum in nine-dimensions is shown in {\bf Table \ref{0Anine}}. Therefore, reducing the candidate type-$0$A spectrum \eqref{routeI_RR1}
on $S^1_a/\mathbb Z_2$ with the Ho\v{r}ava--Witten parity inherited from M-theory
gives the nine-dimensional spectrum:
\begin{equation}
g_{\mu\nu}
\;\oplus\;
\big(C^+_{\mu\nu a},\,C^-_{\mu\nu a}\big)
\;\oplus\;
\big(C^+_\mu,\,C^-_\mu,\,B_{\mu a}\big)
\;\oplus\;
\big(\phi,\,T,\,g_{aa}\big) ,
\label{finalboxed0Ato9d}
\end{equation}
with the tachyon $T$ surviving as a genuine nine-dimensional scalar. We see that, other than the metric, we have two anti-symmetric B-fields, three $U(1)$ gauge fields and three scalars one of it being the tachyon $T$.

\begin{table}[t]
\centering
\resizebox{\textwidth}{!}{%
\renewcommand{\arraystretch}{2.5}
\begin{tabular}{|c|c|c|c|c|c|}
\hline
\textbf{Field type} & \textbf{9d dof per field} & \textbf{Route I mult.} & \textbf{Route I contribution} & \textbf{Route II mult.} & \textbf{Route II contribution} \\
\hline
metric \(g_{\mu\nu}\) & \(27\) & \(1\) & \(1\times 27=27\) & \(1\) & \(1\times 27=27\) \\
\hline
2-form \(B_{\mu\nu}\) & \(21\) & \(2\) & \(2\times 21=42\) & \(1\) & \(1\times 21=21\) \\
\hline
vector \(A_\mu\) & \(7\) & \(3\) & \(3\times 7=21\) & \(4\) & \(4\times 7=28\) \\
\hline
scalar \(\varphi\) & \(1\) & \(3\) & \(3\times 1=3\) & \(5\) & \(5\times 1=5\) \\
\hline
\textbf{Total} & --- & --- & \(\mathbf{93}\) & --- & \(\mathbf{81}\) \\
\hline
\end{tabular}%
}
\caption{The counting of the degrees of freedom from the two spectra \eqref{finalboxed0Ato9d} and \eqref{finalboxed0HW} for the bosonic on-shell fields in nine space-time dimensions.}
\label{DOFcounting}
\end{table}
\subsubsection*{%
\fbox{Route II: first on $S^1_a/\mathbb Z_2$, then on $S^1_+\vee S^1_-$}
}
We now start from the universal bosonic sector of the ten-dimensional heterotic theory,
$g_{MN}$, $B_{MN}$, and  $\phi$
with
$M,N=0,\dots,9$,
and compactify on the wedge circle $S^1_+\vee S^1_-$. The goal is to compare the
resulting nine-dimensional field content with the nine-dimensional spectrum obtained
earlier from the candidate type-$0$A reduction followed by the $S^1_a/\mathbb Z_2$
projection.

The correct statement is slightly subtle. The heterotic wedge reduction reproduces only a
\emph{subsector} of the nine-dimensional candidate type-$0$A frame. In particular, the
heterotic reduction naturally reproduces the universal NS sector and two branchwise
vector slots, which are then identified with part of the type-$0$A field content through
the proposed local duality dictionary. However, it does \emph{not} reproduce the full
candidate type-$0$A spectrum field by field. In particular, the second type-$0$A two-form
slot and the vector descending from $B_{\mu a}$ on the type-$0$A side do not arise from
the universal heterotic bosonic sector alone.

\begin{table}[t]
\centering
\resizebox{\textwidth}{!}{%
\renewcommand{\arraystretch}{1.35}
\begin{tabular}{|c|c|c|}
\hline
\textbf{Field type} & \textbf{Route I: type \(0\)A \(\to\) 9d} & \textbf{Route II: type \(0\)HW \(\to\) 9d} \\
\hline
Metric
&
$g_{\mu\nu}$
&
$g_{\mu\nu}$
\\
\hline
Two-forms
&
$C^+_{\mu\nu a},\; C^-_{\mu\nu a}$
&
$B_{\mu\nu}$
\\
\hline
Vectors
&
$C^+_\mu,\; C^-_\mu,\; B_{\mu a}$
&
$g_{\mu +},\; g_{\mu -},\; B_{\mu +},\; B_{\mu -}$
\\
\hline
Scalars
&
$\phi,\; T,\; g_{aa}$
&
$\phi,\; R_B,\; T,\; g_{+-},\; B_{+-}$
\\
\hline
Total count
&
$1$ metric $+\,2$ two-forms $+\,3$ vectors $+\,3$ scalars
&
$1$ metric $+\,1$ two-form $+\,4$ vectors $+\,5$ scalars
\\
\hline
\end{tabular}%
}
\caption{Nine-dimensional bulk spectra obtained from the two sequential
compactification routes of M-theory on
$\frac{S^1_a}{\mathbb Z_2}\times (S^1\vee S^1)_b$. The two spectra do not match
field by field before any further projection, truncation, or duality
identification.}
\label{compu9d}
\end{table}

To see this, let the ten-dimensional coordinates be split as
$x^M=(x^\mu,y^+,y^-)$,
with $\mu=0,\dots,8$,
where $y^\pm$ parametrize the two branches of the wedge circle.
The heterotic metric, two-form and the dilaton decompose branchwise as:
\begin{equation}
g_{MN}
\;\longrightarrow\;
\Big\{
g_{\mu\nu},
\;
g_{\mu +},\, g_{\mu -},
\;
g_{++},\, g_{--},
\;
g_{+-}
\Big\}\nonumber
\end{equation}
\begin{equation}
B_{MN}
\;\longrightarrow\;
\Big\{
B_{\mu\nu},
\;
B_{\mu +},\, B_{\mu -},
\;
B_{+-}
\Big\}\nonumber
\label{het_B_branchwise}
\end{equation}
\begin{equation}
\phi \;\longrightarrow\; \phi ,
\label{het_phi_branchwise}
\end{equation}
where we followed a straightforward scheme of keeping all the allowed bulk fields, without taking into account the non-abelian sector. They would surely contribute, in the same way as the Type 0A contributions from the copies of the D8/O8 sectors, but we will worry about them later. The spectrum can now be decomposed in the same vein as  \eqref{finalboxed0Ato9d}:
\begin{equation}
g_{\mu\nu}
\;\oplus\;
B_{\mu\nu}
\;\oplus\;
\big(g_{\mu +},\,g_{\mu -},\,B_{\mu +},\,B_{\mu -}\big)
\;\oplus\;
\big(\phi,\,R_B,\,T,\,g_{+-},\,B_{+-}\big) ,
\label{finalboxed0HW}
\end{equation}
with $R_B$ and $T$ defined in the same way as \eqref{RBTdef0A}. The spectrum now consists of one metric, one B-field, four $U(1)$ gauge fields and five scalars. This clearly {\it does not} match with the corresponding Type 0A spectrum in nine dimensions. (See {\bf Table \ref{compu9d}}.) Moreover there is something highly unusual about the two spectra \eqref{finalboxed0Ato9d} and \eqref{finalboxed0HW}: a naive degrees of freedom counting, shown in {\bf Table \ref{DOFcounting}}, 
show that there is a mismatch\footnote{This mismatch persists even after including the gauge DOFs from the D8-branes in the Type 0A side and the vector bundles from 0HW side.}: for route I we get 93 DOFs and for route II we get 81 DOFs. This clearly {\it cannot} be right as two different ways of reducing M-theory to nine dimensions cannot give us two different counts of the degrees of freedom!

Recall that the counting of 81 DOFs is done following the procedure adopted in {\bf Table \ref{DOFcounting}} adopting the multiplicity $(1, 1, 4, 5)$. What if we allow for a  different multiplicity? For example, let $n_1 $ be the multiplicity for the 2-form field $B_{\mu\nu}$, ${n_2\over 2}$ be the multiplicity of the number of gauge fields from the dimensional reduction of the $10d$ metric, ${n_3\over 2}$ be the number of gauge fields from the dimensional reduction of the $10d$ B-field, and $5+n_4$ be the number of the scalar fields. Then the spectrum balance can happen if the following equation is satisfied:
\begin{equation}\label{intisol}
21n_1 + 7(n_2 + n_3) + n_4 = 61 ,
\end{equation}
where $n_i \in \mathbb{Z}_+$. The half-integer choice is kept to allow for some combinations to remain non-dynamical. We can ask how many positive integer solutions are possible for \eqref{intisol}. For $n_4 = 0$ there are {\it no} integer solutions possible, implying that if we take the five scalars at face-value, the spectrum in nine-dimensions {\it cannot} be matched. For $n_4 > 0$, there are eleven solutions as shown in {\bf Table \ref{positivesol}}. However for $n_4 > 5$, the choices do not look reasonable because $R_B$ and $T$ appearing in \eqref{finalboxed0HW} are already in some sense {\it doubled}. Even for 
$0 < n_4 \le 5$, assuming that $(\phi, R_B, T)$ do not get further doubled so that $n_4 = 5$ is accounted from within the remaining two scalars $(g_{+-}, B_{+-})$, the choices with $n_2 = 3$ and $n_3 = 3$ do not look reasonable either. The remaining three choices for $(n_1, n_2, n_3)$ namely $(1, 1, 4), (1, 4, 1)$ and $(2, 1, 1)$ would make sense only if we can properly account for the five extra scalars with either doubling of the B-fields or quadrupling of the gauge fields. Clearly these are not very satisfactory scenarios either.

\begin{table}[t]
\centering
\renewcommand{\arraystretch}{1.25}
\begin{tabular}{|c|c|c|c|}
\hline
${\bf n_1}$ & ${\bf n_2}$ & ${\bf n_3}$ & ${\bf n_4}$ \\
\hline
$1$ & $1$ & $1$ & $26$ \\
\hline
$1$ & $1$ & $2$ & $19$ \\
\hline
$1$ & $2$ & $1$ & $19$ \\
\hline
$1$ & $1$ & $3$ & $12$ \\
\hline
$1$ & $2$ & $2$ & $12$ \\
\hline
$1$ & $3$ & $1$ & $12$ \\
\hline
$1$ & $1$ & $4$ & $5$ \\
\hline
$1$ & $2$ & $3$ & $5$ \\
\hline
$1$ & $3$ & $2$ & $5$ \\
\hline
$1$ & $4$ & $1$ & $5$ \\
\hline
$2$ & $1$ & $1$ & $5$ \\
\hline
\end{tabular}
\caption{Positive integer solutions of $21n_1+7(n_2+n_3)+n_4=61$.}
\label{positivesol}
\end{table}

So what's going on? One thing is clear: the mismatch is a diagnostic that some of the fields we
listed are not all independent propagating nine-dimensional degrees of freedom. The counting in {\bf Table \ref{DOFcounting}}  treated every listed object as an ordinary propagating $9d$ massless field.
That is almost certainly too naive for the wedge reduction. In particular, the following types
of objects need not be independent propagating bulk fields:
\begin{equation}
g_{+-},
\qquad
B_{+-},
\qquad
B_{\mu +}-B_{\mu -},
\qquad
g_{\mu +}-g_{\mu -},
\label{branch_odd_auxiliary}
\end{equation}
as we saw in section \ref{sec001};  and similarly the would-be doubled RR fields on the \(0\)A side may not all be independent
once one imposes the singular wedge identifications and the correct projection. So the issue is that the two routes were being compared at the level of an
\emph{over-complete parametrization}, not at the level of the reduced physical phase space.

There are three standard reasons why two raw Kaluza--Klein field lists can look different
while still describing the same lower-dimensional theory: (a) Some fields could be auxiliary or constrained:
A field appearing in the reduction ansatz need not correspond to an independent dynamical mode.
For wedge reductions, branch-odd combinations and branch-mixing quantities often behave as
junction data rather than propagating bulk fields.
(b) Some fields could be related by dualization:
In nine dimensions, different \(p\)-form descriptions can represent the same physical degrees
of freedom. So a mismatch in the \emph{type} of fields does not automatically imply a mismatch
in physical states. (c) The intermediate \(0{\rm A}\) and \(0{\rm HW}\) descriptions are not literal
KK reductions on smooth manifolds:
They are effective field dictionaries adapted to different local frames. Therefore some fields
that appear naturally in one frame may correspond to composite, constrained, or projected data
in the other. This means, if $x$ and $y$ represent the DOFs in respectively in route I and route II  that are not dynamical, then we expect:
\begin{equation}
x - y = 12.
\end{equation}
Thus {\it if} there is any match, it must be much more subtle than what appears at the face value. A true resolution of the problem is unfortunately beyond the scope of this work. We can however try the following: Instead of comparing the spectra between the two raw field lists, we could compare the spectra
between their common nine-dimensional massless sectors. We define this common
sector by imposing the same projection rules on both routes:

\begin{enumerate}[label=(\roman*)]
\item only node-invariant zero modes are retained as common massless bulk
fields;
\item components with simultaneous \(+\) and \(-\) internal legs are treated
as resolved-junction data rather than bulk zero modes of the strict wedge;
\item fields with one leg along an interval direction are retained only if
they define an independent gauge-invariant interval zero mode after the
boundary conditions and form-gauge redundancies are imposed;
\item the scalar comparison is performed in a fixed branch-even volume frame,
so the physical branch-even scalar is the nine-dimensional dilaton
\(\Phi_{\rm 9d}\), while the orthogonal breathing mode is fixed or lifted.
\end{enumerate}

\noindent Let us see how the aforementioned set of rules help us fix the spectra in nine-dimensions. First, on the candidate type-$0$A side the two two-form slots
\(C^{(+)}_{\mu\nu a}\) and \(C^{(-)}_{\mu\nu a}\) should be decomposed into
even and odd combinations:
\begin{equation}\label{ccombination}
C_{\mu\nu a}^{(e)}
=
\frac12\left(C_{\mu\nu a}^{(+)}+C_{\mu\nu a}^{(-)}\right),
\qquad
C_{\mu\nu a}^{(o)}
=
\frac12\left(C_{\mu\nu a}^{(+)}-C_{\mu\nu a}^{(-)}\right).
\end{equation}
A massless zero mode is constant on each branch. In the common node-invariant truncation used for the comparison, only the
symmetric zero mode is retained. Thus the branch-odd constant mode satisfies
\begin{equation}
C_{\mu\nu a}^{(o)}\big|_{\rm common\;zero\;mode}=0.
\end{equation}
Thus only \(C_{\mu\nu a}^{(e)}\) is retained in the common massless sector.
This removes the \(21\) degrees of freedom of a nine-dimensional two-form. The more subtle one is the vector \(B_{\mu a}\). It appears in the parity-allowed raw type-$0$A list, and
carries one leg along the interval direction. One might impose interval
boundary conditions and two-form gauge redundancy to eliminate it from the spectrum. Unfortunately this doesn't quite work in a simple way. For example, we could however impose the following boundary condition:
\begin{equation}\label{onlyifbnd}
B_{\mu a}\big|_{\partial I}= {1\over 2}(C_{\mu a +} + C_{\mu a -})\bigg|_{\partial I} =  0,
\end{equation}
but such a criterion is hard to justify in the present set-up. Therefore we cannot declare that in the common sector we
keep only the gauge-invariant vectors that match the branch-even vector slots
of the heterotic wedge reduction. The \(B_{\mu a}\) slot is therefore retained as an independent vector in the Type 0A side, but is conspicuously absent in the Type 0HW side.
{\it Only in a formulation
where the interval boundary condition \eqref{onlyifbnd} is imposed directly, the constant
zero mode of \(B_{\mu a}\) is absent}. Under this condition \(7\) degrees of freedom from the
common-sector count can be removed.

On the other hand, the $U(1)$ vectors on the Type 0HW side are under better control. This is because, on the heterotic wedge side, the branchwise vectors are decomposed into
even and odd combinations. The node-invariant sector retains only the even
zero modes. For a generic branch vector
$A_\mu^{(\pm)}(x,y_\pm)=a_\mu^{(\pm)}(x)$,
the node condition gives \(a_\mu^{(+)}=a_\mu^{(-)}\), hence:
\begin{equation}
A_\mu^{(o)}
=
\frac12\left(A_\mu^{(+)}-A_\mu^{(-)}\right)
=0
\end{equation}
for the massless zero mode. Therefore the odd metric and \(B\)-field vectors
\(A_\mu^{(g,o)}\) and \(A_\mu^{(B,o)}\) are absent from the common massless
sector. This removes \(7+7=14\) degrees of freedom.

There is also the subtlety about the mixed scalar sector. These scalars \(g_{+-}\) and \(B_{+-}\) require simultaneous
\(+\) and \(-\) internal legs. The strict wedge \(S^1_+\vee S^1_-\) is
one-dimensional away from the node, and at the node there is no smooth
two-dimensional tangent plane spanned by \(dy^+\) and \(dy^-\). Hence these
mixed components are not bulk zero modes of the strict wedge graph. They are
resolved-junction data. This removes \(1+1=2\) degrees of freedom.

The scalar comparison between the type-$0$A route and the type-$0$HW route
requires some care.  In each individual perturbative frame, the
nine-dimensional dilaton is obtained by the standard Kaluza--Klein reduction
formula:
\begin{equation}
e^{-2\Phi_{\rm 9d}}
=
e^{-2\phi}\,V_{\rm int},
\qquad
\Phi_{\rm 9d}
=
\phi-\frac{1}{2}\log V_{\rm int}.
\label{single_frame_Phi9}
\end{equation}
Here $\phi$ is the ten-dimensional dilaton of the corresponding perturbative
description, and $V_{\rm int}$ is the volume of the internal one-dimensional
space measured in that same frame.

However, \eqref{single_frame_Phi9} should not be interpreted as a direct
identification of the type-$0$A scalar and the type-$0$HW scalar.  It is only
the definition of the nine-dimensional dilaton within a given frame.  When one
compares the two routes,
the resulting type-$0$A and type-$0$HW descriptions are not related by a
trivial equality of perturbative dilatons.  Even in the supersymmetric limit,
the corresponding type-IIA and heterotic descriptions are related by a
non-trivial U-duality transformation.  Therefore the scalar map must allow for
mixing between the dilaton and the radion moduli.
Thus, in the type-$0$A frame one may define:
\begin{equation}
\Phi_{\rm 9d}^{(0A)}
=
\phi_{0A}
-
\frac{1}{2}\log V_a^{(0A)},
\label{Phi9_0A}
\end{equation}
where $V_a^{(0A)}$ denotes the interval volume of $S^1_a/\mathbb Z_2$ in the
type-$0$A frame. (This equals $l_a = \pi R_a$ that we used in \eqref{delta_pushforward}.)  Similarly, in the type-$0$HW frame one defines:
\begin{equation}
\Phi_{\rm 9d}^{(0{\rm HW})}
=
\phi_{0{\rm HW}}
-
\frac{1}{2}\log V_b^{(0{\rm HW})},
\label{Phi9_0HW}
\end{equation}
where $V_b^{(0{\rm HW})}$ denotes the branch-even volume of the wedge circle
$S^1_+\vee S^1_-$ in the type-$0$HW frame.
The two quantities
$\Phi_{\rm 9d}^{(0A)}$
and 
$\Phi_{\rm 9d}^{(0{\rm HW})}$
should not be set equal.  Rather, they are related by a duality-frame
like transformation of the schematic form:
\begin{equation}
\begin{pmatrix}
\Phi_{\rm 9d}^{(0A)}
\\[2pt]
\sigma_{0A}
\end{pmatrix}
=
{\cal U}
\begin{pmatrix}
\Phi_{\rm 9d}^{(0{\rm HW})}
\\[2pt]
\sigma_{0{\rm HW}}
\end{pmatrix}
+
\text{possible branch-odd corrections},
\label{scalar_duality_map_general}
\end{equation}
where $\sigma$ denotes the relevant branch-even radion or breathing-mode
scalar, and ${\cal U}$ is the U-duality like transformation acting on the
nine-dimensional scalar moduli. Quantifying $\mathcal U$ here is more subtle because it is not clear whether these two descriptions are U-dual to each other away from the two supersymmetric end-points. Thus $\mathcal U$ may be interpreted as a functional map between the scalar sectors of the two theories that match the standard U-duality map at the two end-points.
In particular this means, near a supersymmetric endpoint, one expects the map to reduce to
the usual heterotic/type-IIA duality relation.  Schematically, this involves an
inversion of the coupling:
\begin{equation}
(g_{0A}\to g_{IIA})
\sim
\frac{1}{(g_{0{\rm HW}}\to g_{HW})}
\end{equation}
in the corresponding supersymmetric limit, up to moduli-dependent factors.
Equivalently, at the level of effective dilatons one expects a relation of the
form:
\begin{equation}
\Phi_{\rm 9d}^{(0A)}
\stackrel{?}{=}
-\Phi_{\rm 9d}^{(0{\rm HW})}
+
{\cal M}_{\rm rad}
+
\cdots ,
\label{Phi_dual_schematic}
\end{equation}
where ${\cal M}_{\rm rad}$ denotes radion-dependent corrections required by
the precise nine-dimensional duality frame, and the ellipsis denotes corrections
that vanish or become controlled at the supersymmetric endpoint. The relation \eqref{Phi_dual_schematic}, as in \eqref{scalar_duality_map_general}, is subjective to the condition that the two theories have some remnants of the U-duality away from the supersymmetric end-points. 
Thus the correct scalar comparison is not
\eqref{Phi_dual_schematic}, but 
a more accurate statement of the form:
\begin{equation}
\Phi_{\rm 9d}^{(0A)}
\quad
\longleftrightarrow
\quad
\Phi_{\rm 9d}^{(0{\rm HW})}
\end{equation}
as corresponding scalar slots in the common nine-dimensional effective
description, with the understanding that the actual relation between them is a {\it conditional} 
duality-frame map rather than a literal equality up to an overall sign plus corrections.

The branch-odd scalar should be treated separately.  On the wedge side the
branch radii may be written as
$R_\pm
=
R_B\pm T $. The scalar $T$ is branch-odd under
$S^1_+
\longleftrightarrow
S^1_-$,
and is naturally identified with the effective type-$0$ tachyonic scalar.  The
branch-even combination $R_B$ participates in the nine-dimensional
dilaton/radion system, while the branch-odd combination $T$ is a separate
non-supersymmetric deformation.
Therefore a more appropriate scalar dictionary takes the following form:
\begin{equation}
T^{(0{\rm HW})}
\quad
\longleftrightarrow
\quad
T^{(0A)} \nonumber
\label{T_common_dictionary}
\end{equation}
\begin{equation}
\left(
\Phi_{\rm 9d}^{(0{\rm HW})},
\sigma_{0{\rm HW}}
\right)
\quad
\xleftrightarrow{\ \ {\cal U}\ \ }
\quad
\left(
\Phi_{\rm 9d}^{(0A)},
\sigma_{0A}
\right) ,
\label{even_scalar_duality_dictionary}
\end{equation}
where $\sigma$ denotes the branch-even breathing mode, but $\mathcal U$, as mentioned above, {\it does not imply that the two theories are U-dual to each other away from the supersymmetric end points}. As our analysis show, any dualities between the two pictures should be taken with extreme caution.  In a fixed-volume
comparison frame one may set one linear combination of $\Phi_{\rm 9d}$ and
$\sigma$ to a fixed value, but this fixing should be imposed only after the
duality-frame relation has been acknowledged. Under this condition we can lose one degree of freedom from both sides.
With these projections the counts become:
\begin{equation}
93-(21+ {\color{red}{7}} +1)\stackrel{?}= 64,
\qquad
81-(14+2+1)=64 ,
\label{matched_64_count}
\end{equation}
where $\color{red}{7}$ denotes the seven degrees of freedom that are frozen via \eqref{onlyifbnd}. This is conditional equation, and is not clear whether it can be always imposed. In the latter case we cannot claim a match like \eqref{matched_64_count}. The conclusion from the above analysis then appears to be the following.
The raw mismatch in {\bf Table~\ref{DOFcounting}} is a mismatch of the spectra at the pinch/wedge, and not of the common reduced massless sector. However the common massless sector appears to have a good match, but only if certain conditions like \eqref{onlyifbnd} are imposed. Therefore to summarize: motivated by the wedge analysis used earlier, we now form branch-even and
branch-odd combinations. (See also {\bf Table \ref{0hetnine}}). For the
metric vectors, B-field vectors and branchwise radii, we have respectively:
\begin{equation}
A_\mu^{(g,e)}
\equiv
\frac{1}{2}\big(g_{\mu +}+g_{\mu -}\big),
\qquad
A_\mu^{(g,o)}
\equiv
\frac{1}{2}\big(g_{\mu +}-g_{\mu -}\big)\nonumber
\label{het_metric_vectors}
\end{equation}
\begin{equation}
A_\mu^{(B,e)}
\equiv
\frac{1}{2}\big(B_{\mu +}+B_{\mu -}\big),
\qquad
A_\mu^{(B,o)}
\equiv
\frac{1}{2}\big(B_{\mu +}-B_{\mu -}\big)\nonumber
\label{het_B_vectors}
\end{equation}
\begin{equation}
R_\pm \equiv \sqrt{g_{\pm\pm}},
\qquad
R_B \equiv \frac{1}{2}(R_+ + R_-),
\qquad
T \equiv \frac{1}{2}(R_+ - R_-).
\label{het_radii_tachyon}
\end{equation}
\begin{table}[t]
\centering
\resizebox{\textwidth}{!}{%
\renewcommand{\arraystretch}{3.0}
\begin{tabular}{|c|c|c|c|}
\hline
\textbf{10d heterotic field} & \textbf{Branchwise 9d decomposition} & \textbf{Even/odd split} & \textbf{Physical role} \\
\hline
$g_{MN}$
&
$g_{\mu\nu},\ g_{\mu +},\ g_{\mu -},\ g_{++},\ g_{--},\ g_{+-}$
&
$\big(g_{\mu\nu},\,A_\mu^{(g,e)},\,A_\mu^{(g,o)},\,R_B,\,T,\,g_{+-}\big)$
&
$g_{\mu\nu},\ A_\mu^{(g,e)},\ \Phi,\ T$; odd pieces auxiliary
\\
\hline
$B_{MN}$
&
$B_{\mu\nu},\ B_{\mu +},\ B_{\mu -},\ B_{+-}$
&
$\big(B_{\mu\nu},\,A_\mu^{(B,e)},\,A_\mu^{(B,o)},\,B_{+-}\big)$
&
$B_{\mu\nu},\ A_\mu^{(B,e)}$; odd/mixed pieces auxiliary
\\
\hline
$\phi$
&
$\phi$
&
$\Phi$
&
$\Phi$
\\
\hline
\end{tabular}%
}
\caption{Branchwise reduction of the universal heterotic bosonic sector on
$S^1_+\vee S^1_-$. The branch-even combinations provide the physical effective
nine-dimensional fields, while the branch-odd and branch-mixed combinations are
treated as auxiliary wedge/junction data.}
\label{0hetnine}
\end{table}
\noindent As before, the branch-odd scalar modulus $T$ is identified with the tachyonic degree
of freedom in the effective type-$0$ description. The physical branch-even bosonic fields are:
$g_{\mu\nu}$,
$B_{\mu\nu}$,
$\Phi$
$T$ and the two set of gauge fields $A_\mu^{(g, e)}$ and $A_\mu^{(B, e)}$; 
where $\Phi$ denotes the branch-even scalar combination obtained from $\phi$ and the
even radion $R_B$ after the usual frame redefinition. The branch-odd quantities:
\begin{equation}
A_\mu^{(g,o)},
\qquad
A_\mu^{(B,o)},
\qquad
g_{+-},
\qquad
B_{+-}
\label{het_odd_aux}
\end{equation}
are not counted as independent fields in the common nine-dimensional massless
sector. The odd vectors are anti-invariant under the node projection, while
the mixed scalars require a resolved junction geometry. They may still be
useful localized wedge variables in a branch-resolved or resolved-pinch
description, but they are not part of the common bulk zero-mode spectrum. More precisely, away from the pinch point, the branch-odd vector combinations:
\[
A_\mu^{(g,o)}
=
\frac12(g_{\mu+}-g_{\mu-}),
\qquad
A_\mu^{(B,o)}
=
\frac12(B_{\mu+}-B_{\mu-})
\]
are projected out of the node-invariant zero-mode sector. The junction
condition keeps only the branch-even zero modes, while the branch-odd
vectors should be treated as massive or auxiliary odd-sector data rather than
as massless gauge bosons in the common nine-dimensional theory.

This is a tree-level statement about the wedge boundary condition and the
node projection, not a one-loop effect. The odd vectors may still appear in
a branch-resolved or smoothed-junction description, but their effects belong
to the massive threshold sector and do not modify the common massless
spectrum counted in~\eqref{matched_64_count}. Thus the comparison between
the candidate Type $0$A route and the Type $0$HW route should be performed
after removing these odd vector slots from the propagating massless sector.
Putting everything together, the heterotic wedge reduction gives the following natural nine-dimensional
effective field content:
\begin{equation}
\big(
g_{\mu\nu},
\;
B_{\mu\nu},
\;
\Phi,
\;
T,
\;
A_\mu^{(g,e)},
\;
A_\mu^{(B,e)}
\big) ~~~~\oplus
\quad
\text{branch-odd auxiliary data ,}
\label{het_9d_raw}
\end{equation}
as shown in {\bf Table \ref{0hetnine}}.
We can now compare with the candidate Type 0A spectrum.
Earlier, the nine-dimensional spectrum obtained from the candidate Type $0$A frame
after the $S^1_a/\mathbb Z_2$ projection was given in \eqref{finalboxed0Ato9d}.
At this stage one should be careful. The candidate Type $0$A side contains two
two-form slots,
$C_{\mu\nu a}^{(+)}$,
and 
$C_{\mu\nu a}^{(-)}$,
or equivalently their even/odd combinations \eqref{ccombination}
whereas the heterotic wedge reduction produces only one universal physical
two-form \(B_{\mu\nu}\). The common-sector projection identifies this
heterotic two-form with the even combination \(C_{\mu\nu a}^{(e)}\). The
odd combination \(C_{\mu\nu a}^{(o)}\) has no node-invariant zero mode and
is not part of the common massless sector.
Similarly, the candidate Type $0$A side contains the vector
\(A_\mu^{(B)}\equiv B_{\mu a}\). This vector is present in the raw
Type $0$A parametrization but can only be removed from the common universal
nine-dimensional sector if a condition like \eqref{onlyifbnd} is imposed.

Therefore the heterotic wedge reduction does not reproduce the full raw
candidate Type $0$A spectrum field by field. Rather, after imposing the
same massless-sector projection on both sides, including \eqref{onlyifbnd}, the two descriptions might share
common \(64\)-DOF sector. The local dictionary should
therefore be stated more weakly as:
\begin{equation}
g_{\mu\nu}^{\rm(het)}
\;\longleftrightarrow\;
g_{\mu\nu}^{(0A)},
\qquad
\Phi^{\rm(het)}
\;\longleftrightarrow\;
\Phi^{(0A)},
\qquad
T^{\rm(het)}
\;\longleftrightarrow\;
T^{(0A)},
\label{dictionary_common_scalars_metric}
\end{equation}
where one should note that the relations should not be viewed as literal equalities between the fields. As an example, once we reach the supersymmetric end points, the Type IIA and Heterotic theories are U-dual to each other (involving one intermediate S-duality), where the two dilatons are related by a relative minus sign up-to moduli dependent corrections. Away from the supersymmetric points, the relations are more non-trivial and is not clear even there exist any duality between the two theories. In a similar vein
the two heterotic branch-even vectors are mapped to the two
effective type-$0$A vector slots:
\begin{equation}
A_\mu^{(g,e)}
\;\longleftrightarrow\;
C_\mu^{(+)},
\qquad
A_\mu^{(B,e)}
\;\longleftrightarrow\;
C_\mu^{(-)} ,
\label{dictionary_vectors_corrected}
\end{equation}
The remaining raw type-$0$A vector \(B_{\mu a}\) is not part of the common
projected sector, as discussed above.
For the two-form sector, the heterotic theory provides one distinguished
two-form slot,
\begin{equation}
B_{\mu\nu}^{\rm(het)}
\;\longleftrightarrow\;
C_{\mu\nu a}^{(e)} ,
\label{dictionary_single_2form}
\end{equation}
but not the branch-odd combination \(C_{\mu\nu a}^{(o)}\). Thus the
two-form comparison is a comparison of the common node-invariant sector,
not of the full raw branch-resolved field list. The detailed slot-by-slot comparison between the heterotic wedge reduction and the
candidate Type $0$A nine-dimensional description is summarized in
{\bf Table \ref{comp9d}}.

\begin{table}[t]
\centering
\renewcommand{\arraystretch}{1.35}
\begin{tabular} {|p{0.25\textwidth}|p{0.32\textwidth}|p{0.32\textwidth}|}
\hline
\textbf{9d candidate Type $0$A field} & \textbf{Heterotic wedge counterpart?} & \textbf{Comment} \\
\hline
$g_{\mu\nu}$ & $g_{\mu\nu}$ & common metric sector \\
\hline
$\Phi$ & $\Phi$ & common scalar sector \\
\hline
$T$ & $T$ & common tachyonic/branch-odd scalar \\
\hline
$C_\mu^{(+)}$ & $A_\mu^{(g,e)}$ & effective vector-slot matching \\
\hline
$C_\mu^{(-)}$ & $A_\mu^{(B,e)}$ & effective vector-slot matching \\
\hline
$C_{\mu\nu a}^{(+)}$, $C_{\mu\nu a}^{(-)}$ & only one $B_{\mu\nu}$ & heterotic has only one universal 2-form \\
\hline
$B_{\mu a}$ & no universal counterpart & missing in minimal heterotic bosonic sector \\
\hline
$g_{aa}$ & no direct universal counterpart & may enter the scalar dictionary only indirectly \\
\hline
\end{tabular}
\caption{Comparison between the nine-dimensional candidate type-$0$A
reduction and the wedge reduction of the universal heterotic bosonic sector.
The heterotic side matches only a subsector of the candidate type-$0$A field
content. We will however soon provide a more precise map between the DOFs in nine-dimensions for the two theories in {\bf Table \ref{spectramatch}}.}
\label{comp9d}
\end{table}

Unfortunately the condition \eqref{onlyifbnd} and consequently the weak equivalence between the 64 DOFs for the two theories from \eqref{matched_64_count} suggest that the underlying picture may be more subtle that what appears from the naive countings. A better way to compare the spectrum would be to first incorporate the {\it non-dynamical} DOFs from the Type 0A side. These non-dynamical DOFs do not affect the bulk spectrum but appear from the highly localized UV DOFs at the pinch that we discussed in section \ref{sec001}. Since it will be impossible to quantify the full spectrum of the localized UV DOFs at the pinch, we will only take the set from \eqref{nonphys0A}. Thus in the Type 0A side, the ten-dimensional spectrum consists of the set from \eqref{routeI_RR1} and \eqref{nonphys0A}, {\it i.e.}:
\begin{equation}
\big(g_{mn}, B_{mn}, \Phi, T\big)_{\rm NS}
~~\oplus~~
\big(C_1^{(\pm)}, C_3^{(\pm)}\big)_{\rm RR} ~~ \oplus ~~ \big(g_{+-}, C_{m+-}, B^{(o)}_{mn}\big),
\end{equation}
which is now compactified on the interval $S_a^1/\mathbb{Z}_2$. This is an orientifold operation and therefore acts on the three- and the two-forms by flipping their signs. In fact even in the localized pinch sector we expect $C_{m+-} \to -C_{m+-}$ and $B_{mn}^{(o)} \to - B_{mn}^{(o)}$, as they both originate from M-theory three-form. The nine-dimensional spectrum now becomes:
\begin{equation}
g_{\mu\nu} \;\oplus\;
\big(C^+_{\mu\nu a},\,C^-_{\mu\nu a}\big)
\;\oplus\;
\big(C^+_\mu,\,C^-_\mu,\,B_{\mu a}, \, B^{(o)}_{ma}\big)
\;\oplus\;
\big(\phi,\,T,\,g_{aa}, \, g_{+-}, \, C_{a+-}\big), 
\end{equation}
implying that we now have one metric, two anti-symmetric B-fields, four abelian gauge fields and five scalars compared to what we had in \eqref{finalboxed0Ato9d}. Comparing this to the Type 0HW spectrum in nine-dimensions from \eqref{finalboxed0HW} we see that the two spectrum {\it almost} match with the exception that Type 0A has {\it two} anti-symmetric B-fields whereas Type 0HW has only one. In fact this is exactly parallel to the situation we encountered when comparing the spectrum of M-theory on $S^1_+\vee S^1_-$ and Type 0A: the former has only {\it one} three-form field whereas the latter has {\it two}. The resolution therein was simple but instructive: the pinch contributes a {\it massless} three-form DOF in addition to other localized DOFs. Could the same thing happen here too? 

\begin{table}[t]
\centering
\renewcommand{\arraystretch}{1.25}
\begin{tabular}{|c|c|c|c|}
\hline
$n_1$ & $n_2$ & $n_3$ & $n_4$ \\
\hline
$1$ & $1$ & $1$ & $35$ \\
\hline
$1$ & $1$ & $2$ & $28$ \\
\hline
$1$ & $1$ & $3$ & $21$ \\
\hline
$1$ & $1$ & $4$ & $14$ \\
\hline
$1$ & $1$ & $5$ & $7$ \\
\hline
$1$ & $2$ & $1$ & $28$ \\
\hline
$1$ & $2$ & $2$ & $21$ \\
\hline
$1$ & $2$ & $3$ & $14$ \\
\hline
$1$ & $2$ & $4$ & $7$ \\
\hline
$1$ & $3$ & $1$ & $21$ \\
\hline
$1$ & $3$ & $2$ & $14$ \\
\hline
$1$ & $3$ & $3$ & $7$ \\
\hline
$1$ & $4$ & $1$ & $14$ \\
\hline
$1$ & $4$ & $2$ & $7$ \\
\hline
$1$ & $5$ & $1$ & $7$ \\
\hline
$\color{red}{2}$ & $\color{red}{1}$ & $\color{red}{1}$ & $\color{red}{14}$ \\
\hline
$\color{red}{2}$ & $\color{red}{1}$ & $\color{red}{2}$ & $\color{red}{7}$ \\
\hline
$\color{red}{2}$ & $\color{red}{2}$ & $\color{red}{1}$ & $\color{red}{7}$ \\
\hline
\end{tabular}
\caption{Positive integer solutions of $21n_1+7(n_2+n_3)+n_4=70$. The entries in $\color{red}{\rm red}$ show the precise spectrum match conditions.}
\label{finalmatch}
\end{table}

To see this, let us count the number of DOFs in the Type 0A side. With the additional sector inserted in, the total number becomes $27 + 2\times 21 + 4\times 7 + 5\times 1 = {\bf 102}$. Expectedly this still doesn't match with the ${\bf 81}$ DOFs from the type 0HW side, but now we are no longer required to keep the same bulk DOFs from \eqref{finalboxed0HW}, as the pinch may add in new massless DOFs. Following the same counting scheme as in \eqref{intisol}, the DOF matching condition now leads to:
\begin{equation}
21n_1 + 7(n_2 + n_3) + n_4 = 70,~~~~~ n_i \in \mathbb{Z}_+ ,    
\end{equation}
with {\it two} important differences: one, $n_4$ now represents any additional massless DOF from the pinch and not just scalar DOF from earlier, and two, since we do not want to lose (or hide) any DOFs, we put the constraint that $n_2 + n_3 + n_4 \ge 4$. There are now {\bf 18}
non-negative positive-integer solutions to the four-tuples shown in {\bf Table \ref{finalmatch}} out of which the important ones are shown in $\color{red}{\rm red}$. If we demand that $n_4$ counts the number of massless abelian gauge fields, then the three choices $(2, 1, 1, 14), (2, 1, 2, 7)$ and $(2, 2, 1, 7)$  all points to the presence of {\it two} anti-symmetric B-fields and {\it four} abelian gauge fields. Comparing this to the spectrum in \eqref{finalboxed0HW}, our analysis predicts that the {\it pinch must contribute one massless anti-symmetric two-form field to the Type 0HW spectrum in nine-dimensions}, {\it i.e.}:
\begin{equation}\label{splituu}
B_{\mu\nu}^{(\pm)} = B_{\mu\nu} \pm (\check{\Xi}_2)_{\mu\nu},
\end{equation}
similar we what we had in \eqref{chotolookai} and \eqref{chotolook} for the doubling of Type 0A three-form. At the supersymmetric end-points, and exactly as in the three-form case \eqref{chotolookai}, the massless contribution $\check{\Xi}_2$ should either go to zero, or become a pure gauge. This way the spectrum on both sides would match precisely in the supersymmetric and the non-supersymmetric regimes.

\begin{table}[t]
\centering
\renewcommand{\arraystretch}{1.25}
\begin{tabular}{|c|c|}
\hline
${\bf Type~0A}$ & ${\bf Type~0HW}$\\
\hline
$g_{\mu\nu}$ & ${\rm g}_{\mu\nu}$ \\
\hline
$C^+_{\mu\nu a},\, C^-_{\mu\nu a}$ & $ \mathcal B_{\mu\nu}^{(+)} = \mathcal B_{\mu\nu} + (\check{\Xi}_2)_{\mu\nu}, \, \mathcal B_{\mu\nu}^{(-)} = \mathcal B_{\mu\nu} - (\check{\Xi}_2)_{\mu\nu}$  \\
\hline
$ C^+_\mu,\,C^-_\mu,\,B_{\mu a}, \, B^{(o)}_{ma}$ & $ {\rm g}_{\mu +},\,{\rm g}_{\mu -},\,\mathcal B_{\mu +},\, \mathcal B_{\mu -}  $  \\
\hline
$ \phi,\,T,\,g_{aa}, \, g_{+-}, \, C_{a+-}   $ & $ \varphi,\,\mathcal R_B,\,\mathcal T,\,{\rm g}_{+-},\,\mathcal B_{+-}  $  \\
\hline
\end{tabular}
\caption{Precise matching of the Type 0A spectrum with the Type 0HW spectrum in nine-dimensions.}
\label{spectramatch}
\end{table}

The equality of the \({\bf 102}\) DOF sectors should not be read
as a proof of a full duality away from the supersymmetric endpoints.  It is a
more limited, but still nontrivial, statement: after imposing the same node,
junction, interval, and scalar-frame projections, including the split in
\eqref{splituu}, the two local descriptions contain the same common universal
massless bosonic sector in nine dimensions.
This matching shows that the Type 0A and Type 0{\rm HW} frames can be
organized so as to describe the same effective nine-dimensional field content,
even though the microscopic interpretation of the fields is different in the
two pictures.  In the Type 0A frame, part of this sector is naturally
described in terms of branch-resolved RR fields.  In the Type 0{\rm HW}
frame, the same number of degrees of freedom is instead accounted for by the
branchwise metric, scalar, gauge, junction-sensitive sectors, and branch resolved two-forms
arising after the interval and wedge projections. 

Thus the spectrum matching should be interpreted as a controlled equality of
effective massless bosonic degrees of freedom, not as an identification of the
underlying microscopic variables.  In particular, it does not imply that the
perturbative heterotic description literally contains the Type 0A RR
fields.  Rather, the two reductions provide different local organizations of a
common nine-dimensional sector.

This distinction is important.  Equality of the spectrum is a necessary
consistency check for any proposed relation between the two frames, but it is
not sufficient to establish a non-supersymmetric U-duality.  A genuine duality
would require additional data to match, including the localized junction
degrees of freedom, their interactions, anomaly or inflow constraints, flux
quantization, the precise map of moduli, and the gauge-sector couplings.
Therefore the result should be viewed as evidence for a precise effective
correspondence between the two descriptions, rather than as a proof of a full
microscopic duality in the non-supersymmetric regime.


\subsubsection{The Type $0$HW torsion classes in ten and  nine-dimensions \label{comparisoon0}}

The two routes that we studied above give similar nine-dimensional structures, but the origin of the
terms is different.  In Route I, the nine-dimensional result is obtained by
taking the already-pinch-corrected Type $0$A torsion classes and then imposing
the interval projection.  In Route II, one first obtains the interval-reduced
type-$0$HW torsion classes, and only then adds the wedge-induced corrections.
In other words, the two routes are:
\begin{equation}
{\rm Route\ A:}
\qquad
{\rm M\ theory}
\xrightarrow{\;S^1_+\vee S^1_-\;}
0A
\xrightarrow{\;S^1_a/\mathbb Z_2\;}
9d \nonumber
\label{route_0A_again}
\end{equation}
\begin{equation}
{\rm Route\ HW:}
\qquad
{\rm M\ theory}
\xrightarrow{\;S^1_a/\mathbb Z_2\;}
0{\rm HW}
\xrightarrow{\;S^1_+\vee S^1_-\;}
9d .
\label{2routes}
\end{equation}
In Route A, the first reduction is already on the singular wedge circle
$S^1_+\vee S^1_-$.  Therefore the type-$0$A torsion classes already contain
the branch-odd wedge and pinch data before the subsequent reduction on
$S^1_a/\mathbb Z_2$.  In Route HW, however, the first reduction is on the
Ho\v{r}ava--Witten interval $S^1_a/\mathbb Z_2$.  The wedge circle enters only
in the second step, when the type-$0$HW branch is compactified further to nine
dimensions.  Consequently, the ten-dimensional type-$0$HW torsion classes
should be kept in their simpler interval-reduction form, while the more
expanded $T$, $dT$, $d\varepsilon$, and branch-odd terms should be interpreted
as nine-dimensional wedge-reduction corrections. This will clarify the issue that we raised earlier in \eqref{W5full}.  In the following let us elaborate the story.

The main point is that, in the Type 0HW case, the torsion classes listed in \eqref{W5full} should be interpreted carefully. In fact this will also help us to figure out the dotted terms therein.
  The key point is that there
are two different stages in the type-$0$HW route, and the localized
wedge-junction terms should be assigned to the correct stage.
The two reduction routes are in \eqref{2routes}, and as mentioned above,
the order of reduction matters because the singular wedge circle is used in
the first step of Route A, but only in the second step of Route HW.

Therefore, in the type-$0$A route, the pinch and branch-odd wedge data are
already present in the type-$0$A torsion classes.  In the type-$0$HW route,
the bare ten-dimensional type-$0$HW torsion classes should not yet contain
wedge-node terms such as:
\begin{equation}
T\,\delta_{\Sigma_5^{\rm jct}}\,\xi_i,
\qquad
dT\wedge \mathbf y_i^{(o)},
\qquad
d\varepsilon,
\qquad
\varepsilon\,w_i^{(\varepsilon)} ,
\end{equation}
where the parameters will be defined soon.
These terms arise after the second step, namely after compactifying the
type-$0$HW branch on
$S^1_+\vee S^1_-$. This means the torsion classes in \eqref{W5full} should {\it not} be interpreted to be completely a ten-dimensional result.

Immediately after reducing M-theory on the Ho\v{r}ava--Witten interval
$S^1_a/\mathbb Z_2$, the wedge circle has not yet been used as a
compactification direction.  Thus the clean ten-dimensional type-$0$HW torsion
classes should be written as:
\begin{equation}
W_1^{(0{\rm HW},10d)}
=
w_1^{\rm fib}
+
w_1^{\rm flux} \nonumber
\label{W1_0HW_10d_clean}
\end{equation}
\begin{equation}
W_2^{(0{\rm HW},10d)}
=
w_2^{\rm fib}
+
w_2^{\rm flux} \nonumber
\label{W2_0HW_10d_clean}
\end{equation}
\begin{equation}
W_3^{(0{\rm HW},10d)}
=
w_3^{\rm fib}
+
w_3^{\rm flux} \nonumber
\label{W3_0HW_10d_clean}
\end{equation}
\begin{equation}
W_4^{(0{\rm HW},10d)}
=
w_4^{\rm fib}
+
w_4^{\rm flux},
\qquad
W_5^{(0{\rm HW},10d)}
=
w_5^{\rm fib}
+
w_5^{\rm flux} ,
\label{W45_0HW_10d_clean}
\end{equation}
where $w_i^{\rm fib}$ denotes the fibration contribution, while
$w_i^{\rm flux}$ denotes the flux-induced contribution in the
interval-reduced type-$0$HW frame. This looks distinctively simpler than \eqref{Wi_total_final_combined} and \eqref{W_flux_pinch_combined_final}, and rightly so because the pinch (or the wedge-junction) effects have not yet been inserted in.
The important observation is that the localized wedge-junction terms
\begin{equation}
\kappa_2\,T\,\delta_{\Sigma_5^{\rm jct}}\,\xi_2,
\qquad
\kappa_3\,T\,\delta_{\Sigma_5^{\rm jct}}\,\xi_3
\end{equation}
should not be included in
$W_2^{(0{\rm HW},10d)}$,
and $W_3^{(0{\rm HW},10d)}$
in the strict order-of-reduction picture.  They are localized data associated
with the wedge node, and the wedge node appears only after the second
compactification on $S^1_+\vee S^1_-$.  Thus the earlier schematic formula in \eqref{W5full}:
\begin{equation}
W_2^{(0{\rm HW},10d)}
=
w_2^{\rm fib}
+
w_2^{\rm flux}
+
\kappa_2\,T\,\delta_{\Sigma_5^{\rm jct}}\,\xi_2
\end{equation}
should be reinterpreted as a nine-dimensional wedge-reduced expression, not
as the bare ten-dimensional type-$0$HW expression.  The same applies to the
corresponding $W_3$ formula.

The above discussion explains all the terms in \eqref{W5full}, but does not yet provide a full derivation of the dotted terms. Plus the comparison to the torsion classes in the Type 0A side is still lacking. In the following let us elaborate this by studying the 
nine-dimensional type-$0$HW torsion classes after the wedge compactification. In other words, after the second step:
\begin{equation}
0{\rm HW}
\xrightarrow{\;S^1_+\vee S^1_-\;}
9d ,
\end{equation}
the wedge geometry is part of the compactification data.  One can then
introduce the branch-even and branch-odd combinations of the wedge radii:
\begin{equation}
R_B
=
\frac{1}{2}
\left(
R_+ + R_-
\right),
\qquad
T
=
\frac{1}{2}
\left(
R_+ - R_-
\right) ,
\label{RB_T_def_0HW}
\end{equation}
and this is where the information of the tachyon $T$ and its subsequent effects kick in. Recall that  
the scalar $T$ is branch-odd under the exchange
$S^1_+
\longleftrightarrow
S^1_-$.
If the wedge node is smoothed or resolved, one may also introduce a local
resolution parameter
$\varepsilon$. For all practical purposes, $\varepsilon$ could be chosen to be the same as the one from \eqref{global_eps_endpoint}.
It vanishes at the symmetric point $R_+=R_-$ and transforms as
$\varepsilon\mapsto -\varepsilon$ under branch exchange.  In a fixed
branch-even volume frame it is a nonlinear function of the branch-odd modulus
$T$, so that near $T=0$,
\begin{equation}\label{epstacion}
\varepsilon
=
-4e^{-1}\frac{T}{R_B}
+
O\left(\frac{T^3}{R_B^3}\right) ,
\end{equation}
as we saw earlier. An alternative option of choosing a different resolution parameter than the one from \eqref{global_eps_endpoint} is possible but does not serve any useful purpose here. Therefore we will stick with $\varepsilon$ defined as in \eqref{global_eps_endpoint} and 
\eqref{epstacion}. Keeping this in mind, the nine-dimensional type-$0$HW torsion classes should be
organized as:
\begin{equation}
W_i^{(0{\rm HW},9d)}
=
W_i^{(0{\rm HW},10d)}
+
\Delta_{\rm wedge}W_i^{(0{\rm HW}\to 9d)} ,
\label{Wi_0HW_9d_split_unified}
\end{equation}
where $W_i^{(0{\rm HW},10d)}$ are the ten-dimensional torsion classes from \eqref{W45_0HW_10d_clean}. This split is important and crucial to our discussion because all the wedge-induced corrections are trapped in the second term. We can quantify 
the wedge-induced corrections in the following way:
\begin{equation}
\begin{aligned}
\Delta_{\rm wedge}W_1^{(0{\rm HW}\to 9d)}
&=
T\,\mathbf w_1^{(o)}
+
\iota_{V_T}\mathbf y_1^{(o)}
+
O(T^2),
\\[4pt]
\Delta_{\rm wedge}W_2^{(0{\rm HW}\to 9d)}
&=
\kappa_2\,T\,\delta_{\Sigma_5^{\rm jct}}\,\xi_2
+
T\,\mathbf w_2^{(o)}
+
dT\wedge\mathbf y_2^{(o)}
+
O(T^2),
\\[4pt]
\Delta_{\rm wedge}W_3^{(0{\rm HW}\to 9d)}
&=
\kappa_3\,T\,\delta_{\Sigma_5^{\rm jct}}\,\xi_3
+
T\,\mathbf w_3^{(o)}
+
dT\wedge\mathbf y_3^{(o)}
+
O(T^2),
\\[4pt]
\Delta_{\rm wedge}W_4^{(0{\rm HW}\to 9d)}
&=
\hat{a}_4^{(o)}\,\mathbf x_\perp^{(o)}
+
\hat{c}_4^{(o)}\,dT
+
T\,\mathbf w_4^{(o)}
+
\hat{b}_4\,d\varepsilon
+
\varepsilon\,w_4^{(\varepsilon)}
+
O(T^2,\varepsilon^2),
\\[4pt]
\Delta_{\rm wedge}W_5^{(0{\rm HW}\to 9d)}
&=
\hat{a}_5^{(o)}\,\mathbf x_\perp^{(o)}
+
\hat{c}_5^{(o)}\,dT
+
T\,\mathbf w_5^{(o)}
+
\hat{b}_5\,d\varepsilon
+
\varepsilon\,w_5^{(\varepsilon)}
+
O(T^2,\varepsilon^2) ,
\end{aligned}
\label{Delta_wedge_0HW_unified}
\end{equation}
where note that the corrections are all packaged in the odd-branch parameters and take exactly similar form as in \eqref{W_flux_pinch_combined_final}. Moreover the two terms, $ \kappa_2\,T\,\delta_{\Sigma_5^{\rm jct}}\,\xi_2$ and $ \kappa_3\,T\,\delta_{\Sigma_5^{\rm jct}}\,\xi_3$, that we originally motivated from the ${\bf 27}$ of $G_2$, now appears rightly in the nine-dimensional $W_2$ and $W_3$. Moreover, with the definition of $\varepsilon$ from \eqref{global_eps_endpoint}, the terms
$\hat{b}_4\,d\varepsilon
+
\varepsilon\,w_4^{(\varepsilon)}$
and
$\hat{b}_5\,d\varepsilon
+
\varepsilon\,w_5^{(\varepsilon)}$
in the Lee-form torsion classes are smooth branch-odd resolved-wedge
corrections.  The derivative term $d\varepsilon$ is a one-form and therefore
can contribute naturally to $W_4$ and $W_5$ as 
$d\varepsilon\in \Lambda^1$.
Thus the full nine-dimensional type-$0$HW torsion classes are:
\begin{equation}
\begin{aligned}
W_1^{(0{\rm HW},9d)}
&=
w_1^{\rm fib}
+
w_1^{\rm flux}
+
T\,\mathbf w_1^{(o)}
+
\iota_{V_T}\mathbf y_1^{(o)}
+
O(T^2),
\\[4pt]
W_2^{(0{\rm HW},9d)}
&=
w_2^{\rm fib}
+
w_2^{\rm flux}
+
\kappa_2\,T\,\delta_{\Sigma_5^{\rm jct}}\,\xi_2
+
T\,\mathbf w_2^{(o)}
+
dT\wedge\mathbf y_2^{(o)}
+
O(T^2),
\\[4pt]
W_3^{(0{\rm HW},9d)}
&=
w_3^{\rm fib}
+
w_3^{\rm flux}
+
\kappa_3\,T\,\delta_{\Sigma_5^{\rm jct}}\,\xi_3
+
T\,\mathbf w_3^{(o)}
+
dT\wedge\mathbf y_3^{(o)}
+
O(T^2),
\\[4pt]
W_4^{(0{\rm HW},9d)}
&=
w_4^{\rm fib}
+
w_4^{\rm flux}
+
\hat{a}_4^{(o)}\,\mathbf x_\perp^{(o)}
+
\hat{c}_4^{(o)}\,dT
+
T\,\mathbf w_4^{(o)}
+
\hat{b}_4\,d\varepsilon
+
\varepsilon\,w_4^{(\varepsilon)}
+
O(T^2,\varepsilon^2),
\\[4pt]
W_5^{(0{\rm HW},9d)}
&=
w_5^{\rm fib}
+
w_5^{\rm flux}
+
\hat{a}_5^{(o)}\,\mathbf x_\perp^{(o)}
+
\hat{c}_5^{(o)}\,dT
+
T\,\mathbf w_5^{(o)}
+
\hat{b}_5\,d\varepsilon
+
\varepsilon\,w_5^{(\varepsilon)}
+
O(T^2,\varepsilon^2) ,
\end{aligned}
\label{W0HW_9d_full_unified}
\end{equation}
which, as mentioned above, is the expression that contains the localized junction terms in $W_2$ and
$W_3$, as well as the smooth wedge terms in $W_4$ and $W_5$.  It is a
nine-dimensional expression, not the bare ten-dimensional type-$0$HW
torsion formula. The distinction between the ten-dimensional and nine-dimensional type-$0$HW
expressions are summarized in {\bf Table \ref{comp1011d}}. This table makes clear that the localized junction terms and the smooth
resolution terms are not part of the bare ten-dimensional interval reduction.
They belong to the wedge correction
$\Delta_{\rm wedge}W_i^{(0{\rm HW}\to 9d)}$ quantified in \eqref{Delta_wedge_0HW_unified}.

\begin{table}[t]
\centering
\resizebox{\textwidth}{!}{%
\renewcommand{\arraystretch}{1.35}
\begin{tabular}{|c|c|c|c|}
\hline
\text{\bf term}
&
\text{\bf form degree}
&
\text{\bf origin in the type 0HW route}
&
\text{\bf appears in}
\\[2pt]
\hline
$w_i^{\rm fib}$
&
\text{same as }$W_i$
&
\text{interval-reduced fibration data}
&
$W_i^{(0{\rm HW},10d)}$
\\[4pt]\hline
$w_i^{\rm flux}$
&
\text{same as }$W_i$
&
\text{interval-reduced flux data}
&
$W_i^{(0{\rm HW},10d)}$
\\[4pt]\hline
$T\,\mathbf w_1^{(o)}$
&
$\Lambda^0$
&
\text{branch-odd wedge scalar correction}
&
$\Delta_{\rm wedge}W_1^{(0{\rm HW}\to 9d)}$
\\[4pt]\hline
$\iota_{V_T}\mathbf y_1^{(o)}$
&
$\Lambda^0$
&
\text{wedge-gradient contraction}
&
$\Delta_{\rm wedge}W_1^{(0{\rm HW}\to 9d)}$
\\[4pt]\hline
$\kappa_2T\delta_{\Sigma_5^{\rm jct}}\xi_2$
&
$\Lambda_{\rm prim}^{(1,1)}$
&
\text{localized wedge-junction source}
&
$\Delta_{\rm wedge}W_2^{(0{\rm HW}\to 9d)}$
\\[4pt]\hline
$\kappa_3T\delta_{\Sigma_5^{\rm jct}}\xi_3$
&
$\Lambda_{\rm prim}^{(2,1)+(1,2)}$
&
\text{localized wedge-junction source}
&
$\Delta_{\rm wedge}W_3^{(0{\rm HW}\to 9d)}$
\\[4pt]\hline
$T\,\mathbf w_2^{(o)}$
&
$\Lambda_{\rm prim}^{(1,1)}$
&
\text{smooth branch-odd wedge correction}
&
$\Delta_{\rm wedge}W_2^{(0{\rm HW}\to 9d)}$
\\[4pt]\hline
$dT\wedge\mathbf y_2^{(o)}$
&
$\Lambda_{\rm prim}^{(1,1)}$
&
\text{gradient of branch asymmetry}
&
$\Delta_{\rm wedge}W_2^{(0{\rm HW}\to 9d)}$
\\[4pt]\hline
$T\,\mathbf w_3^{(o)}$
&
$\Lambda_{\rm prim}^{(2,1)+(1,2)}$
&
\text{smooth branch-odd wedge correction}
&
$\Delta_{\rm wedge}W_3^{(0{\rm HW}\to 9d)}$
\\[4pt]\hline
$dT\wedge\mathbf y_3^{(o)}$
&
$\Lambda_{\rm prim}^{(2,1)+(1,2)}$
&
\text{gradient of branch asymmetry}
&
$\Delta_{\rm wedge}W_3^{(0{\rm HW}\to 9d)}$
\\[4pt]\hline
$\mathbf x_\perp^{(o)}$
&
$\Lambda^1$
&
\text{branch-odd Lee-form data}
&
$\Delta_{\rm wedge}W_{4,5}^{(0{\rm HW}\to 9d)}$
\\[4pt]\hline
$dT$
&
$\Lambda^1$
&
\text{one-form from branch-odd scalar}
&
$\Delta_{\rm wedge}W_{4,5}^{(0{\rm HW}\to 9d)}$
\\[4pt]\hline
$d\varepsilon$
&
$\Lambda^1$
&
\text{smooth resolution one-form}
&
$\Delta_{\rm wedge}W_{4,5}^{(0{\rm HW}\to 9d)}$
\\[4pt]\hline
$\varepsilon\,w_{4,5}^{(\varepsilon)}$
&
$\Lambda^1$
&
\text{algebraic smoothing correction}
&
$\Delta_{\rm wedge}W_{4,5}^{(0{\rm HW}\to 9d)}$\\
\hline
\end{tabular}%
}
\caption{Corrected comparison between the ten- and the nine-dimensional candidate Type 0HW theory. From here one may directly compare the terms to \eqref{Wi_total_final_combined} and \eqref{W_flux_pinch_combined_final} that we got for the Type 0A case.}
\label{comp1011d}
\end{table}

We are now ready to compare with the Type 0A route. The difference between the two routes should be clear. In the Type 0A case, we have a complicated theory compactified on a simpler six-dimensional torsional manifold, whereas in the Type 0HW case, we have a simple theory compactified on a more complicated six-dimensional torsional manifold. Graphically,
the type-$0$A route has the opposite order:
\begin{equation}
{\rm M\ theory}
\xrightarrow{\;S^1_+\vee S^1_-\;}
0A
\xrightarrow{\;S^1_a/\mathbb Z_2\;}
9d .
\end{equation}
Here the wedge compactification is performed in the first step.  Therefore the
branch-odd and pinch data are already built into the type-$0$A torsion
classes.  One can write
\begin{equation}
W_i^{(0A)}
=
W_i^{\rm fib}
+
W_i^{\rm flux+pinch},
\qquad
i=1,\ldots,5 .
\label{Wi_0A_total_unified}
\end{equation}
The fibration piece depends only on the fibration geometry, {\it i.e.}
$W_i^{\rm fib}
=
W_i^{\rm fib}[{\rm fibration}]$.
Since no T-duality is being performed in this step, the fibration sector does
not mix with the flux-plus-pinch sector.  The latter is already given in \eqref{W_flux_pinch_combined_final}, 
where note that 
in contrast with the type-$0$HW route, the symbols
$T, dT, d\varepsilon$ and $\varepsilon$
are already part of the type-$0$A torsion data after the first compactification,
because the first compactification was on the wedge circle itself. We can now make quantitative comparison of the two nine-dimensional descriptions. This can be written in a clean form in the following way.
For the 
Type $0$A route:
\begin{equation}
W_i^{(0A,9d)}
=
W_i^{\rm fib}
+
W_i^{\rm flux+pinch} ,
\label{box_0A_9d_torsion}
\end{equation}
which we already had in \eqref{Wi_total_final_combined}, but now we interpret this a nine-dimensional result because in the 0A side the reduction over the orientifold only projects out certain field components. This just implies that $\mathcal H_3^{(s)}$ and $G_4^{(27, s)}$ appearing there are the components that survive the orientifold projections. 
For the type-$0$HW route, we have:
\begin{equation}
W_i^{(0{\rm HW},9d)}
=
W_i^{(0{\rm HW},10d)}
+
\Delta_{\rm wedge}W_i^{(0{\rm HW}\to 9d)}
=
w_i^{\rm fib}
+
w_i^{\rm flux}
+
\Delta_{\rm wedge}W_i^{(0{\rm HW}\to 9d)}, 
\label{box_0HW_9d_torsion}
\end{equation}
A detailed comparison of the five torsion classes between Type 0A and Type 0HW is presented in {\bf Table \ref{torucomp99}}.
This table should be read structurally, not as a term-by-term equality.  The
two columns use different reduction frames.  The symbols
${\bf W}_i^{(o)}$ and 
$\mathbf w_i^{(o)}$
both denote branch-odd torsion data in the relevant $SU(3)$ representation,
but they need not be literally equal.  Similarly,
$X_\perp^{(o)}$
and 
$\mathbf x_\perp^{(o)}$
are both branch-odd one-form data contributing to the Lee-form channels
$W_4$ and $W_5$, but they are related only after specifying the
nine-dimensional field/frame map.

\begin{table}[t]
\centering
\resizebox{\textwidth}{!}{%
\renewcommand{\arraystretch}{2.4}
\begin{tabular}{|c|c|c|}
\hline
\textbf{Torsion class}
&
\textbf{Route A: ${\rm M}/(S^1_+\vee S^1_-)\to 0A$}
&
\textbf{Route HW: ${\rm M}/(S^1_a/\mathbb Z_2)\to 0{\rm HW}\to 9d$}
\\
\hline
$W_1$
&
$
\begin{aligned}
&
\Pi_{(3,0)+(0,3)}\!\left(\mathcal H_3^{(e)}\right)
+
T{\bf W}_1^{(o)}
+
\iota_{V_T}{\bf Y}_1^{(o)}
\end{aligned}
$
&
$
\begin{aligned}
&
w_1^{\rm fib}
+
w_1^{\rm flux}
+
T\mathbf w_1^{(o)}
+
\iota_{V_T}\mathbf y_1^{(o)}
\end{aligned}
$
\\
\hline
$W_2$
&
$
\begin{aligned}
&
{\rm L}_J^{-1}
\!\left[
\Pi_{\rm prim}^{(2,2)}
\left((d\Omega)_{\rm flux,e}\right)
\right]
+
\widetilde a_2
\Pi_{\rm prim}^{(1,1)}
\!\left(\iota_{v_e}*_7G_4^{(27,e)}\right)
\\
&\quad
+
\widetilde a_2^{(o)}
\Pi_{\rm prim}^{(1,1)}
\!\left(\iota_{v_o}*_7G_4^{(27,o)}\right)
+
T{\bf W}_2^{(o)}
+
dT\wedge{\bf Y}_2^{(o)}
\end{aligned}
$
&
$
\begin{aligned}
&
w_2^{\rm fib}
+
w_2^{\rm flux}
+
\kappa_2T\delta_{\Sigma_5^{\rm jct}}\xi_2
\\
&\quad
+
T\mathbf w_2^{(o)}
+
dT\wedge\mathbf y_2^{(o)}
\end{aligned}
$
\\
\hline
$W_3$
&
$
\begin{aligned}
&
\Pi_{\rm prim}^{(2,1)+(1,2)}
\!\left(\mathcal H_3^{(e)}\right)
+
\Pi_{\rm prim}^{(2,1)+(1,2)}
\!\left(\mathcal H_3^{(o)}\right)
\\
&\quad
+
T{\bf W}_3^{(o)}
+
dT\wedge{\bf Y}_3^{(o)}
\end{aligned}
$
&
$
\begin{aligned}
&
w_3^{\rm fib}
+
w_3^{\rm flux}
+
\kappa_3T\delta_{\Sigma_5^{\rm jct}}\xi_3
\\
&\quad
+
T\mathbf w_3^{(o)}
+
dT\wedge\mathbf y_3^{(o)}
\end{aligned}
$
\\
\hline
$W_4$
&
$
\begin{aligned}
&
a_4X_\perp^{(e)}
+
a_4^{(o)}X_\perp^{(o)}
+
c_4^{(o)}dT
+
T{\cal W}_4^{(o)}
\\
&\quad
+
b_4d\varepsilon
+
\varepsilon W_4^{(\varepsilon,{\rm pinch})}
\end{aligned}
$
&
$
\begin{aligned}
&
w_4^{\rm fib}
+
w_4^{\rm flux}
+
\hat{a}_4^{(o)}\mathbf x_\perp^{(o)}
+
\hat{c}_4^{(o)}dT
\\
&\quad
+
T\mathbf w_4^{(o)}
+
\hat{b}_4d\varepsilon
+
\varepsilon w_4^{(\varepsilon)}
\end{aligned}
$
\\
\hline
$W_5$
&
$
\begin{aligned}
&
a_5X_\perp^{(e)}
+
a_5^{(o)}X_\perp^{(o)}
+
c_5^{(o)}dT
+
T{\cal W}_5^{(o)}
\\
&\quad
+
b_5d\varepsilon
+
\varepsilon W_5^{(\varepsilon,{\rm pinch})}
\end{aligned}
$
&
$
\begin{aligned}
&
w_5^{\rm fib}
+
w_5^{\rm flux}
+
\hat{a}_5^{(o)}\mathbf x_\perp^{(o)}
+
\hat{c}_5^{(o)}dT
\\
&\quad
+
T\mathbf w_5^{(o)}
+
\hat{b}_5d\varepsilon
+
\varepsilon w_5^{(\varepsilon)}
\end{aligned}
$
\\
\hline
\end{tabular}%
}
\caption{Comparison of the five $SU(3)$-structure torsion classes in the two
nine-dimensional reduction routes.  In Route A, M-theory is first compactified
on the wedge circle $S^1_+\vee S^1_-$, so the type-$0$A torsion classes already
contain the flux-plus-pinch contributions before the subsequent
$S^1_a/\mathbb Z_2$ reduction.  In Route HW, M-theory is first compactified on
the Ho\v{r}ava--Witten interval $S^1_a/\mathbb Z_2$; the wedge-dependent terms
enter only after the further compactification of the type-$0$HW branch on
$S^1_+\vee S^1_-$.  The table is therefore a structural comparison of
representation channels, not a term-by-term equality of the two frames.}
\label{torucomp99}
\end{table}

\subsubsection{Refined comparison of the type 0A and type $0{\rm HW}$ torsion sectors \label{comparisoon}}

We now refine the comparison between the Type $0A$ and Type $0{\rm HW}$
torsion sectors by separating smooth branch-odd corrections from localized
distributional descendants of the parent $G_2$ pinch source.  This separation
is useful because some of the compact notation used earlier hides localized
terms inside the total branch-odd coefficients.

The two reduction routes are in \eqref{2routes}.
In Route A, the wedge reduction is performed first, so the pinch data are
already encoded in the type-$0A$ torsion classes.  In Route HW, the interval
reduction is performed first, so the bare ten-dimensional type-$0{\rm HW}$
torsion classes are simpler, and the wedge-induced terms appear only after the
second compactification. The parent localized pinch source is:
\begin{equation}
\tau_3^{\rm pinch}
=
T\,\delta_{\Sigma_6^{\rm jct}}\,\Xi_{27}
+
O(T^2).
\label{tau3_pinch_refined}
\end{equation}
Although this is a $\mathbf{27}$ of $G_2$, after reducing to the
six-dimensional $SU(3)$-structure manifold one has the schematic branching of ${\bf 27}$ as in \eqref{27decomp}.
The relevant pieces for the torsion classes are:
\begin{equation}
{\bf 8}
\longrightarrow
W_2,
\qquad
{\bf 6}\oplus\overline{\bf 6}
\longrightarrow
W_3 .
\label{G2_27_to_W2_W3_refined}
\end{equation}
Therefore the localized $\tau_3^{\rm pinch}$ data may have direct descendants
in both the $W_2$ and $W_3$ torsion classes. To see this we need to revisit the Type 0A torsion classes \eqref{W_flux_pinch_combined_final}
and express the smooth versus localized pieces in the type 0A notation.
Concretely this means the following. Previously
the odd flux and pinch terms in the type-$0A$ $W_2$ channel were
combined as \eqref{mapW2b_combined}. 
The refined point is that the pinch coefficient itself may have a smooth part
and a localized distributional part defined in the following way:
\begin{equation}
W_2^{(T,{\rm pinch})}
=
W_{2,{\rm smooth}}^{(T,{\rm pinch})}
+
\kappa_2^{(0A)}
\delta_{\Sigma_5^{\rm jct}}
\zeta_2^{(0A)} ,
\label{W2_pinch_split_refined}
\end{equation}
which basically splits contribution in two parts. But this more than just a split, because it tells us that even after we remove the localized (here delta-function) contribution from the pinch, the pinch continues to contribute non-zero smooth contribution to the $W_2$ torsion class.
Thus the total branch-odd coefficient may be decomposed as:
\begin{equation}
{\bf W}_2^{(o)}
=
{\bf W}_{2,{\rm smooth}}^{(o)}
+
\kappa_2^{(0A)}
\delta_{\Sigma_5^{\rm jct}}
\zeta_2^{(0A)} = {\cal W}_2^{(o)}
+
W_{2,{\rm smooth}}^{(T,{\rm pinch})} + \kappa_2^{(0A)}
\delta_{\Sigma_5^{\rm jct}}
\zeta_2^{(0A)}\nonumber
\label{W2_bold_split_refined}
\end{equation}
\begin{equation}
T{\bf W}_2^{(o)}
=
T{\bf W}_{2,{\rm smooth}}^{(o)}
+
\kappa_2^{(0A)}
T\delta_{\Sigma_5^{\rm jct}}
\zeta_2^{(0A)} .
\label{T_W2_bold_split_refined}
\end{equation}
This is precisely why $T{\bf W}_2^{(o)}$ contains more information:    one may either write the localized term explicitly or
hide it inside $T{\bf W}_2^{(o)}$.
Similarly, for $W_3$ we previously wrote \eqref{mapW3b_combined}. Following the above strategy, we can define a refined splitting of the following form:
\begin{equation}
{\bf W}_3^{(o)}
=
{\bf W}_{3,{\rm smooth}}^{(o)}
+
\kappa_3^{(0A)}
\delta_{\Sigma_5^{\rm jct}}
\zeta_3^{(0A)} = {\cal W}_3^{(o)}
+
W_{3,{\rm smooth}}^{(T,{\rm pinch})} + \kappa_3^{(0A)}
\delta_{\Sigma_5^{\rm jct}}
\zeta_3^{(0A)}\nonumber
\label{W2_bold_split_refined}
\end{equation}
\begin{equation}
T{\bf W}_3^{(o)}
=
T{\bf W}_{3,{\rm smooth}}^{(o)}
+
\kappa_3^{(0A)}
T\delta_{\Sigma_5^{\rm jct}}
\zeta_3^{(0A)} .
\label{T_W3_bold_split_refined}
\end{equation}
The notation $\zeta_i^{(0A)}$ is used here for the localized Type 0A
representative.  It can be identified with $\xi_i^{(0A)}$ if one wants a single
symbol, {\it i.e.}
$\zeta_i^{(0A)}
\equiv
\xi_i^{(0A)}$, but this is not necessary. Now with all the aforementioned refinements at hand, we can rewrite the Type 0A torsion classes by displaying the localized terms more explicitly.  
The Type 0A flux-plus-pinch
torsion classes then take the following form:
\begin{equation}
\begin{aligned}
W_1^{\rm flux+pinch}
&=
\Pi_{(3,0)+(0,3)}
\big(\mathcal H_3^{(e)}\big)
+
T\,{\bf W}_1^{(o)}
+
\iota_{V_T}{\bf Y}_1^{(o)}
+
O(T^2),
\\[4pt]
W_2^{\rm flux+pinch}
&=
{\rm L}_J^{-1}
\left[
\Pi_{\rm prim}^{(2,2)}
\big(d\Omega\big)_{\rm flux,e}
\right]
+
\widetilde a_2
\Pi_{\rm prim}^{(1,1)}
\big(
\iota_{v_e} *_7G_4^{(27,e)}
\big)
\\
&\quad
+
\widetilde a_2^{(o)}
\Pi_{\rm prim}^{(1,1)}
\big(
\iota_{v_o} *_7G_4^{(27,o)}
\big)
+
\kappa_2^{(0A)}
T\,\delta_{\Sigma_5^{\rm jct}}\zeta_2^{(0A)}
\\
&\quad
+
T\,{\bf W}_{2,{\rm smooth}}^{(o)}
+
dT\wedge{\bf Y}_2^{(o)}
+
O(T^2),
\\[4pt]
W_3^{\rm flux+pinch}
&=
\Pi_{\rm prim}^{(2,1)+(1,2)}
\big(\mathcal H_3^{(e)}\big)
+
\Pi_{\rm prim}^{(2,1)+(1,2)}
\big(\mathcal H_3^{(o)}\big)
\\
&\quad
+
\kappa_3^{(0A)}
T\,\delta_{\Sigma_5^{\rm jct}}\zeta_3^{(0A)}
+
T\,{\bf W}_{3,{\rm smooth}}^{(o)}
+
dT\wedge{\bf Y}_3^{(o)}
+
O(T^2),
\\[4pt]
W_4^{\rm flux+pinch}
&=
a_4X_\perp^{(e)}
+
a_4^{(o)}X_\perp^{(o)}
+
c_4^{(o)}dT
+
T{\cal W}_4^{(o)}
+
b_4d\varepsilon
+
\varepsilon W_4^{(\varepsilon,{\rm pinch})}
+
O(T^2,\varepsilon^2),
\\[4pt]
W_5^{\rm flux+pinch}
&=
a_5X_\perp^{(e)}
+
a_5^{(o)}X_\perp^{(o)}
+
c_5^{(o)}dT
+
T{\cal W}_5^{(o)}
+
b_5d\varepsilon
+
\varepsilon W_5^{(\varepsilon,{\rm pinch})}
+
O(T^2,\varepsilon^2) ,
\end{aligned}
\label{W0A_flux_pinch_refined}
\end{equation}
which may be compared to \eqref{W_flux_pinch_combined_final}. They are both identical and one can return to the compact notation by absorbing the localized
terms into ${\bf W}_2^{(o)}$ and ${\bf W}_3^{(o)}$ using \eqref{T_W3_bold_split_refined} and 
\eqref{T_W3_bold_split_refined} respectively. In a similar vein one can study the refined Type $0{\rm HW}$ torsion classes. 
The Type $0{\rm HW}$ route first gives the ten-dimensional interval-reduced
torsion classes as in \eqref{W45_0HW_10d_clean}.
After the further compactification on $S^1_+\vee S^1_-$, the nine-dimensional
torsion classes are:
\begin{equation}
\begin{aligned}
W_1^{(0{\rm HW},9d)}
&=
w_1^{\rm fib}
+
w_1^{\rm flux}
+
T\,\mathbf w_1^{(o)}
+
\iota_{V_T}\mathbf y_1^{(o)}
+
O(T^2),
\\[4pt]
W_2^{(0{\rm HW},9d)}
&=
w_2^{\rm fib}
+
w_2^{\rm flux}
+
\kappa_2^{(0{\rm HW})}
T\,\delta_{\Sigma_5^{\rm jct}}\xi_2^{(0{\rm HW})}
+
T\,\mathbf w_{2,{\rm smooth}}^{(o)}
+
dT\wedge\mathbf y_2^{(o)}
+
O(T^2),
\\[4pt]
W_3^{(0{\rm HW},9d)}
&=
w_3^{\rm fib}
+
w_3^{\rm flux}
+
\kappa_3^{(0{\rm HW})}
T\,\delta_{\Sigma_5^{\rm jct}}\xi_3^{(0{\rm HW})}
+
T\,\mathbf w_{3,{\rm smooth}}^{(o)}
+
dT\wedge\mathbf y_3^{(o)}
+
O(T^2),
\\[4pt]
W_4^{(0{\rm HW},9d)}
&=
w_4^{\rm fib}
+
w_4^{\rm flux}
+
\hat{a}_4^{(o)}\,\mathbf x_\perp^{(o)}
+
\hat{c}_4^{(o)}\,dT
+
T\,\mathbf w_4^{(o)}
+
\hat{b}_4\,d\varepsilon
+
\varepsilon\,w_4^{(\varepsilon)}
+
O(T^2,\varepsilon^2),
\\[4pt]
W_5^{(0{\rm HW},9d)}
&=
w_5^{\rm fib}
+
w_5^{\rm flux}
+
\hat{a}_5^{(o)}\,\mathbf x_\perp^{(o)}
+
\hat{c}_5^{(o)}\,dT
+
T\,\mathbf w_5^{(o)}
+
\hat{b}_5\,d\varepsilon
+
\varepsilon\,w_5^{(\varepsilon)}
+
O(T^2,\varepsilon^2) ,
\end{aligned}
\label{W0HW_9d_refined}
\end{equation}
where the only difference from \eqref{W0HW_9d_full_unified} is notational: ${\bf w}_i^{(o)}$ is replaced by ${\bf w}^{(o)}_{i, {\rm smooth}}$ and $\xi_i$ is replaced by $\xi_i^{({\rm 0HW})}$. No other structural refinements have been implemented.
If desired, the localized $W_2$ and $W_3$ terms can again be absorbed into
total boldface coefficients:
\begin{equation}\label{patcline}
\widetilde{\mathbf w}_2^{(o)}
\equiv 
\mathbf w_{2,{\rm smooth}}^{(o)}
+
\kappa_2^{(0{\rm HW})}
\delta_{\Sigma_5^{\rm jct}}\xi_2^{(0{\rm HW})}, ~~~~~
\widetilde{\mathbf w}_3^{(o)}
=
\mathbf w_{3,{\rm smooth}}^{(o)}
+
\kappa_3^{(0{\rm HW})}
\delta_{\Sigma_5^{\rm jct}}\xi_3^{(0{\rm HW})}.
\end{equation}
Therefore \eqref{W0A_flux_pinch_refined} and \eqref{W0HW_9d_refined} are our final answers for the Type 0A and Type 0HW torsion classes in nine-dimensions respectively.  We can now compare all the five torsion class channels in the following way.

\paragraph{Comparison of $W_1$ channel.}

The torsion class $W_1$ is a complex scalar
$W_1\in \Lambda^0\otimes\mathbb C$.
It is extracted from the singlet component of $dJ$ {\it i.e.}
$W_1
=
-\frac{i}{12}\,\overline\Omega\lrcorner dJ$.
Since the direct localized $G_2$ pinch source sits in the $\mathbf{27}$ and
its relevant direct $SU(3)$ descendants lie in the $\mathbf{8}$ and
$\mathbf{6}\oplus\overline{\mathbf 6}$ channels, it does not directly generate
a scalar singlet contribution to $W_1$.  Therefore the branch-odd pinch
dependence of $W_1$ is induced rather than direct. In the type-$0A$ route one may write:
\begin{equation}
W_1^{(0A)}
=
W_1^{\rm fib}
+
\Pi_{(3,0)+(0,3)}
\big(\mathcal H_3^{(e)}\big)
+
T\,{\bf W}_1^{(o)}
+
\iota_{V_T}{\bf Y}_1^{(o)}
+
O(T^2) ,
\end{equation}
where the first non-fibration term is the branch-even flux contribution, while
the last two terms are induced branch-odd corrections.  These terms include the
smooth response of $J(T)$ and $\Omega(T)$ to the wedge asymmetry.
In the type-$0{\rm HW}$ route, the corresponding nine-dimensional expression is:
\begin{equation}
W_1^{(0{\rm HW},9d)}
=
w_1^{\rm fib}
+
w_1^{\rm flux}
+
T\,\mathbf w_1^{(o)}
+
\iota_{V_T}\mathbf y_1^{(o)}
+
O(T^2) ,
\end{equation}
where the first two terms are the interval-reduced ten-dimensional type-$0{\rm HW}$
contribution; and the last two terms arise only after compactifying the
type-$0{\rm HW}$ branch on $S^1_+\vee S^1_-$.  Thus the structural match is:
\begin{equation}
\Pi_{(3,0)+(0,3)}
\big(\mathcal H_3^{(e)}\big)
\quad
\leftrightarrow
\quad
w_1^{\rm flux}\nonumber
\end{equation}
\begin{equation}
T\,{\bf W}_1^{(o)}
+
\iota_{V_T}{\bf Y}_1^{(o)}
\quad
\leftrightarrow
\quad
T\,\mathbf w_1^{(o)}
+
\iota_{V_T}\mathbf y_1^{(o)}.
\end{equation}
This is a representation-level comparison of scalar torsion data, not a
literal equality of coefficients. For more detail see {\bf Table \ref{W1tabletop}}.

\begin{table}[t]
\centering
\resizebox{\textwidth}{!}{%
\renewcommand{\arraystretch}{1.7}
\begin{tabular}{|c|c|c|}
\hline
\textbf{$W_1$~Data}
&
\textbf{0A}
&
\textbf{0HW}
\\
\hline
singlet flux
&
$
\Pi_{(3,0)+(0,3)}
\left(
{\cal H}_3^{(e)}
\right)
$
&
$
w_1^{\rm flux}
$
\\
\hline
direct localized $\tau_3$ descendant
&
$
0
$
&
$
0
$
\\
\hline
induced branch-odd scalar
&
$
T{\bf W}_1^{(o)}
$
&
$
T\mathbf w_1^{(o)}
$
\\
\hline
induced gradient/contraction scalar
&
$
\iota_{V_T}{\bf Y}_1^{(o)}
$
&
$
\iota_{V_T}\mathbf y_1^{(o)}
$
\\
\hline
\end{tabular}%
}
\caption{Comparison of the $W_1$ torsion-channel data in the type-$0A$ and
type-$0{\rm HW}$ nine-dimensional reduction routes.  The direct localized
descendant of the parent $G_2$ $\tau_3^{\rm pinch}$ source does not contribute
to the scalar singlet channel.}
\label{W1tabletop}
\end{table}

\paragraph{Comparison of $W_2$ channel.}

The torsion class $W_2$ is a primitive $(1,1)$ form:
$W_2\in \Lambda_{\rm prim}^{(1,1)}$.
It is extracted from $d\Omega$ through the following standard procedure:
\begin{equation}
J\wedge W_2
=
\Pi_{\rm prim}^{(2,2)}(d\Omega),
\qquad
W_2
=
L_J^{-1}
\left[
\Pi_{\rm prim}^{(2,2)}(d\Omega)
\right],
\qquad
L_J(\alpha)=J\wedge\alpha.
\end{equation}
Since $W_2$ transforms in the $\mathbf{8}$ of $SU(3)$, it can receive a direct
localized descendant from the $\mathbf{8}$ component in the branching of the
parent $G_2$ $\mathbf{27}$, {\it i.e.}
${\bf 27}_{G_2}
\supset
{\bf 8}_{SU(3)}$ implies that 
$W_2^{\rm pinch,direct}
\neq 0$.
We denote this direct localized contribution by
$W_2^{\rm pinch,direct}
=
\kappa_2\,T\,\delta_{\Sigma_5^{\rm jct}}\,\xi_2$,
with 
$\xi_2\in\Lambda_{\rm prim}^{(1,1)}$.
In the type-$0A$ route, a refined form of the $W_2$ torsion class is given by:
\begin{equation}
\begin{aligned}
W_2^{(0A)}
&=
W_2^{\rm fib}
+
{\rm L}_J^{-1}
\left[
\Pi_{\rm prim}^{(2,2)}
\big(d\Omega\big)_{\rm flux,e}
\right]
+
\widetilde a_2\,
\Pi_{\rm prim}^{(1,1)}
\big(
\iota_{v_e}*_7G_4^{(27,e)}
\big)
\\
&\quad
+
\widetilde a_2^{(o)}
\Pi_{\rm prim}^{(1,1)}
\big(
\iota_{v_o}*_7G_4^{(27,o)}
\big)
+
\kappa_2^{(0A)}
T\,\delta_{\Sigma_5^{\rm jct}}\zeta_2^{(0A)}
\\
&\quad
+
T\,{\bf W}_{2,{\rm smooth}}^{(o)}
+
dT\wedge{\bf Y}_2^{(o)}
+
O(T^2).
\end{aligned}
\end{equation}
The localized piece can also be hidden inside the total branch-odd coefficient.
Indeed, if one writes \eqref{mapW2b_combined}, 
then the refined decomposition is given by \eqref{W2_pinch_split_refined}
so that we get \eqref{T_W2_bold_split_refined}, implying that 
the explicit and compact notations are equivalent.
In the type-$0{\rm HW}$ route, the corresponding nine-dimensional expression is:
\begin{equation}
\begin{aligned}
W_2^{(0{\rm HW},9d)}
&=
w_2^{\rm fib}
+
w_2^{\rm flux}
+
\kappa_2^{(0{\rm HW})}
T\,\delta_{\Sigma_5^{\rm jct}}\xi_2^{(0{\rm HW})}
\\
&\quad
+
T\,\mathbf w_{2,{\rm smooth}}^{(o)}
+
dT\wedge\mathbf y_2^{(o)}
+
O(T^2).
\end{aligned}
\end{equation}
Equivalently, one may absorb the localized term into the total boldface
coefficient by defining the tilde variables as in \eqref{patcline}, although this is not absolutely essential for us here.
Thus the $W_2$ comparison may be summarized as:
\begin{equation}
\begin{aligned}
&{\rm Type}\;0A:
\quad
\left[
{\rm flux\;primitive}\;(2,2)
\right]
+
\left[
G_4^{(27)}\to{\bf 8}
\right]
+
\left[
{\rm localized}\;{\bf 8}\;{\rm pinch}
\right]
+
\left[
{\rm smooth\;odd}
\right]
\\
&{\rm Type}\;0{\rm HW}:
\quad
\left[
w_2^{\rm flux}
\right]
+
\left[
{\rm localized}\;{\bf 8}\;{\rm pinch}
\right]
+
\left[
{\rm smooth\;odd}
\right] \nonumber
\end{aligned}
\end{equation}
The two descriptions agree at the level of $SU(3)$ representation content, even
though the coefficients and representatives need not coincide. More details may be inferred from {\bf Table \ref{W2tabletop}}.

\begin{table}[t]
\centering
\renewcommand{\arraystretch}{1.8}
\begin{tabular}{|p{0.18\textwidth}|p{0.38\textwidth}|p{0.38\textwidth}|}
\hline
\textbf{$W_2$~Data}
&
\textbf{0A}
&
\textbf{0HW}
\\
\hline
primitive $(2,2)$ flux
&
$
L_J^{-1}
\Pi_{\rm prim}^{(2,2)}
\left(
(d\Omega)_{\rm flux,e}
\right)
$
&
$
w_2^{\rm flux}
$
\\
\hline
$G_4^{(27)}$ primitive $(1,1)$ channel
&
$
\begin{aligned}
&
\widetilde a_2
\Pi_{\rm prim}^{(1,1)}
\left(
\iota_{v_e}*_7G_4^{(27,e)}
\right)
\\
&\quad
+
\widetilde a_2^{(o)}
\Pi_{\rm prim}^{(1,1)}
\left(
\iota_{v_o}*_7G_4^{(27,o)}
\right)
\end{aligned}
$
&
$
\text{included, if present, inside }w_2^{\rm flux}
$
\\
\hline
direct localized $\tau_3$ descendant from ${\bf 8}_{SU(3)}$
&
$
\kappa_2^{(0A)}
T\delta_{\Sigma_5^{\rm jct}}\zeta_2^{(0A)}
$
&
$
\kappa_2^{(0{\rm HW})}
T\delta_{\Sigma_5^{\rm jct}}\xi_2^{(0{\rm HW})}
$
\\
\hline
smooth part of the total odd coefficient
&
$
T{\bf W}_{2,{\rm smooth}}^{(o)}=~T\left({\cal W}_2^{(o)}+W_{2,{\rm smooth}}^{(T,{\rm pinch})}
\right)
$
&
$
T\mathbf w_{2,{\rm smooth}}^{(o)}
$
\\
\hline
gradient-induced smooth correction
&
$
dT\wedge{\bf Y}_2^{(o)}
$
&
$
dT\wedge\mathbf y_2^{(o)}
$
\\
\hline
compact notation
&
$
T{\bf W}_2^{(o)}
=
T{\bf W}_{2,{\rm smooth}}^{(o)}
+
\kappa_2^{(0A)}
T\delta_{\Sigma_5^{\rm jct}}\zeta_2^{(0A)}
$
&
$
T\mathbf w_2^{(o)}
=~
T\mathbf w_{2,{\rm smooth}}^{(o)}
+
\kappa_2^{(0{\rm HW})}
T\delta_{\Sigma_5^{\rm jct}}\xi_2^{(0{\rm HW})}
$
\\
\hline
\end{tabular}
\caption{Comparison of the $W_2$ torsion-channel data in the type-$0A$ and
type-$0{\rm HW}$ nine-dimensional reduction routes.  The $G_2$ $\mathbf{27}$
pinch source contains an $SU(3)$ $\mathbf 8$ component, which can contribute
directly to $W_2$.  In the compact notation this localized contribution is
absorbed into the total branch-odd coefficient ${\bf W}_2^{(o)}$ or
$\mathbf w_2^{(o)}$; in the refined notation it is displayed separately.}
\label{W2tabletop}
\end{table}

\paragraph{Comparison of $W_3$ channel.}

The torsion class $W_3$ is a primitive three-form:
$W_3\in \Lambda_{\rm prim}^{(2,1)+(1,2)}$.
It is extracted from $dJ$ as
$W_3
=
\Pi_{\rm prim}^{(2,1)+(1,2)}(dJ)$.
Since the parent $G_2$ $\mathbf{27}$ contains the
$\mathbf{6}\oplus\overline{\mathbf 6}$ of $SU(3)$, the localized
$\tau_3^{\rm pinch}$ source also has a direct descendant in $W_3$. In other words, since 
${\bf 27}_{G_2}
\supset
{\bf 6}\oplus\overline{\bf 6}$, this implies that 
$W_3^{\rm pinch,direct}
\neq 0$.
We can write this contribution as
$W_3^{\rm pinch,direct}
=
\kappa_3\,T\,\delta_{\Sigma_5^{\rm jct}}\,\xi_3$, with 
$\xi_3\in \Lambda_{\rm prim}^{(2,1)+(1,2)}$.
In the Type 0A route, the refined $W_3$ expression is:
\begin{equation}
\begin{aligned}
W_3^{(0A)}
&=
W_3^{\rm fib}
+
\Pi_{\rm prim}^{(2,1)+(1,2)}
\big(\mathcal H_3^{(e)}\big)
+
\Pi_{\rm prim}^{(2,1)+(1,2)}
\big(\mathcal H_3^{(o)}\big)
\\
&\quad
+
\kappa_3^{(0A)}
T\,\delta_{\Sigma_5^{\rm jct}}\zeta_3^{(0A)}
+
T\,{\bf W}_{3,{\rm smooth}}^{(o)}
+
dT\wedge{\bf Y}_3^{(o)}
+
O(T^2).
\end{aligned}
\end{equation}
Again, the localized piece may be absorbed into the total branch-odd
coefficient.  If \eqref{mapW3b_combined} holds, then one can easily show that \eqref{T_W3_bold_split_refined} also holds, implying equivalency between the two descriptions.
In the type-$0{\rm HW}$ route, the corresponding nine-dimensional expression is:
\begin{equation}
\begin{aligned}
W_3^{(0{\rm HW},9d)}
&=
w_3^{\rm fib}
+
w_3^{\rm flux}
+
\kappa_3^{(0{\rm HW})}
T\,\delta_{\Sigma_5^{\rm jct}}\xi_3^{(0{\rm HW})}
\\
&\quad
+
T\,\mathbf w_{3,{\rm smooth}}^{(o)}
+
dT\wedge\mathbf y_3^{(o)}
+
O(T^2) ,
\end{aligned}
\end{equation}
where as before one can go to the tilde variables as in \eqref{patcline} to cloak the singular piece in a function. Doing this obscures the map between the torsion classes, so we will avoid this.
From here the $W_3$ comparison between the two pictures is:
\begin{equation}
\begin{aligned}
&{\rm Type}\;0A:
\quad
\left[
{\rm primitive\;three\mbox{-}form\;flux}
\right]
+
\left[
{\rm localized}\;({\bf 6}\oplus\overline{\bf 6})\;{\rm pinch}
\right]
+
\left[
{\rm smooth\;odd}
\right]
\\
&{\rm Type}\;0{\rm HW}:
\quad
\left[
w_3^{\rm flux}
\right]
+
\left[
{\rm localized}\;({\bf 6}\oplus\overline{\bf 6})\;{\rm pinch}
\right]
+
\left[
{\rm smooth\;odd}
\right] \nonumber
\end{aligned}
\end{equation}
The direct localized contribution to $W_3$ is therefore not optional: it is the
natural descendant of the $\mathbf{6}\oplus\overline{\mathbf 6}$ part of the
parent $G_2$ $\mathbf{27}$. More details can be found in {\bf Table \ref{W3tabletop}}.
\begin{table}[t]
\centering
\renewcommand{\arraystretch}{1.8}
\begin{tabular} {|p{0.18\textwidth}|p{0.38\textwidth}|p{0.38\textwidth}|}
\hline
\textbf{$W_3$~Data}
&
\textbf{0A}
&
\textbf{0HW}
\\
\hline
primitive three-form flux
&
$
\begin{aligned}
&
\Pi_{\rm prim}^{(2,1)+(1,2)}
\left(
{\cal H}_3^{(e)}
\right)
\\
&\quad
+
\Pi_{\rm prim}^{(2,1)+(1,2)}
\left(
{\cal H}_3^{(o)}
\right)
\end{aligned}
$
&
$
w_3^{\rm flux}
$
\\
\hline
direct localized $\tau_3$ descendant from
${\bf 6}\oplus\overline{\bf 6}$
&
$
\kappa_3^{(0A)}
T\delta_{\Sigma_5^{\rm jct}}\zeta_3^{(0A)}
$
&
$
\kappa_3^{(0{\rm HW})}
T\delta_{\Sigma_5^{\rm jct}}\xi_3^{(0{\rm HW})}
$
\\
\hline
smooth part of the total odd coefficient
&
$
T{\bf W}_{3,{\rm smooth}}^{(o)}
=~
T\left(
{\cal W}_3^{(o)}
+
W_{3,{\rm smooth}}^{(T,{\rm pinch})}
\right)
$
&
$
T\mathbf w_{3,{\rm smooth}}^{(o)}
$
\\
\hline
gradient-induced smooth correction
&
$
dT\wedge{\bf Y}_3^{(o)}
$
&
$
dT\wedge\mathbf y_3^{(o)}
$
\\
\hline
compact notation
&
$
T{\bf W}_3^{(o)}
=
T{\bf W}_{3,{\rm smooth}}^{(o)}
+
\kappa_3^{(0A)}
T\delta_{\Sigma_5^{\rm jct}}\zeta_3^{(0A)}
$
&
$
T\mathbf w_3^{(o)}
=
T\mathbf w_{3,{\rm smooth}}^{(o)}
+
\kappa_3^{(0{\rm HW})}
T\delta_{\Sigma_5^{\rm jct}}\xi_3^{(0{\rm HW})}
$
\\
\hline
\end{tabular}
\caption{Comparison of the $W_3$ torsion-channel data in the type-$0A$ and
type-$0{\rm HW}$ nine-dimensional reduction routes.  The $G_2$ $\mathbf{27}$
pinch source contains an $SU(3)$ $\mathbf 6\oplus\overline{\mathbf 6}$
component, which contributes directly to the primitive
$(2,1)+(1,2)$ torsion class $W_3$.  In the compact notation this localized
contribution is absorbed into the total branch-odd coefficient.}
\label{W3tabletop}
\end{table}

\paragraph{Comparison of $W_4$ and $W_5$ channels.}

The torsion classes $W_4$ and $W_5$ are one-forms:
$W_4,W_5\in \Lambda^1$,
and they are extracted from
$W_4
=
\frac12 J\lrcorner dJ$,
and
$W_5
=
\frac12{\rm Re}
\left(
\overline\Omega^{-1}\lrcorner d\Omega
\right)$.
The localized $\mathbf{8}$ and
$\mathbf{6}\oplus\overline{\mathbf 6}$ descendants of the parent
$\tau_3^{\rm pinch}$ source are assigned to $W_2$ and $W_3$, respectively.
The Lee-form torsion classes instead receive smooth branch-odd and
resolved-node corrections.
In the Type 0A route, one may write:

{\footnotesize
\begin{equation}
\begin{aligned}
W_4^{(0A)}
&=
W_4^{\rm fib}
+
a_4X_\perp^{(e)}
+
a_4^{(o)}X_\perp^{(o)}
+
c_4^{(o)}dT
+
T{\cal W}_4^{(o)}
+
b_4d\varepsilon
+
\varepsilon W_4^{(\varepsilon,{\rm pinch})}
+
O(T^2,\varepsilon^2),
\\[4pt]
W_5^{(0A)}
&=
W_5^{\rm fib}
+
a_5X_\perp^{(e)}
+
a_5^{(o)}X_\perp^{(o)}
+
c_5^{(o)}dT
+
T{\cal W}_5^{(o)}
+
b_5d\varepsilon
+
\varepsilon W_5^{(\varepsilon,{\rm pinch})}
+
O(T^2,\varepsilon^2),
\end{aligned}
\end{equation}}
where $X_\perp^{(e)}$ and $X_\perp^{(o)}$ are even and odd Lee-form-type flux
data, while $dT$, $d\varepsilon$, and the algebraic terms proportional to $T$
and $\varepsilon$ encode the smooth branch-odd and resolved-node corrections. In the Type $0{\rm HW}$ route, the corresponding expressions may be laid out in the following way:

{\footnotesize
\begin{equation}
\begin{aligned}
W_4^{(0{\rm HW},9d)}
&=
w_4^{\rm fib}
+
w_4^{\rm flux}
+
\hat{a}_4^{(o)}\mathbf x_\perp^{(o)}
+
\hat{c}_4^{(o)}dT
+
T\mathbf w_4^{(o)}
+
\hat{b}_4d\varepsilon
+
\varepsilon w_4^{(\varepsilon)}
+
O(T^2,\varepsilon^2),
\\[4pt]
W_5^{(0{\rm HW},9d)}
&=
w_5^{\rm fib}
+
w_5^{\rm flux}
+
\hat{a}_5^{(o)}\mathbf x_\perp^{(o)}
+
\hat{c}_5^{(o)}dT
+
T\mathbf w_5^{(o)}
+
\hat{b}_5d\varepsilon
+
\varepsilon w_5^{(\varepsilon)}
+
O(T^2,\varepsilon^2) ,
\end{aligned}
\end{equation}}
where the first two terms are the interval-reduced type-$0{\rm HW}$ data, while the
remaining terms are generated by the later compactification on
$S^1_+\vee S^1_-$.  The comparison is therefore:
\begin{equation}
a_{4,5}X_\perp^{(e)}
+
a_{4,5}^{(o)}X_\perp^{(o)}
\quad
\leftrightarrow
\quad
w_{4,5}^{\rm flux}
+
\hat{a}_{4,5}^{(o)}\mathbf x_\perp^{(o)}
\end{equation}
\begin{equation}
c_{4,5}^{(o)}dT
+
T{\cal W}_{4,5}^{(o)}
+
b_{4,5}d\varepsilon
+
\varepsilon W_{4,5}^{(\varepsilon,{\rm pinch})}
\quad
\leftrightarrow
\quad
\hat{c}_{4,5}^{(o)}dT
+
T\mathbf w_{4,5}^{(o)}
+
\hat{b}_{4,5}d\varepsilon
+
\varepsilon w_{4,5}^{(\varepsilon)}.\nonumber
\end{equation}
Again, this is not a term-by-term equality.  It is a matching of the one-form
Lee-channel structure after both routes have been reduced to nine dimensions. For more details see {\bf Table \ref{W45tabletop}}. A purely group theoretic summary of the various contributions to the $SU(3)$ torsion classes may be summarized in the  following way.

\begin{table}[t]
\centering
\resizebox{\textwidth}{!}{%
\renewcommand{\arraystretch}{1.0}
\begin{tabular}{|c|c|c|}
\hline
\textbf{$W_{4, 5}$~Data}
&
\textbf{0A}
&
\textbf{0HW}
\\
\hline
even Lee-form flux
&
$
a_{4,5}X_\perp^{(e)}
$
&
$
w_{4,5}^{\rm flux}
$
\\
\hline
odd Lee-form flux
&
$
a_{4,5}^{(o)}X_\perp^{(o)}
$
&
$
\hat{a}_{4,5}^{(o)}\mathbf x_\perp^{(o)}
$
\\
\hline
direct localized $\tau_3$ descendant
&
$
0
$
&
$
0
$
\\
\hline
branch-gradient one-form
&
$
c_{4,5}^{(o)}dT
$
&
$
\hat{c}_{4,5}^{(o)}dT
$
\\
\hline
algebraic odd correction
&
$
T{\cal W}_{4,5}^{(o)}
$
&
$
T\mathbf w_{4,5}^{(o)}
$
\\
\hline
resolved-node derivative
&
$
b_{4,5}d\varepsilon
$
&
$
\hat{b}_{4,5}d\varepsilon
$
\\
\hline
resolved-node algebraic term
&
$
\varepsilon W_{4,5}^{(\varepsilon,{\rm pinch})}
$
&
$
\varepsilon w_{4,5}^{(\varepsilon)}
$
\\
\hline
\end{tabular}%
}
\caption{Comparison of the Lee-form torsion-channel data in the Type 0A and
Type $0{\rm HW}$ nine-dimensional reduction routes.  The direct localized
descendants of the parent $G_2$ $\mathbf{27}$ pinch source are assigned to
$W_2$ and $W_3$.  The Lee-form classes $W_4$ and $W_5$ receive smooth
branch-odd and resolved-node corrections.}
\label{W45tabletop}
\end{table}

\begin{enumerate}

\item The torsion class $W_1$ is a complex scalar, therefore 
$W_1\in \Lambda^0\otimes\mathbb C$.
The direct localized descendants of the $G_2$ $\mathbf{27}$ pinch source do
not sit in this scalar singlet channel.  Thus the $W_1$ pinch dependence is
only through smooth induced branch-odd corrections.

\item The torsion class $W_2$ is a primitive $(1,1)$ form, implying that 
$W_2\in\Lambda_{\rm prim}^{(1,1)}$.
Equivalently, it lies in the $SU(3)$ $\mathbf 8$.  Since
${\bf 27}_{G_2}\supset {\bf 8}_{SU(3)}$,
the parent $\tau_3^{\rm pinch}$ source can have a direct localized descendant
in $W_2$.

\item The torsion class $W_3$ is a primitive three-form, implying that 
$W_3\in\Lambda_{\rm prim}^{(2,1)+(1,2)}$.
Equivalently, it lies in the
${\bf 6}\oplus\overline{\bf 6}$
of $SU(3)$.  Since
${\bf 27}_{G_2}
\supset
{\bf 6}\oplus\overline{\bf 6}$,
the parent $\tau_3^{\rm pinch}$ source can also have a direct localized
descendant in $W_3$.

\item The torsion classes $W_4$ and $W_5$ are one-forms, and therefore
$W_4,W_5\in\Lambda^1$.
The localized $\mathbf 8$ and
${\bf 6}\oplus\overline{\bf 6}$
descendants of the parent $\tau_3^{\rm pinch}$ source are naturally assigned
to $W_2$ and $W_3$, respectively.  The Lee-form channels receive smooth
branch-odd and resolved-node corrections.

\end{enumerate}

\noindent Therefore we can conclude the following: Although the type-$0$HW branch does not see the lower-dimensional pinch in the
same way as the type-$0$A reduction, it is still sensitive to the same upstairs
branch-odd wedge data. The matching of the singlet $W_1$ channel is simple. In the Lee-form channels $W_4$ and $W_5$, this data
appears through branch-odd one-forms such as $\mathbf{x}_\perp^{(o)}$, $dT$,
and $d\varepsilon$, together with localized algebraic corrections proportional to $T$ for the $W_2$ and $W_3$ torsion classes.
The $W_{4, 5}$ terms have the same representation structure as the
odd/pinch corrections on the type-$0$A side, because both are one-form
corrections to the $SU(3)$ Lee-form torsion classes. In a similar vein the $W_{2, 3}$ torsion classes are correlated. The difference is that,
on the type-$0$A side, they are naturally interpreted as explicit pinch
corrections, whereas on the type-$0$HW side they should be regarded as inherited
branch-odd wedge or resolved-junction data, becoming explicit pinch terms only
after the further compactification on $S^1_+\vee S^1_-$. A final structural comparison between the two theories appears in {\bf Table \ref{W12345}}.

\subsubsection{Type $0{\rm HW}$ gauge theory sector and comparison to Type 0A \label{sec424}}

We now compare the gauge sectors of the Type 0A and Type $0{\rm HW}$
branches in a way that parallels the torsion-class comparison discussed above.
The important point is again the order of reductions that we presented in \eqref{2routes}.
In Route A, the wedge circle is used already in the first reduction, so the
branch-even, branch-odd, and wedge-deformation data are already naturally
present in the Type 0A gauge-sector analysis.  In Route HW, the first
reduction is instead the Ho\v{r}ava--Witten interval reduction.  Therefore the
bare ten-dimensional Type $0{\rm HW}$ gauge sector is the interval-reduced wall
gauge sector.  The wedge-dependent branch resolution and possible wedge-induced
mass terms enter only after the subsequent compactification on
$S^1_+\vee S^1_-$.
Thus the comparison is structurally analogous to the torsion-class comparison: (i) in Type 0A, the 
wedge data enter already in the first reduction, and (ii) in Type 0HW,
wedge data enter only after the second reduction.
This means that the two gauge-sector formulae can have the same logical
architecture without having the same microscopic origin or identical
coefficients.

The starting gauge sectors are different.  In the Type 0A route, the gauge
sector is tied to the orientifold/D-brane construction and to the doubled
bulk Type 0A RR gauge fields.  Schematically, the relevant higher-dimensional
gauge algebra may be denoted
$\mathfrak g_{8d}^{(0A)}$.
This algebra can include gauge fields supported by the D8/O8 system and
effective gauge fields descending from the doubled Type 0A RR sector, as discussed in detail in section \ref{sosiebrake}.

In the Type $0{\rm HW}$ route, the starting point is different.  The ordinary
Ho\v{r}ava--Witten interval reduction has two end-of-the-world walls carrying
two $E_8$ gauge sectors given by the following Lie algebra decomposition:
\begin{equation}
\mathfrak g_{\rm HW}
=
\mathfrak e_{8,L}
\oplus
\mathfrak e_{8,R}.
\end{equation}
After the later compactification on the wedge circle, each wall sees the
branch structure of
$(S^1_+\vee S^1_-)_b$.
In the Type $0$ branch, the branch-resolved current algebra is reduced from
$E_8$ to $SO(16)$ on each branch.  Thus one writes
$E_{8,L}
\longrightarrow
SO(16)_{L}^{(+)}
\oplus
SO(16)_{L}^{(-)}$, and 
$E_{8,R}
\longrightarrow
SO(16)_{R}^{(+)}
\oplus
SO(16)_{R}^{(-)}$, which are the expected doubled enhancement of the gauge groups. 
At the level of Lie algebras this gives:
\begin{equation}
\mathfrak g_{\rm wall}^{(0{\rm HW})}
=
\mathfrak{so}(16)_{L}^{(+)}
\oplus
\mathfrak{so}(16)_{L}^{(-)}
\oplus
\mathfrak{so}(16)_{R}^{(+)}
\oplus
\mathfrak{so}(16)_{R}^{(-)} .
\label{gwall0HW_gauge_compare}
\end{equation}
This is the effective $SO(16)^4$ gauge algebra of the branch-resolved
Type $0{\rm HW}$ wall sector. Therefore the first essential distinction is: (i) 
$\mathfrak g_{8d}^{(0A)}$
is an orientifold/D-brane and doubled-RR gauge sector, and (ii)
$\mathfrak g_{\rm wall}^{(0{\rm HW})}
=
\mathfrak{so}(16)^4$
is a branch-resolved Ho\v{r}ava--Witten wall gauge sector. This is elaborated further in {\bf Table \ref{gaugobeta01}}.

\begin{table}[t]
\centering
\renewcommand{\arraystretch}{2.8}
\begin{tabular} {|p{0.18\textwidth}|p{0.38\textwidth}|p{0.38\textwidth}|}
\hline
\textbf{Theory}
&
\textbf{First reduction}
&
\textbf{Starting gauge data}
\\
\hline
Type 0A
&
$
{\rm M\ theory}
\xrightarrow{\;S^1_+\vee S^1_-\;}
0A
$
&
orientifold/D-brane gauge data and doubled type $0A$ RR gauge fields
\\
\hline
Type $0{\rm HW}$
&
$
{\rm M\ theory}
\xrightarrow{\;S^1_a/\mathbb Z_2\;}
0{\rm HW}
$
&
Ho\v{r}ava--Witten wall data
$
\mathfrak e_{8,L}\oplus\mathfrak e_{8,R}
$
before branch resolution
\\
\hline
Type $0{\rm HW}$ after wedge reduction
&
$
0{\rm HW}
\xrightarrow{\;S^1_+\vee S^1_-\;}
9d
$
&
branch-resolved wall algebra
$
\mathfrak{so}(16)_{L}^{(+)}
\oplus
\mathfrak{so}(16)_{L}^{(-)}
\oplus
\mathfrak{so}(16)_{R}^{(+)}
\oplus
\mathfrak{so}(16)_{R}^{(-)}
$
\\
\hline
\end{tabular}
\caption{Starting gauge data in the two reduction routes.  In the type $0A$
route the wedge reduction is performed first, so the branch structure is already
built into the type $0A$ gauge-sector analysis.  In the type $0{\rm HW}$ route
the first reduction is the Ho\v{r}ava--Witten interval reduction; the
branch-resolved $SO(16)^4$ structure appears only after the subsequent
compactification on $S^1_+\vee S^1_-$.}
\label{gaugobeta01}
\end{table}

For the Type 0A branch, the four-dimensional gauge algebra is obtained by
requiring a generator to survive the D-brane bundle holonomies, the fibration
monodromies, the flux-induced St\"uckelberg mass matrix, and the wedge
deformation data.  Schematically, this is given by \eqref{G4d_commutant_0A_rewritten}. 
At the level of Lie algebras, one may equivalently write this as
\eqref{frakg8d}. More explicitly, a generator
$X_A\in \mathfrak g_{8d}^{(0A)}$
survives if it satisfies
\begin{equation}
[U_\gamma,X_A]=0,~~~
[\rho(M_\eta),X_A]=0,~~~
X_A\in \ker(M_{\rm Stuck}^{2}),~~~
[{\cal W}^{(0A)}_r,X_A]=0 ,
\label{0A_wedge_condition}
\end{equation}
where $U_\gamma$ is a D8-brane bundle holonomy, $\rho(M_\eta)$ is the induced
action of the fibration monodromy on charge or gauge data, $M_{\rm St\"uck.}^2$
is the flux-induced St\"uckelberg mass matrix, and ${\cal W}_r^{(0A)}$ denotes
the wedge-deformation data in the Type $0A$ frame. (See section \ref{sosiebrake} for more details.)

For the Type $0{\rm HW}$ branch, the starting algebra is not
$\mathfrak g_{8d}^{(0A)}$, but rather the branch-resolved Ho\v{r}ava--Witten
wall algebra given via:
\begin{equation}
\mathfrak g_{\rm wall}^{(0{\rm HW})}
=
\bigoplus_{\alpha\in\{L^\pm,R^\pm\}}
\mathfrak{so}(16)_\alpha ,
\end{equation}
implying that each branch-resolved wall factor may carry a bundle
$V_\alpha$, where
$\alpha\in\{L^\pm, R^\pm\}$. The $\pm$ resolution is precisely due to the compactification over the wedge $S^1_+ \vee S^1_-$ to nine dimensions.
The holonomy projection is therefore:
\begin{equation}
\mathfrak g_{4d}^{\rm hol}
=
\bigoplus_{\alpha}
{\rm Com}_{SO(16)_\alpha}
\left(
{\rm Hol}(V_\alpha)
\right) ,
\label{HW_hol_gauge_compare}
\end{equation}
where the notation
${\rm Com}_{SO(16)_\alpha}({\cal S})$
means the commutant, or centralizer, of the set ${\cal S}$ inside the group $SO(16)$ for any $\alpha$. This is described in details in \eqref{commdeff}.
Equivalently, if
$U_{\gamma,\alpha}
=
{\cal P}
\exp
\left(
i\oint_\gamma A_\alpha
\right)$
is a holonomy of $V_\alpha$, then a generator
$X_A^{(\alpha)}\in \mathfrak{so}(16)_\alpha$
survives only if:
\begin{equation}
[U_{\gamma,\alpha},X_A^{(\alpha)}]=0, 
\end{equation}
for all relevant cycles $\gamma$. In a similar vein,  the monodromy projection is parallel to the Type 0A case, but now the
monodromy acts on wall gauge or lattice data.  If a loop
$\zeta\in H_1(\Sigma_3,\mathbb Z)$
induces a monodromy $M_\zeta$, and if this monodromy is embedded into the
wall gauge data through a representation $\rho_\alpha$, the survival condition
is:
\begin{equation}
[\rho_\alpha(M_\zeta),X_A^{(\alpha)}]=0
\qquad
\forall\,\zeta\in H_1(\Sigma_3,\mathbb Z),
\qquad
\forall\,\alpha .
\label{HW_monodromy_gauge_compare}
\end{equation}
If the monodromy is purely geometric and does not act on the wall gauge bundle
or gauge lattice, then
$\rho_\alpha(M_\zeta)={\bf 1}$
on the gauge algebra, and the monodromy condition is trivial.

\begin{table}[t]
\centering
\renewcommand{\arraystretch}{1.9}
\begin{tabular} {|p{0.25\textwidth}|p{0.32\textwidth}|p{0.32\textwidth}|}
\hline
\textbf{Projection condition}
&
\textbf{Type $0A$}
&
\textbf{Type $0{\rm HW}$}
\\
\hline
parent generator
&
$
X_A\in\mathfrak g_{8d}^{(0A)}
$
&
$
X_A^{(\alpha)}\in \mathfrak{so}(16)_\alpha,
\quad
\alpha=L^+,L^-,R^+,R^-
$
\\
\hline
holonomy projection
&
$
[U_\gamma,X_A]=0
$
&
$
[U_{\gamma,\alpha},X_A^{(\alpha)}]=0
$
\\
\hline
monodromy projection
&
$
[\rho(M_\gamma),X_A]=0
$
&
$
[\rho_\alpha(M_\gamma),X_A^{(\alpha)}]=0
$
\\
\hline
masslessness condition
&
$
X_A\in\ker(M_{\rm St\"uck.}^2)
$
&
$
X_A\in\ker(M_{\rm wedge}^2)
$
\\
\hline
wedge-deformation projection
&
$
[{\cal W}^{(0A)}_r,X_A]=0
$
&
$
[{\cal W}^{(0{\rm HW})}_r,X_A]=0
$
\\
\hline
surviving algebra
&
$
\mathfrak g_{4d}^{(0A)}
$
&
$
\mathfrak g_{4d}^{(0{\rm HW})}
$
\\
\hline
\end{tabular}
\caption{Projection conditions for four-dimensional gauge generators in the
type $0A$ and type $0{\rm HW}$ routes.  The equations are structurally
parallel, but they act on different parent gauge algebras and on different
physical data.}
\label{gaugocomp008}
\end{table}

The wedge deformation also acts differently from the Type 0A case because it
acts on the branch-resolved wall algebra.  If the wedge descendants are
represented by algebra-valued elements
${\cal W}^{(0{\rm HW})}_r
\in
\mathfrak g_{\rm wall}^{(0{\rm HW})}$,
then the wedge projection is:
\begin{equation}
[{\cal W}^{(0{\rm HW})}_r,X_A]=0
\qquad
\forall r,
\label{HW_wedge_comm_condition}
\end{equation}
or equivalently
$X_A
\in
\ker\left(
{\rm ad}_{{\rm Def}_{\rm wedge}^{(0{\rm HW})}}
\right)$.
The mass-lifting sector should be treated more cautiously in the type
$0{\rm HW}$ frame.  Unlike the Type 0A case, where the doubled RR sector and
flux axions naturally produce a St\"uckelberg mass matrix, the Type
0HW wall gauge bosons require an explicit local junction model to
derive a detailed mass matrix.  Thus the safest statement is a formal kernel
condition:
$X_A
\in
\ker(M_{\rm wedge}^2)$,
where $M_{\rm wedge}^2$ denotes possible wedge-induced, junction-induced, or
branch-odd mass mixing.
Therefore the four-dimensional type $0{\rm HW}$ gauge algebra is:

{\footnotesize
\begin{equation}
\mathfrak g_{4d}^{(0{\rm HW})}
=
\mathfrak g_{\rm wall}^{(0{\rm HW})}
\cap
\ker\left({\rm ad}_{{\rm Hol}(V)}\right)
\cap
\ker\left({\rm ad}_{{\rm Mon}_{\rm fib}}\right)
\cap
\ker(M_{\rm wedge}^2)
\cap
\ker\left({\rm ad}_{{\rm Def}_{\rm wedge}^{(0{\rm HW})}}\right) ,
\label{g4d0HW_gauge_compare}
\end{equation}}
which looks somewhat similar to the Type 0A result in \eqref{frakg8d}, so it will be instructive to compare them carefully. (See {\bf Table \ref{gaugocomp008}}.) In fact the gauge-sector comparison follows the same logic as the torsion comparison.
For the torsion classes, the Type 0A and Type 0HW sides naturally take the following forms in nine-dimensions:
\begin{equation}
W_i^{(0A,9d)}
=
W_i^{\rm fib}
+
W_i^{\rm flux+pinch}\nonumber
\end{equation}
\begin{equation}
W_i^{(0{\rm HW},9d)}
=
W_i^{(0{\rm HW},10d)}
+
\Delta_{\rm wedge}W_i,
\end{equation}
because for Type 0A the wedge reduction is performed first whereas for the type $0{\rm HW}$ side the wedge is introduced only in the second reduction.
The gauge-sector analogue is almost similar in spirit:
\begin{table}[t]
\centering
\renewcommand{\arraystretch}{2.5}
\begin{tabular} {|p{0.25\textwidth}|p{0.32\textwidth}|p{0.32\textwidth}|}
\hline
\textbf{Ingredient}
&
\textbf{Type $0A$}
&
\textbf{Type $0{\rm HW}$}
\\
\hline
parent gauge algebra
&
$
\mathfrak g_{8d}^{(0A)}
$
from D8/O8 and doubled type $0A$ RR data
&
$
\mathfrak g_{\rm wall}^{(0{\rm HW})}
=
\displaystyle
\bigoplus_{\alpha=L^\pm,R^\pm}
\mathfrak{so}(16)_\alpha
$
\\
\hline
bundle holonomy
&
$
{\rm Hol}(V_{\rm D8})
$
&
$
{\rm Hol}(V_\alpha),
\quad
\alpha=~ L^+,L^-,R^+,R^-
$
\\
\hline
monodromy action
&
$
\rho(M_\gamma)
$
acting on D-brane, charge, or doubled-RR data
&
$
\rho_\alpha(M_\gamma)
$
acting on wall gauge or Narain/lattice data
\\
\hline
mass-lifting condition
&
$
X_A\in\ker(M_{\rm St\"uck.}^2)
$
from flux-axion St\"uckelberg couplings
&
$
X_A\in\ker(M_{\rm wedge}^2)
$
from possible wedge or junction couplings
\\
\hline
wedge projection
&
$
{\rm Def}_{\rm wedge}^{(0A)}
$
acting in the type $0A$ gauge frame
&
$
{\rm Def}_{\rm wedge}^{(0{\rm HW})}
$
acting on the branch-resolved wall algebra
\\
\hline
stage at which wedge data enter
&
first reduction
&
second reduction
\\
\hline
interpretation
&
wedge-first orientifold/D-brane and doubled-RR gauge sector
&
interval-first wall gauge sector plus later wedge projection
\\
\hline
\end{tabular}
\caption{Structural comparison of the gauge-sector projections in the type
$0A$ and type $0{\rm HW}$ routes.  The two analyses have the same algebraic
architecture, namely holonomy, monodromy, mass-kernel, and wedge-deformation
projections.  However, the parent gauge algebras and the physical origin of
the projections are different.}
\label{gaugocomp007}
\end{table}
\begin{equation}
\mathfrak g_{4d}^{(0A)}
=
\mathfrak g_{8d}^{(0A)}
\cap
\left(
\text{projections already including wedge-frame data}
\right)
\end{equation}
\begin{equation}
\mathfrak g_{4d}^{(0{\rm HW})}
=
\mathfrak g_{\rm wall}^{(0{\rm HW})}
\cap
\left(
\text{interval-reduced projections}
\right)
\cap
\left(
\text{later wedge projections}
\right), \nonumber
\end{equation}
implying that 
the common structure is a sequence of projections, but the parent gauge
algebras and the stage at which wedge data enter are different. We can make this more quantitative by allowing a more explicit comparison of the two commutant formulae. The Type 0A and Type 0HW Lie algebra structures may be written as:
\begin{equation}
\mathfrak g_{4d}^{(0A)}
=
\mathfrak g_{8d}^{(0A)}
\cap
\ker\left({\rm ad}_{{\rm Hol}(V_{\rm D8})}\right)
\cap
\ker\left({\rm ad}_{{\rm Mon}_{\rm fib}}\right)
\cap
\ker(M_{\rm Stuck.}^2)
\cap
\ker\left({\rm ad}_{{\rm Def}_{\rm wedge}^{(0A)}}\right)
\label{g4d0A_final_commutant}
\end{equation}
\begin{equation}
\mathfrak g_{4d}^{(0{\rm HW})}
=
\mathfrak g_{\rm wall}^{(0{\rm HW})}
\cap
\ker\left({\rm ad}_{{\rm Hol}(V)}\right)
\cap
\ker\left({\rm ad}_{{\rm Mon}_{\rm fib}}\right)
\cap
\ker(M_{\rm wedge}^2)
\cap
\ker\left({\rm ad}_{{\rm Def}_{\rm wedge}^{(0{\rm HW})}}\right), \nonumber
\label{g4d0HW_final_commutant}
\end{equation}
where all the terms in the Type 0A construction have been elaborated from \eqref{frakg8d} and for the Type 0HW, they were explained above.
The two equations are formally parallel, but the meanings of the factors differ. For example 
$\mathfrak g_{8d}^{(0A)}
\neq
\mathfrak g_{\rm wall}^{(0{\rm HW})}$,
$M_{\rm Stuck.}^2
\neq
M_{\rm wedge}^2$ in general, and 
${\rm Def}_{\rm wedge}^{(0A)}
\neq
{\rm Def}_{\rm wedge}^{(0{\rm HW})}$
as concrete operators.
They are analogous structures, not identical data.  This is the same lesson as
for the torsion classes: the two descriptions match by representation and
projection structure after the nine-dimensional reduction, not by term-by-term
identity. For more details see {\bf Table \ref{gaugocomp007}}.

However there is one special point in the non-supersymmetric moduli space where we can make some quantitative analysis and this is the symmetric branch point.
Recall that the symmetric branch point is characterized by vanishing branch-odd data:
\begin{equation}
T=0,
\qquad
{\rm Def}_{\rm wedge}^{(0{\rm HW})}=0,
\qquad
{\rm Def}_{\rm wedge}^{({\rm 0A})}=0 .
\end{equation}
In this limit, the wedge-induced projections and branch-odd mass mixings are
removed.  However, the interpretation of the symmetric limit differs in the two
frames.

On the Type 0A side, the symmetric limit removes the branch-odd wedge
deformations and leaves the even Type 0A sector, which can approach an
ordinary IIA-like organization at the appropriate endpoint.  On the type
0{HW} side, the restoration of the full $E_8$ gauge algebra is not
automatic from the low-energy commutant formula.  This is because,
as an $\mathfrak{so}(16)$ representation:
\begin{equation}
\mathfrak e_8
\cong
\mathfrak{so}(16)
\oplus
{\bf 128}_{\rm spin} ~~~~ {\rm or}~~~ {\bf 248}
\longrightarrow
{\bf 120}
\oplus
{\bf 128} ,
\end{equation}
where
the ${\bf 120}$ is the adjoint of $SO(16)$, while the ${\bf 128}$ is the
spinorial current-algebra sector; an enhancement
$SO(16)\longrightarrow E_8$
requires the restoration of the spinorial states.  It is not produced by the
commutant formula alone. Thus, for each wall branch,
\begin{equation}
\mathfrak{so}(16)_\alpha
\oplus
{\bf 128}_\alpha
\longrightarrow
\mathfrak e_{8,\alpha}
\end{equation}
{\it only} when the branch-resolved Type $0$ projection is removed and the spinorial
states return.  Without these extra states, the low-energy gauge algebra
remains $\mathfrak{so}(16)_\alpha$. This is detailed in {\bf Table \ref{gaugocomp009}}.

\begin{table}[t]
\centering
\renewcommand{\arraystretch}{2.5}
\begin{tabular} {|p{0.25\textwidth}|p{0.32\textwidth}|p{0.32\textwidth}|}
\hline
\textbf{Limit or effect}
&
\textbf{Type $0A$}
&
\textbf{Type $0{\rm HW}$}
\\
\hline
branch-symmetric point
&
$
T=0,\quad
{\rm Def}_{\rm wedge}^{(0A)}=0
$
&
$
T=0,\quad
{\rm Def}_{\rm wedge}^{(0{\rm HW})}=0
$
\\
\hline
effect on wedge projections
&
branch-odd wedge projections vanish
&
branch-odd wedge projections vanish
\\
\hline
even-sector interpretation
&
approaches an ordinary IIA-like even sector at the appropriate endpoint
&
approaches the ordinary HW structure only if spinorial current states are restored
\\
\hline
gauge enhancement issue
&
depends on the type $0A$/IIA endpoint and D-brane/orientifold data
&
$
SO(16)\to E_8
$
requires restoration of the
$
{\bf 128}_{\rm spin}
$
sector
\\
\hline
low-energy commutant alone
&
does not determine full endpoint enhancement
&
does not restore $E_8$ by itself
\\
\hline
\end{tabular}
\caption{Comparison of the symmetric-limit interpretation in the type $0A$ and
type $0{\rm HW}$ gauge-sector analyses.  In the type $0{\rm HW}$ route, the
enhancement from $SO(16)$ to $E_8$ requires spinorial current-algebra states
and is not obtained from the low-energy commutant formula alone.}
\label{gaugocomp009}
\end{table}

Therefore, the gauge-sector comparison between the two theories should be read
as follows.  While, both analyses have the same mathematical architecture, the parent gauge algebras and the physical meanings of the projections
are quite different. In Type 0A we have 
orientifold/D-brane and doubled RR gauge data,
whereas
in Type 
0HW we have 
branch-resolved Ho\v{r}ava--Witten wall gauge data.
Thus one should not identify the two gauge sectors term by term.  Rather, after
both compactification routes have been reduced to the same dimension, one
compares their surviving gauge algebras by the common projection logic:
\begin{equation}
\begin{aligned}
\mathfrak g_{4d}
 &=
\text{parent algebra}
\cap
\text{holonomy commutant}\\
&~~~\cap
\text{monodromy commutant}
\cap
\text{massless kernel}
\cap
\text{wedge commutant} \nonumber
\end{aligned}
\end{equation}
The Type 0A and Type $0{\rm HW}$ frames realize this same structure with
different microscopic inputs, just as their torsion classes realize the same
$SU(3)$ representation channels with different reduction-frame
representatives\footnote{{Interestingly, the branch-exchange projection that we discussed towards the end of section \ref{susyresto} has a parallel interpretation in the
Type $0$ Ho\v{r}ava--Witten frame. It retains the diagonal combinations of
the branch-resolved wall degrees of freedom. At the level of the adjoint
gauge fields this gives diagonal $SO(16)$ factors. Recovery of the complete
supersymmetric $E_8\times E_8$ theory additionally requires the restoration
of the spinorial $\mathbf{128}$ sectors in:
\begin{equation}
\mathbf{248}
=
\mathbf{120}\oplus\mathbf{128}. \nonumber
\end{equation}
After this enhancement, a supersymmetric heterotic compactification is
obtained only when the internal geometry and gauge bundle satisfy the
appropriate Killing-spinor, Hermitian Yang--Mills, and anomaly-cancellation
conditions. Thus the branchwise projection reproduces the correct
effective organization of the supersymmetric heterotic frame, but it does
not replace the usual geometric and bundle consistency conditions.}}.

\subsection{Torsion-frame maps near the supersymmetric endpoints \label{sec4343}}

The discussion of a strong--weak map between the Type 0A and Type $0{\rm HW}$
torsion classes should be formulated with an important qualification.  A genuine
U-duality between the two descriptions is well motivated only at the
supersymmetric Type IIA endpoints.  Away from those endpoints, a comparison of
torsion classes, spectra, and gauge-sector projection data is not sufficient to
establish a microscopic duality.  What can be defined away from the endpoints is
a \emph{formal frame map} between the $SU(3)$-structure data, useful for
organizing the comparison, but not by itself a proof of U-duality.

The wedge variables are $R_B$ and $T$.
At fixed branch-even radius $R_B$, the symmetric Type $0$ point and the two
Type IIA endpoints are respectively at $T = 0, R_+ = R_-$ and $T = \pm R_B, R_\pm = 0, R_\mp > 0$.
Thus the supersymmetric Type IIA limits occur when the branch-odd field has
condensed to one of the endpoint values and 
not at the symmetric point. With this in mind, let us define a notation:
\begin{equation}\label{duamap}
{\cal D}_{\rm s/w}
\end{equation}
which we shall call the {\it duality map}. There is however an immediate caution accompanying this map: it should not be interpreted as a proven
strong--weak duality map throughout the full wedge moduli space.  Rather, it is
best regarded as an endpoint map, or as a formal comparison map, whose physical
status is secure only when the branch-odd modulus has reached one of the
type-IIA endpoints. Thus \eqref{duamap}
is physically justified only at
$T=\pm R_B$
or in a controlled neighborhood thereof.
A careful way to write the statement is therefore:
\begin{equation}
{\cal D}_{\rm s/w}^{\rm endpoint}
\equiv
{\cal P}_{SU(3)}^{(0{\rm HW})}
\circ
{\cal C}_{A,\vartheta}
\circ
{\cal R}_{\rm s/w}^{*}
\circ
{\cal M}_{\rm s/w}\nonumber
\label{Dsw_endpoint_composition}
\end{equation}
\begin{equation}\label{duamap2}
{\cal D}_{\rm s/w}^{\rm endpoint}
:
\left.
\left(
J_{\rm 0A},\Omega_{\rm 0A}
\right)
\right|_{T=\pm R_B}
\longmapsto
\left.
\left(
J_{0{\rm HW}},\Omega_{0{\rm HW}}
\right)
\right|_{T=\pm R_B} ,
\end{equation}
where ${\cal M}_{\rm s/w}$ denotes the moduli and coupling map,
${\cal R}_{\rm s/w}^{*}$ denotes the pullback between the internal coframes,
${\cal C}_{A,\vartheta}$ denotes the Weyl rescaling and possible phase
rotation of the $SU(3)$ structure, and ${\cal P}_{SU(3)}^{(0{\rm HW})}$ denotes
projection onto the $SU(3)$ representation channels in the type-$0{\rm HW}$
frame.  The crucial point is that \eqref{Dsw_endpoint_composition} is an
endpoint map, not a derived duality of the full non-supersymmetric interior. At the level of the $SU(3)$ structure forms, the endpoint comparison may be
written as:
\begin{equation}
\left.
J_{0{\rm HW}}
\right|_{T=\pm R_B}
=
e^{2A_{\rm s/w}}\,
{\cal R}_{\rm s/w}^{*}
\left.
J_{0A}
\right|_{T=\pm R_B}\nonumber
\label{J_Dsw_endpoint}
\end{equation}
\begin{equation}
\left.
\Omega_{0{\rm HW}}
\right|_{T=\pm R_B}
=
e^{3A_{\rm s/w}+i\vartheta_{\rm s/w}}\,
{\cal R}_{\rm s/w}^{*}
\left.
\Omega_{0A}
\right|_{T=\pm R_B} ,
\label{Omega_Dsw_endpoint}
\end{equation}
where $A_{\rm s/w}$ is the Weyl factor needed to compare the two frames in a
common normalization, while $\vartheta_{\rm s/w}$ is a possible phase rotation
of the holomorphic three-form\footnote{The two equations in \eqref{Omega_Dsw_endpoint}
are precisely the action of the factor
${\cal C}_{A,\vartheta}\circ{\cal R}_{\rm s/w}^{*}$
on the $SU(3)$-structure forms.
More explicitly, the endpoint map was written as \eqref{duamap2}.
The factor
${\cal M}_{\rm s/w}$
acts first on the moduli, radii, coupling, flux labels, and endpoint data.  In
particular, it tells us how the Type 0A endpoint moduli are to be rewritten in
the Type $0{\rm HW}$ endpoint frame.  Schematically:
\begin{equation}
{\cal M}_{\rm s/w}:
\left(
R_a^{(0A)},R_B^{(0A)},\Phi_{0A},T_{0A},\ldots
\right)
\longmapsto
\left(
R_a^{(0{\rm HW})},R_B^{(0{\rm HW})},\Phi_{0{\rm HW}},T_{0{\rm HW}},\ldots
\right). \nonumber
\end{equation}
At the endpoint this includes 
$T_{0A}=\pm R_B^{(0A)}
\longmapsto
T_{0{\rm HW}}=\pm R_B^{(0{\rm HW})}$,
with the appropriate strong--weak relation among the remaining moduli.
After this moduli map has been made, one applies
${\cal R}_{\rm s/w}^{*}$,
which is the pullback on forms induced by the change of internal frame.  For
example,
${\cal R}_{\rm s/w}^{*}:
\Lambda^p(X_{0A})
\longrightarrow
\Lambda^p(X_{0{\rm HW}})$.
Thus
${\cal R}_{\rm s/w}^{*}J_{0A}$
is the Type 0A two-form rewritten as a two-form in the
Type $0{\rm HW}$ internal coframe, and similarly
${\cal R}_{\rm s/w}^{*}\Omega_{\rm 0A}$
is the pulled-back three-form in the type-$0{\rm HW}$ frame.
The next factor is
${\cal C}_{A,\vartheta}$.
This is the Weyl rescaling and phase rotation of the $SU(3)$ structure.  It is
defined by its action:

{\scriptsize
\begin{equation}
{\cal C}_{A,\vartheta}(J)
=
e^{2A_{\rm s/w}}J,~~~~~
{\cal C}_{A,\vartheta}(\Omega)
=
e^{3A_{\rm s/w}+i\vartheta_{\rm s/w}}\Omega \nonumber
\end{equation}
\begin{equation}
({\cal C}_{A,\vartheta}\circ{\cal R}_{\rm s/w}^{*})(J_{0A})
=
e^{2A_{\rm s/w}}
{\cal R}_{\rm s/w}^{*}J_{0A},~~~
({\cal C}_{A,\vartheta}\circ{\cal R}_{\rm s/w}^{*})(\Omega_{0A})
=
e^{3A_{\rm s/w}+i\vartheta_{\rm s/w}}
{\cal R}_{\rm s/w}^{*}\Omega_{0A}.\nonumber
\end{equation}
\begin{equation}
\left.
J_{0{\rm HW}}
\right|_{T=\pm R_B}
=
\left.
\left(
{\cal C}_{A,\vartheta}
\circ
{\cal R}_{\rm s/w}^{*}
\circ
{\cal M}_{\rm s/w}
\right)
J_{0A}
\right|_{T=\pm R_B},~
\left.
\Omega_{0{\rm HW}}
\right|_{T=\pm R_B}
=
\left.
\left(
{\cal C}_{A,\vartheta}
\circ
{\cal R}_{\rm s/w}^{*}
\circ
{\cal M}_{\rm s/w}
\right)
\Omega_{0A}
\right|_{T=\pm R_B} , \nonumber
\end{equation}}
which are exactly the two equations in 
\eqref{Omega_Dsw_endpoint}.
The reason the displayed equations only show
${\cal C}_{A,\vartheta}$ and 
${\cal R}_{\rm s/w}^{*}$
is that
${\cal M}_{\rm s/w}$
has already been absorbed into the meaning of the endpoint variables and the
functions
$A_{\rm s/w},
\vartheta_{\rm s/w}$ and 
${\cal R}_{\rm s/w}^{*}$.
Similarly, the final projection
${\cal P}_{SU(3)}^{(0{\rm HW})}$
does not appear explicitly in
\eqref{Omega_Dsw_endpoint} because these equations are written for
the defining $SU(3)$-structure forms themselves.  Once
$J_{0{\rm HW}}$,
and $\Omega_{0{\rm HW}}$
have been obtained, the structure group has already been fixed to the
Type $0{\rm HW}$ $SU(3)$ frame.  The projection
${\cal P}_{SU(3)}^{(0{\rm HW})}$
becomes important when extracting the torsion classes from
$dJ_{0{\rm HW}}$,
and 
$d\Omega_{0{\rm HW}}$,
because one must project the resulting forms onto
$\Lambda^0,
\Lambda_{\rm prim}^{(1,1)},
\Lambda_{\rm prim}^{(2,1)+(1,2)}$ and 
$\Lambda^1$.
For example:
\begin{equation}
W_2^{(0{\rm HW})}
=
{\cal P}_{\rm prim}^{(1,1)}
\left[
\text{the relevant component of }d\Omega_{0{\rm HW}}
\right] \nonumber
\end{equation}
\begin{equation}
W_3^{(0{\rm HW})}
=
{\cal P}_{\rm prim}^{(2,1)+(1,2)}
\left[
\text{the relevant component of }dJ_{0{\rm HW}}
\right] \nonumber
\end{equation}
Thus the clean distinction can be stated in the following way: (i)
${\cal M}_{\rm s/w}$
acts on moduli and labels, (ii) 
${\cal R}_{\rm s/w}^{*}$
acts on differential forms by changing the internal frame, (iii)
${\cal C}_{A,\vartheta}$
rescales and rephases the $SU(3)$ structure, and (iv)
${\cal P}_{SU(3)}^{(0{\rm HW})}$
extracts the $SU(3)$ torsion representations after differentiation.
Therefore, the equations for $J$ and $\Omega$ display only the middle two
operations because they are equations for the defining forms themselves, {\it i.e.}
${\cal R}_{\rm s/w}^{*}$
followed by
${\cal C}_{A,\vartheta}$.
The moduli map is implicit in the endpoint variables, and the projection map is
used only after taking $dJ$ and $d\Omega$ to read off the torsion classes.}. Both should be evaluated in the endpoint
duality frame. Using $dJ$ and $d\Omega$ from \eqref{dJtorsion}
one obtains the corresponding endpoint transformation of torsion classes:
\begin{equation}
\begin{aligned}
&\left.
W_1^{(0{\rm HW})}
\right|_{T=\pm R_B}
=
e^{-A_{\rm s/w}+i\vartheta_{\rm s/w}}\,
{\cal R}_{\rm s/w}^{*}
\left.
W_1^{(0A)}\right|_{T=\pm R_B}\\
&\left.
W_2^{(0{\rm HW})}
\right|_{T=\pm R_B}
=
e^{A_{\rm s/w}+i\vartheta_{\rm s/w}}\,
{\cal P}_{\rm prim}^{(1,1)}
{\cal R}_{\rm s/w}^{*}
\left.
W_2^{(0A)}
\right|_{T=\pm R_B} \\
& \left.
W_3^{(0{\rm HW})}
\right|_{T=\pm R_B}
=
e^{2A_{\rm s/w}}\,
{\cal P}_{\rm prim}^{(2,1)+(1,2)}
{\cal R}_{\rm s/w}^{*}
\left.
W_3^{(0A)}
\right|_{T=\pm R_B} \\
& \left.
W_4^{(0{\rm HW})}
\right|_{T=\pm R_B}
=
{\cal R}_{\rm s/w}^{*}
\left.
W_4^{(0A)}
\right|_{T=\pm R_B}
+
2\,dA_{\rm s/w} \\
& \left.
W_5^{(0{\rm HW})}
\right|_{T=\pm R_B}
=
{\cal R}_{\rm s/w}^{*}
\left.
W_5^{(0A)}
\right|_{T=\pm R_B}
+
3\,dA_{\rm s/w}
+
i\,d\vartheta_{\rm s/w} ,
\end{aligned}
\end{equation}
where these equations should be interpreted as the transformation of torsion classes under
the endpoint strong--weak frame change, and {\it not} as a proof
that the two non-supersymmetric interiors are U-dual. Putting everything together we see that, at the supersymmetric end-points where the two theories can be U-dual to each other, the torsion classes may be related via:
\begin{equation}
\left.
W_i^{(0{\rm HW})}
\right|_{T=\pm R_B}
=
{\cal D}_{\rm s/w}^{\rm endpoint}
\left[
\left.
W_i^{(0A)}
\right|_{T=\pm R_B}
\right] ,
\label{endpoint_torsion_duality_summary}
\end{equation}
where ${\cal D}_{\rm s/w}^{\rm endpoint}$ is the end-point map given by \eqref{duamap2}.
By contrast, away from the endpoints one should write only a formal comparison
map:
\begin{equation}
W_i^{(0{\rm HW})}
\overset{\rm formal}{\sim}
{\cal D}_{\rm frame}
\left[
W_i^{(0A)}
\right],
\qquad
|T|\neq R_B,
\label{formal_frame_map_nonendpoint}
\end{equation}
where ${\cal D}_{\rm frame}$ is a useful bookkeeping map between representation
channels and frame conventions, but {\it not} an established duality map. This distinction matters especially for the localized pinch terms.  For
example, the direct localized $W_2$ descendants in the two frames may be
written as
$W_{2,{\rm loc}}^{(0A)}
=
\kappa_2^{(0A)}
T_{0A}
\delta_{\Sigma_5^{\rm jct}}
\zeta_2^{(0A)}$
and
$W_{2,{\rm loc}}^{(0{\rm HW})}
=
\kappa_2^{(0{\rm HW})}
T_{0{\rm HW}}
\delta_{\Sigma_5^{\rm jct}}
\xi_2^{(0{\rm HW})}$, and similarly the localized $W_3$, where we use superscripts to distinguish the various parameters in the two theories.
Near the endpoints one may impose the strong--weak matching conditions:
\begin{equation}
\begin{aligned}
&
\lim_{T\to \pm R_B}
\left[
\kappa_2^{(0{\rm HW})}
T_{0{\rm HW}}
\delta_{\Sigma_5^{\rm jct}}
\xi_2^{(0{\rm HW})}
\right]
\\
&\hspace{1.2cm}
=
\lim_{T\to \pm R_B}
\left[
e^{A_{\rm s/w}+i\vartheta_{\rm s/w}}\,
{\cal P}_{\rm prim}^{(1,1)}
{\cal R}_{\rm s/w}^{*}
\left(
\kappa_2^{(0A)}
T_{0A}
\delta_{\Sigma_5^{\rm jct}}
\zeta_2^{(0A)}
\right)
\right],
\\[6pt]
&
\lim_{T\to \pm R_B}
\left[
\kappa_3^{(0{\rm HW})}
T_{0{\rm HW}}
\delta_{\Sigma_5^{\rm jct}}
\xi_3^{(0{\rm HW})}
\right]
\\
&\hspace{1.2cm}
=
\lim_{T\to \pm R_B}
\left[
e^{2A_{\rm s/w}}\,
{\cal P}_{\rm prim}^{(2,1)+(1,2)}
{\cal R}_{\rm s/w}^{*}
\left(
\kappa_3^{(0A)}
T_{0A}
\delta_{\Sigma_5^{\rm jct}}
\zeta_3^{(0A)}
\right)
\right] ,
\end{aligned}
\label{localized_near_endpoint_matching}
\end{equation}
where these equations compare how the localized torsion currents approach the
endpoint from the non-supersymmetric wedge interior.  They should not be read
as saying that the type-IIA endpoint itself contains a nonzero pinch source. At the exact endpoint the wedge singularity is absent, and the
localized current must either vanish, decouple, or be reorganized into smooth
endpoint-adapted data. Away from the endpoints, however, these equations are not consequences of a
known U-duality.  They are at most possible matching conditions that would have
to be derived from a microscopic junction theory.
The same caveat applies to the Lee-form terms.  Near the endpoints one may compare
$W_{4,\varepsilon}^{(0A)}$
with 
$W_{4,\varepsilon}^{(0{\rm HW})}$
through the endpoint frame maps:
\begin{equation}
\begin{aligned}
& \left.
W_{4,\varepsilon}^{(0{\rm HW})}
\right|_{T \to\pm R_B}
=
{\cal R}_{\rm s/w}^{*}
\left[
\left.
W_{4,\varepsilon}^{(0A)}
\right|_{T\to\pm R_B}
\right]
+
2dA_{\rm s/w}\\
& \left.
W_{5,\varepsilon}^{(0{\rm HW})}
\right|_{T\to\pm R_B}
=
{\cal R}_{\rm s/w}^{*}
\left[
\left.
W_{5,\varepsilon}^{(0A)}
\right|_{T\to\pm R_B}
\right]
+
3dA_{\rm s/w}
+
id\vartheta_{\rm s/w}.
\end{aligned}
\end{equation}
These formulae do not imply
$b_{4,5}^{(0A)}
=
b_{4,5}^{(0{\rm HW})}$.
The coefficients are frame-dependent.  What may be compared at the endpoint is
the full Lee-form torsion representative after the appropriate change of frame. Therefore away from $T=\pm R_B$, one should instead state
that the Type 0A and Type $0{\rm HW}$ torsion sectors exhibit a controlled
structural correspondence, not a demonstrated U-duality
In particular, the comparison of torsion data alone cannot establish a
microscopic duality in the non-supersymmetric interior of the wedge moduli
space.  Such a claim would require additional evidence, including the matching
of localized junction modes, their interactions, anomaly constraints, flux
quantization, and the full gauge-sector map.

\subsection{Fate of the other degrees of freedom at the pinch and torsion classes \label{fate}}

So far in our analysis we took a simplified picture in which we have a geometric pinch and the degrees of freedom at the pinch act as sources of the doubled fields that we studied in detail in section \ref{sec001}. However there could be other degrees of freedom that are highly localized at the pinch that do not influence any of the  bulk degrees of freedom\footnote{One example being the set of fields in \eqref{nonphys0A}, or more generically the ones studied in \eqref{haramibob} and \eqref{Sjct_light_UV_split}.}. They are clearly not {\it geometric} and they are invisible at the level of the doubled fields, but could they be absorbed in the ${\bf 27}$? In general this is not possible, but there could be special cases that allow such a scenario. In the following let us discuss this in some details. 

The same representation-theoretic reasoning that places the geometric pinch
deformation in the $\mathbf{27}$ of $G_2$ does not automatically imply that all
localized degrees of freedom at the junction must also lie in the
$\mathbf{27}$.  It implies only that the localized modes which source the same
geometric channel as the pinch must have a nonzero projection onto the
$\mathbf{27}$.

The geometric pinch contribution is special because it is a localized
deformation of the $G_2$ structure which is neither an overall singlet
rescaling, nor a Lee-form deformation, nor an adjoint-valued two-form
obstruction.  It is therefore naturally encoded in the primitive traceless
component of the intrinsic torsion:
\begin{equation}
\tau_3^{\rm pinch}
=
T\,\delta_{\Sigma^{\rm jct}}\,\Xi^{(3)}_{27}
+
O(T^2),
\qquad
\Xi^{(3)}_{27}\in\Lambda^3_{27}(M_7).
\end{equation}
This is a statement about the representation content of the localized
\emph{geometric} deformation. By contrast, genuine junction degrees of freedom may carry several possible
quantum numbers and representation components.  If we denote these DOFs as $\chi_{\rm jct} \equiv \{\chi\} = \chi_{\rm UV}$ or $\chi_{\rm heavy}$ from \eqref{Sjct_light_UV_split} or \eqref{haramibob} respectively, then the localized junction source should be written more cleanly as:
\begin{equation}
{\cal J}_{\rm jct}(\chi)
=
{\cal J}_{\rm jct}^{[0]}(\chi)
+
{\cal J}_{\rm jct}^{[1]}(\chi)
+
{\cal J}_{\rm jct}^{[2]}(\chi)
+
{\cal J}_{\rm jct}^{[3]}(\chi)
+
{\cal J}_{\rm jct}^{[4]}(\chi)
+
\cdots ,
\end{equation}
where
${\cal J}_{\rm jct}^{[p]}(\chi)$
denotes a $p$-form current localized at the junction.  For example,
${\cal J}_{\rm jct}^{[0]}(\chi) $
is a localized scalar source,
${\cal J}_{\rm jct}^{[1]}(\chi)$
is a localized one-form current,
${\cal J}_{\rm jct}^{[2]}(\chi)$
is a localized two-form current, and so on. The dots stand for any additional localized defect operators that may be
present at the pinch but are not explicitly written.  These can include
higher-degree localized currents, derivatives of localized fields,
fermion-bilinear sources, gauge-current contributions, anomaly-inflow terms,
higher-derivative operators, and nonlinear products of the lower-degree
currents, captured by say \eqref{Sjct_light_UV_split}.  In other words, the dots do not represent additional irreducible
$G_2$ representations of a fixed form.  They represent additional possible
operators in the localized junction effective theory.
A more explicit schematic expansion would be:
\begin{equation}\label{gencurrent}
{\cal J}_{\rm jct}(\chi)
=
{\cal J}_{\rm jct}^{[0]}(\chi) + ..
+
{\cal J}_{\rm jct}^{[4]}(\chi)
+
{\cal J}_{\rm jct}^{\rm der}(\chi)
+
{\cal J}_{\rm jct}^{\rm gauge}(\chi)
+
{\cal J}_{\rm jct}^{\rm ferm}(\chi)
+
{\cal J}_{\rm jct}^{\rm inflow}(\chi)
+
\cdots ,
\end{equation}
where ${\cal J}_{\rm jct}^{\rm der}(\chi)$ denotes derivative and higher-derivative
localized operators, ${\cal J}_{\rm jct}^{\rm gauge}(\chi)$ denotes localized gauge
currents, ${\cal J}_{\rm jct}^{\rm ferm}(\chi)$ denotes possible fermion bilinears, and 
${\cal J}_{\rm jct}^{\rm inflow}(\chi)$ denotes anomaly-inflow or Chern--Simons-type
localized terms.

Once a fixed form degree is selected, one then decomposes that object into
$G_2$ irreducible representations.  For instance, a two- and a three-form currents may be decomposed in the following way:
\begin{equation}
{\cal J}_{\rm jct}^{[2]}(\chi)
=
{\cal J}_{\rm jct}^{[2,{\bf 7}]}(\chi)
+
{\cal J}_{\rm jct}^{[2,{\bf 14}]}(\chi) \nonumber
\end{equation}
\begin{equation}
{\cal J}_{\rm jct}^{[3]}(\chi)
=
{\cal J}_{\rm jct}^{[3,{\bf 1}]}(\chi)
+
{\cal J}_{\rm jct}^{[3,{\bf 7}]}(\chi)
+
{\cal J}_{\rm jct}^{[3,{\bf 27}]}(\chi) ,
\end{equation}
from where we see that only the component
${\cal J}_{\rm jct}^{[3,{\bf 27}]}(\chi)$
can be absorbed into the same effective $\tau_3$ source as the geometric
pinch contribution.  Thus one may write:
\begin{equation}\label{tau3jct}
\tau_3^{\rm jct}
=
\delta_{\Sigma^{\rm jct}}
\Big(
T\,\Xi^{(3)}_{27}
+
{\cal J}_{\rm jct}^{[3,{\bf 27}]}(\chi)
\Big),
\end{equation}
implying that, if the localized degrees of freedom are assumed to describe only the
same primitive traceless deformation of the $G_2$ structure as the pinch, then
one may consistently take their contribution to lie in the $\mathbf{27}$.
In that restricted sense, the same reasoning applies.  But if the junction
sector contains more general defect fields, gauge modes, localized matter, or
source terms, then those modes need not lie purely in the $\mathbf{27}$.
Their $\mathbf{27}$ projection feeds $\tau_3$, while other projections may feed
other torsion classes or other localized constraints. In other words
we can rewrite \eqref{tau3jct} as:
\begin{equation}
\tau_3^{\rm jct}
=
T\,\delta_{\Sigma^{\rm jct}}\Xi_{27}^{\rm eff},
\qquad
\Xi_{27}^{\rm eff}
=
\Xi^{(3)}_{27}
+
T^{-1}{\cal J}_{\rm jct}^{[3,{\bf 27}]}(\chi) ,
\end{equation}
implying that the other components of ${\cal J}_{\rm jct}(\chi)$ should be kept separately,
because they may source other torsion channels, localized Bianchi identities,
or additional junction constraints\footnote{Only in the special case where
\(\mathcal J^{[3,27]}_{\rm jct}(\chi)=T\,\widehat{\mathcal J}^{[3,27]}_{\rm jct}(\chi)\) may one absorb it into an effective \(\Xi_{27}^{\rm eff}\), as a genuine localized defect current may exist independently of the branch-odd geometric modulus $T$.}.

The more general junction picture implies that the localized pinch sector can
source more than the minimal geometric descendants of the parent $G_2$
torsion.  In the minimal geometric truncation, the leading pinch deformation
was treated as a primitive traceless $G_2$ deformation and was therefore
encoded mainly through the $\mathbf{27}$ channel.  After reduction to an
$SU(3)$ structure, this produced localized contributions to the primitive
$(1,1)$ and primitive $(2,1)+(1,2)$ torsion classes, namely $W_2$ and $W_3$,
together with induced smooth corrections to the remaining classes.

If, however, the pinch supports a more general localized junction sector as in \eqref{gencurrent}, then
each fixed-degree current must be decomposed into irreducible
representations of the reduced $SU(3)$ structure.  Only the components that
land in the appropriate $SU(3)$ representation contribute to a given torsion
class.  Thus the localized junction sector should be projected as:
\begin{equation}
{\cal J}_{\rm jct}
\quad
\longrightarrow
\quad
\left(
{\cal J}_{W_1},
{\cal J}_{W_2},
{\cal J}_{W_3},
{\cal J}_{W_4},
{\cal J}_{W_5}
\right),
\end{equation}
where each ${\cal J}_{W_i}$ denotes the part of the localized current that lies
in the representation appropriate to $W_i$. Recall that the five $SU(3)$ torsion classes have the representation content
$W_1\in \Lambda^0\otimes \mathbb C,
W_2\in \Lambda_{\rm prim}^{(1,1)},
W_3\in \Lambda_{\rm prim}^{(2,1)+(1,2)}$ and 
$W_4,W_5\in \Lambda^1$.
Therefore a generic localized junction sector can contribute in the following way:
\begin{equation}
{\cal J}_{W_1}(\chi)
=
\Pi_{(3,0)+(0,3)}
\left[
{\cal J}_{\rm jct}^{[3]}(\chi)
\right]
+
\Pi_{\bf 1}
\left[
{\cal J}_{\rm jct}^{[0]}(\chi)
\right] \nonumber
\end{equation}
\begin{equation}
{\cal J}_{W_2}(\chi)
=
L_J^{-1}
\left[
\Pi_{\rm prim}^{(2,2)}
\left(
{\cal J}_{\rm jct}^{[4]}(\chi)
\right)
\right]
+
\Pi_{\rm prim}^{(1,1)}
\left[
{\cal J}_{\rm jct}^{[2]}(\chi)
\right] \nonumber
\end{equation}
\begin{equation}
~~~ {\cal J}_{W_3}(\chi)
=
\Pi_{\rm prim}^{(2,1)+(1,2)}
\left[
{\cal J}_{\rm jct}^{[3]}(\chi)
\right],~~~~~~ {\cal J}_{W_4}(\chi),
\ {\cal J}_{W_5}(\chi)
\in
\Lambda^1 ,
\end{equation}
where for the last two cases, they 
may be sourced by localized one-form currents, gradients of localized scalars,
or contractions of higher-degree localized currents with the reduced structure
forms. Thus the full torsion classes should be written schematically 
as:
\begin{equation}
W_i
=
W_i^{\rm fib}
+
W_i^{\rm flux}
+
W_i^{\rm pinch,geom}
+
W_i^{\rm jct},
\qquad
i=1,\ldots,5 ,
\end{equation}
where $ W_i^{\rm fib}$
and 
$W_i^{\rm flux}$ denote the fibration and the flux contributions that we developed earlier, and $W_i^{\rm pinch,geom}$ denotes the minimal geometric pinch contribution. From nine-dimensional point of view they are exactly the ones we had for both the Type 0A and Type 0HW cases discussed here. The new contribution is 
$W_i^{\rm jct}$. This denotes the extra contribution from genuine localized
junction degrees of freedom that do not directly influence any of the doubled fluxes that we studied here. More explicitly, one may write:
\begin{equation}
W_1^{\rm jct}
=
\delta_{\Sigma_5^{\rm jct}}\,
{\cal J}_{W_1},~~~
W_2^{\rm jct}
=
\delta_{\Sigma_5^{\rm jct}}\,
{\cal J}_{W_2},~~~
W_3^{\rm jct}
=
\delta_{\Sigma_5^{\rm jct}}\,
{\cal J}_{W_3}\nonumber
\end{equation}
\begin{equation}
~~~ W_4^{\rm jct}
=
\delta_{\Sigma_5^{\rm jct}}\,
{\cal J}_{W_4}
+
{\cal J}_{W_4}^{\rm smooth},
\qquad
W_5^{\rm jct}
=
\delta_{\Sigma_5^{\rm jct}}\,
{\cal J}_{W_5}
+
{\cal J}_{W_5}^{\rm smooth} ,
\end{equation}
where the smooth terms may include branch-odd gradients such as $dT$ and
$d\varepsilon$, while the distributional terms represent genuinely localized
junction currents. In this more general language, the earlier minimal pinch formulas become a
special truncation:
\begin{equation}
W_2^{\rm jct}
=
\kappa_2\,T\,\delta_{\Sigma_5^{\rm jct}}\,\xi_2,
\qquad
W_3^{\rm jct}
=
\kappa_3\,T\,\delta_{\Sigma_5^{\rm jct}}\,\xi_3,
\end{equation}
with $W_1^{\rm jct}
=
W_4^{\rm jct}
=
W_5^{\rm jct}
=
0$
at the level of direct localized sources.  The Lee-form classes $W_4$ and
$W_5$ were then affected only indirectly through smooth branch-odd data such as
$dT$ and $d\varepsilon$. The generic junction theory is less restrictive.  It allows:
\begin{equation}
W_1^{\rm jct}\neq 0,
\qquad
W_4^{\rm jct}\neq 0,
\qquad
W_5^{\rm jct}\neq 0
\end{equation}
if the localized degrees of freedom contain the appropriate scalar or one-form
currents.  Therefore the minimal geometric-pinch result should be understood as
a controlled ans\"atze, not as the most general consequence of having localized
degrees of freedom at the pinch.

In summary, the geometric pinch source naturally enters the $SU(3)$ torsion
classes through the descendants of the parent $G_2$ $\mathbf{27}$ channel.
However, a genuinely localized junction sector may contain additional
$SU(3)$ representation components, and these must be projected separately onto
the five torsion classes.  Therefore the restriction of localized sources to
the $W_2$ and $W_3$ channels should be understood as a simplifying truncation,
valid only after imposing a specific microscopic assumption about the junction
theory.

This broader viewpoint opens up a richer and potentially more informative
description of the junction sector than the minimal geometric-pinch ansatz
alone.  In particular, it suggests that the localized physics at the pinch can
be studied in a controlled representation-theoretic way.  From this perspective
the torsion-class framework becomes especially powerful: it provides a unified
language for organizing branchwise fluxes, singular geometric data, and
localized defect contributions, while keeping track of how each ingredient
feeds into the appropriate $SU(3)$ torsion channel.

\section{Conclusion and outlook \label{sec5}}

The main conclusion of our analysis is that the Type 0A and Type 0HW
descriptions should be viewed as two different effective organizations of the
same underlying M-theory data.  The two routes differ in the order in which one
compactifies on the wedge circle and on the Ho\v{r}ava--Witten interval:
\begin{equation}\label{decom1110}
{\rm M\ theory}
\xrightarrow{\;S^1_+\vee S^1_-\;}
0A
\xrightarrow{\;S^1_a/\mathbb Z_2\;}
9d,
\qquad
{\rm M\ theory}
\xrightarrow{\;S^1_a/\mathbb Z_2\;}
0{\rm HW}
\xrightarrow{\;S^1_+\vee S^1_-\;}
9d.
\end{equation}
This difference in the order of reductions is important.  In the Type 0A
route, the wedge data are present already in the first reduction, whereas in
the Type 0HW route the wedge data enter only after the subsequent
compactification on $S^1_+\vee S^1_-$.  Consequently, one should not expect a
term-by-term equality between the two sets of torsion classes or gauge-sector
data.  The correct comparison is a comparison of representation channels and
effective structures after both routes have been reduced to the same
nine-dimensional setting.

A central point in the analysis is that the doubled branchwise structure becomes
geometrically useful once it is encoded through the branchwise decomposition of
the M-theory flux and through the pinched coframe. This is where our motivation differs slightly from that of \cite{BDV}.
We treat the doubled field configurations as part of the effective
branchwise description, and then ask how this description can be reconciled
with the field-dependent resolution properties emphasized in \cite{BDV}.
In our approach, the branchwise decomposition is not merely a convenient
rewriting of the fields; it is also intended to encode information about the
pinch singularity and the degrees of freedom localized at the junction.  Concretely, one introduces
branchwise data of the form:
\begin{equation}
G^{\rm br}_4
=
\eta_+\wedge H_3^{(+)}
+
\eta_-\wedge H_3^{(-)}
+
F_4^{(+)}
+
F_4^{(-)},
\qquad
H_3^{(\pm)}=\iota_{v_\pm}G_4,
\end{equation}
where $\pm$ denote the doubled decomposition of {\it all} the low-energy DOFs that we elaborated in section \ref{sec001};
together with a wedge one-form that contains both the branch-even radius and
the branch-odd imbalance mode:
\begin{equation}
e^4_{\rm wedge}
=
R_B(\cdots)
+
T(\cdots),
\qquad
R_B=\frac12(R_++R_-),
\qquad
T=\frac12(R_+-R_-).
\end{equation}
Once these are accepted as part of the effective description, the $G_2$ and
$SU(3)$ torsion classes become a very efficient language for the problem.  They
do not, by themselves, derive the full microscopic doubled spectrum.  Rather,
they provide a powerful geometric bookkeeping device for organizing the
branchwise smooth data, the flux data, and the pinch-induced intrinsic torsion. Equivalently, given the singular wedge geometry, the branchwise flux decomposition, and the
fibration data as input, the $G_2$ and $SU(3)$ torsion-class analysis is a
valid effective analysis of the intrinsic torsion on the smooth strata together
with localized junction contributions.  It should not be interpreted as a
smooth global $G_2$-holonomy construction, but it is a natural and useful
description of the singular effective geometry.

This qualification is important.  The torsion classes are not being used as a
proof that the doubled sectors must exist microscopically.  Instead, the logic
is conditional: once the doubled branchwise flux decomposition and the pinched
coframe are built into the effective description, the torsion classes become
the natural variables for describing the resulting geometry.  They convert the
branchwise data into intrinsic torsion data of the relevant $G_2$ and
$SU(3)$ structures.

The reason this works is simple.  Torsion classes are sensitive to the failure
of the defining forms to be closed.  The fibration, the fluxes, and the wedge
pinch all contribute precisely to such failures of closure.  In particular, the
non-closure of the pinched one-form schematically contains smooth branch-even
terms, smooth branch-odd terms, and junction-supported terms:
\begin{equation}
de^4_{\rm wedge}
\sim
dR_B\wedge(\cdots)
+
dT\wedge(\cdots)
+
\text{junction-supported terms}.
\end{equation}
Thus the intrinsic torsion naturally records three types of information:
smooth fibration data, branchwise flux data, and localized pinch or junction
data. For example, on the seven-dimensional side, the $G_2$ torsion classes provide the most
compact way to organize the parent M-theory geometry.  The decomposition of
$d\varphi$ and $d\psi$ into:
\begin{equation}
\tau_0,\qquad \tau_1,\qquad \tau_2,\qquad \tau_3
\end{equation}
is not a supersymmetry assumption.  It is a decomposition of intrinsic torsion.
Supersymmetry would impose additional restrictions on these classes, but in the
present non-supersymmetric wedge/pinch setting the point is precisely that the
torsion is nonzero and carries the physical information.  The localized pinch
source is naturally described as a contribution to the $\mathbf{27}$ component,
schematically:
\begin{equation}
\tau_3^{\rm pinch}
=
T\,\delta_{\Sigma_6^{\rm jct}}\Xi^{(3)}_{27}
+
\cdots .
\end{equation}
After reducing to the six-dimensional $SU(3)$-structure space, this $G_2$
$\mathbf{27}$ decomposes into several $SU(3)$ representations.  In particular,
the $\mathbf 8$ component contributes to the primitive $(1,1)$ torsion class
$W_2$, while the $\mathbf 6\oplus\overline{\mathbf 6}$ component contributes to
the primitive $(2,1)+(1,2)$ torsion class $W_3$.  This explains why the direct
localized descendants of the parent $G_2$ pinch data can appear in both $W_2$
and $W_3$.  The remaining classes, especially $W_1$, $W_4$, and $W_5$, receive
pinch-dependent corrections more indirectly, through the deformation of the
reduced structure forms $J(T)$ and $\Omega(T)$ and through regular
resolved-node data such as $\varepsilon$.

Dimensional reduction via two different ways, shown in \eqref{decom1110}, lead to either Type 0A and Type 0HW compactified on two different non-K\"ahler manifolds for which the $SU(3)$ torsion classes succinctly capture the important geometrical features.
In the Type 0A route, the torsion classes are naturally written in the form:
\begin{equation}
W_i^{(0A)}
=
W_i^{\rm fib}
+
W_i^{\rm flux+pinch}.
\end{equation}
This is because the wedge circle is already part of the first reduction.
Accordingly, the Type 0A torsion classes already contain the branch-even flux
data, the branch-odd flux data, the localized pinch descendants, and the
induced smooth corrections.  The localized contributions to $W_2$ and $W_3$
may either be displayed explicitly or absorbed into the total branch-odd
coefficients.  The refined notation is useful because it separates the smooth
odd corrections from the distributional junction-supported pieces.

In the Type $0{\rm HW}$ route, shown in \eqref{decom1110}, the organization is different.  The first step
is the Ho\v{r}ava--Witten interval reduction, so before compactifying on the
wedge circle the torsion classes have the simpler form:
\begin{equation}
W_i^{(0{\rm HW},10d)}
=
w_i^{\rm fib}
+
w_i^{\rm flux}.
\end{equation}
Only after the further compactification on $S^1_+\vee S^1_-$ do the
wedge-induced corrections appear.  Therefore the nine-dimensional torsion
classes take the schematic form:
\begin{equation}
W_i^{(0{\rm HW},9d)}
=
w_i^{\rm fib}
+
w_i^{\rm flux}
+
\Delta_{\rm wedge}W_i.
\end{equation}
The direct localized descendants of the parent $G_2$ pinch source again appear
in $W_2$ and $W_3$, while the Lee-form classes receive smooth branch-odd and
resolved-node corrections.

This also clarifies the comparison between the two pictures.  The Type 0A
and Type $0{\rm HW}$ torsion classes are not equal term by term.  The
coefficients, the representatives, and even the natural way of grouping the
terms are frame-dependent.  What matches is the representation-theoretic
structure.  Both sides organize the data into the same five $SU(3)$ torsion
channels:
\begin{equation}
W_1,\qquad W_2,\qquad W_3,\qquad W_4,\qquad W_5.
\end{equation}
The Type 0A frame reaches these channels by reducing first on the wedge,
while the Type $0{\rm HW}$ frame reaches them by reducing first on the interval
and adding the wedge data later.

The strong--weak relation between the supersymmetric endpoints should be
understood in the same way.  It is not a statement that the individual torsion
coefficients in the two frames are equal.  Rather, the full torsion data are
related by a duality-frame map involving a moduli map, a pullback of the
internal coframe, a Weyl rescaling, and possible phase rotations of
$\Omega$.  In particular, coefficients such as $b_4$ and $b_5$ in the Lee-form
channels should be written separately in the two frames, for example
$b_{4,5}$ and $\hat{b}_{4,5}$ respectively as shown in section \ref{comparisoon}.  They multiply analogous
one-form structures, but they are not equal unless an additional matching
prescription is imposed. However, away from the two supersymmetric endpoints, a comparison of the spectra, gauge sectors, and torsion classes is not by itself sufficient to establish a U-duality between the two descriptions.  Indeed, our analysis of the spectrum shows that, after the appropriate branch projection and treatment of the pinch-localized modes, the Type 0A compactification on $S^1/\mathbb{Z}_2$ and the Type 0HW compactification on $S_+^1\vee S_-^1$ give the same physical spectrum. In this precise sense, the two descriptions exhibit a manifest spectral matching. However, this matching by itself does not establish a full duality between the two constructions. Thus the result should be interpreted as strong evidence for a common localized-spectrum sector, rather than as a proof of complete equivalence between the two theories.

It remains possible that a more subtle duality exists away from the supersymmetric limits. However, such a claim would require considerable caution and substantially more evidence. In particular, one would need to perform additional checks involving the localized junction modes, the matching of interactions, anomaly constraints, flux quantization, and the precise mapping of moduli and gauge-sector data before asserting a genuine duality.

The gauge-sector comparison follows the same logic.  In the Type 0A route,
the gauge sector is tied to the D8/O8 system, the D-brane bundle data, and the
doubled Type 0A RR gauge fields.  The four-dimensional gauge algebra is
obtained by imposing holonomy, monodromy, flux-induced St\"uckelberg, and wedge
commutant conditions on the parent type-$0A$ gauge algebra.  In the
Type $0{\rm HW}$ route, the starting point is instead the Ho\v{r}ava--Witten
wall gauge sector.  After branch resolution, the effective wall algebra is
schematically:
\begin{equation}
\mathfrak{so}(16)_{L}^{(+)}
\oplus
\mathfrak{so}(16)_{L}^{(-)}
\oplus
\mathfrak{so}(16)_{R}^{(+)}
\oplus
\mathfrak{so}(16)_{R}^{(-)}.
\end{equation}
The same kinds of projections are then imposed: bundle holonomy, fibration
monodromy, possible wedge-induced mass terms, and wedge-deformation
commutants.  Thus the two gauge-sector analyses have the same mathematical
architecture, but they begin from different parent gauge algebras and have
different physical origins.

In particular, the Type $0{\rm HW}$ gauge-sector formula should be understood
as a branch-resolved effective description.  Recovering the full $E_8$ factors
at the supersymmetric endpoint requires the restoration of the spinorial
current-algebra states, since:
\begin{equation}
{\bf 248}
\longrightarrow
{\bf 120}
\oplus
{\bf 128}
\end{equation}
under $SO(16)$.  The commutant formula alone retains the adjoint
$SO(16)$ part; the enhancement back to $E_8$ is an additional symmetric-limit
or non-perturbative restoration of the spinorial sector.

The main conceptual outcome is therefore the following.  Without the doubled
branchwise ansatz, the torsion classes by themselves are not enough to derive
the full doubled story.  With the doubled ansatz and the pinched coframe built
in, however, the $G_2$ and $SU(3)$ torsion classes become one of the best tools
for analyzing the effective geometry.  They unify the fibration data, the
branchwise flux data, the branch-odd tachyonic imbalance, and the localized
junction contributions in a single geometric framework (see {\bf Table \ref{W12345}}).  Their limitation is
not a weakness of the torsion formalism; it is simply the statement that the
torsion classes encode the effective geometric organization of the doubled
setup rather than proving its microscopic origin.

{Additionally, the analysis here points towards an effective construction
with three logically distinct regimes. On the smooth strata away from the
pinch, the Wilsonian bulk theory and its equations of motion apply. At the
pinch, the bulk theory must be supplemented by localized junction degrees
of freedom and matching conditions. Finally, the two supersymmetric Type
IIA limits arise in different ways: either through the dynamical collapse
of one wedge branch or through the perturbative
$0{\rm A}/(-1)^{G_L}$ projection at the symmetric point. In either case,
recovery of the supersymmetric spectrum must be supplemented by the
Killing-spinor, flux, bundle, and Bianchi-identity conditions appropriate
to the surviving compactification.}

In this sense, the final picture is controlled but deliberately modest.  The
Type 0A and Type $0{\rm HW}$ constructions do not provide identical
term-by-term descriptions.  Rather, after both compactification routes are
brought to nine dimensions, they display the same structural organization of
torsion and gauge data.  The torsion classes package the geometric and flux
information into $SU(3)$ representation channels, while the gauge sector is
organized by holonomy, monodromy, mass-kernel, and wedge-projection conditions.
This is the appropriate effective language for comparing the two routes at
the level of representation channels, torsion data, and gauge projections,
while leaving the microscopic equivalence question open away from the
supersymmetric endpoints.

\section*{Acknowledgements}

We would like to thank Zihni Baykara and Cumrun Vafa for helpful discussions; and especially Joydeep Chakravarty for many helpful comments that helped us to improve the draft. {We would also like to thank the anonymous referee for several valuable comments that helped us clarify our construction and relate it more directly to certain established dualities.} The work of KD is supported in part by a
Discovery Grant from the Natural Sciences and Engineering Research Council of Canada (NSERC). The work of RT is carried out with the support of the STFC Consolidated Grant ST/X000699/1.

\begin{table}[H]
\centering
\renewcommand{\arraystretch}{2.0}
\begin{tabular}  {|p{0.18\textwidth}|p{0.38\textwidth}|p{0.38\textwidth}|}
\hline
\textbf{Feature}
&
\textbf{Type 0A route}
&
\textbf{Type 0HW route}
\\
\hline
First reduction
&
$S^1_+\vee S^1_-$
&
$S^1_a/\mathbb Z_2$
\\
\hline
Second reduction to $9d$
&
$S^1_a/\mathbb Z_2$
&
$S^1_+\vee S^1_-$
\\
\hline
When wedge data enter
&
Already in the first reduction
&
Only in the second reduction
\\
\hline
Generic torsion structure
&
$
W_i^{(0A,9d)}
=
W_i^{\rm fib}
+
W_i^{\rm flux+pinch}
$
&
$
W_i^{(0{\rm HW},9d)}
=
W_i^{(0{\rm HW},10d)}
+
\Delta_{\rm wedge}W_i
$
\\
\hline
Bare ten-dimensional torsion classes
&
Already wedge-sensitive
&
$
W_i^{(0{\rm HW},10d)}
=
w_i^{\rm fib}
+
w_i^{\rm flux}
$
\\
\hline
Parent localized source
&
$
\tau_3^{\rm pinch}\in{\bf 27}_{G_2}
$
&
$
\tau_3^{\rm pinch}\in{\bf 27}_{G_2}
$
\\
\hline
$G_2\to SU(3)$ branching relevant for direct descendants
&
$
{\bf 27}
\supset
{\bf 8}
\oplus
{\bf 6}
\oplus
\overline{\bf 6}
$
&
$
{\bf 27}
\supset
{\bf 8}
\oplus
{\bf 6}
\oplus
\overline{\bf 6}
$
\\
\hline
Direct localized $W_2$ descendant
&
$
\kappa_2^{(0A)}
T\delta_{\Sigma_5^{\rm jct}}\zeta_2^{(0A)}
$
&
$
\kappa_2^{(0{\rm HW})}
T\delta_{\Sigma_5^{\rm jct}}\xi_2^{(0{\rm HW})}
$
\\
\hline
Direct localized $W_3$ descendant
&
$
\kappa_3^{(0A)}
T\delta_{\Sigma_5^{\rm jct}}\zeta_3^{(0A)}
$
&
$
\kappa_3^{(0{\rm HW})}
T\delta_{\Sigma_5^{\rm jct}}\xi_3^{(0{\rm HW})}
$
\\
\hline
$W_1,W_4,W_5$ pinch dependence
&
Induced smooth corrections
&
Induced smooth corrections after wedge compactification
\\
\hline
Smooth $dT$ terms
&
Present in the flux-plus-pinch sector
&
Appear after compactification on $S^1_+\vee S^1_-$
\\
\hline
Smooth $d\varepsilon$ terms
&
Present as resolved-pinch corrections
&
Present after wedge compactification or after resolving the node
\\
\hline
Interpretation of matching
&
Wedge-first torsion data
&
Interval-first torsion data plus later wedge correction
\\
\hline
\end{tabular}
\caption{Structural comparison between the Type 0A and
Type $0{\rm HW}$ nine-dimensional torsion sectors.  The localized parent
$G_2$ $\mathbf{27}$ pinch source has direct $SU(3)$ descendants in both
$W_2$ and $W_3$, corresponding respectively to the $\mathbf 8$ and
$\mathbf 6\oplus\overline{\mathbf 6}$ components.  These localized pieces may
be displayed explicitly or absorbed into the total branch-odd coefficients.}
\label{W12345}
\end{table}


\end{document}